# Nonlinear and Linear Elastodynamics Transformation Cloaking

Arash Yavari*[1,2] and Ashkan Golgoon[1]

[1] *School of Civil and Environmental Engineering, Georgia Institute of Technology, Atlanta, GA 30332, USA*
[2] *The George W. Woodruff School of Mechanical Engineering, Georgia Institute of Technology, Atlanta, GA 30332, USA*

July 26, 2018

**Abstract**

In this paper we formulate the problems of nonlinear and linear elastodynamics transformation cloaking in a geometric framework. In particular, it is noted that a cloaking transformation is neither a spatial nor a referential change of frame (coordinates); a cloaking transformation maps the boundary-value problem of an isotropic and homogeneous elastic body (virtual problem) to that of an anisotropic and inhomogeneous elastic body with a hole surrounded by a cloak that is to be designed (physical problem). The virtual body has a desired mechanical response while the physical body is designed to mimic the same response outside the cloak using a cloaking transformation. We show that nonlinear elastodynamics transformation cloaking is not possible while nonlinear elastostatics transformation cloaking may be possible for special deformations, e.g., radial deformations in a body with either a cylindrical or a spherical cavity. In the case of linear elastodynamics, in agreement with the previous observations in the literature, we show that the elastic constants in the cloak are not fully symmetric; they do not possess the minor symmetries. We prove that elastodynamics transformation cloaking is not possible regardless of the shape of the hole and the cloak. We next show that linear elastodynamics transformation cloaking cannot be achieved for gradient elastic solids either; similar to classical linear elasticity the balance of angular momentum is the obstruction to transformation cloaking. We finally prove that transformation cloaking is not possible for linear elastic generalized Cosserat solids in dimension two for any shape of the hole and the cloak. In particular, in dimension two transformation cloaking cannot be achieved in linear Cosserat elasticity. We also show that transformation cloaking for a spherical cavity covered by a spherical cloak is not possible in the setting of linear elastic generalized Cosserat solids. We conjecture that this result is true for a cavity of any shape.

**Keywords:** Cloaking, Nonlinear Elasticity, Gradient Elasticity, Cosserat Elasticity, Elastic Waves.

## Contents

*Corresponding author, e-mail: arash.yavari@ce.gatech.edu

# 1 Introduction

Invisibility has been a dream for centuries. Making objects invisible to electromagnetic waves has been a subject of intense research in recent years. Pendry et al. [2006] and Leonhardt [2006] independently showed the possibility of electromagnetic cloaking. This was later experimentally verified by Schurig et al. [2006], Liu et al. [2009], and Ergin et al. [2010]. In many references, including [Pendry et al., 2006], it is argued that the main idea of cloaking in electromagnetism is the invariance of Maxwell's equations under coordinate transformations (covariance). The covariance of Maxwell's equations has been known for a long time [Post, 1997]. However, one should note that covariance of Maxwell's equations is not the direct underlying principle of transformation cloaking. In electromagnetic cloaking one maps one problem to another problem with some desirable response. For example, a domain with a hole surrounded by a cloak with unknown physical properties is mapped to a domain without a hole (or with a very small one) made of an isotropic and homogeneous material. Then one tries to find the transformed fields such that both problems satisfy Maxwell's equations [Kohn et al., 2008]. This will then determine the physical properties of the cloak. In particular, the transformed quantities are not necessarily what one would expect under a coordinate transformation, i.e., the two problems are not related by push-forward or pull-back using the cloaking map.

More specifically, the idea of cloaking for electromagnetism is as follows. Suppose one is given a body (domain) $\Omega$ with a hole $\mathcal{H}$ surrounded by a cloaking region $\mathcal{C}$ (see Fig.1). The hole can be of any shape and the one shown in Fig.1 is assumed to be circular (spherical in 3D). Suppose the physical properties in $\Omega \setminus \mathcal{C}$ are uniform and isotropic. One is interested in designing the cloaking region $\mathcal{C}$ such that an electromagnetic wave passing through $\Omega$ would not interact with $\mathcal{H}$. In other words, $\mathcal{C}$ redirects the waves such that the boundary measurements are identical to those of another body with the same outer boundary as $\Omega$ without the hole and made of the same homogeneous (and isotropic) material. Let us consider a smooth mapping $\psi_\rho : \Omega \to \Omega$ such that $\psi_\rho|_{\Omega\setminus\mathcal{C}} = id$ and it shrinks the hole $\mathcal{H}$ to a small circle (sphere) of radius $\rho > 0$ (see Fig.1(b)). Note that this map is not unique, and hence, many possibilities for a cloak $\mathcal{C}$. One then transforms the physical fields such that the two problems satisfy Maxwell's equations. This usually makes the physical properties of the cloaking region $\mathcal{C}$ both inhomogeneous and anisotropic. The perfect cloaking case (the limit $\rho \to 0$) may require singular physical properties and should be studied carefully [Kohn et al., 2008].

Cloaking in the context of conductivity [Greenleaf et al., 2003a,b], electrical impedance tomography, and electromagnetism has been studied rigorously and is well understood [Bryan and Leise, 2010, Greenleaf et al., 2007, 2009a,b]. Interestingly, the idea of cloaking has been explored in many other fields of science and engineering, e.g., acoustics [Chen and Chan, 2007, Farhat et al., 2008, 2009a,b, Norris, 2008, Cummer et al., 2008a,b, Zhou et al., 2008], optics [Leonhardt and Philbin, 2012], thermodynamics (design of thermal cloaks) [Guenneau et al., 2012], diffusion [Guenneau and Puvirajesinghe, 2013], quantum mechanics [Zhang et al., 2008], and elastodynamics [Milton et al., 2006]. The recent reviews [Kadic et al., 2013, 2015] discuss these applications in some detail. The least understood among these applications is elastodynamics. In our opinion, the main problem is that none of the existing works in the literature has formulated the problem of elastodynamics cloaking properly. In particular, boundary and continuity conditions and the restrictions they impose on cloaking maps have not been discussed. In this paper, we formulate both the nonlinear and linearized cloaking problems in a mathematically precise form. One should note that cloaking is an inherently geometric problem. This is explained in the case of optical cloaking and invisibility in the recent book [Leonhardt and Philbin, 2012]. We will see that this is the case for elastodynamics cloaking as well; geometry plays a critical role in a proper formulation of elastodynamics transformation cloaking.

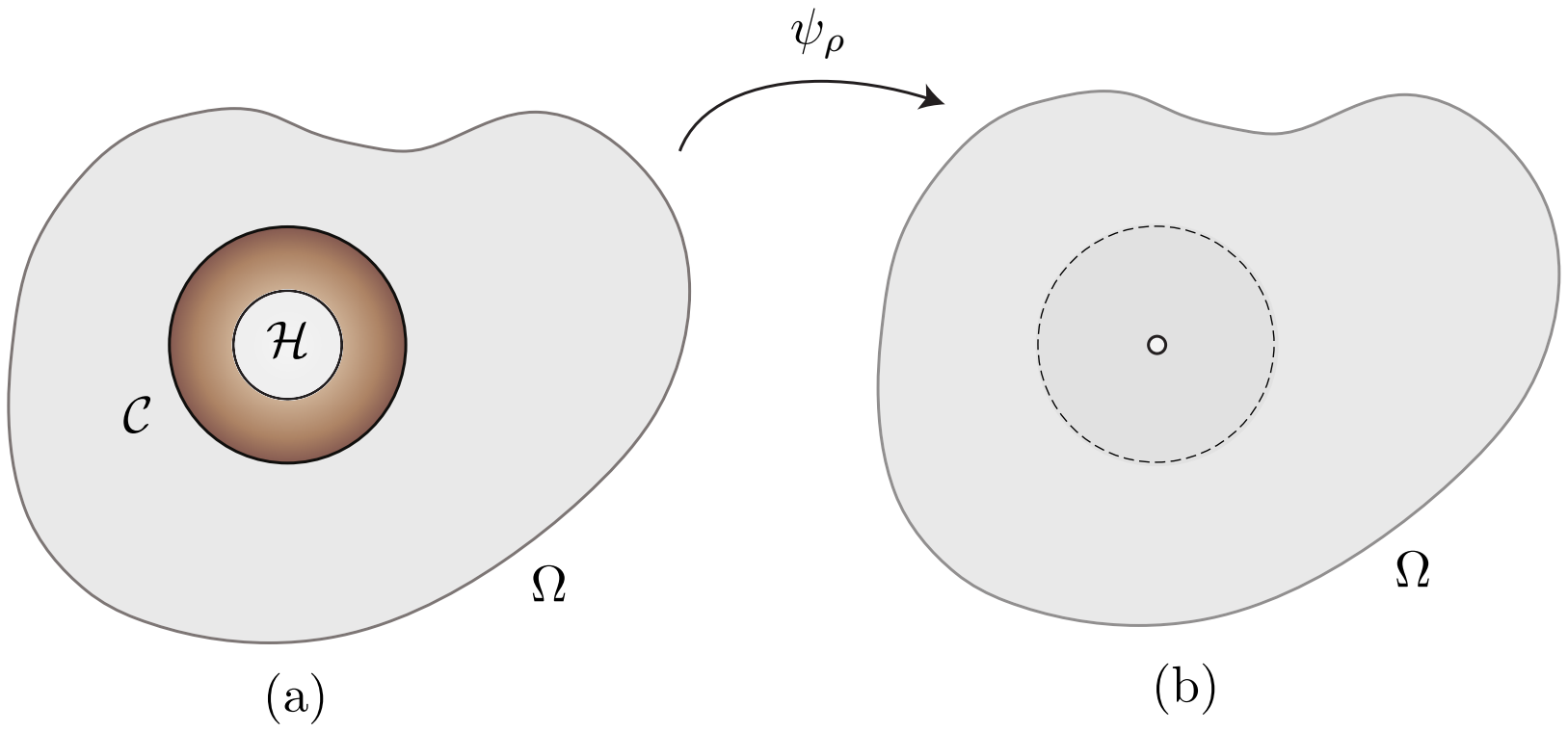

Figure 1: *Cloaking an object in a hole $\mathcal{H}$ by a cloak $\mathcal{C}$. The system (b) has uniform physical properties. The cloaking transformation is the identity map outside the cloaking region $\mathcal{C}$.*

The first ideas related to cloaking in elasticity go back to the 1930s and 1940s in the works of Gurney [1938] and Reissner and Morduchow [1949] on reinforced holes in elastic sheets in the framework of linear elasticity. The first systematic study of cloaking in linear elasticity is due to Mansfield [1953] who introduced the concept of neutral holes. Mansfield considered a sheet (a plane problem) under a given (far-field) load. For the same far-field load applied to an uncut sheet one knows (or can calculate) the corresponding Airy stress function $\phi = \phi(x, y)$. He then put a hole(s) in the sheet and asked if the hole(s) can be reinforced such that the stress field outside the hole(s) is identical to that of the uncut sheet. In other words, the reinforcement hides the hole(s) from the stress field. Mansfield [1953] showed that the boundary of a neutral hole is given by the equation $\phi(x, y) + ax + by + c = 0$, where $a, b, c$ are constants. The reinforcement is a pre-stressed axially-loaded member (no bending stiffness) that may have a non-uniform cross sectional area. Design of such neutral holes depend on the external loading. However, unlike the cloaking problem in which a hole is to be hidden from arbitrary elastic waves, the shape of a neutral hole and the characteristics of its reinforcement explicitly depend on the stress field of the uncut sheet. In other words, a reinforced hole neutral under one far-field load may not be neutral under another one. In this paper we will present a nonlinear analogue of Mansfield's neutral holes for radial deformations in §4.2.

The main difference between electromagnetic (and optical) cloaking and elastodynamics cloaking is that the governing equations of elasticity are written with respect to two frames and that they are tensor-valued. In elasticity, one writes the governing equations with respect to a reference and a current configuration; this leads to two-point tensors in the governing equations. If formulated properly, both nonlinear and linearized elasticity are spatially covariant, i.e., their governing equations are invariant under arbitrary time-dependent changes of frame (or coordinate transformations if viewed passively) [Steigmann, 2007, Yavari and Ozakin, 2008]. Invariance under referential changes of frame is more subtle as can be seen in the work of Yavari et al. [2006], who showed that the balance of energy is not invariant under an arbitrary time-dependent referential diffeomorphism. They obtained the transformed balance of energy that has some new terms corresponding to the velocity of the referential change of frame. Mazzucato and Rachele [2006] showed that the balance of linear momentum is invariant under any time-independent change in the reference configuration. In this paper, we will show that the results of Yavari et al. [2006] and Mazzucato and Rachele [2006] are consistent and the balance of energy and all the governing equations of nonlinear elasticity are invariant under arbitrary time-independent referential coordinate transformations. The goal in elastodynamics cloaking is to make a hole (cavity) invisible to elastic waves. This may be achieved by covering the boundary of the cavity by a cloak that has inhomogeneous and anisotropic elastic properties, in general. The cloak will deflect the elastic waves resulting in elastic measurements away from the cavity (more specifically, outside the cloak) identical to those when the cavity is absent. Pulling back the homogeneous material properties using a cloaking transformation (that will be defined later), one would obtain the desired inhomogeneous and anisotropic mechanical properties of the cloak. To achieve design of an elastic cloak one would need to answer the following questions: i. Are the governing equations of nonlinear (and linear) elasticity invariant under coordinate transformations? First, one must define what a coordinate transformation means in nonlinear elasticity. There are two types of transformations that have very different physical meanings. ii. Can cloaking be achieved using a spatial or referential coordinate transformation? Or a cloaking transformation is more than a change of coordinates? Related to question i, note that any properly formulated physical field theory has to be covariant. This was Einstein's idea in the theory of general relativity. No coordinate system can be a distinguished one; nature does not discriminate between different observers and they all see the same physical laws. In nonlinear elasticity covariance is understood as invariance of the governing equations under arbitrary time-dependent coordinate transformations in the ambient space [Hughes and Marsden, 1977, Marsden and Hughes, 1983, Yavari et al., 2006]. Regarding question ii we will show that a cloaking transformation is neither a spatial nor a referential change of frame (coordinates); a cloaking transformation maps the boundary-value problem of an isotropic and homogeneous elastic body (virtual problem) to that of an anisotropic and inhomogeneous elastic body with a hole surrounded by a cloak that is to be designed (physical problem).

Traditionally, many workers in solid mechanics start from linear elasticity. This is appropriate for many practical applications, and linear elasticity has been quite successful in the past. The governing equations of linear elasticity are linear partial differential equations, and hence, superposition is applicable, one can use Green's functions, etc. However, there are many problems for which linearized elasticity is not appropriate. The first practical application of nonlinear elasticity was in the rubber industry in the 1940s and 1950s, which motivated Rivlin's seminal contributions [Rivlin, 1948,a,b,c, 1949a,b, Rivlin and Saunders, 1951]. In recent years, nonlinear elasticity has been revived motivated by the biomechanics applications in which biological tissues undergo large strains [Goriely, 2017]. However, the presence of large strains is not the only reason to work with nonlinear elasticity. Unlike electromagnetism with only one configuration (ambient space), in nonlinear elasticity, there are two inherently different configurations: reference and current. Linear elasticity does not distinguish between these two configurations, and this has been a source of confusion in the recent literature of elastodynamics transformation cloaking. In the reference configuration the body is stress free[1] and any measure of strain is defined with respect to this configuration. Consequently, the stored energy of an elastic body explicitly depends on the reference configuration as well. In the classical formulation of nonlinear elasticity, it is well understood that coordinate transformations in the reference and current configurations are very different. Local referential transformations are related to material symmetries, while the global transformations of the ambient space (current configuration) are related to objectivity (or material frame indifference). This implies that any cloaking study, even when strains are small, should be formulated in the framework of nonlinear elasticity.

A classic example of an improper use of the governing equations of linear elasticity can be seen in almost all the existing discussions on the objectivity of linear elasticity. It has long been argued that linearized elasticity is not objective, i.e., its governing equations are not invariant under rigid body translations and rotations of the ambient space (see [Steigmann, 2007] for references). This is unnatural and accepting it, one, at least implicitly, is assuming that linear elasticity is a "special" field theory. This cannot be true and linear elasticity, like any other field theory, has to be objective (and, more generally, spatially covariant) if it is properly formulated. This problem was revisited independently by Steigmann [2007] and Yavari and Ozakin [2008]. These authors showed that if formulated and interpreted properly, linear elasticity is objective (covariant) as expected. In short, Navier's equations are written with respect to one coordinate system and do not have the proper geometric structure to be used in studying the transformation properties of the balance of linear momentum in linear elasticity.

[1] We should mention that there are recent geometric developments using non-Euclidean reference configurations that allow for sources of residual stress [Ozakin and Yavari, 2010, Yavari, 2010, Yavari and Goriely, 2012a,b,c, 2014, Sadik and Yavari, 2015, Efrati et al., 2013, Golgoon et al., 2016, Golgoon and Yavari, 2017, 2018b, Sadik et al., 2016, Sozio and Yavari, 2017, Golgoon and Yavari, 2018a].

**The work of Lodge [1952, 1955].** Arthur S. Lodge showed that the static equilibrium solutions of certain anisotropic homogeneous linear elastic bodies can be mapped to those of isotropic homogenous linear elastic bodies using affine transformations of position and displacement vectors. In other words, knowing an equilibrium solution for a homogeneous isotropic linear elastic body, new equilibrium solutions can be generated for certain anisotropic bodies. We should point out that in [Lodge, 1952, 1955] and [Ostrosablin, 2006] the matrices of the coordinate and displacement transformations are inverses of each other. In particular, Lodge [1952] in his first paper considered the following transformations for coordinates and displacements: $(x, y, z) \to (x', y', z') = (x, y, \nu^{-\frac{1}{2}} z)$ and $(u, v, w) \to (u', v', w') = (u, v, \nu^{\frac{1}{2}} z)$, for some $\nu > 0$. He showed that the governing equations of linear elasticity are invariant under these transformations. Note that these transformations map the equilibrium solution of an isotropic elastic body to that of another elastic body that is anisotropic. This should not be confused with the transformation of the governing equations of one given body under referential or spatial coordinate transformations. In other words, Lodge [1955] (see also Lang [1956]) finds an equilibrium solution for an anisotropic body using that of another elastic body, which is isotropic. The position and displacement vectors are linearly related. However, the two problems are not related by a coordinate transformation. Olver [1988] showed that any planar anisotropic linear elastic solid is equivalent to an orthotropic solid through some linear transformations of coordinates and displacements that are independent.

Lodge [1955]'s idea of mapping the boundary-value problem of an anisotropic linearly elastic body to that of an isotropic body can be summarized as follows. The position vector and the displacement field are transformed homogeneously as

$$\mathbf{x}' = \mathbf{A}\mathbf{x}, \quad \mathbf{u}' = \mathbf{A}^{-\mathsf{T}}\mathbf{u}\,. \tag{1.1}$$

We will refer to the transformation $(\mathbf{x}', \mathbf{u}') = (\mathbf{A}\mathbf{x}, \mathbf{A}^{-\mathsf{T}}\mathbf{u})$ as a *Lodge transformation*. Using this pair of linear transformations, strain is transformed as $\boldsymbol{\epsilon}' = \mathbf{A}^{-\mathsf{T}}\boldsymbol{\epsilon}\mathbf{A}^{-1}$. Assuming that under (1.1), $\boldsymbol{\sigma}' : \boldsymbol{\epsilon}' = \boldsymbol{\sigma} : \boldsymbol{\epsilon}$, stress is transformed as $\boldsymbol{\sigma}' = \mathbf{A}\boldsymbol{\sigma}\mathbf{A}^{\mathsf{T}}$. Lodge [1955] assumed that body force transforms like a vector, i.e., $\mathbf{b}' = \mathbf{A}\mathbf{b}$. Implicitly, he assumed that mass density transforms like a scalar, i.e., it remains unchanged: $\rho' = \rho$. This is similar to the way mass density transforms under a change of spatial frame. Under these assumptions one can show that

$$\mathrm{div}'\,\boldsymbol{\sigma}' + \rho'\mathbf{b}' = \mathbf{A}\left(\mathrm{div}\,\boldsymbol{\sigma} + \rho\mathbf{b}\right)\,. \tag{1.2}$$

The inertial force is transformed as $\rho'\mathbf{a}' = \mathbf{A}^{-\mathsf{T}}(\rho\mathbf{a})$. Therefore, starting from $\mathrm{div}\,\boldsymbol{\sigma} + \rho\mathbf{b} = \rho\mathbf{a}$, one obtains the balance of linear momentum for the transformed body as

$$\mathrm{div}'\,\boldsymbol{\sigma}' + \rho'\mathbf{b}' = \mathbf{A}\mathbf{A}^{\mathsf{T}}\rho'\mathbf{a}\,. \tag{1.3}$$

Lodge then concluded that the balance of linear momentum is invariant under the transformation (1.1) only when inertial forces are ignored. Under a Lodge transformation elastic constants, and hence, the anisotropy type transform. Lodge finally calculated the matrix $\mathbf{A}$ such that the transformed body is isotropic. Using this transformation, one can generate equilibrium solutions for certain anisotropic bodies having the corresponding solutions for an isotropic linearly elastic body.

**Remark 1.1.** Fifty years later, not being aware of the work of Lodge, Milton et al. [2006] for a completely different purpose used the Lodge transformations (1.1) but with a position-dependent $\mathbf{A}$. They assumed a harmonic time dependence and ignored the body forces. Let us examine their Eq.(2.4) for a constant matrix $\mathbf{A}$. Their transformed wave equation in this special case reads

$$\mathrm{div}'\,\boldsymbol{\sigma}' = -\omega^2\boldsymbol{\rho}'\mathbf{u}'\,, \tag{1.4}$$

where their matrix-valued mass density is defined as $\boldsymbol{\rho}' = \frac{1}{\det \mathbf{A}}\rho\mathbf{A}\mathbf{A}^{\mathsf{T}}$. They justify a matrix-valued mass density arguing that it has been observed for composites. Two comments are in order here: i) Note that the factor $\frac{1}{\det \mathbf{A}}$ appears when one uses the Piola identity as we will discuss in §4.1 and §4.3. When $\mathbf{A}$ is a constant matrix, $\mathrm{div}'\,\boldsymbol{\sigma}'$ is $\frac{1}{\det \mathbf{A}}$ times that of Lodge's. ii) If the body force term $\rho\mathbf{b}$ is kept, this matrix-valued mass density would not work unless one assumes that $\mathbf{b}' = \mathbf{A}^{-\mathsf{T}}\mathbf{b}$, which is different from Lodge's original transformation.

**The work of Milton et al. [2006] on elastodynamics cloaking.** The first theoretical study of elastodynamics transformation cloaking is due to Milton et al. [2006]. They observed that the governing equations of linear elasticity are not invariant under coordinate transformations (or what they called "curvilinear transformations"), and hence, cloaking of elastic waves cannot be achieved using coordinate transformations. One should note that Navier's equations are written with respect to one coordinate system, and hence, do not have the proper geometric structure to be used in studying the transformation properties of the governing equations of linear elasticity under coordinate (or cloaking) transformations. Milton et al. [2006] start with the wave equation in the setting of linear elasticity, i.e., $\nabla \cdot \boldsymbol{\sigma} + \omega^2 \rho \mathbf{u} = \mathbf{0}$, where $\boldsymbol{\sigma} = \mathsf{C}\nabla\mathbf{u}$, and $\mathsf{C}$ is the elasticity tensor, and consider mappings $\mathbf{x} \to \mathbf{x}'(\mathbf{x})$ and $\mathbf{u}(\mathbf{x}) \to \mathbf{u}'(\mathbf{x}')$. They pointed out that the two mappings can be chosen freely. Denoting the derivative of the map $\mathbf{x}'(\mathbf{x})$ by $\mathbf{A}(\mathbf{x})$, i.e., $A_{i'j} = \partial x'_i/\partial x_j$, they assume that instead of $\mathbf{u}' = \mathbf{A}(\mathbf{x})\mathbf{u}$ (assuming that the change of coordinates is a spatial coordinate transformation), displacement is transformed as $\mathbf{u}' = \mathbf{A}^{-\mathsf{T}}(\mathbf{x})\mathbf{u}$. Milton et al. [2006] point out that these (Lodge-type) transformations preserve the symmetries of the elasticity tensor. They finally show that under these changes of variables the Cauchy stress and the wave equation transform as

$$\mathrm{div}'\,\boldsymbol{\sigma}' = -\omega^2 \boldsymbol{\rho}'\mathbf{u}' + \mathbf{D}'\nabla'\mathbf{u}' \qquad \text{and} \qquad \boldsymbol{\sigma}' = \mathsf{C}'\nabla'\mathbf{u}' + \mathbf{S}'\mathbf{u}', \tag{1.5}$$

where ($J = \det \mathbf{A}$) $C'_{pqrs} = \frac{1}{J}\frac{\partial x'^p}{\partial x^i}\frac{\partial x'^q}{\partial x^j}\frac{\partial x'^r}{\partial x^k}\frac{\partial x'^s}{\partial x^l}C_{ijkl}$, $S'_{pqr} = \frac{1}{J}\frac{\partial x'^p}{\partial x^i}\frac{\partial x'^q}{\partial x^j}\frac{\partial^2 x'^r}{\partial x^k \partial x^l}C_{ijkl} = S'_{qrp}$, $D'_{pqr} = S'_{qrp}$, and the matrix-valued mass density[2] $\rho'_{pq} = \frac{\rho}{J}\frac{\partial x'_p}{\partial x_i}\frac{\partial x'_q}{\partial x_i} + \frac{1}{J}\frac{\partial^2 x'_p}{\partial x_i \partial x_j}C_{ijkl}\frac{\partial^2 x'_q}{\partial x_k \partial x_l}$. Banerjee [2011] discusses this in more detail and concludes that: "Therefore, unlike Maxwell's equations the equations of elastodynamics change form under a coordinate transformation." This conclusion is, unfortunately, incorrect; governing equations of linear elasticity are form-invariant under both spatial and referential coordinate transformations. However, as we will show cloaking transformations are not coordinate transformations. It is, nevertheless, correct that transformation cloaking is not possible in classical linear elastodynamics.

Several authors have looked at in-plane waves arguing that in this particular case the governing equations are invariant under coordinate transformations (however, it is not clear what type of coordinate transformations is being considered), and hence, transformation cloaking can be acheived. In particular, Brun et al. [2009] observed that for in-plane elastic waves, the balance of linear momentum is invariant under an arbitrary change of coordinates (they do not distinguish between referential and spatial transformations), and following the ideas in electromagnetic cloaking, introduced a cylindrical cloak. They considered an annular cloak of inner and outer radii $r_0$ and $r_1$, respectively, and used Pendry et al. [2006]'s coordinate transformation $(r, \theta) \to (r', \theta') = (r'(r), \theta)$, where

$$r'(r) = \begin{cases} r_0 + \frac{r_1 - r_0}{r_1} r, & r \leq r_1, \\ r, & r \geq r_1. \end{cases} \tag{1.6}$$

This (singular) transformation maps a disk of radius $r_1$ to an annulus of inner and outer radii $r_0$ and $r_1$, respectively, and is the identity transformation outside the disk. These authors observed that assuming $\mathbf{u}' = \mathbf{u}$, Navier's equations are form-invariant, and mass density transforms as $\rho'(r) = \frac{r - r_0}{r}\left(\frac{r_1}{r_1 - r_0}\right)^2 \rho(r)$. In the transformed coordinates, they obtained an elasticity tensor that does not possess the minor symmetries. Then their recourse is to argue that the transformed body is made of a Cosserat solid.[3] We will show in this paper that the map (1.6), which has been borrowed from the literature of electromagnetic cloaking is not admissible for elastodynamics cloaking; the derivative of this map is not the identity on the outer boundary of the cloak, i.e., $f'(r_1) \neq 1$ (see §5.2). Instead of assuming that the cloak is made of a Cosserat solid without even discussing its elastic constants, we believe one should assume that both the physical and virtual bodies (that will be defined in §4) are made of Cosserat solids and see if a Cosserat cloak can be designed while all the balance laws are respected in both bodies. We will show that the minor symmetries of the elastic constants are not preserved under cloaking transformations (that will be defined in §4.1). Our conclusion is that classical linear elasticity is not flexible enough to allow for cloaking. This would force one to start from some kind of a solid with microstructure. In addition to finding the classical elastic constants in the cloak, the non-classical elastic constants must be calculated as well. This has not been discussed in the literature to this date. This is a non-trivial calculation that will be discussed in §5. In the literature, it has been implicitly assumed that the material outside the cloak is a classical linear elastic solid. We will see that this is not possible (Remark 5.5).

[2] There is a typo in their transformed mass density. The second term does not have the correct physical dimension. The similar expression in [Banerjee, 2011] has the correct dimension with a factor $1/\omega^2$ but has the opposite sign.

[3] Surprisingly, in none of the works that accept non-symmetric Cauchy stresses in the cloak is there any mention of the balance of angular momentum and the distribution of couple stresses. There is also no discussion on what the extra elastic constants of the Cosserat cloak should be.

Starting from linear elasticity, Norris and Shuvalov [2011] tried to find the governing equations in a transformed domain using Lodge-type transformations. Similar to the work of Milton et al. [2006], they assumed that displacement field does not transform the way it does under a spatial coordinate transformation. They refer to this as a linear gauge change. Unlike that of Milton et al. [2006], their displacement transformation is completely independent of the coordinate transformation. This is similar to the transformations that had earlier been used by Olver [1988]. They discussed several possibilities for the displacement transformation and observed a loss of minor symmetries of the elastic constants in the cloak and assumed that it is made of a Cosserat solid. However, the balance of angular momentum and the calculation of the non-classical elastic constants of the Cosserat elastic cloak were not discussed.

Olsson and Wall [2011] studied time-harmonic cloaking a finite rigid body that is fixed, i.e., cannot move, embedded in an elastic matrix. In the case of both a rigid circular disk and a rigid spherical ball they assumed that the matrix is an elastic medium with inextensible radial fibers. To our best knowledge, this is the first paper on elastic cloaking that actually discusses boundary conditions. In the physical and virtual bodies (that we will define in §4) in both 2D and 3D they assumed the following relation between displacement vectors: $\mathbf{U}/R = \tilde{\mathbf{U}}/\tilde{R}$. They observed that the balance of linear momentum of the two configurations have the same form, and that elastic constants retain their full symmetries. Khlopotin et al. [2015] investigated cloaking a finite rigid body embedded in an elastic medium. Their motivation was cloaking an object in a soft matrix from surface elastic waves. They assumed that the matrix is a micropolar solid. They discussed boundary conditions. However, in this work there is no discussion on how the non-classical elastic constants of the cloak should be calculated. In particular, there is no mention of how the couple stiffness tensor and the couple stress tensor of the micropolar medium are transformed under a cloaking transformation. We believe that a proper formulation of linear elastodynamics cloaking should consider both the physical and virtual bodies to be made of generalized Cosserat solids and all the elastic constants of the cloak must be calculated. This will be discussed in §5.

Parnell [2012] (see also [Norris and Parnell, 2012]) first considered antiplane deformations of an isotropic linear elastic solid for which the displacement field in cylindrical coordinates has the form $(0, 0, W(R, \Theta))$. He concluded that the balance of linear momentum (wave equation) is form-invariant under coordinate transformations. The transformed mass density is identical to that of Brun et al. [2009]. Next, Parnell considered the wave equation in the small-on-large theory of Green et al. [1952], which is simply linearization about a finitely-deformed (and stressed) configuration [Ogden, 1997]. He considered a body with a small hole made of an incompressible neo-Hookean solid. Using a (static) applied internal pressure the hole is inflated. Parnell showed that the incremental wave equation with respect to this pre-stressed configuration has anisotropic shear moduli. However, they are different from those of a cloak. One should also note that inflating an initially small hole in a body the entire body would deform. In other words, the small-on-large elasticity of such a body cannot be identical to that of a stress-free and homogeneous linear elastic body outside any finite region. In this sense, the idea of pre-stress cannot be useful in the context of transformation cloaking.

There have been other efforts in the literature on guiding elastic waves in structures. We should mention Amirkhizi et al. [2010] who proposed the idea of redirecting stress waves by smoothly changing anisotropy of a structure. In particular, they experimentally and numerically showed that when the direction of a stress wave is known and is fixed its propagation in a structure made of a transversely isotropic material with a varying axis of anisotropy can be guided. Of course, their construction is restricted and useful only for one specific direction of wave propagation.

**Contributions of this paper.** In this paper, we investigate the problem of hiding a hole from elastic waves in both nonlinear and linear elastic solids. We start by discussing the invariance of the governing equations of nonlinear and linearized (with respect to any finitely-deformed configuration) elasticity under both arbitrary time-dependent spatial changes of frame (coordinate transformations) and arbitrary time-independent referential changes of frame. We, however, note that cloaking cannot be achieved using either (or both) spatial or referential coordinate transformations. We define a cloaking map to be a mapping that transforms the boundary-value problem of an elastic body with a hole reinforced by a cloak (physical body) to that of a homogeneous and isotropic body with an infinitesimal hole (virtual body). The cloak needs to be designed while the loads and boundary conditions in the virtual body are not known a priori. We define a cloaking transformation to be a map between the boundary-value problems of an elastic body to be designed and a virtual elastic body that has some desired mechanical response. The main contributions of this work can be summarized as follows:

- We provide a geometric formulation of transformation cloaking in nonlinear elasticity. It is shown that nonlinear elastodynamics transformation cloaking is not possible (Proposition 4.6).
- It is shown that nonlinear elastostatics transformation cloaking may be possible for special deformations. This is somehow a nonlinear analogue of Mansfield's neutral holes. We provide one such example, namely radial deformations in an infinitely long solid cylinder with a cylindrical hole or a finite spherical ball with a spherical cavity.
- Classical linear elasticity is not flexible enough to allow for transformation cloaking (Proposition 4.8). More specifically, linear elastodynamics cloaking cannot be achieved because the elastic constants in the cloak lose their minor symmetries. This is true for a hole of any shape reinforced by a cloak with an arbitrary shape.
- Assuming that the virtual body is isotropic and centro-symmetric, elastodynamics transformation cloaking is not possible in the setting of gradient elasticity (Proposition 5.3). This result is independent of the shape of the hole.
- Elastodynamics transformation cloaking is not possible in the setting of generalized Cosserat elasticity in dimension two (Proposition 5.7). In particular, in dimension two transformation cloaking cannot be achieved in Cosserat elasticity (with rigid directors) either. This result is independent of the shapes of the hole and the cloak. Elastodynamics transformation cloaking is not possible for a spherical cavity using a spherical cloak in the setting of generalized Cosserat elasticity (Proposition 5.9). We conjecture that this result in dimension three is independent of the shapes of the cavity and the cloak (Conjecture 5.11).

This paper is structured as follows. In §2, we revisit nonlinear elasticity and discuss the invariance of its governing equations under both arbitrary time-dependent transformations of the ambient space (current configuration) and arbitrary time-independent transformations of the reference configuration. In particular, structural tensors are discussed in some detail. In §3, linearization of nonlinear elasticity is discussed in detail. Then spatial and referential covariance of the governing equations of linearized elasticity are investigated. The problems of cloaking for both nonlinear elastodynamics and elastostatics are formulated in §4.1 and §4.2, respectively. Cloaking transformation in classical linear elasticity is investigated in §4.3. In §5, elastodynamics transformation cloaking in solids with microstructure is investigated. The impossibility of transformation cloaking in linearized gradient elasticity is discussed in §5.1. Elastodynamics transformation cloaking in generalized Cosserat solids is formulated in §5.2.

# 2 Nonlinear Elastodynamics

**Kinematics.** In nonlinear elasticity, motion is a time-dependent mapping between a reference configuration (or natural configuration) and the ambient space (see Fig.2). Geometrically, we write this as $\varphi_t : \mathcal{B} \to \mathcal{S}$, where $(\mathcal{B}, \mathbf{G})$ and $(\mathcal{S}, \mathbf{g})$ are the material and the ambient space Riemannian manifolds, respectively [Marsden and Hughes, 1983]. Here, $\mathbf{G}$ is the material metric (that allows one to measure distances in a natural stress-free configuration) and $\mathbf{g}$ is the background metric of the ambient space. The Levi-Civita connections associated with the metrics $\mathbf{G}$ and $\mathbf{g}$ are denoted as $\nabla^{\mathbf{G}}$ and $\nabla^{\mathbf{g}}$, respectively. The corresponding Christoffel symbols of $\nabla^{\mathbf{G}}$ and $\nabla^{\mathbf{g}}$ in the local coordinate charts $\{X^A\}$ and $\{x^a\}$ are denoted by $\Gamma^A{}_{BC}$ and $\gamma^a{}_{bc}$, respectively. These can be directly expressed in terms of the metric components as

$$\gamma^a{}_{bc} = \frac{1}{2} g^{ak} \left( g_{kb,c} + g_{kc,b} - g_{bc,k} \right) , \qquad \Gamma^A{}_{BC} = \frac{1}{2} G^{AK} \left( G_{KB,C} + G_{KC,B} - G_{BC,K} \right) . \tag{2.1}$$

The deformation gradient $\mathbf{F}$ is the tangent map of $\varphi_t$, which is defined as $\mathbf{F}(X,t) = T\varphi_t(X) : T_X\mathcal{B} \to T_{\varphi_t(X)}\mathcal{S}$. The transpose of $\mathbf{F}$ is denoted by $\mathbf{F}^\mathsf{T}$, where

$$\mathbf{F}^\mathsf{T}(X,t) : T_{\varphi_t(X)}\mathcal{S} \to T_X\mathcal{B} \,, \qquad \langle\!\langle \mathbf{W}, \mathbf{F}^\mathsf{T}\mathbf{w} \rangle\!\rangle_{\mathbf{G}} = \langle\!\langle \mathbf{F}\mathbf{W}, \mathbf{w} \rangle\!\rangle_{\mathbf{g}}, \quad \forall\, \mathbf{W} \in T_X\mathcal{B},\, \mathbf{w} \in T_{\varphi_t(X)}\mathcal{S} \,. \tag{2.2}$$

In components, $(F^\mathsf{T})^A{}_a = G^{AB} F^b{}_B g_{ab}$. The right Cauchy-Green deformation tensor[4] is defined as $\mathbf{C} = \mathbf{F}^\mathsf{T}\mathbf{F} : T_X\mathcal{B} \to T_X\mathcal{B}$, which in components reads $C^A{}_B = F^a{}_L F^b{}_B g_{ab} G^{AL}$.

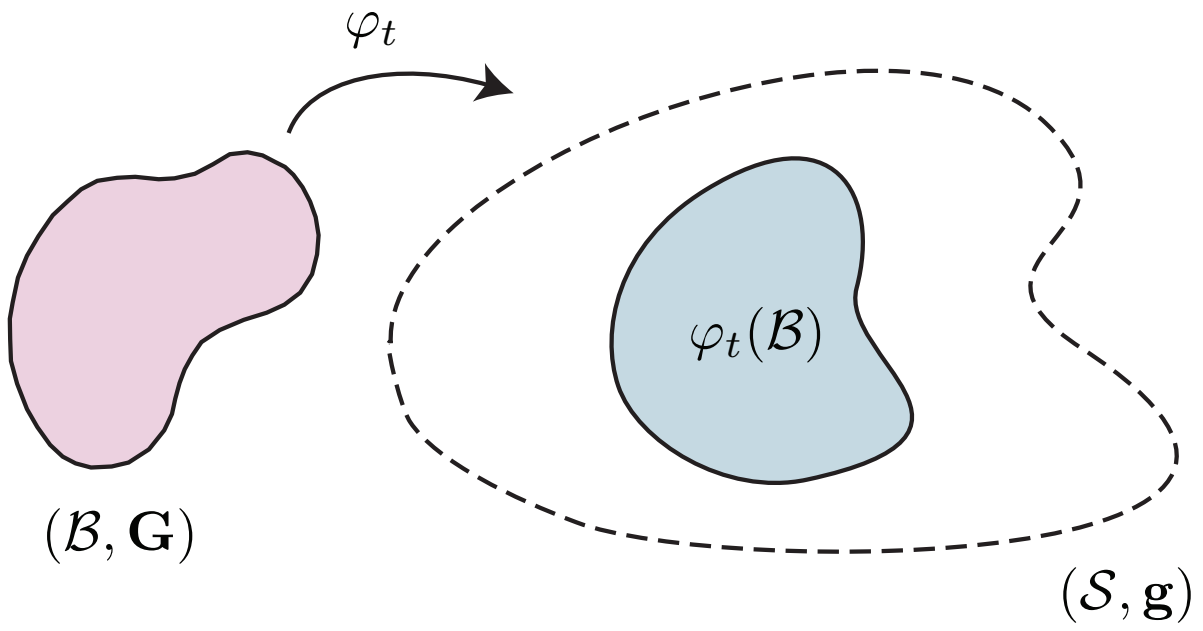

Figure 2: *Motion in nonlinear elasticity is a time-dependent mapping between two manifolds.*

The material velocity of the motion is the mapping $\mathbf{V} : \mathcal{B} \times \mathbb{R}^+ \to T\mathcal{S}$, where $\mathbf{V}(X,t) \in T_{\varphi_t(X)}\mathcal{S}$, and in components, $V^a(X,t) = \frac{\partial \varphi^a}{\partial t}(X,t)$. The spatial velocity is defined as $\mathbf{v} : \varphi_t(\mathcal{B}) \times \mathbb{R}^+ \to T\mathcal{S}$ such that $\mathbf{v}_t(x) = \mathbf{V}_t \circ \varphi_t^{-1}(x) \in T_x\mathcal{S}$, where $x = \varphi_t(X)$. The convected velocity is defined as $\boldsymbol{\mathscr{V}}_t = \varphi_t^*(\mathbf{v}_t) = T\varphi_t^{-1} \circ \mathbf{v}_t \circ \varphi_t = \mathbf{F}^{-1} \cdot \mathbf{V}$. The material acceleration is a mapping $\mathbf{A} : \mathcal{B} \times \mathbb{R}^+ \to T\mathcal{S}$ defined as $\mathbf{A}(X,t) := D_t^{\mathbf{g}}\mathbf{V}(X,t) = \nabla^{\mathbf{g}}_{\mathbf{V}(X,t)}\mathbf{V}(X,t) \in T_{\varphi_t(X)}\mathcal{S}$, where $D_t^{\mathbf{g}}$ denotes the covariant derivative along the curve $\varphi_t(X)$ in $\mathcal{S}$. In components, $A^a = \frac{\partial V^a}{\partial t} + \gamma^a{}_{bc} V^b V^c$. To motivate this definition note that the time derivative of the kinetic energy density is calculated as[5]

$$\frac{d}{dt}\frac{1}{2}\rho_0(X)\langle\!\langle \mathbf{V}(X,t), \mathbf{V}(X,t) \rangle\!\rangle_{\mathbf{g}} = \rho_0(X)\langle\!\langle \mathbf{V}(X,t), D_t^{\mathbf{g}}\mathbf{V}(X,t) \rangle\!\rangle_{\mathbf{g}} = \rho_0(X)\langle\!\langle \mathbf{V}(X,t), \mathbf{A}(X,t) \rangle\!\rangle_{\mathbf{g}}. \tag{2.3}$$

Therefore, $\mathbf{A}(X,t)$ is the covariant time derivative of the velocity vector field. The spatial acceleration is defined as $\mathbf{a} : \varphi_t(\mathcal{B}) \times \mathbb{R}^+ \to T\mathcal{S}$ such that $\mathbf{a}_t(x) = \mathbf{A}_t \circ \varphi_t^{-1}(x) \in T_x\mathcal{S}$. In components, it reads $a^a = \frac{\partial v^a}{\partial t} + \frac{\partial v^a}{\partial x^b} v^b + \gamma^a{}_{bc} v^b v^c$. The spatial acceleration can also be expressed as the material time derivative of $\mathbf{v}$, i.e., $\mathbf{a} = \dot{\mathbf{v}} = \frac{\partial \mathbf{v}}{\partial t} + \nabla^{\mathbf{g}}_{\mathbf{v}}\mathbf{v}$. The convected acceleration is defined as [Simo et al., 1988]

$$\boldsymbol{\mathscr{A}}_t = \varphi_t^*(\mathbf{a}_t) = \frac{\partial \boldsymbol{\mathscr{V}}_t}{\partial t} + \nabla^{\varphi_t^*\mathbf{g}}_{\boldsymbol{\mathscr{V}}_t}\boldsymbol{\mathscr{V}}_t = \frac{\partial \boldsymbol{\mathscr{V}}_t}{\partial t} + \nabla^{\mathbf{C}^\flat}_{\boldsymbol{\mathscr{V}}_t}\boldsymbol{\mathscr{V}}_t \,. \tag{2.4}$$

**Balance laws.** Balance of linear momentum in spatial and material forms reads

$$\operatorname{div}_{\mathbf{g}} \boldsymbol{\sigma} + \rho\mathbf{b} = \rho\mathbf{a}, \quad \operatorname{Div} \mathbf{P} + \rho_0\mathbf{B} = \rho_0\mathbf{A}, \tag{2.5}$$

[4]Note that $\mathbf{C}^\flat$ agrees with the pull-back of the ambient space metric by $\varphi_t$, i.e., $\mathbf{C}^\flat = \varphi_t^*\mathbf{g}$.

[5]Note that if a connection $\nabla$ is $\mathbf{G}$-compatible, then $\frac{d}{dt}\langle\!\langle \mathbf{X}, \mathbf{Y}(X,t) \rangle\!\rangle_{\mathbf{G}} = \langle\!\langle D_t\mathbf{X}, \mathbf{Y} \rangle\!\rangle_{\mathbf{G}} + \langle\!\langle \mathbf{X}, D_t\mathbf{Y} \rangle\!\rangle_{\mathbf{G}}$, where $D_t$ is the covariant time derivative.

where $\boldsymbol{\sigma}$ and $\mathbf{P}$ are the Cauchy stress and the first Piola-Kirchhoff stress, respectively. $\rho_0$, $\mathbf{B}$, and $\mathbf{A}$ are the material mass density, material body force, and material acceleration, respectively, and $\rho$, $\mathbf{b}$, and $\mathbf{a}$ are their corresponding spatial counterparts. Note that $\operatorname{div}_{\mathbf{g}} \boldsymbol{\sigma}$ and $\operatorname{Div} \mathbf{P}$ have the following coordinate expressions

$$\begin{aligned} \operatorname{div}_{\mathbf{g}} \boldsymbol{\sigma} &= \sigma^{ab}{}_{|b} \frac{\partial}{\partial x^a} = \left( \frac{\partial \sigma^{ab}}{\partial x^b} + \sigma^{ac} \gamma^b{}_{cb} + \sigma^{cb} \gamma^a{}_{cb} \right) \frac{\partial}{\partial x^a}, \\ \operatorname{Div} \mathbf{P} &= P^{aA}{}_{|A} \frac{\partial}{\partial x^a} = \left( \frac{\partial P^{aA}}{\partial X^A} + P^{aB} \Gamma^A{}_{AB} + P^{cA} F^b{}_A \gamma^a{}_{bc} \right) \frac{\partial}{\partial x^a}. \end{aligned} \tag{2.6}$$

In coordinates, $J\sigma^{ab} = F^a{}_A P^{bA}$, where $J$ is the Jacobian of deformation that relates the deformed and undeformed Riemannian volume elements as $dv(x, \mathbf{g}) = J dV(X, \mathbf{G})$, and

$$J = \sqrt{\frac{\det \mathbf{g}}{\det \mathbf{G}}} \det \mathbf{F}. \tag{2.7}$$

One can pull back the balance of linear momentum to the reference configuration, i.e., $\varphi_t^* (\operatorname{div}_{\mathbf{g}} \boldsymbol{\sigma}) + \varphi_t^* (\rho \mathbf{b}) = \varphi_t^* (\rho \mathbf{a})$. This can be written as

$$\operatorname{div}_{\mathbf{C}^\flat} \boldsymbol{\Sigma} + \varrho \mathscr{B}_t = \varrho \mathscr{A}_t, \tag{2.8}$$

where $\boldsymbol{\Sigma} = \varphi_t^* \boldsymbol{\sigma}$ is the convected stress, $\mathscr{B}_t = \varphi_t^* \mathbf{b}$ is the convected body force, and $\varrho = \rho \circ \varphi_t$.

Identifying a material point with its position in the material manifold $X \in \mathcal{B}$, we have $x = \varphi_t(X)$. When the ambient space is Euclidean one defines the material displacement field as $\mathbf{U} = \varphi_t(X) - X$. The spatial displacement field is denoted by $\mathbf{u} = \mathbf{U} \circ \varphi_t^{-1}$.

Balance of angular momentum in local form reads $\boldsymbol{\sigma}^{\mathsf{T}} = \boldsymbol{\sigma}$ or $\mathbf{F}\mathbf{P}^\star = \mathbf{P}\mathbf{F}^\star$, where $\mathbf{P}^\star$ and $\mathbf{F}^\star$ are duals of $\mathbf{P}$ and $\mathbf{F}$, respectively, and are defined as

$$\begin{aligned} \mathbf{F} &= F^a{}_A \frac{\partial}{\partial x^a} \otimes dX^A, & \mathbf{F}^\star &= F^a{}_A dX^A \otimes \frac{\partial}{\partial x^a}, \\ \mathbf{P} &= P^{aA} \frac{\partial}{\partial x^a} \otimes \frac{\partial}{\partial X^A}, & \mathbf{P}^\star &= P^{aA} \frac{\partial}{\partial X^A} \otimes \frac{\partial}{\partial x^a}. \end{aligned} \tag{2.9}$$

Note that $\mathbf{F}^\star : T^*_{\varphi_t(X)} \mathcal{S} \to T^*_X \mathcal{B}$, where $T^*_{\varphi_t(X)} \mathcal{S}$ and $T^*_X \mathcal{B}$ denote the cotangent spaces of $T_{\varphi_t(X)} \mathcal{S}$ and $T_X \mathcal{B}$, respectively.

Conservation of mass implies that $\rho dv = \rho_0 dV$ or $\rho J = \rho_0$, where $\rho_o$ and $\rho$ denote the material and spatial mass densities, respectively. In terms of Lie derivatives, conservation of mass can be written as $\mathbf{L}_{\mathbf{v}} \rho = 0$ [Marsden and Hughes, 1983].

**Constitutive equations.** In nonlinear elasticity, the energy function (per unit undeformed volume) of an inhomogeneous anisotropic hyperelastic material at a material point $X$ is written in the following general form

$$W = \hat{W}(X, \mathbf{C}^\flat, \mathbf{G}, \boldsymbol{\zeta}_1, \ldots, \boldsymbol{\zeta}_n), \tag{2.10}$$

where $\boldsymbol{\zeta}_i, i = 1, \ldots, n$ are a collection of the so called *structural tensors* characterizing the material symmetry group at the point $X$. The inclusion of the structural tensors, along with $\mathbf{C}^\flat$ in the energy function as shown in (2.10) constructs an isotropic function, i.e., it is invariant under the orthogonal group [Liu, 1982]. Therefore, (2.10) can be treated as the energy function of an isotropic material, and hence, the second Piola-Kirchhoff stress tensor is given as

$$\mathbf{S} = 2 \frac{\partial \hat{W}}{\partial \mathbf{C}^\flat}. \tag{2.11}$$

Alternatively, by the Doyle-Ericksen formula [Doyle and Ericksen, 1956], the Cauchy, the first Piola-Kirchhoff, and the convected stress tensors are expressed as

$$\boldsymbol{\sigma} = \frac{2}{J} \frac{\partial \hat{W}}{\partial \mathbf{g}}, \quad \mathbf{P} = \mathbf{g}^\sharp \frac{\partial \hat{W}}{\partial \mathbf{F}}, \quad \boldsymbol{\Sigma} = \frac{2}{J} \frac{\partial \hat{W}}{\partial \mathbf{C}^\flat}, \tag{2.12}$$

where, with a slight abuse of notation, one may write

$$\hat{W}(x, \mathbf{G}\circ\varphi^{-1}, \mathbf{g}, \mathbf{F}, \boldsymbol{\zeta}_1\circ\varphi^{-1}, \ldots, \boldsymbol{\zeta}_n\circ\varphi^{-1}) = \hat{W}(X, \mathbf{G}, \mathbf{g}\circ\varphi, \mathbf{F}, \boldsymbol{\zeta}_1, \ldots, \boldsymbol{\zeta}_n) = \hat{W}(X, \mathbf{G}, \mathbf{C}^\flat, \boldsymbol{\zeta}_1, \ldots, \boldsymbol{\zeta}_n)\,. \tag{2.13}$$

For an incompressible solid the relations in (2.12) are modified to read

$$\boldsymbol{\sigma} = -p\mathbf{g}^\sharp + \frac{2}{J}\frac{\partial \hat{W}}{\partial \mathbf{g}}\,, \quad \mathbf{P} = -pJ\mathbf{g}^\sharp\mathbf{F}^{-\star} + \mathbf{g}^\sharp\frac{\partial \hat{W}}{\partial \mathbf{F}}, \quad \boldsymbol{\Sigma} = -p\mathbf{C}^{-1} + \frac{2}{J}\frac{\partial \hat{W}}{\partial \mathbf{C}^\flat}\,. \tag{2.14}$$

According to Hilbert's theorem, for any finite number of tensors, there exist a finite number of isotropic invariants forming a basis called *integrity basis* for the space of isotropic invariants of the collection of tensors.[6] Thus, if $I_j, j = 1, \ldots, m$, form an integrity basis for the set of tensors in (2.10), one has $W = W(X, I_1, ..., I_m)$. Hence, using (2.11), one obtains

$$\mathbf{S} = \sum_{j=1}^{j=m} 2W_{I_j}\frac{\partial I_j}{\partial \mathbf{C}^\flat}\,, \quad W_{I_j} := \frac{\partial W}{\partial I_j}\,, \quad j = 1, \ldots, m\,. \tag{2.15}$$

If the material is isotropic, i.e., $W = W(X, I_1, I_2, I_3)$, where $I_1 = \operatorname{tr}\mathbf{C}$, $I_2 = \det\mathbf{C}\operatorname{tr}\mathbf{C}^{-1}$, and $I_3 = \det\mathbf{C}$ are the principal invariants of the right Cauchy-Green deformation tensor, it follows from (2.15) that

$$\mathbf{S} = 2\left\{W_{I_1}\mathbf{G}^\sharp + W_{I_2}(I_2\mathbf{C}^{-1} - I_3\mathbf{C}^{-2}) + W_{I_3}I_3\mathbf{C}^{-1}\right\}\,. \tag{2.16}$$

If the material is incompressible, i.e., $I_3 = 1$, one writes $\mathbf{S} = 2\{W_{I_1}\mathbf{G}^\sharp - W_{I_2}\mathbf{C}^{-2}\} - p\mathbf{C}^{-1}$, where $p$ is the Lagrange multiplier associated with the incompressibility constraint $J = 1$.

## 2.1 Spatial Covariance of the Governing Equations of Nonlinear Elasticity

It turns out that in continuum mechanics (and even discrete systems) one can obtain all the balance laws using the energy balance and postulating its invariance under some groups of transformations. This idea was introduced by Green and Rivlin [1964] in the case of Euclidean ambient spaces and was later extended to manifolds by Hughes and Marsden [1977]. See also Marsden and Hughes [1983], Simo and Marsden [1984], Yavari et al. [2006], Yavari and Ozakin [2008], Yavari [2008], Yavari and Marsden [2009a,b] for applications of covariance ideas in different continuous and discrete systems.

Consider an arbitrary time-dependent spatial diffeomorphism $\xi_t : \mathcal{S} \to \mathcal{S}$ (see Fig.3). Denoting the fields in the transformed configuration by primes, we know that [Marsden and Hughes, 1983, Yavari et al., 2006]

$$R' = R, \quad H' = H, \quad \rho_0' = \rho_0, \quad \mathbf{T}' = \xi_{t*}\mathbf{T}, \quad \mathbf{V}' = \xi_{t*}\mathbf{V} + \mathbf{w}\circ\varphi_t, \tag{2.17}$$

where $\mathbf{w} = \frac{\partial}{\partial t}\xi_t\circ\varphi_t$ is the velocity of the change of frame and $\mathbf{T}$ is the traction vector. Note that $\mathbf{A} = \nabla^{\mathbf{g}}_{\mathbf{V}}\mathbf{V}$ and hence

$$\begin{aligned}\mathbf{A}' &= \nabla^{\mathbf{g}'}_{\mathbf{V}'}\mathbf{V}' = \nabla^{\xi_{t*}\mathbf{g}}_{\xi_{t*}\mathbf{V}+\mathbf{w}\circ\varphi_t}\left(\xi_{t*}\mathbf{V} + \mathbf{w}\circ\varphi_t\right)\\ &= \xi_{t*}\left(\nabla^{\mathbf{g}}_{\mathbf{V}}\mathbf{V} + \nabla^{\mathbf{g}}_{\mathbf{W}}\mathbf{W} + \nabla^{\mathbf{g}}_{\mathbf{V}}\mathbf{W} + \nabla^{\mathbf{g}}_{\mathbf{W}}\mathbf{V}\right)\\ &= \xi_{t*}\left(\mathbf{A} + \nabla^{\mathbf{g}}_{\mathbf{W}}\mathbf{W} + 2\nabla^{\mathbf{g}}_{\mathbf{V}}\mathbf{W} + [\mathbf{W},\mathbf{V}]\right).\end{aligned} \tag{2.18}$$

It is assumed that body forces are transformed such that [Marsden and Hughes, 1983] $\mathbf{B}' - \mathbf{A}' = \xi_{t*}(\mathbf{B} - \mathbf{A})$. Similar transformations hold for the spatial quantities. It can be shown that [Marsden and Hughes, 1983, Yavari et al., 2006]

$$\begin{aligned}&\operatorname{div}'\boldsymbol{\sigma}' + \rho'\mathbf{b}' - \rho'\mathbf{a}' = \xi_{t*}\left(\operatorname{div}\boldsymbol{\sigma} + \rho\mathbf{b} - \rho\mathbf{a}\right),\\ &\operatorname{Div}'\mathbf{P}' + \rho_0'\mathbf{B}' - \rho_0'\mathbf{A}' = \xi_{t*}\left(\operatorname{Div}\mathbf{P} + \rho_0\mathbf{B} - \rho_0\mathbf{A}\right),\end{aligned} \tag{2.19}$$

i.e., the balance of linear momentum is spatially covariant provided that the Doyle-Ericksen formula $(2.12)_1$ is satisfied. This in turn restricts the body to be isotropic. A way out to have a covariant elasticity theory for anisotropic bodies is to include structural tensors in the energy function. This was discussed earlier. Note that the balance of angular momentum (symmetry of the Cauchy stress) is covariant as well, i.e., $\boldsymbol{\sigma}'^{\mathsf{T}} = \boldsymbol{\sigma}'$.

[6]See [Spencer, 1971] for a detailed discussion on integrity basis for a finite set of tensors.

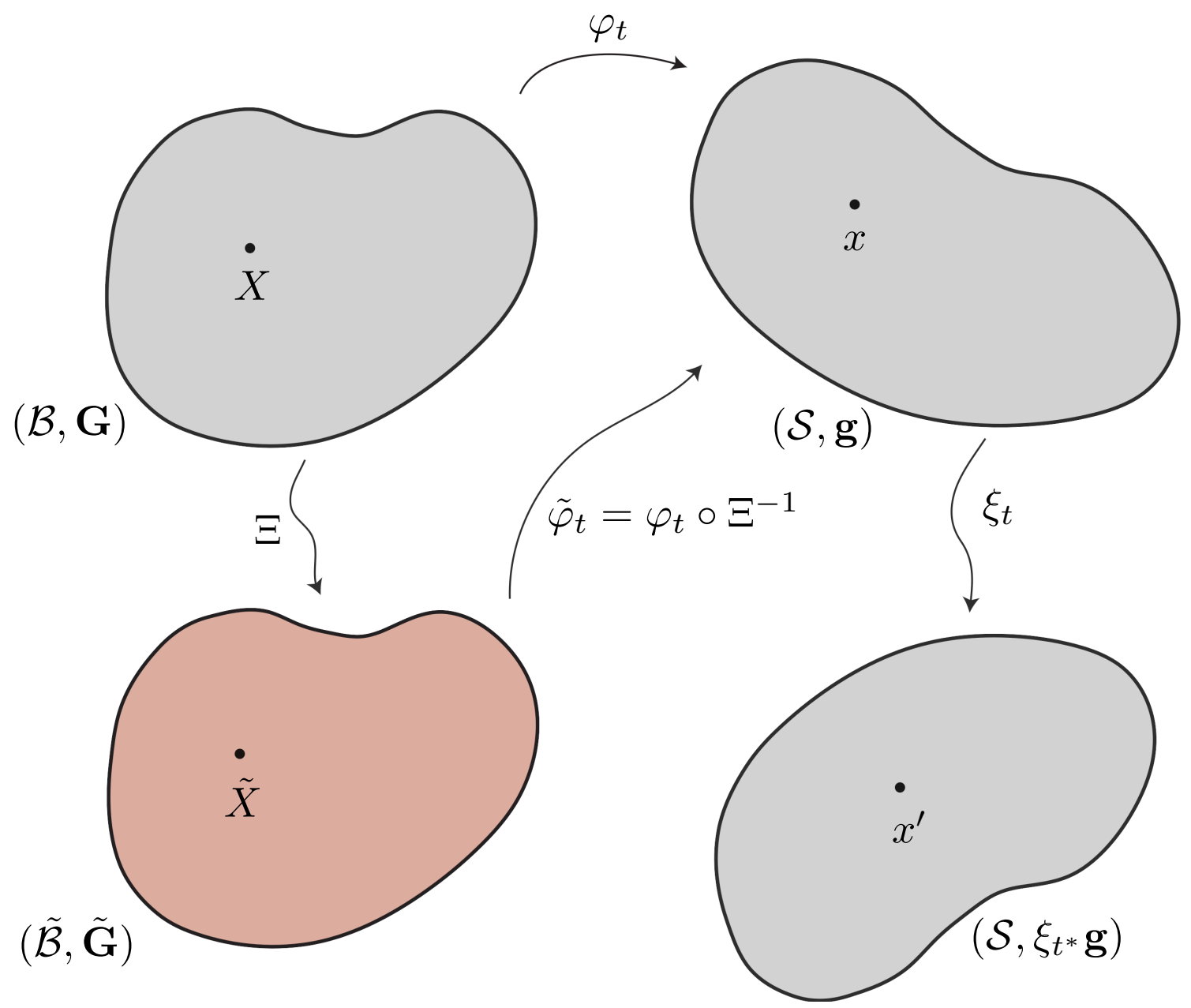

Figure 3: *Motion of a nonlinear elastic body and spatial and material changes of frame. $\xi_t$ is a time-dependent spatial change of frame and $\Xi$ is a time-independent material (referential) change of frame.*

The main idea in covariant elasticity is that in the ambient space the physical laws (here, the balance of energy or the first law of thermodynamics) should be observer-independent.[7] Yavari et al. [2006] investigated the possibility of the covariance of the energy balance under diffeomorphisms of the reference configuration. Their motivation was to see if there was any connection between material covariance and balance of the so-called configurational forces. It was observed that the energy balance is not invariant under material diffeomorphisms, in general. They obtained a transformation equation for the balance of energy. This is discussed next.

## 2.2 Material Covariance of the Governing Equations of Nonlinear Elasticity

Note that a spatial diffeomorphism is nothing but a change of observer. In other words, given an elastic body in (dynamic) equilibrium, a spatial diffeomorphism is simply representing the same configuration in another frame. For cloaking applications, one needs to know what the elastic properties of the cloak should be in order to make a given cavity invisible to elastic waves. This means that given a reference configuration, one should be looking at material (referential) diffeomorphisms. This is the motivation for the following discussion.

In this section, we discuss the transformation of the governing equations of nonlinear elasticity for an anisotropic and inhomogenous body under a time-independent material diffeomorphism. More specifically, consider a diffeomorphism $\Xi : \mathcal{B} \to \tilde{\mathcal{B}}$. We use the coordinate charts $\{X^A\}$, $\{\tilde{X}^{\tilde{A}}\}$, and $\{x^a\}$ for the reference configuration, the transformed reference configuration, and the ambient space, respectively, see Fig 3. The deformation map $\varphi_t : \mathcal{B} \to \mathcal{S}$ is transformed under the material diffeomorphism to $\tilde{\varphi}_t : \mathcal{B} \to \mathcal{S}$, where $\tilde{\varphi}_t = \varphi_t \circ \Xi^{-1}$. Material velocity with respect to the new reference configuration reads

$$\tilde{\mathbf{V}}(\tilde{X}, t) = \frac{\partial}{\partial t}\tilde{\varphi}(\tilde{X}, t) = \frac{\partial}{\partial t}\varphi(\Xi^{-1}(\tilde{X}), t) = \mathbf{V}(\Xi^{-1}(\tilde{X}), t). \tag{2.20}$$

[7] This is also known as the principle of material objectivity (see, e.g., [Noll, 1958]).

Thus, $\tilde{\mathbf{V}} = \mathbf{V} \circ \Xi^{-1}$. Material acceleration with respect to the new reference configuration reads

$$\tilde{\mathbf{A}}(\tilde{X}, t) = \nabla^{\mathbf{g}}_{\frac{\partial}{\partial t}} \tilde{\mathbf{V}}(\tilde{X}, t) = \nabla^{\mathbf{g}}_{\frac{\partial}{\partial t}} \mathbf{V}(\Xi^{-1}(\tilde{X}), t) = \nabla^{\mathbf{g}}_{\frac{\partial}{\partial t}} \mathbf{V}_t \circ \Xi^{-1}(\tilde{X}) = \mathbf{A}_t \circ \Xi^{-1}(\tilde{X}). \tag{2.21}$$

The deformation gradient $\mathbf{F} = T\varphi_t : T_X\mathcal{B} \to T_{\varphi_t(X)}\mathcal{S}$, under $\Xi$ is transformed to $\tilde{\mathbf{F}} = T\tilde{\varphi}_t : T_{\tilde{X}}\mathcal{B} \to T_{\tilde{\varphi}_t(\tilde{X})}\mathcal{S} = T_{\varphi_t(X)}\mathcal{S}$, where

$$\tilde{\mathbf{F}} = T\tilde{\varphi}_t = T(\varphi_t \circ \Xi^{-1}) = T\varphi_t \circ T\Xi^{-1} = \mathbf{F} \circ \bar{\bar{\bar{\mathbf{F}}}}^{-1}, \tag{2.22}$$

and $\bar{\bar{\bar{\mathbf{F}}}} = T\Xi$. Moreover, note that $\tilde{\mathbf{G}} = \Xi_* \mathbf{G}$. In coordinates, $\tilde{G}_{\tilde{A}\tilde{B}} = (\bar{\bar{\bar{F}}}^{-1})^A{}_{\tilde{A}} (\bar{\bar{\bar{F}}}^{-1})^B{}_{\tilde{B}} G_{AB}$. The right Cauchy-Green deformation tensor is written as $\mathbf{C}^\flat = {\varphi_t}^* \mathbf{g}$. Hence, one notes that

$$\tilde{\mathbf{C}}^\flat = {\tilde{\varphi}_t}^* \mathbf{g} = \left(\varphi_t \circ \Xi^{-1}\right)^* \mathbf{g} = \Xi_* \circ {\varphi_t}^* \mathbf{g} = \Xi_* \left({\varphi_t}^* \mathbf{g}\right) = \Xi_* \mathbf{C}^\flat . \tag{2.23}$$

In coordinates, $\tilde{C}_{\tilde{A}\tilde{B}} = (\bar{\bar{\bar{F}}}^{-1})^A{}_{\tilde{A}} (\bar{\bar{\bar{F}}}^{-1})^B{}_{\tilde{B}} C_{AB}$.

**Material symmetry.** The material symmetry group $\mathcal{G}_X$ associated with an elastic body made of a simple material[8] with the response function $\mathscr{R}$[9] at a point $X$ with respect to the reference configuration $(\mathcal{B}, \mathbf{G})$ is defined as

$$\mathscr{R}\left(\mathbf{FK}\right) = \mathscr{R}\left(\mathbf{F}\right) , \qquad \forall\, \mathbf{K} \in \mathcal{G}_X , \tag{2.24}$$

for all deformation gradients $\mathbf{F}$, where $\mathbf{K} : T_X\mathcal{B} \to T_X\mathcal{B}$ is an invertible linear transformation. For a hyperelastic solid, objectivity requires that the energy function depend on the deformation through the right Cauchy-Green deformation tensor $\mathbf{C}^\flat$, i.e., $W = W(X, \mathbf{C}^\flat, \mathbf{G})$ at a referential point $X$. Therefore, material symmetry group $\mathcal{G}_X$ for a hyperelastic solid is defined to be the subgroup of $\mathbf{G}$-orthogonal transformations $\mathrm{Orth}(\mathbf{G})$ such that[10] [Ehret and Itskov, 2009]

$$W(X, \mathbf{Q}^{-\star}\mathbf{C}^\flat\mathbf{Q}^{-1}, \mathbf{G}) = W(X, \mathbf{C}^\flat, \mathbf{G}) , \qquad \forall\, \mathbf{Q} \in \mathcal{G}_X \leqslant \mathrm{Orth}\left(\mathbf{G}\right) . \tag{2.25}$$

The symmetry group of the material relative to a transformed reference configuration $(\tilde{\mathcal{B}}, \tilde{\mathbf{G}})$ is denoted by $\widetilde{\mathcal{G}}$. According to Noll's rule [Noll, 1958, Coleman and Noll, 1959, 1963, 1964], one can write

$$\widetilde{\mathcal{G}} = \Xi_* \mathcal{G} = \bar{\bar{\bar{\mathbf{F}}}}\, \mathcal{G}\, \bar{\bar{\bar{\mathbf{F}}}}^{-1} . \tag{2.26}$$

In other words, in the sense of group theory, at each material point $X$, $\mathcal{G}$ and $\widetilde{\mathcal{G}}$ are conjugate subgroups of the general linear group, and hence, isomorphic ($\widetilde{\mathcal{G}} \cong \mathcal{G}$). Note that if $\bar{\bar{\bar{\mathbf{F}}}} \in \mathcal{G}$, then $\mathcal{G} = \widetilde{\mathcal{G}}$. Also, the symmetry group is not affected by a change of reference configuration ($\widetilde{\mathcal{G}} = \mathcal{G}$) if $\bar{\bar{\bar{\mathbf{F}}}} = \alpha \mathbf{I}$ (pure dilatation) for some positive scalar $\alpha$ [Ogden, 1997]. More generally, $\widetilde{\mathcal{G}} = \mathcal{G}$ if and only if $\bar{\bar{\bar{\mathbf{F}}}}$ belongs to the normalizer group[11] of $\mathcal{G}$ within the general linear group. It is straightforward to see that (2.26) is satisfied if and only if it holds for all the generators of the group $\mathcal{G}$. Therefore, if $\mathcal{G}$ is finitely generated, the elements of the generating sets of $\mathcal{G}$ and $\widetilde{\mathcal{G}}$ denoted by $\{\mathbf{Q}_1, \ldots, \mathbf{Q}_m\}$ and $\{\tilde{\mathbf{Q}}_1, \ldots, \tilde{\mathbf{Q}}_m\}$, respectively, are related as $\tilde{\mathbf{Q}}_j = \bar{\bar{\bar{\mathbf{F}}}}\mathbf{Q}_j\bar{\bar{\bar{\mathbf{F}}}}^{-1}$, $j = 1, \ldots, m$. The symmetry group can be characterized using a finite collection of structural tensors[12] $\boldsymbol{\zeta}_i$ of order $\mu_i$, $i = 1, \ldots, n$, as follows [Liu, 1982, Boehler, 1987, Zheng and Spencer, 1993a, Zheng, 1994, Lu and Papadopoulos, 2000, Mazzucato and Rachele, 2006]

$$\mathbf{Q} \in \mathcal{G} \leqslant \mathrm{Orth}\left(\mathbf{G}\right) \iff \langle \mathbf{Q} \rangle_{\mu_1} \boldsymbol{\zeta}_1 = \boldsymbol{\zeta}_1 , \ldots , \langle \mathbf{Q} \rangle_{\mu_n} \boldsymbol{\zeta}_n = \boldsymbol{\zeta}_n , \tag{2.27}$$

[8] The response of a *simple* material at any material point depends only on the first deformation gradient (and its evolution) at that point [Noll, 1958].

[9] Here we assume that $\mathscr{R}$ is the energy function. Response function may be any measure of stress as well.

[10] Note that $\mathrm{Orth}(\mathbf{G}) = \left\{\mathbf{Q} : T_X\mathcal{B} \to T_X\mathcal{B} \;\mid\; \mathbf{Q}^\top = \mathbf{Q}^{-1}\right\}$. We use the notation $\mathscr{G} \leqslant \mathscr{H}$ when $\mathscr{G}$ is a subgroup of $\mathscr{H}$.

[11] The normalizer group $\mathcal{N}_{\mathscr{G}}(\mathcal{Q})$ of a subgroup $\mathcal{Q}$ of $\mathscr{G}$ ($\mathcal{Q} \leqslant \mathscr{G}$) is defined as $\mathcal{N}_{\mathscr{G}}(\mathcal{Q}) = \{g_i \in \mathscr{G} : g_i \mathcal{Q} g_i^{-1} = \mathcal{Q}\}$.

[12] Note that such a collection forms a basis for the space of tensors that are invariant under the action of $\mathcal{G}$.

where the $\mu$-th power Kronecker product $\langle \mathbf{Q} \rangle_\mu$ of a $\mathbf{G}$-orthogonal transformation $\mathbf{Q}$ for any $\mu$-th order tensor $\boldsymbol{\zeta}$ is defined as[13]

$$(\langle \mathbf{Q} \rangle_\mu \zeta)^{\bar{A}_1 \ldots \bar{A}_\mu} = Q^{\bar{A}_1}{}_{A_1} \ldots Q^{\bar{A}_\mu}{}_{A_\mu} \zeta^{A_1 \ldots A_\mu} . \tag{2.28}$$

Note that (2.27) suggests that the material symmetry group $\mathcal{G}$ is the invariance group of the set of the structural tensors $\boldsymbol{\zeta}_i$, $i = 1, \ldots, n$. Using (2.26) and (2.27), one obtains the following relation for the transformed structural tensors under the material diffeomorphism, which characterize the transformed symmetry group $\widetilde{\mathcal{G}}$

$$\tilde{\mathbf{Q}} \in \widetilde{\mathcal{G}} \leqslant \mathrm{Orth}(\tilde{\mathbf{G}}) \iff \langle \tilde{\mathbf{Q}} \rangle_{\mu_1} \tilde{\boldsymbol{\zeta}}_1 = \tilde{\boldsymbol{\zeta}}_1 , \ldots , \langle \tilde{\mathbf{Q}} \rangle_{\mu_n} \tilde{\boldsymbol{\zeta}}_n = \tilde{\boldsymbol{\zeta}}_n , \tag{2.29}$$

where $\tilde{\mathbf{Q}} = \Xi_* \mathbf{Q} = \bar{\bar{\mathbf{F}}} \, \mathbf{Q} \, \bar{\bar{\mathbf{F}}}^{-1}$ and $\tilde{\boldsymbol{\zeta}}_i = \Xi_* \boldsymbol{\zeta}_i$, $i = 1, \ldots, n$. Therefore, the type of the symmetry group of the material is preserved under a material change of frame $\Xi$.

**Balance of energy.** Yavari et al. [2006] showed that the balance of energy is not invariant under an arbitrary time-dependent material diffeomorphism. However, it can be shown that the balance of energy is always invariant under time-independent material diffeomorphisms. The strain energy function satisfies (2.25) if and only if it is represented as an isotropic function (invariant under the special orthogonal group) of the structural tensors and $\mathbf{C}^\flat$ at a material point $X$ (see (2.30)) [Boehler, 1979, Zhang and Rychlewski, 1990, Ehret and Itskov, 2009]. Thus, we write the general form of the energy function (per unit undeformed volume) of an inhomogeneous anisotropic hyperelastic material with a set of structural tensors $\boldsymbol{\zeta}_i$, $i = 1, \ldots, n$ (cf. (2.27)) characterizing the material symmetry group (at a referential point $X$) as

$$W = \hat{W}(X, \mathbf{C}^\flat, \mathbf{G}, \boldsymbol{\zeta}_1, \ldots, \boldsymbol{\zeta}_n) . \tag{2.30}$$

Similarly, using (2.23) and (2.29), one obtains

$$\tilde{W} = \widetilde{\hat{W}}(\tilde{X}, \tilde{\mathbf{C}}^\flat, \tilde{\mathbf{G}}, \tilde{\boldsymbol{\zeta}}_1, \ldots, \tilde{\boldsymbol{\zeta}}_n) = \hat{W}(\Xi(X), \Xi_* \mathbf{C}^\flat, \Xi_* \mathbf{G}, \Xi_* \boldsymbol{\zeta}_1, \ldots, \Xi_* \boldsymbol{\zeta}_n) = \Xi_* W . \tag{2.31}$$

Using the theory of invariants [Spencer, 1971, 1982], the energy function can be represented in terms of a finite set of isotropic invariants. It can be shown that these invariants do not change under material diffeomorphisms, which in turn implies the material covariance of the energy function, i.e., (see also [Lu and Papadopoulos, 2000])

$$\tilde{W} = W \circ \Xi^{-1} . \tag{2.32}$$

**Balance of linear momentum.** Balance of linear momentum in terms of the first Piola-Kirchhoff stress reads $\operatorname{Div} \mathbf{P} + \rho_0 \mathbf{B} = \rho_0 \mathbf{A}$. We now examine the transformed first and second Piola-Kirchhoff stress tensors, denoted, respectively, by $\tilde{\mathbf{P}}$ and $\tilde{\mathbf{S}}$, and the material acceleration $\tilde{\mathbf{A}}$ pertaining to the transformed configuration $(\tilde{\mathcal{B}}, \tilde{\mathbf{G}})$ with the deformation $\tilde{\varphi}_t = \varphi_t \circ \Xi^{-1}$. The second Piola-Kirchhoff stress tensor is written as $\mathbf{S} = 2 \frac{\partial \hat{W}}{\partial \mathbf{C}^\flat}$. Therefore, using (2.31), one obtains

$$\tilde{\mathbf{S}} = 2 \frac{\partial \widetilde{\hat{W}}}{\partial \tilde{\mathbf{C}}^\flat} = 2 \frac{\partial (\Xi_* \hat{W})}{\partial (\Xi_* \mathbf{C}^\flat)} . \tag{2.33}$$

Moreover, employing (2.32) and (2.33), one can write

$$\begin{aligned} (\Xi_* S)^{\tilde{A}\tilde{B}} &= \bar{\bar{F}}^{\tilde{A}}{}_A \bar{\bar{F}}^{\tilde{B}}{}_B \, S^{AB} \circ \Xi^{-1} = 2 \bar{\bar{F}}^{\tilde{A}}{}_A \bar{\bar{F}}^{\tilde{B}}{}_B \, \frac{\partial \hat{W}}{\partial C_{AB}} \circ \Xi^{-1} \\ &= 2 \bar{\bar{F}}^{\tilde{A}}{}_A \bar{\bar{F}}^{\tilde{B}}{}_B \frac{\partial \widetilde{\hat{W}}}{\partial \tilde{C}_{\tilde{D}\tilde{H}}} \frac{\partial \tilde{C}_{\tilde{D}\tilde{H}} \circ \Xi}{\partial C_{AB}} \\ &= 2 \bar{\bar{F}}^{\tilde{A}}{}_A \bar{\bar{F}}^{\tilde{B}}{}_B (\bar{\bar{F}}^{-1})^A{}_{\tilde{D}} (\bar{\bar{F}}^{-1})^B{}_{\tilde{H}} \, \frac{\partial \widetilde{\hat{W}}}{\partial \tilde{C}_{\tilde{D}\tilde{H}}} = 2 \frac{\partial \widetilde{\hat{W}}}{\partial \tilde{C}_{\tilde{A}\tilde{B}}} = \tilde{S}^{\tilde{A}\tilde{B}} . \end{aligned} \tag{2.34}$$

[13]Note that $\langle \mathbf{Q} \rangle_m (\mathbf{v}_1 \otimes \ldots \otimes \mathbf{v}_m) = \mathbf{Q}\mathbf{v}_1 \otimes \ldots \otimes \mathbf{Q}\mathbf{v}_m$, where $\mathbf{v}_i \in T_X \mathcal{B}$, $i = 1, \ldots, m$, are arbitrary vectors.

Hence, it immediately follows that $\tilde{\mathbf{S}} = \Xi_* \mathbf{S}$. Under the diffeomorphism $\Xi : \mathcal{B} \to \tilde{\mathcal{B}}$, the two-point tensor $\mathbf{P}$ is transformed to $\Xi_* \mathbf{P}$, where in components

$$\tilde{P}^{a\tilde{A}} = \frac{\partial \tilde{X}^{\tilde{A}}}{\partial X^A} P^{aA} = \bar{\bar{F}}^{\tilde{A}}{}_A P^{aA} \circ \Xi^{-1}. \tag{2.35}$$

We note that, using (2.22), one has $\tilde{\mathbf{P}} = \tilde{\mathbf{F}}\tilde{\mathbf{S}} = (\mathbf{F} \circ \bar{\bar{\mathbf{F}}}^{-1})\tilde{\mathbf{S}}$. Therefore, in components (cf. (2.35) and $\tilde{\mathbf{S}} = \Xi_* \mathbf{S}$)

$$\begin{aligned} \tilde{P}^{a\tilde{A}} =& F^a{}_A (\bar{\bar{F}}^{-1})^A{}_{\tilde{B}} \, \tilde{S}^{\tilde{B}\tilde{A}} = F^a{}_A (\bar{\bar{F}}^{-1})^A{}_{\tilde{B}} \bar{\bar{F}}^{\tilde{B}}{}_C \bar{\bar{F}}^{\tilde{A}}{}_B \, S^{CB} \circ \Xi^{-1} \\ =& F^a{}_A \bar{\bar{F}}^{\tilde{A}}{}_B \, S^{AB} \circ \Xi^{-1} = \bar{\bar{F}}^{\tilde{A}}{}_B P^{aB} \circ \Xi^{-1} = (\Xi_* P)^{a\tilde{A}} \,, \end{aligned} \tag{2.36}$$

that is, $\tilde{\mathbf{P}} = \Xi_* \mathbf{P}$. Note that $(\operatorname{Div} \mathbf{P})^a = \partial P^{aA}/\partial X^A + \Gamma^A{}_{AB} P^{aB} + (\gamma^a{}_{bc} \circ \varphi_t) \, F^b{}_A P^{cA}$. Thus, $(\widetilde{\operatorname{Div}} \tilde{\mathbf{P}})^a = \partial \tilde{P}^{a\tilde{A}}/\partial \tilde{X}^{\tilde{A}} + \tilde{\Gamma}^{\tilde{A}}{}_{\tilde{A}\tilde{B}} \tilde{P}^{a\tilde{B}} + (\gamma^a{}_{bc} \circ \tilde{\varphi}_t) \, \tilde{F}^b{}_{\tilde{A}} \tilde{P}^{c\tilde{A}}$. Note that $\tilde{\mathbf{F}} = T\tilde{\varphi}_t = T\left(\varphi_t \circ \Xi^{-1}\right) = T\varphi_t \circ (T\Xi)^{-1} = \mathbf{F} \circ \bar{\bar{\mathbf{F}}}^{-1}$. Thus, $\tilde{\mathbf{F}} \circ \tilde{\mathbf{P}} = \mathbf{F} \circ \mathbf{P} \circ \Xi^{-1}$. It can be shown that $\partial \tilde{P}^{a\tilde{A}}/\partial \tilde{X}^{\tilde{A}} = \partial P^{aA}/\partial X^A \circ \Xi^{-1}$. Using the transformation of connection coefficients, it is straightforward to show that $\tilde{\Gamma}^{\tilde{A}}{}_{\tilde{A}\tilde{B}} = (\bar{\bar{F}}^{-1})^B{}_{\tilde{B}} \Gamma^A{}_{AB} \circ \Xi^{-1}$. Therefore, $\widetilde{\operatorname{Div}} \tilde{\mathbf{P}} = \operatorname{Div} \mathbf{P} \circ \Xi^{-1}$. Note that $\rho_0 dV = \tilde{\rho}_0 d\tilde{V} = \tilde{\rho}_0 J_\Xi dV$, where

$$J_\Xi = \sqrt{\frac{\det \tilde{\mathbf{G}}}{\det \mathbf{G}}} \det \bar{\bar{\mathbf{F}}}. \tag{2.37}$$

Note also that $\tilde{\mathbf{G}} = \Xi_* \mathbf{G}$ and $\det \tilde{\mathbf{G}} = \det \mathbf{G} \, (\det \bar{\bar{\mathbf{F}}})^{-2}$. Hence, $J_\Xi = 1$, and therefore, $\tilde{\rho}_0 = \rho_0 \circ \Xi^{-1}$. For the spatial mass density, $\tilde{\rho} = \rho$.

Body force is a vector field in the ambient space, i.e., $\mathbf{B}_X \in T_{\varphi_t(X)} \mathcal{S}$, and hence, $\tilde{\mathbf{B}} = \Xi_* \mathbf{B} = \mathbf{B} \circ \Xi^{-1}$. For a time-independent material diffeomprphism, $\tilde{\mathbf{V}} = \mathbf{V} \circ \Xi^{-1}$, and hence, acceleration transforms as $\tilde{\mathbf{A}} = \mathbf{A} \circ \Xi^{-1}$. Therefore, the balance of linear momentum is invariant under the diffeomorphism $\Xi : \mathcal{B} \to \tilde{\mathcal{B}}$, i.e.,

$$\widetilde{\operatorname{Div}} \tilde{\mathbf{P}} + \tilde{\rho}_0 \tilde{\mathbf{B}} - \tilde{\rho}_0 \tilde{\mathbf{A}} = \Xi_* \left( \operatorname{Div} \mathbf{P} + \rho_0 \mathbf{B} - \rho_0 \mathbf{A} \right) = \left( \operatorname{Div} \mathbf{P} + \rho_0 \mathbf{B} - \rho_0 \mathbf{A} \right) \circ \Xi^{-1}. \tag{2.38}$$

This is identical to what Mazzucato and Rachele [2006] proved.

**Balance of angular momentum.** Balance of angular momentum in terms of the first Piola-Kirchhoff stress reads $P^{aA} F^b{}_A - P^{bA} F^a{}_A = 0$. It is straightforward to show that $\tilde{P}^{a\tilde{A}} \tilde{F}^b{}_{\tilde{A}} = P^{aA} F^b{}_A \circ \Xi^{-1}$. Therefore, $\tilde{P}^{a\tilde{A}} \tilde{F}^b{}_{\tilde{A}} - \tilde{P}^{b\tilde{A}} \tilde{F}^a{}_{\tilde{A}} = \left( P^{aA} F^b{}_A - P^{bA} F^a{}_A \right) \circ \Xi^{-1}$, i.e., the balance of angular momentum is invariant under material diffeomorphisms.

**Conservation of mass.** Conservation of mass in local form reads $\rho_0 - J\rho \circ \varphi = 0$, where the Jacobian is written as $J = \sqrt{\frac{\det \mathbf{g}}{\det \mathbf{G}}} \det \mathbf{F}$. Thus

$$\tilde{J} = \sqrt{\frac{\det \tilde{\mathbf{g}}}{\det \tilde{\mathbf{G}}}} \det \tilde{\mathbf{F}} = J \circ \Xi^{-1}. \tag{2.39}$$

Therefore, $\tilde{\rho}_0 - \tilde{J} \tilde{\rho} \circ \tilde{\varphi} = \Xi_* (\rho_0 - J\rho \circ \varphi) = (\rho_0 - J\rho \circ \varphi) \circ \Xi^{-1}$, i.e., conservation of mass is materially covariant.

We next observe that under the diffeomorphism $\Xi : \mathcal{B} \to \tilde{\mathcal{B}}$, the Cauchy stress tensor remains unchanged and is transformed as $\tilde{\boldsymbol{\sigma}} = \boldsymbol{\sigma} \circ \Xi^{-1}$. To see this, note that (cf. (2.35) and (2.39))

$$\tilde{\boldsymbol{\sigma}} = \tilde{J}^{-1} \tilde{\mathbf{F}} \tilde{\mathbf{P}}^\star = (J^{-1} \mathbf{F} \bar{\bar{\mathbf{F}}}^{-1} \bar{\bar{\mathbf{F}}} \mathbf{P}^\star) \circ \Xi^{-1} = \boldsymbol{\sigma} \circ \Xi^{-1} \,. \tag{2.40}$$

## 3 Linearized Elastodynamics

Classical linear elasticity can be derived from nonlinear elasticity if one linearizes the governing equations of nonlinear elasticity with respect to a stress-free equilibrium configuration. More generally, nonlinear elasticity can be linearized with respect to any stressed and finitely-deformed (in either static or dynamic equilibrium) configuration. This is the so-called small-on-large theory of Green et al. [1952]. In the language of geometric mechanics, Marsden and Hughes [1983] presented a geometric linearization of nonlinear elasticity. In particular, in their formulation elastic constants are properly defined in terms of two-point tensors. See also Yavari and Ozakin [2008] for a discussion on covariance in linearized elasticity.

Variation of a map $\varphi_t : \mathcal{B} \to \mathcal{S}$ is a map $\Phi_t : \mathcal{B} \times I$, where $I = (-a, a)$ is some interval, such that $\Phi_t(X, 0) = \mathring{\varphi}_t$, which we call the reference motion. Let us denote $\varphi_{t,\epsilon}(X) = \Phi_t(X, \epsilon)$, and hence, $\varphi_{t,0} = \mathring{\varphi}_t$. Variation field is defined as

$$\delta\varphi_t(X) = \mathbf{U}(X,t) = \frac{d}{d\epsilon}\Big|_{\epsilon=0}\varphi_{t,\epsilon}(X). \tag{3.1}$$

The spatial variation (displacement) field is defined as $\mathbf{u} = \mathbf{U} \circ \mathring{\varphi}_t^{-1}$ or $u^a(x,t) = U^a(X,t)$. Note that $\delta\varphi \in \Gamma(\mathring{\varphi}^{-1}T\mathcal{S})$ (see the appendix for the definition of the induced bundle). The tangent space $T_{(X,\epsilon)}(\mathcal{B} \times I)$ of the product manifold $\mathcal{B} \times I$ at $(X, \epsilon)$ is identified with $T_X\mathcal{B} \otimes T_\epsilon I$.

**Linearization of the material velocity and acceleration.** Material velocity is defined as $\mathbf{V}_\epsilon(X,t) = \frac{\partial\varphi_{t,\epsilon}(X)}{\partial t}$. Note that $\mathbf{V}_\epsilon(X,t) \in T_{\varphi_{t,\epsilon}(X)}\mathcal{S}$, i.e., for different values of $\epsilon$, velocity lies in different tangent spaces, and hence, a covariant derivative along the curve $\epsilon \mapsto \varphi_{t,\epsilon}(X)$ should be used to find the linearization of velocity [Marsden and Hughes, 1983]. Therefore

$$\delta\mathbf{V}(X,t) = \nabla_{\frac{\partial}{\partial\epsilon}}\frac{\partial\varphi_{t,\epsilon}(X)}{\partial t}\Big|_{\epsilon=0} = \nabla_{\frac{\partial}{\partial t}}\frac{\partial\varphi_{t,\epsilon}(X)}{\partial \epsilon}\Big|_{\epsilon=0} = \nabla_t\delta\varphi_t(X) = D_t\mathbf{U}(X,t) = \nabla_{\mathring{\mathbf{V}}}\mathbf{U} =: \dot{\mathbf{U}}, \tag{3.2}$$

i.e., variation of the velocity field is the covariant time derivative of the displacement field. In the above calculation in the second equality the symmetry lemma of Riemannian geometry [Lee, 1997] was used. In components, $(D_t\mathbf{U}(X,t))^a = \dot{U}^a = \frac{\partial U^a}{\partial t} + \gamma^a{}_{bc}\mathring{V}^bU^c$. Note that $\frac{\partial U^a}{\partial t} = \frac{\partial u^a}{\partial t} + \frac{\partial u^a}{\partial x^b}\frac{\partial x^b}{\partial t}$. Therefore, $\delta\mathbf{v} = \frac{\partial\mathbf{u}}{\partial t} + \nabla^{\mathbf{g}}_{\mathring{\mathbf{v}}}\mathbf{u} =: \dot{\mathbf{u}}$.

Material acceleration is the covariant time derivative of velocity, i.e., $\mathbf{A} = D_t\mathbf{V} = \nabla_{\mathbf{V}}\mathbf{V}$, which in coordinates reads [Marsden and Hughes, 1983] $A^a = \frac{\partial V^a}{\partial t} + \gamma^a{}_{bc}V^bV^c$. Therefore

$$\begin{aligned}
\delta\mathbf{A}(X,t) &= \nabla_{\frac{\partial}{\partial\epsilon}}\nabla_{\frac{\partial}{\partial t}}\frac{\partial\varphi_{t,\epsilon}(X)}{\partial t}\Big|_{\epsilon=0}\\
&= \nabla_{\frac{\partial}{\partial t}}\nabla_{\frac{\partial}{\partial \epsilon}}\frac{\partial\varphi_{t,\epsilon}(X)}{\partial t}\Big|_{\epsilon=0} + \nabla_{[\frac{\partial}{\partial\epsilon},\frac{\partial}{\partial t}]}\frac{\partial\varphi_{t,\epsilon}(X)}{\partial t}\Big|_{\epsilon=0} + \boldsymbol{\mathcal{R}}_{\mathbf{g}}\left(\frac{\partial}{\partial\epsilon},\frac{\partial}{\partial t}\right)\frac{\partial\varphi_{t,\epsilon}(X)}{\partial t}\Big|_{\epsilon=0}\\
&= \nabla_{\frac{\partial}{\partial t}}\nabla_{\frac{\partial}{\partial t}}\frac{\partial\varphi_{t,\epsilon}(X)}{\partial \epsilon}\Big|_{\epsilon=0} + \nabla_{[\mathbf{U},\mathring{\mathbf{V}}]}\mathring{\mathbf{V}} + \boldsymbol{\mathcal{R}}_{\mathbf{g}}\left(\mathbf{U},\mathring{\mathbf{V}},\mathring{\mathbf{V}}\right)\\
&= D_tD_t\mathbf{U} + \nabla_{[\mathbf{U},\mathring{\mathbf{V}}]}\mathring{\mathbf{V}} + \boldsymbol{\mathcal{R}}_{\mathbf{g}}(\mathbf{U},\mathring{\mathbf{V}},\mathring{\mathbf{V}})\\
&= \nabla_{\mathring{\mathbf{V}}}\nabla_{\mathring{\mathbf{V}}}\mathbf{U} + \nabla_{[\mathbf{U},\mathring{\mathbf{V}}]}\mathring{\mathbf{V}} + \boldsymbol{\mathcal{R}}_{\mathbf{g}}(\mathbf{U},\mathring{\mathbf{V}},\mathring{\mathbf{V}})\\
&= \ddot{\mathbf{U}} + \nabla_{[\mathbf{U},\mathring{\mathbf{V}}]}\mathring{\mathbf{V}} + \boldsymbol{\mathcal{R}}_{\mathbf{g}}(\mathbf{U},\mathring{\mathbf{V}},\mathring{\mathbf{V}}),
\end{aligned} \tag{3.3}$$

where $\boldsymbol{\mathcal{R}}_{\mathbf{g}}$ is the curvature tensor of the metric $\mathbf{g}$ and $\ddot{\mathbf{U}} = D_tD_t\mathbf{U}$ is the second covariant time derivative of the displacement field. Note that $\delta\mathbf{a}_t = \delta\mathbf{A}_t \circ \mathring{\varphi}_t^{-1} = \ddot{\mathbf{U}} \circ \mathring{\varphi}_t^{-1} + \nabla^{\mathbf{g}}_{[\mathbf{u},\mathring{\mathbf{v}}]}\mathring{\mathbf{v}} + \boldsymbol{\mathcal{R}}_{\mathbf{g}}\left(\mathbf{u},\mathring{\mathbf{v}},\mathring{\mathbf{v}}\right)$.

**Linearization of the right Cauchy-Green strain, the Jacobian, and the deformation gradient.** The right Cauchy-Green strain of the motion $\varphi_{t,\epsilon}$ is defined as $\mathbf{C}^\flat_\epsilon = \varphi^*_{t,\epsilon}\mathbf{g} \circ \varphi_{t,\epsilon}$. Note that $\mathbf{C}^\flat_\epsilon \in \Gamma(\mathcal{B}, T^*\mathcal{B} \otimes T^*\mathcal{B})$ for all $\epsilon$, where $\Gamma(\mathcal{B}, T^*\mathcal{B} \otimes T^*\mathcal{B})$ is the set of $\binom{0}{2}$-tensors on $\mathcal{B}$. Linearization of $\mathbf{C}^\flat$ is calculated as

$$\delta\mathbf{C}^\flat = \frac{d}{d\epsilon}\mathbf{C}^\flat_\epsilon\Big|_{\epsilon=0} = \frac{d}{d\epsilon}\varphi^*_{t,\epsilon}\mathbf{g}\Big|_{\epsilon=0} = \mathring{\varphi}^*_t\left(\mathbf{L}_{\mathbf{u}}\mathbf{g}\right) = \mathring{\varphi}^*_t\left(\mathfrak{L}_{\mathbf{u}}\mathbf{g}\right) = \mathring{\varphi}^*_t\left(\nabla^{\mathbf{g}}\mathbf{u}^\flat + (\nabla^{\mathbf{g}}\mathbf{u}^\flat)^\star\right) = 2\mathring{\varphi}^*_t\boldsymbol{\epsilon}, \tag{3.4}$$

where $\mathbf{L_u}$ is Lie derivative with respect to the vector field $\mathbf{u}$ and $\mathfrak{L}_\mathbf{u}$ is the autonomous Lie derivative with respect to $\mathbf{u}$ (see the appendix). The linearized strain $\boldsymbol{\epsilon}$ has the components $2\epsilon_{ab} = u_{a|b} + u_{b|a}$.

The Jacobian for the perturbed motion is written as

$$J_\epsilon = \sqrt{\frac{\det \mathbf{g}\circ\varphi_{t,\epsilon}}{\det \mathbf{G}}}\det \mathbf{F}_\epsilon = \sqrt{\frac{\det \mathbf{C}^\flat_\epsilon}{\det \mathbf{G}}}. \tag{3.5}$$

Hence

$$\delta J = \frac{d}{d\epsilon}J_\epsilon\Big|_{\epsilon=0} = \frac{1}{2\sqrt{\det \mathbf{G}\det \mathring{\mathbf{C}}^\flat}}\frac{d}{d\epsilon}\det \mathbf{C}^\flat_\epsilon\Big|_{\epsilon=0}. \tag{3.6}$$

Using Jacobi's formula we know that

$$\frac{d}{d\epsilon}\det \mathbf{C}^\flat_\epsilon\Big|_{\epsilon=0} = \det \mathring{\mathbf{C}}^\flat \operatorname{tr}\left[(\mathring{\mathbf{C}}^\flat)^{-1} 2\mathring{\varphi}_t^*\boldsymbol{\epsilon}\right]. \tag{3.7}$$

Note that $(\mathring{\mathbf{C}}^\flat)^{-1} = \mathring{\varphi}_t^*\mathring{\mathbf{g}}^\sharp$, and hence, the right-hand side of (3.7) is simplified to read $(\det \mathring{\mathbf{C}}^\flat)\mathbf{g}^\sharp : \boldsymbol{\epsilon}$. Therefore, $\delta J = \mathring{J}\mathbf{g}^\sharp : \boldsymbol{\epsilon} = \mathring{J}\operatorname{div}_\mathbf{g}\mathbf{u}$.

Deformation gradient of the map $\varphi_{t,\epsilon}$ is defined as $\mathbf{F}_{t,\epsilon} = \frac{\partial\varphi_{t,\epsilon}}{\partial X}$. Consider the vector fields $\left(\frac{\partial}{\partial X^A}, 0\right)$ and $\left(0, \frac{\partial}{\partial\epsilon}\right)$ on $\mathcal{B}\times I$, and note that $\left[\left(\frac{\partial}{\partial X^A}, 0\right), \left(0, \frac{\partial}{\partial\epsilon}\right)\right] = 0$. Thus, from (A.8) one can write

$$\nabla_{\left(0,\frac{\partial}{\partial\epsilon}\right)}\varphi_{t,\epsilon *}\left(\frac{\partial}{\partial X^A}, 0\right) = \nabla_{\left(\frac{\partial}{\partial X^A}, 0\right)}\varphi_{t,\epsilon *}\left(0, \frac{\partial}{\partial\epsilon}\right). \tag{3.8}$$

Or

$$\nabla_{\frac{\partial}{\partial\epsilon}}\frac{\partial\varphi^a_\epsilon}{\partial X^A} = \nabla_{\frac{\partial}{\partial X^A}}\frac{\partial\varphi^a_\epsilon}{\partial\epsilon}. \tag{3.9}$$

Therefore, $\delta F^a{}_A = \varphi^a{}_{|A} = U^a{}_{|A} = F^b{}_A U^a{}_{|b}$. Note that $U^a{}_{|b} = \frac{\partial U^b}{\partial x^b} + \gamma^a{}_{bc}U^c$.

**Linearization of the first and the second Piola-Kirchhoff stresses.** For a hyperelastic solid, given an energy function $W = W(X, \mathbf{F}, \mathbf{G}, \mathbf{g}\circ\varphi, \boldsymbol{\zeta}_1, \ldots, \boldsymbol{\zeta}_n)$, the first Piola-Kirchhoff stress is given as $\mathbf{P} = \mathbf{g}^\sharp\frac{\partial W}{\partial\mathbf{F}}$. In components, $P^{aA} = g^{ab}\frac{\partial W}{\partial F^b{}_A}$. For the perturbed motion, $W_\epsilon = W(X, \mathbf{F}_\epsilon, \mathbf{G}, \mathbf{g}\circ\varphi_\epsilon, \boldsymbol{\zeta}_1, \ldots, \boldsymbol{\zeta}_n)$, and hence

$$\delta\mathbf{P} = \mathbf{g}^\sharp\nabla_{\frac{\partial}{\partial\epsilon}}\frac{\partial W_\epsilon}{\partial\mathbf{F}_\epsilon}\Big|_{\epsilon=0} = \mathbf{g}^\sharp\frac{\partial^2 W_\epsilon}{\partial\mathbf{F}_\epsilon\partial\mathbf{F}_\epsilon} : \nabla_{\frac{\partial}{\partial\epsilon}}\mathbf{F}_\epsilon\Big|_{\epsilon=0} + \mathbf{g}^\sharp\frac{\partial^2 W_\epsilon}{\partial\mathbf{g}\circ\varphi_\epsilon\partial\mathbf{F}_\epsilon} : \nabla_{\frac{\partial}{\partial\epsilon}}\mathbf{g}\circ\varphi_\epsilon\Big|_{\epsilon=0} = \mathbf{g}^\sharp\frac{\partial^2 W}{\partial\mathbf{F}\partial\mathbf{F}} : \nabla\mathbf{U}. \tag{3.10}$$

In the above derivation, in the second equality the geometric $\omega$-lemma [Marsden and Hughes, 1983], and in the last equality the metric compatibility of the Levi-Civita connection ($\nabla^\mathbf{g}\mathbf{g} = \mathbf{0}$) was used. Therefore, $\delta\mathbf{P} = \mathsf{A} : \nabla\mathbf{U}$, where $\mathsf{A}$ is the first elasticity tensor [Marsden and Hughes, 1983] with components

$$\mathsf{A}^{aA}{}_b{}^B = \frac{\partial P^{aA}}{\partial F^b{}_B} = g^{ac}\frac{\partial^2 W}{\partial F^c{}_A\partial F^b{}_B}. \tag{3.11}$$

Thus, in components $\delta P^{aA} = \mathsf{A}^{aA}{}_b{}^B U^b{}_B$.

The second Piola-Kirchhoff stress is defined as $S^{AB} = (F^{-1})^A{}_a P^{aB}$. Using the material frame-indifference (objectivity) the energy function is written as $W = \hat{W}(X, \mathbf{G}, \mathbf{C}^\flat, \boldsymbol{\zeta}_1, \ldots, \boldsymbol{\zeta}_n)$. The second Piola-Kirchhoff stress is written as $\mathbf{S} = 2\frac{\partial\hat{W}}{\partial\mathbf{C}^\flat}$. For the one-parameter family of motions $\varphi_{t,\epsilon}$, one has $\mathbf{S}_\epsilon = 2\frac{\partial\hat{W}_\epsilon}{\partial\mathbf{C}^\flat_\epsilon}$, where $\hat{W}_\epsilon = \hat{W}(X, \mathbf{G}, \mathbf{C}^\flat_\epsilon, \boldsymbol{\zeta}_1, \ldots, \boldsymbol{\zeta}_n)$. Therefore

$$\delta\mathbf{S} = 2\frac{\partial^2\hat{W}_\epsilon}{\partial\mathbf{C}^\flat_\epsilon\partial\mathbf{C}^\flat_\epsilon} : \frac{d}{d\epsilon}\mathbf{C}^\flat_\epsilon\Big|_{\epsilon=0} = 4\frac{\partial^2\hat{W}}{\partial\mathring{\mathbf{C}}^\flat\partial\mathring{\mathbf{C}}^\flat} : \mathring{\varphi}^*\boldsymbol{\epsilon} = \mathsf{C} : \mathring{\varphi}^*\boldsymbol{\epsilon}, \tag{3.12}$$

where $\mathsf{C}$ is the second elasticity tensor with components

$$\mathsf{C}^{ABCD} = 4\frac{\partial^2\hat{W}}{\partial C_{AB}\partial C_{CD}}. \tag{3.13}$$

In components, $\delta S^{AB} = \mathsf{C}^{ABCD}(\mathring{\varphi}^*\epsilon)_{CD}$. It is straightforward to show that the first and the second elasticity tensors are related as

$$\mathsf{A}_a{}^A{}_b{}^B = \mathsf{C}^{AMBN}\mathring{F}^m{}_M\mathring{F}^n{}_N g_{am}g_{bn} + \mathring{S}^{AB}g_{ab}. \tag{3.14}$$

Or

$$\mathsf{A}^{aAbB} = \mathsf{C}^{AMBN}\mathring{F}^a{}_M\mathring{F}^b{}_N + \mathring{S}^{AB}g^{ab}. \tag{3.15}$$

Similarly, the spatial elasticity tensors are defined as $\mathsf{a} = \frac{1}{J}\varphi_*\mathsf{A}$, and $\mathsf{c} = \frac{1}{J}\varphi_*\mathsf{C}$. In components

$$\mathsf{a}^{ac}{}_b{}^d = \frac{1}{J}F^c{}_AF^d{}_B\mathsf{A}^{aA}{}_b{}^B\,, \quad \mathsf{c}^{abcd} = \frac{\partial\sigma^{ab}}{\partial g_{cd}} = \frac{1}{J}F^a{}_AF^b{}_BF^c{}_CF^d{}_D\mathsf{C}^{ABCD}\,. \tag{3.16}$$

The relation analogous to (3.15) is given by $\mathsf{a}^{ac}{}_b{}^d = \sigma^{cd}\delta^a_b + \mathsf{c}^{aced}g_{eb}$. Note that the spatial elasticity tensors have the following symmetries $\mathsf{a}^{acbd} = \mathsf{a}^{bdac}$, and $\mathsf{c}^{abcd} = \mathsf{c}^{bacd} = \mathsf{c}^{abdc} = \mathsf{c}^{cdab}$.

For an incompressible solid the variation of the first Piola-Kirchhoff stress is written as $\delta\mathbf{P} = \delta\mathbf{P}_{\text{const.}} - \delta\left(p\mathbf{g}^\sharp\mathbf{F}^{-\star}\right)$, where $\delta\mathbf{P}_{\text{const.}}$ is the linearization of the constitutive part of the stress, which is given in (3.10). In components

$$\delta P^{aA} = \left[\mathsf{A}^{aA}{}_b{}^B + \mathring{p}\ g^{am}(\mathring{F}^{-1})^A{}_b(\mathring{F}^{-1})^B{}_m\right]U^b{}_{|B} - \delta p\ g^{ab}(\mathring{F}^{-1})^A{}_b. \tag{3.17}$$

Similarly, the linearized second Piola-Kirchhoff stress in components reads

$$\delta S^{AB} = \left[\mathsf{C}^{ABCD} + 2\mathring{p}\ (\mathring{C}^{-1})^{AC}(\mathring{C}^{-1})^{BD}\right](\mathring{\varphi}^*\epsilon)_{CD} - \delta p\ (\mathring{C}^{-1})^{AB}. \tag{3.18}$$

For an incompressible solid the relation (3.15) is modified to read

$$\mathsf{A}_a{}^A{}_b{}^B = \mathsf{C}^{AMBN}\mathring{F}^m{}_M\mathring{F}^n{}_N g_{am}g_{bn} + \mathring{S}^{AB}_{\text{const.}}\ g_{ab}, \tag{3.19}$$

where

$$\mathring{S}^{AB}_{\text{const.}} = \mathring{S}^{AB} + \mathring{p}\mathring{C}^{AB}. \tag{3.20}$$

**Linearization of conservation of mass.** For the family of motions $\varphi_{t,\epsilon}$, conservation of mass is locally written as $\rho_0(X) = J_\epsilon\ \rho\circ\varphi_{t,\epsilon}(X)$. Taking derivatives of both sides with respect to $\epsilon$ and evaluating at $\epsilon = 0$, one obtains $0 = \delta J\ \rho\circ\mathring{\varphi}_t(X) + \mathring{J}\ \mathbf{d}\mathring{\rho}\cdot\delta\varphi$, where $\mathbf{d}\mathring{\rho}$ is the exterior derivative of mass density of the reference motion (a 1-form). Thus, $\mathbf{d}\mathring{\rho}\cdot\mathbf{u} + \mathring{\rho}\mathbf{g}^\sharp\!:\!\boldsymbol{\epsilon} = 0$. Note that $\mathbf{g}^\sharp\!:\!\boldsymbol{\epsilon} = \mathbf{g}^\sharp\!:\!\nabla\mathbf{u}^\flat$. Hence, the linearized conservation of mass can be written as, $\mathbf{d}\mathring{\rho}\cdot\mathbf{u} + \mathring{\rho}\mathbf{g}^\sharp\!:\!\nabla^{\mathbf{g}}\mathbf{u}^\flat = 0$. Therefore

$$\delta\rho = -\mathring{\rho}\,\boldsymbol{\epsilon}\!:\!\mathbf{g}^\sharp = -\mathring{\rho}\,\mathrm{div}_{\mathbf{g}}\,\mathbf{u}. \tag{3.21}$$

**Linearization of the balance of linear momentum.** Balance of linear momentum in terms of the first Piola-Kirchhoff stress reads $\mathrm{Div}\,\mathbf{P} + \rho_0\mathbf{B} = \rho_0\mathbf{A}$. Note that the operator Div depends on both $\mathbf{G}$ and $\mathbf{g}$ and is defined through the Piola identity (see the appendix) : $\mathrm{Div}\,\mathbf{P} = J\,\mathrm{div}_{\mathbf{g}}\,\boldsymbol{\sigma}$. Balance of linear momentum in terms of the Cauchy stress reads $\mathrm{div}_{\mathbf{g}}\,\boldsymbol{\sigma} + \rho\mathbf{b} = \rho\mathbf{a}$. Using the Doyle-Ericksen formula, the Cauchy stress is written as [Doyle and Ericksen, 1956] $\boldsymbol{\sigma} = \frac{2}{J}\frac{\partial W}{\partial\mathbf{g}}$, where $W = W(X,\mathbf{F},\mathbf{g}\circ\varphi,\mathbf{G},\boldsymbol{\zeta}_1,\ldots,\boldsymbol{\zeta}_n)$ is the energy density per unit undeformed volume. The convected stress tensor $\boldsymbol{\Sigma} = \varphi_t^*\boldsymbol{\sigma}$ can be written as [Simo et al., 1988] $\boldsymbol{\Sigma} = \frac{2}{J}\frac{\partial\hat{W}}{\partial\mathbf{C}^\flat}$, where $W = \hat{W}(X,\mathbf{C}^\flat,\mathbf{G},\boldsymbol{\zeta}_1,\ldots,\boldsymbol{\zeta}_n)$.

**Linearization of the convected balance of linear momentum.** For the one-parameter family of motions $\varphi_{t,\epsilon}$, one has

$$\mathrm{div}_{\mathbf{C}^\flat_\epsilon}\,\boldsymbol{\Sigma}_\epsilon + \varrho_\epsilon\boldsymbol{\mathscr{B}}_\epsilon = \varrho_\epsilon\boldsymbol{\mathscr{A}}_\epsilon. \tag{3.22}$$

Therefore, the linearized balance of linear momentum is defined as

$$\frac{d}{d\epsilon}\left[\mathrm{div}_{\mathbf{C}^\flat_\epsilon}\,\boldsymbol{\Sigma}_\epsilon\right]\Big|_{\epsilon=0} + \frac{d}{d\epsilon}\left[\varrho_\epsilon\boldsymbol{\mathscr{B}}_\epsilon\right]\Big|_{\epsilon=0} = \frac{d}{d\epsilon}\left[\varrho_\epsilon\boldsymbol{\mathscr{A}}_\epsilon\right]\Big|_{\epsilon=0}. \tag{3.23}$$

Note that for $X \in \mathcal{B}$, all the terms in (3.22) lie in the same tangent space $T_X\mathcal{B}$, and hence, linearizing the balance of linear momentum is straightforward when using the convected stress.

It is a simple calculation to show that the linearized convected stress reads

$$\delta\boldsymbol{\Sigma} = \frac{4}{\mathring{J}} \frac{\partial^2 \hat{W}}{\partial \mathring{\mathbf{C}}^\flat \partial \mathring{\mathbf{C}}^\flat} : \mathring{\varphi}_t^* \boldsymbol{\epsilon} - (\boldsymbol{\epsilon} : \mathbf{g}^\sharp) \mathring{\boldsymbol{\Sigma}}. \tag{3.24}$$

Note that $\delta\boldsymbol{\sigma} = \mathring{\varphi}_{t*}\delta\boldsymbol{\Sigma}$, and hence

$$\delta\boldsymbol{\sigma} = \frac{4}{\mathring{J}} \frac{\partial^2 W}{\partial \mathbf{g} \partial \mathbf{g}} : \boldsymbol{\epsilon} - (\boldsymbol{\epsilon} : \mathbf{g}^\sharp) \mathring{\boldsymbol{\sigma}}. \tag{3.25}$$

We define the following fourth-order elasticity tensor

$$\mathbb{C} := \frac{4}{\mathring{J}} \frac{\partial^2 W}{\partial \mathbf{g} \partial \mathbf{g}} \,, \tag{3.26}$$

which in components reads $\mathbb{C}^{abcd} = \frac{4}{\mathring{J}} \frac{\partial^2 W}{\partial g_{ab} \partial g_{cd}}$. Note that $\mathbb{C} = \mathring{\varphi}_* \mathsf{C} / \mathring{J}$. We now expand the linearization of each term in (3.23). For the body force term one can write

$$\begin{aligned} \frac{d}{d\epsilon} \left[ \varrho_\epsilon \mathscr{B}_\epsilon \right] \Big|_{\epsilon=0} &= \frac{d}{d\epsilon} \left[ (\rho \circ \varphi_{t,\epsilon}) \varphi_{t,\epsilon}^* \mathbf{b} \right] \Big|_{\epsilon=0} = \delta\rho \circ \mathring{\varphi}_t \, \mathring{\mathscr{B}} + \mathring{\varrho} \frac{d}{d\epsilon} \left[ (\varphi_{t,\epsilon}^* \mathbf{b}) \right] \Big|_{\epsilon=0} \\ &= -(\mathbf{g}^\sharp : \boldsymbol{\epsilon}) \circ \mathring{\varphi}_t \; \mathring{\varrho} \mathring{\mathscr{B}} + \mathring{\varrho} \; \mathring{\varphi}_t^* \Big( \mathbf{L}_{\mathbf{u}} \mathring{\mathbf{b}} \Big) \,. \end{aligned} \tag{3.27}$$

Note that $\partial \mathring{\mathbf{b}} / \partial \epsilon = \mathbf{0}$, and hence, $\mathbf{L}_{\mathbf{u}} \mathring{\mathbf{b}} = \mathfrak{L}_{\mathbf{u}} \mathring{\mathbf{b}}$. Therefore, $\delta(\varrho\mathscr{B}) = -(\mathbf{g}^\sharp : \boldsymbol{\epsilon}) \circ \mathring{\varphi}_t \; \mathring{\varrho} \mathring{\mathscr{B}} + \mathring{\varrho} \; \mathring{\varphi}_t^* \Big( \mathfrak{L}_{\mathbf{u}} \mathring{\mathbf{b}} \Big)$. Also note that $\mathfrak{L}_{\mathbf{u}} \mathring{\mathbf{b}} = \nabla^{\mathbf{g}}_{\mathbf{u}} \mathring{\mathbf{b}} - \nabla^{\mathbf{g}}_{\mathring{\mathbf{b}}} \mathbf{u}$. Hence

$$\delta(\varrho\mathscr{B}) = -(\mathbf{g}^\sharp : \boldsymbol{\epsilon}) \circ \mathring{\varphi}_t \; \mathring{\varrho} \mathring{\mathscr{B}} + \mathring{\varrho} \; \mathring{\varphi}_t^* \Big( \nabla^{\mathbf{g}}_{\mathbf{u}} \mathring{\mathbf{b}} - \nabla^{\mathbf{g}}_{\mathring{\mathbf{b}}} \mathbf{u} \Big) \,. \tag{3.28}$$

The convected acceleration is linearized as follows

$$\begin{aligned} \frac{d}{d\epsilon} [\mathscr{A}_\epsilon] \Big|_{\epsilon=0} &= \varphi_{t,\epsilon}^* (\mathbf{L}_{\mathbf{u}} \mathbf{a}_\epsilon) \Big|_{\epsilon=0} = \varphi_{t,\epsilon}^* (\mathfrak{L}_{\mathbf{u}} \mathbf{a}_\epsilon) \Big|_{\epsilon=0} = \varphi_{t,\epsilon}^* \left[ \nabla^{\mathbf{g}}_{\mathbf{u}} \mathbf{a}_\epsilon - \nabla^{\mathbf{g}}_{\mathbf{a}_\epsilon} \mathbf{u} \right] \Big|_{\epsilon=0} \\ &= T\varphi_{t,\epsilon}^{-1} \circ \left[ \nabla_{\mathbf{U}} \mathbf{A}_\epsilon - \nabla_{\mathbf{A}_\epsilon} \mathbf{U} \right] \Big|_{\epsilon=0} = T\varphi_{t,\epsilon}^{-1} \circ \left[ D_\epsilon \mathbf{A}_\epsilon - \nabla_{\mathbf{A}_\epsilon} \mathbf{U} \right] \Big|_{\epsilon=0} \\ &= T\varphi_{t,\epsilon}^{-1} \circ \left[ D_\epsilon D_t \mathbf{V}_\epsilon - \nabla_{\mathbf{A}_\epsilon} \mathbf{U} \right] \Big|_{\epsilon=0} \\ &= T\varphi_{t,\epsilon}^{-1} \circ \left[ D_t D_\epsilon \frac{\partial \varphi_{t,\epsilon}}{\partial t} + \nabla_{[\mathbf{U}, \mathbf{V}_\epsilon]} \mathbf{V}_\epsilon + \boldsymbol{\mathcal{R}}_{\mathbf{g}}(\mathbf{U}, \mathbf{V}_\epsilon, \mathbf{V}_\epsilon) - \nabla_{\mathbf{A}_\epsilon} \mathbf{U} \right] \Bigg|_{\epsilon=0} \\ &= \mathring{\mathbf{F}}_t^{-1} \cdot \left[ D_t D_t \mathbf{U} + \nabla_{[\mathbf{U}, \mathring{\mathbf{V}}]} \mathring{\mathbf{V}} + \boldsymbol{\mathcal{R}}_{\mathbf{g}}(\mathbf{U}, \mathring{\mathbf{V}}, \mathring{\mathbf{V}}) - \nabla_{\mathring{\mathbf{A}}} \mathbf{U} \right] \,. \end{aligned} \tag{3.29}$$

Therefore

$$\delta(\varrho\mathscr{A}) = -(\mathbf{g}^\sharp : \boldsymbol{\epsilon}) \circ \mathring{\varphi}_t \; \mathring{\varrho} \mathring{\mathscr{A}} + \mathring{\varrho} \mathring{\mathbf{F}}_t^{-1} \cdot \left[ \ddot{\mathbf{U}} + \nabla_{[\mathbf{U}, \mathring{\mathbf{V}}]} \mathring{\mathbf{V}} + \boldsymbol{\mathcal{R}}_{\mathbf{g}}(\mathbf{U}, \mathring{\mathbf{V}}, \mathring{\mathbf{V}}) - \nabla_{\mathring{\mathbf{A}}} \mathbf{U} \right] \,. \tag{3.30}$$

To calculate $\delta \left( \operatorname{div}_{\mathbf{C}^\flat} \boldsymbol{\Sigma} \right)$, Sadik and Yavari [2016] expanded $\operatorname{div}_{\mathbf{C}^\flat_\epsilon} \boldsymbol{\Sigma}_\epsilon$ in a coordinate chart and then took derivatives with respect to $\epsilon$ and finally evaluated at $\epsilon = 0$. They also assumed a flat ambient space. Here, we do not assume a flat ambient space. Note that

$$\delta(\operatorname{div}_{\mathbf{C}^\flat} \boldsymbol{\Sigma}) = \frac{d}{d\epsilon} \left[ \operatorname{div}_{\mathbf{C}^\flat_\epsilon} \boldsymbol{\Sigma}_\epsilon \right] \Bigg|_{\epsilon=0} = \frac{d}{d\epsilon} \varphi_{t,\epsilon}^* \left[ \operatorname{div}_{\mathbf{g}} \boldsymbol{\sigma}_\epsilon \right] \Big|_{\epsilon=0} = \varphi_{t,\epsilon}^* \mathbf{L}_{\mathbf{u}} \left[ \operatorname{div}_{\mathbf{g}} \boldsymbol{\sigma}_\epsilon \right] \Big|_{\epsilon=0} = \varphi_{t,\epsilon}^* \mathfrak{L}_{\mathbf{u}} \left[ \operatorname{div}_{\mathbf{g}} \boldsymbol{\sigma}_\epsilon \right] \Big|_{\epsilon=0} . \tag{3.31}$$

We know that [Yano, 1957, Marsden and Hughes, 1983]

$$\mathfrak{L}_{\mathbf{u}} \nabla^{\mathbf{g}} = \nabla^{\mathbf{g}} \nabla^{\mathbf{g}} \mathbf{u} + \mathbf{u} \cdot \boldsymbol{\mathcal{R}}_{\mathbf{g}} \,, \tag{3.32}$$

where in components, $(\mathfrak{L}_{\mathbf{u}}\nabla^{\mathbf{g}})^a{}_{bc} = u^a{}_{|bc} + \mathcal{R}^a{}_{ebc}u^e$. We also know that [Yano, 1957]

$$\begin{aligned}\mathfrak{L}_{\mathbf{u}}\sigma^{ab}{}_{|b} &= (\mathfrak{L}_{\mathbf{u}}\sigma^{ab})_{|b} + (\mathfrak{L}_{\mathbf{u}}\gamma^a{}_{bd})\sigma^{bd} + (\mathfrak{L}_{\mathbf{u}}\gamma^b{}_{bd})\sigma^{ad} \\ &= (\mathfrak{L}_{\mathbf{u}}\sigma^{ab})_{|b} + (u^a{}_{|bd} + \mathcal{R}^a{}_{cbd}u^c)\sigma^{bd} + (u^b{}_{|bd} + \mathcal{R}^b{}_{cbd}u^c)\sigma^{ad}.\end{aligned} \tag{3.33}$$

Note that $\mathcal{R}^a{}_{cbd} = -\mathcal{R}^a{}_{cdb}$, and hence, $\mathcal{R}^a{}_{cbd}\sigma^{bd} = 0$. Also, $\mathcal{R}^b{}_{cbd} = \mathsf{Ric}_{cd}$ is the Ricci curvature. Thus

$$\mathfrak{L}_{\mathbf{u}}\sigma^{ab}{}_{|b} = (\mathfrak{L}_{\mathbf{u}}\sigma^{ab})_{|b} + u^a{}_{|bd}\sigma^{bd} + \left[(u^b{}_{|b})_{,d} + u^c\mathsf{Ric}_{cd}\right]\sigma^{ad}. \tag{3.34}$$

Therefore

$$\mathfrak{L}_{\mathbf{u}}\left(\operatorname{div}_{\mathbf{g}}\mathring{\boldsymbol{\sigma}}\right) = \operatorname{div}_{\mathbf{g}}\left(\mathfrak{L}_{\mathbf{u}}\mathring{\boldsymbol{\sigma}}\right) + \nabla^{\mathbf{g}}\nabla^{\mathbf{g}}\mathbf{u} : \mathring{\boldsymbol{\sigma}} + \mathbf{d}(\boldsymbol{\epsilon}\!:\!\mathbf{g}^\sharp)\cdot\mathring{\boldsymbol{\sigma}} + \mathring{\boldsymbol{\sigma}}\cdot\mathbf{Ric}_{\mathbf{g}}\cdot\mathbf{u}. \tag{3.35}$$

Note that

$$\mathfrak{L}_{\mathbf{u}}\mathring{\boldsymbol{\sigma}} = \mathfrak{L}_{\mathbf{u}}\left(\frac{2}{\mathring{J}}\frac{\partial W}{\partial \mathbf{g}}\right) = -\frac{2}{\mathring{J}^2}\frac{\partial W}{\partial \mathbf{g}}\mathfrak{L}_{\mathbf{u}}\mathring{J} + \frac{2}{\mathring{J}}\left(\frac{\partial^2 W}{\partial\mathbf{g}\partial\mathbf{g}}\!:\!\mathfrak{L}_{\mathbf{u}}\mathbf{g} + \frac{\partial^2 W}{\partial\mathbf{F}\partial\mathbf{g}}\!:\!\mathfrak{L}_{\mathbf{u}}\mathbf{F}\right). \tag{3.36}$$

However, $\mathfrak{L}_{\mathbf{u}}\mathbf{F} = \mathbf{0}$, because for any vector $\mathbf{W}(X) \in T_X\mathcal{B}$, one can write

$$\begin{aligned}\mathfrak{L}_{\mathbf{u}}\mathbf{F}\cdot\mathbf{W} &= \mathfrak{L}_{\mathbf{u}}\left(\mathbf{F}\cdot\mathbf{W}\right) = \varphi_{t*}\frac{d}{d\epsilon}\left[\varphi^*_{t,\epsilon}\left(\mathbf{F}_{t,\epsilon}\cdot\mathbf{W}\right)\right]\Big|_{\epsilon=0} \\ &= \varphi_{t*}\frac{d}{d\epsilon}\left[\varphi^*_{t,\epsilon}\left(\varphi_{t,\epsilon*}\circ\mathbf{W}\right)\right]\Big|_{\epsilon=0} = \varphi_{t*}\frac{d}{d\epsilon}\left[\mathbf{W}\right]\Big|_{\epsilon=0} = \mathbf{0}.\end{aligned} \tag{3.37}$$

Hence

$$\mathfrak{L}_{\mathbf{u}}\mathring{\boldsymbol{\sigma}} = \frac{4}{\mathring{J}}\frac{\partial^2 W}{\partial\mathbf{g}\partial\mathbf{g}}\!:\!\boldsymbol{\epsilon} - (\boldsymbol{\epsilon}\!:\!\mathbf{g}^\sharp)\mathring{\boldsymbol{\sigma}}. \tag{3.38}$$

Therefore

$$\delta(\operatorname{div}_{\mathbf{C}^\flat}\boldsymbol{\Sigma}) = \mathring{\varphi}_t^*\left[\nabla^{\mathbf{g}}\nabla^{\mathbf{g}}\mathbf{u} : \mathring{\boldsymbol{\sigma}} + \mathbf{d}(\boldsymbol{\epsilon}\!:\!\mathbf{g}^\sharp)\cdot\mathring{\boldsymbol{\sigma}} + \mathring{\boldsymbol{\sigma}}\cdot\mathbf{Ric}_{\mathbf{g}}\cdot\mathbf{u} + \operatorname{div}_{\mathbf{g}}\left(\frac{4}{\mathring{J}}\frac{\partial^2 W}{\partial\mathbf{g}\partial\mathbf{g}}\!:\!\boldsymbol{\epsilon} - (\boldsymbol{\epsilon}\!:\!\mathbf{g}^\sharp)\mathring{\boldsymbol{\sigma}}\right)\right]. \tag{3.39}$$

Or

$$\delta(\operatorname{div}_{\mathbf{C}^\flat}\boldsymbol{\Sigma}) = \mathring{\varphi}_t^*\left[\nabla^{\mathbf{g}}\nabla^{\mathbf{g}}\mathbf{u} : \mathring{\boldsymbol{\sigma}} - (\boldsymbol{\epsilon}\!:\!\mathbf{g}^\sharp)\operatorname{div}_{\mathbf{g}}\mathring{\boldsymbol{\sigma}} + \mathring{\boldsymbol{\sigma}}\cdot\mathbf{Ric}_{\mathbf{g}}\cdot\mathbf{u} + \operatorname{div}_{\mathbf{g}}\left(\mathbb{C}\!:\!\boldsymbol{\epsilon}\right)\right]. \tag{3.40}$$

Now the push-forward of the balance of linear momentum by $\mathring{\varphi}_t$ using (3.28), (3.30), and (3.40) reads

$$\begin{aligned}&\nabla^{\mathbf{g}}\nabla^{\mathbf{g}}\mathbf{u} : \mathring{\boldsymbol{\sigma}} - (\boldsymbol{\epsilon}\!:\!\mathbf{g}^\sharp)\operatorname{div}_{\mathbf{g}}\mathring{\boldsymbol{\sigma}} + \mathring{\boldsymbol{\sigma}}\cdot\mathbf{Ric}_{\mathbf{g}}\cdot\mathbf{u} + \operatorname{div}_{\mathbf{g}}\left(\mathbb{C}\!:\!\boldsymbol{\epsilon}\right) - \mathbf{g}^\sharp\!:\!\boldsymbol{\epsilon}\;\mathring{\rho}\mathring{\mathbf{b}} + \mathring{\rho}\left(\nabla^{\mathbf{g}}_{\mathbf{u}}\mathring{\mathbf{b}} - \nabla^{\mathbf{g}}_{\mathring{\mathbf{b}}}\mathbf{u}\right) \\ &= -\mathbf{g}^\sharp\!:\!\boldsymbol{\epsilon}\;\mathring{\rho}\mathring{\mathbf{a}} + \mathring{\rho}\left[\ddot{\mathbf{U}}\circ\mathring{\varphi}_t^{-1} + \nabla^{\mathbf{g}}_{[\mathbf{u},\mathring{\mathbf{v}}]}\mathring{\mathbf{v}} + \boldsymbol{\mathcal{R}}_{\mathbf{g}}(\mathbf{u},\mathring{\mathbf{v}},\mathring{\mathbf{v}}) - \nabla^{\mathbf{g}}_{\mathring{\mathbf{a}}}\mathbf{u}\right].\end{aligned} \tag{3.41}$$

We assume the reference motion is "physical," i.e., it satisfies all the balance laws. In particular, $\operatorname{div}_{\mathbf{g}}\mathring{\boldsymbol{\sigma}} + \mathring{\rho}\mathring{\mathbf{b}} = \mathring{\rho}\mathring{\mathbf{a}}$. Thus, we have the following two identities

$$\begin{aligned}&-\mathbf{g}^\sharp\!:\!\boldsymbol{\epsilon}\;\mathring{\rho}\mathring{\mathbf{b}} + \mathbf{g}^\sharp\!:\!\boldsymbol{\epsilon}\;\mathring{\rho}\mathring{\mathbf{a}} = (\mathbf{g}^\sharp\!:\!\boldsymbol{\epsilon})\operatorname{div}_{\mathbf{g}}\mathring{\boldsymbol{\sigma}} \\ &-\mathring{\rho}\nabla^{\mathbf{g}}_{\mathring{\mathbf{b}}}\mathbf{u} + \mathring{\rho}\nabla^{\mathbf{g}}_{\mathring{\mathbf{a}}}\mathbf{u} = \nabla^{\mathbf{g}}_{-\mathring{\rho}(\mathring{\mathbf{b}}-\mathring{\mathbf{a}})}\mathbf{u} = \nabla^{\mathbf{g}}_{\operatorname{div}_{\mathbf{g}}\mathring{\boldsymbol{\sigma}}}\mathbf{u} = \nabla^{\mathbf{g}}\mathbf{u}\cdot\operatorname{div}_{\mathbf{g}}\mathring{\boldsymbol{\sigma}}.\end{aligned} \tag{3.42}$$

Therefore, the linearized balance of linear momentum is simplified to read

$$\nabla^{\mathbf{g}}\nabla^{\mathbf{g}}\mathbf{u} : \mathring{\boldsymbol{\sigma}} + \nabla^{\mathbf{g}}\mathbf{u}\cdot\operatorname{div}_{\mathbf{g}}\mathring{\boldsymbol{\sigma}} + \mathring{\boldsymbol{\sigma}}\cdot\mathbf{Ric}_{\mathbf{g}}\cdot\mathbf{u} + \operatorname{div}_{\mathbf{g}}\left(\mathbb{C}\!:\!\boldsymbol{\epsilon}\right) + \mathring{\rho}\nabla^{\mathbf{g}}_{\mathbf{u}}\mathring{\mathbf{b}} = \mathring{\rho}\left[\ddot{\mathbf{U}}\circ\mathring{\varphi}_t^{-1} + \nabla^{\mathbf{g}}_{[\mathbf{u},\mathring{\mathbf{v}}]}\mathring{\mathbf{v}} + \boldsymbol{\mathcal{R}}_{\mathbf{g}}(\mathbf{u},\mathring{\mathbf{v}},\mathring{\mathbf{v}})\right]. \tag{3.43}$$

Note that $\nabla^{\mathbf{g}}\nabla^{\mathbf{g}}\mathbf{u} : \mathring{\boldsymbol{\sigma}} + \nabla^{\mathbf{g}}\mathbf{u}\cdot\operatorname{div}_{\mathbf{g}}\mathring{\boldsymbol{\sigma}} = \operatorname{div}_{\mathbf{g}}\left(\nabla^{\mathbf{g}}\mathbf{u}\cdot\mathring{\boldsymbol{\sigma}}\right)$. Hence, the linearized balance of linear momentum is written as[14]

$$\operatorname{div}_{\mathbf{g}}\left(\mathbb{C}\!:\!\boldsymbol{\epsilon}\right) + \operatorname{div}_{\mathbf{g}}\left(\nabla^{\mathbf{g}}\mathbf{u}\cdot\mathring{\boldsymbol{\sigma}}\right) + \mathring{\boldsymbol{\sigma}}\cdot\mathbf{Ric}_{\mathbf{g}}\cdot\mathbf{u} + \mathring{\rho}\nabla^{\mathbf{g}}_{\mathbf{u}}\mathring{\mathbf{b}} = \mathring{\rho}\left[\ddot{\mathbf{U}}\circ\mathring{\varphi}_t^{-1} + \nabla^{\mathbf{g}}_{[\mathbf{u},\mathring{\mathbf{v}}]}\mathring{\mathbf{v}} + \boldsymbol{\mathcal{R}}_{\mathbf{g}}(\mathbf{u},\mathring{\mathbf{v}},\mathring{\mathbf{v}})\right]. \tag{3.44}$$

[14]For a flat ambient space this is identical to the corresponding equation in [Sadik and Yavari, 2016]. However, note that even for a flat ambient space this is not identical to what Marsden and Hughes [1983] obtained; they do not have the term $\operatorname{div}_{\mathbf{g}}\left(\nabla^{\mathbf{g}}\mathbf{u}\cdot\mathring{\boldsymbol{\sigma}}\right)$.

**Linearization of the spatial balance of linear momentum.** For the motion $\varphi_{t,\epsilon}$, the balance of linear momentum reads $\operatorname{div}_{\mathbf{g}} \boldsymbol{\sigma}_\epsilon + \rho_\epsilon \mathbf{b}_\epsilon = \rho_\epsilon \mathbf{a}_\epsilon$. Linearization of the body force term is calculated as $\delta(\rho\mathbf{b}) = \delta\rho\,\mathring{\mathbf{b}} + \mathring{\rho}\nabla^{\mathbf{g}}_{\frac{\partial}{\partial\epsilon}}\mathbf{b}_\epsilon\Big|_{\epsilon=0} = -(\mathbf{g}^\sharp\!:\!\boldsymbol{\epsilon})\mathring{\rho}\,\mathring{\mathbf{b}} + \mathring{\rho}\nabla^{\mathbf{g}}_{\mathbf{u}}\mathring{\mathbf{b}}$. Linearization of the inertial force term reads

$$
\begin{aligned}
\delta(\rho\mathbf{a}) &= \delta\rho\,\mathring{\mathbf{a}} + \mathring{\rho}\nabla^{\mathbf{g}}_{\frac{\partial}{\partial\epsilon}}\mathbf{a}_\epsilon\Big|_{\epsilon=0} = -(\mathbf{g}^\sharp\!:\!\boldsymbol{\epsilon})\mathring{\rho}\,\mathring{\mathbf{a}} + \mathring{\rho}\nabla^{\mathbf{g}}_{\frac{\partial}{\partial\epsilon}}\mathbf{a}_\epsilon\Big|_{\epsilon=0} \\
&\quad - (\mathbf{g}^\sharp\!:\!\boldsymbol{\epsilon})\mathring{\rho}\,\mathring{\mathbf{a}} + \mathring{\rho}\left[\ddot{\mathbf{U}}\circ\mathring{\varphi}_t^{-1} + \nabla^{\mathbf{g}}_{[\mathbf{u},\mathring{\mathbf{v}}]}\mathring{\mathbf{v}} + \boldsymbol{\mathcal{R}}_{\mathbf{g}}\left(\mathbf{u},\mathring{\mathbf{v}},\mathring{\mathbf{v}}\right)\right].
\end{aligned}
\tag{3.45}
$$

Note that $\delta(\operatorname{div}_{\mathbf{g}}\boldsymbol{\sigma}) = \nabla^{\mathbf{g}}_{\frac{\partial}{\partial\epsilon}}(\operatorname{div}_{\mathbf{g}}\boldsymbol{\sigma}_\epsilon)\Big|_{\epsilon=0} = \nabla^{\mathbf{g}}_{\mathbf{u}}(\operatorname{div}_{\mathbf{g}}\mathring{\boldsymbol{\sigma}})$. Knowing that the Levi-Civita connection is torsion-free, one can write $\nabla^{\mathbf{g}}_{\mathbf{u}}(\operatorname{div}_{\mathbf{g}}\mathring{\boldsymbol{\sigma}}) = \mathfrak{L}_{\mathbf{u}}(\operatorname{div}_{\mathbf{g}}\mathring{\boldsymbol{\sigma}}) + \nabla^{\mathbf{g}}_{\operatorname{div}_{\mathbf{g}}\mathring{\boldsymbol{\sigma}}}\mathbf{u} = \mathfrak{L}_{\mathbf{u}}(\operatorname{div}_{\mathbf{g}}\mathring{\boldsymbol{\sigma}}) + \nabla^{\mathbf{g}}\mathbf{u}\cdot\operatorname{div}_{\mathbf{g}}\mathring{\boldsymbol{\sigma}}$. Hence, from (3.35), one has $\delta(\operatorname{div}_{\mathbf{g}}\boldsymbol{\sigma}) = \operatorname{div}_{\mathbf{g}}(\mathfrak{L}_{\mathbf{u}}\mathring{\boldsymbol{\sigma}}) + \operatorname{div}_{\mathbf{g}}(\nabla^{\mathbf{g}}\mathbf{u}\cdot\mathring{\boldsymbol{\sigma}}) + \mathbf{d}(\boldsymbol{\epsilon}\!:\!\mathbf{g}^\sharp)\cdot\mathring{\boldsymbol{\sigma}} + \mathring{\boldsymbol{\sigma}}\cdot\mathsf{Ric}_{\mathbf{g}}\cdot\mathbf{u}$. Using (3.38), one obtains

$$
\delta(\operatorname{div}_{\mathbf{g}}\boldsymbol{\sigma}) = \operatorname{div}_{\mathbf{g}}(\nabla^{\mathbf{g}}\mathbf{u}\cdot\mathring{\boldsymbol{\sigma}}) + \mathring{\boldsymbol{\sigma}}\cdot\mathsf{Ric}_{\mathbf{g}}\cdot\mathbf{u} + \operatorname{div}_{\mathbf{g}}(\mathbb{C}\!:\!\boldsymbol{\epsilon}) - (\boldsymbol{\epsilon}\!:\!\mathbf{g}^\sharp)\operatorname{div}_{\mathbf{g}}\mathring{\boldsymbol{\sigma}}\,.
\tag{3.46}
$$

**Remark 3.1.** Note that linearizing a convected vector and pushing it forward to the ambient space is not the same as linearizing the corresponding spatial vector using a covariant derivative. In particular, we note that

$$
\begin{aligned}
\delta(\operatorname{div}_{\mathbf{g}}\boldsymbol{\sigma}) &= \mathring{\varphi}_{t*}\delta(\operatorname{div}_{\mathbf{C}^\flat}\boldsymbol{\Sigma}) + \nabla^{\mathbf{g}}\mathbf{u}\cdot\operatorname{div}_{\mathbf{g}}\mathring{\boldsymbol{\sigma}}, \\
\delta(\rho\mathbf{b}) &= \mathring{\varphi}_{t*}\delta(\varrho\mathscr{B}) + \mathring{\rho}\nabla^{\mathbf{g}}_{\mathring{\mathbf{b}}}\mathbf{u}, \\
\delta(\rho\mathbf{a}) &= \mathring{\varphi}_{t*}\delta(\varrho\mathscr{A}) + \mathring{\rho}\nabla^{\mathbf{g}}_{\mathring{\mathbf{a}}}\mathbf{u}.
\end{aligned}
\tag{3.47}
$$

However, note that $\nabla^{\mathbf{g}}\mathbf{u}\cdot\operatorname{div}_{\mathbf{g}}\mathring{\boldsymbol{\sigma}} + \mathring{\rho}\nabla^{\mathbf{g}}_{\mathring{\mathbf{b}}}\mathbf{u} - \mathring{\rho}\nabla^{\mathbf{g}}_{\mathring{\mathbf{a}}}\mathbf{u} = \nabla^{\mathbf{g}}_{\operatorname{div}_{\mathbf{g}}\mathring{\boldsymbol{\sigma}}+\mathring{\rho}\mathring{\mathbf{b}}-\mathring{\rho}\mathring{\mathbf{a}}}\mathbf{u} = \mathbf{0}$, i.e., the two approaches give the same linearized balance of linear momentum (3.44).

**Linearization of the material balance of linear momentum.** Balance of linear momentum in terms of the first Piola-Kirchhoff stress is linearized as follows. Linearized body and inertial forces are (note that $\rho_0 = \rho_0(X)$ has a vanishing variation) $\delta(\rho_0\mathbf{B}) = \rho_0\nabla^{\mathbf{g}}_{\mathbf{U}}\mathring{\mathbf{B}}$, and $\delta(\rho_0\mathbf{A}) = \rho_0\left[\ddot{\mathbf{U}} + \nabla^{\mathbf{g}}_{[\mathbf{U},\mathring{\mathbf{V}}]}\mathring{\mathbf{V}} + \boldsymbol{\mathcal{R}}_{\mathbf{g}}\left(\mathbf{U},\mathring{\mathbf{V}},\mathring{\mathbf{V}}\right)\right]$. Note that $\delta(\operatorname{Div}\mathbf{P}) = \delta(J\operatorname{div}_{\mathbf{g}}\boldsymbol{\sigma}) = \delta J\operatorname{div}_{\mathbf{g}}\mathring{\boldsymbol{\sigma}} + \mathring{J}\delta(\operatorname{div}_{\mathbf{g}}\boldsymbol{\sigma}) = (\boldsymbol{\epsilon}\!:\!\mathbf{g}^\sharp)\operatorname{Div}\mathring{\mathbf{P}} + \mathring{J}\delta(\operatorname{div}_{\mathbf{g}}\boldsymbol{\sigma})$. From (3.46), we know that

$$
\mathring{J}\delta(\operatorname{div}_{\mathbf{g}}\boldsymbol{\sigma}) = \mathring{J}\operatorname{div}_{\mathbf{g}}(\nabla^{\mathbf{g}}\mathbf{u}\cdot\mathring{\boldsymbol{\sigma}}) + \mathring{J}\mathring{\boldsymbol{\sigma}}\cdot\mathsf{Ric}_{\mathbf{g}}\cdot\mathbf{u} + \mathring{J}\operatorname{div}_{\mathbf{g}}(\mathbb{C}\!:\!\boldsymbol{\epsilon}) - (\boldsymbol{\epsilon}\!:\!\mathbf{g}^\sharp)\mathring{J}\operatorname{div}_{\mathbf{g}}\mathring{\boldsymbol{\sigma}}\,.
\tag{3.48}
$$

We next use the Piola identity and rewrite the divergence terms with respect to the reference configuration. Note that

$$
\begin{aligned}
\mathring{J}\operatorname{div}_{\mathbf{g}}(\nabla^{\mathbf{g}}\mathbf{u}\cdot\mathring{\boldsymbol{\sigma}}) &= \mathring{J}\left(u^a{}_{|b}\mathring{\sigma}^{bc}\right)_{|c}\frac{\partial}{\partial x^a} = \left[u^a{}_{|b}\mathring{J}(\mathring{F}^{-1})^B{}_c\mathring{\sigma}^{bc}\right]_{|B}\frac{\partial}{\partial x^a} \\
&= \left(u^a{}_{|b}\mathring{P}^{bB}\right)_{|B}\frac{\partial}{\partial x^a} = \operatorname{Div}(\nabla^{\mathbf{g}}\mathbf{U}\cdot\mathring{\mathbf{P}})\,.
\end{aligned}
\tag{3.49}
$$

The last term in (3.48) is simplified to read $-(\boldsymbol{\epsilon}:\mathbf{g}^\sharp)\operatorname{Div}\mathring{\mathbf{P}}$. The second term on the right-hand side of (3.48) is simplified as $\mathring{\mathbf{F}}\mathring{\mathbf{P}}^\star\cdot\mathsf{Ric}_{\mathbf{g}}\circ\mathring{\varphi}\cdot\mathbf{U}$. The second divergence term is simplified using the Piola identity as $\mathring{J}\operatorname{div}_{\mathbf{g}}(\mathbb{C}\!:\!\boldsymbol{\epsilon}) = \mathring{J}\left(\mathbb{C}^{abcd}u_{c|d}\right)_{|b}\frac{\partial}{\partial x^a} = \left[\mathring{J}(\mathring{F}^{-1})^B{}_b\mathbb{C}^{abcd}u_{c|d}\right]_{|B}\frac{\partial}{\partial x^a}$. From (3.15), we know that $\mathring{J}(\mathring{F}^{-1})^B{}_b\mathbb{C}^{abcd} = F^a{}_M F^c{}_P F^d{}_Q\mathsf{C}^{BMPQ} = \mathsf{A}^{aBcQ}F^d{}_Q - \mathring{P}^{dB}g^{ac}$. Thus

$$
\left[\mathring{J}(\mathring{F}^{-1})^B{}_b\mathbb{C}^{abcd}u_{c|d}\right]_{|B} = \left(\mathsf{A}^{aBcQ}F^d{}_Q u_{c|d}\right)_{|B} - \left(\mathring{P}^{dB}g^{ac}u_{c|d}\right)_{|B} = \left(\mathsf{A}^{aBcQ}u_{c|Q}\right)_{|B} - \left(\mathring{P}^{dB}u^a{}_{|d}\right)_{|B}.
\tag{3.50}
$$

Hence, $\mathring{J}\operatorname{div}_{\mathbf{g}}(\mathbb{C}\!:\!\boldsymbol{\epsilon}) = \operatorname{Div}(\mathbf{A}\!:\!\nabla\mathbf{U}) - \operatorname{Div}(\nabla\mathbf{U}\cdot\mathring{\mathbf{P}})$. Therefore, $\delta(\operatorname{Div}\mathbf{P}) = \operatorname{Div}(\mathbf{A}\!:\!\nabla\mathbf{U}) + \mathring{\mathbf{F}}\mathring{\mathbf{P}}^\star\cdot\mathsf{Ric}_{\mathbf{g}}\circ\mathring{\varphi}\cdot\mathbf{U}$. In summary, the linearized material balance of linear momentum reads

$$
\operatorname{Div}(\mathbf{A}\!:\!\nabla\mathbf{U}) + \mathring{\mathbf{F}}\mathring{\mathbf{P}}^\star\cdot\mathsf{Ric}_{\mathbf{g}}\circ\mathring{\varphi}\cdot\mathbf{U} + \rho_0\nabla_{\mathbf{U}}\mathring{\mathbf{B}} = \rho_0\left[\ddot{\mathbf{U}} + \nabla_{[\mathbf{U},\mathring{\mathbf{V}}]}\mathring{\mathbf{V}} + \boldsymbol{\mathcal{R}}_{\mathbf{g}}\left(\mathbf{U},\mathring{\mathbf{U}},\mathring{\mathbf{V}}\right)\right].
\tag{3.51}
$$

**Linearization of the balance of angular momentum.** In terms of the first Piola-Kirchhoff stress, the balance of angular momentum in coordinates reads $P^{aA}F^b{}_A = P^{bA}F^a{}_A$. Linearization of this relation reads $\delta P^{aA}\mathring{F}^b{}_A + \mathring{P}^{aA}U^b{}_{|A} = \delta P^{bA}\mathring{F}^a{}_A + \mathring{P}^{bA}U^a{}_{|A}$. Or

$$\delta\mathbf{P}\cdot\mathring{\mathbf{F}}^\star + \mathring{\mathbf{P}}\cdot(\nabla\mathbf{U})^\star = \mathring{\mathbf{F}}\cdot\delta\mathbf{P}^\star + \nabla\mathbf{U}\cdot\mathring{\mathbf{P}}^\star. \tag{3.52}$$

In terms of the second Piola-Kirchhoff stress, $\delta\mathbf{S} = \mathsf{C}:\mathring{\varphi}^*\boldsymbol{\epsilon}$, the balance of angular momentum is equivalent to $\delta\mathbf{S}^\star = \delta\mathbf{S}$, or $\mathsf{C}^{ABCD} = \mathsf{C}^{BACD}$. In terms of the first elasticity tensor, the balance of angular momentum reads

$$\mathsf{A}^{[aA}{}_m{}^M U^m{}_{|M}\mathring{F}^{b]}{}_A + \mathring{P}^{[aA}U^{b]}{}_{|A} = 0. \tag{3.53}$$

We next discuss the invariance of the governing equations of linear elasticity under both time-dependent spatial and time-independent referential changes of coordinates.

## 3.1 Spatial Covariance of Linearized Elasticity

Consider a spatial change of frame (or a coordinate transformation in the current configuration) $\xi_t : \mathcal{S} \to \mathcal{S}$. Under this change of frame, the one-parameter family of deformations $\varphi_{t,\epsilon}$ is transformed to $\varphi'_{t,\epsilon} = \xi_t\circ\varphi_{t,\epsilon} : \mathcal{B}\to\mathcal{S}$. We next study the effect of this change of frame on all the fields and governing equations of linear elasticity. The transformed variation field is defined as

$$\delta\varphi'_t(X) = \mathbf{U}'(X,t) = \frac{d}{d\epsilon}\Big|_{\epsilon=0}\xi_t\circ\varphi_{t,\epsilon}(X) = T\xi_t\cdot\delta\varphi_t = \overset{\xi}{\mathbf{F}}\cdot\mathbf{U}(X,t). \tag{3.54}$$

**Transformation of the linearized velocity, acceleration, and the deformation gradient.** Linearization of the material velocity with respect to the new spatial frame is calculated as $\delta\mathbf{V}'(X,t) = \nabla^{\mathbf{g}'}_{\xi_*\frac{\partial}{\partial\epsilon}}\frac{\partial\varphi'_{t,\epsilon}(X)}{\partial t}\Big|_{\epsilon=0}$, where $\mathbf{g}' = \xi_*\mathbf{g}$. Note that

$$\frac{\partial\varphi'_{t,\epsilon}(X)}{\partial t} = \frac{\partial}{\partial t}\xi_t\circ\varphi_{t,\epsilon}(X) = T\xi_t\cdot\frac{\partial\varphi_{t,\epsilon}(X)}{\partial t} + \frac{\partial\xi_t}{\partial t}\circ\varphi_{t,\epsilon}(X) = \xi_*\frac{\partial\varphi_{t,\epsilon}(X)}{\partial t} + \mathbf{w}_t\circ\varphi_{t,\epsilon}(X), \tag{3.55}$$

where $\mathbf{w}_t$ is the velocity of the change of frame. Thus

$$\delta\mathbf{V}'(X,t) = \left(\nabla_{\xi_*\frac{\partial}{\partial\epsilon}}\xi_*\frac{\partial\varphi_{t,\epsilon}(X)}{\partial t} + \nabla_{\xi_*\frac{\partial}{\partial\epsilon}}\mathbf{w}_t\circ\varphi_{t,\epsilon}(X)\right)\Big|_{\epsilon=0} = \xi_*\left(\nabla_{\frac{\partial}{\partial\epsilon}}\frac{\partial\varphi_{t,\epsilon}(X)}{\partial t} + \nabla_{\mathbf{U}}\mathbf{W}_t\circ\varphi_{t,\epsilon}(X)\right)\Big|_{\epsilon=0}, \tag{3.56}$$

where $\mathbf{W}_t = \xi_t^*\mathbf{w}_t$. Therefore, using (3.2)

$$\begin{aligned}\delta\mathbf{V}'(X,t) &= \xi_*\left(\delta\mathbf{V}(X,t) + \nabla_{\mathbf{U}}\mathbf{W}_t\circ\mathring{\varphi}_t(X)\right)\\ &= \xi_*\left(D_t\mathbf{U}(X,t) + \nabla_{\mathbf{U}}\mathbf{W}_t\circ\mathring{\varphi}_t(X)\right)\\ &= \xi_*\left(\dot{\mathbf{U}}(X,t) + \nabla_{\mathbf{U}}\mathbf{W}_t\circ\mathring{\varphi}_t(X)\right).\end{aligned} \tag{3.57}$$

The transformed linearized acceleration is calculated as

$$\begin{aligned}\delta\mathbf{A}'(X,t) &= \nabla_{\mathring{\mathbf{V}}'}\mathring{\mathbf{V}}' = \nabla_{\xi_*\mathring{\mathbf{V}}+\mathbf{w}\circ\varphi_t}(\xi_*\mathring{\mathbf{V}} + \mathbf{w}\circ\varphi_t)\\ &= \xi_*\left\{\nabla_{\mathring{\mathbf{V}}+\mathbf{W}\circ\varphi_t}(\mathring{\mathbf{V}} + \mathbf{W}\circ\varphi_t)\right\}\\ &= \xi_*\delta\mathbf{A}(X,t) + \xi_*\left(\nabla_{\mathring{\mathbf{V}}}\mathbf{W}\circ\varphi_t + \nabla_{\mathbf{W}\circ\varphi_t}\mathring{\mathbf{V}} + \nabla_{\mathbf{W}\circ\varphi_t}\mathbf{W}\circ\varphi_t\right).\end{aligned} \tag{3.58}$$

We assume that body force is transformed such that $\mathbf{A}'(X,t) - \mathbf{B}'(X,t) = \xi_{t*}(\mathbf{A}(X,t) - \mathbf{B}(X,t))$ [Marsden and Hughes, 1983]. Therefore, $\delta\mathbf{A}(X,t) - \delta\mathbf{B}(X,t) = \xi_{t*}(\delta\mathbf{A}(X,t) - \delta\mathbf{B}(X,t))$. Note that $\mathbf{F}'_{t,\epsilon} = T\varphi'_{t,\epsilon} = T(\xi_t\circ\varphi_{t,\epsilon}) = T\xi_t\cdot T\varphi_{t,\epsilon} = \xi_{t*}T\varphi_{t,\epsilon}$. Therefore

$$\delta\mathbf{F}' = \nabla_{\xi_*\frac{\partial}{\partial\epsilon}}\xi_{t*}T\varphi_{t,\epsilon}\Big|_{\epsilon=0} = \xi_*\nabla_{\frac{\partial}{\partial\epsilon}}T\varphi_{t,\epsilon}\Big|_{\epsilon=0} = \xi_*\delta\mathbf{F} = \xi_*\nabla\mathbf{U} = \overset{\Xi}{\mathbf{F}}\cdot\nabla\mathbf{U}. \tag{3.59}$$

**Transformation of the linearized first and second Piola-Kirchhoff, and Cauchy stresses.** The linearized first and second Piola-Kirchhoff stresses are written as $\delta\mathbf{P} = \mathbf{A}\!:\!\nabla^{\mathbf{g}}\mathbf{U}$, and $\delta\mathbf{S} = \mathbf{C}\!:\!\mathring{\varphi}^*\boldsymbol{\epsilon}$. We know that both $\mathbf{A}$ and $\mathbf{C}$ are tensors and under $\xi_t : \mathcal{S} \to \mathcal{S}$ are transformed as

$$\mathbf{A}' = \xi_{t*}\mathbf{A}, \quad \mathsf{A}'^{a'A}{}_{b'}{}^{B} = \overset{\xi}{F}{}^{a'}{}_{a}(\overset{\xi}{F}{}^{-1})^{b}{}_{b'}\mathsf{A}^{aA}{}_{bB}, \quad \mathbf{C}' = \mathbf{C}. \tag{3.60}$$

Therefore

$$\delta\mathbf{P}' = \xi_{t*}\mathbf{A}\!:\!\nabla^{\xi_{t*}\mathbf{g}}\xi_{t*}\mathbf{U} = \xi_{t*}\left(\mathbf{A}\!:\!\nabla\mathbf{U}\right) = \xi_{t*}\mathbf{P}, \quad \delta\mathbf{S}' = \delta\mathbf{S}. \tag{3.61}$$

In the case of Cauchy stress, $\delta\boldsymbol{\sigma} = \mathbb{C}\!:\!\boldsymbol{\epsilon} - (\mathrm{div}_{\mathbf{g}}\,\mathbf{u})\mathring{\boldsymbol{\sigma}}$, and hence

$$\delta\boldsymbol{\sigma}' = \mathbb{C}'\!:\!\boldsymbol{\epsilon}' - (\mathrm{div}_{\mathbf{g}'}\,\mathbf{u}')\mathring{\boldsymbol{\sigma}}' = \xi_{t*}\mathbb{C}\!:\!\xi_{t*}\boldsymbol{\epsilon} - (\mathrm{div}_{\xi_{t*}\mathbf{g}}\,\xi_{t*}\mathbf{u})\xi_{t*}\mathring{\boldsymbol{\sigma}} = \xi_{t*}\left[\mathbb{C}\!:\!\boldsymbol{\epsilon} - (\mathrm{div}_{\mathbf{g}}\,\mathbf{u})\mathring{\boldsymbol{\sigma}}\right] = \xi_{t*}\delta\boldsymbol{\sigma}. \tag{3.62}$$

Note that the elasticity tensor $\mathbb{C}$ is transformed as $\mathbb{C}'^{a'b'c'd'} = \overset{\xi}{F}{}^{a'}{}_{a}\overset{\xi}{F}{}^{b'}{}_{b}\overset{\xi}{F}{}^{c'}{}_{c}\overset{\xi}{F}{}^{d'}{}_{d}\,\mathbb{C}^{abcd}$.

**Transformation of the linearized balance of linear momentum.** Linearized spatial balance of linear momentum reads $\mathrm{div}_{\mathbf{g}}\left(\mathbb{C}\!:\!\boldsymbol{\epsilon}\right) + \mathrm{div}_{\mathbf{g}}\left(\nabla^{\mathbf{g}}\mathbf{u}\cdot\mathring{\boldsymbol{\sigma}}\right) + \mathring{\boldsymbol{\sigma}}\cdot\mathbf{Ric}_{\mathbf{g}}\cdot\mathbf{u} + \mathring{\rho}(\delta\mathbf{b} - \delta\mathbf{a}) = \mathbf{0}$. Therefore, in the new spatial frame it reads

$$\begin{aligned}
&\mathrm{div}_{\mathbf{g}'}\left(\mathbb{C}'\!:\!\boldsymbol{\epsilon}'\right) + \mathrm{div}_{\mathbf{g}'}\left(\nabla^{\mathbf{g}'}\mathbf{u}'\cdot\mathring{\boldsymbol{\sigma}}'\right) + \mathring{\boldsymbol{\sigma}}'\cdot\mathbf{Ric}'_{\mathbf{g}'}\cdot\mathbf{u}' + \mathring{\rho}'(\delta\mathbf{b}' - \delta\mathbf{a}') \\
&= \mathrm{div}_{\mathbf{g}'}\left(\mathbb{C}'\!:\!\boldsymbol{\epsilon}'\right) + \mathrm{div}_{\xi_*\mathbf{g}}\left(\nabla^{\xi_*\mathbf{g}}\xi_*\mathbf{u}\cdot\xi_*\mathring{\boldsymbol{\sigma}}\right) + \xi_*\mathring{\boldsymbol{\sigma}}\cdot\xi_*\mathbf{Ric}_{\mathbf{g}}\cdot\xi_*\mathbf{u} + (\mathring{\rho}\circ\xi)\xi_*(\delta\mathbf{b} - \delta\mathbf{a}) \\
&= \xi_*\left[\mathrm{div}_{\mathbf{g}}\left(\mathbb{C}\!:\!\boldsymbol{\epsilon}\right) + \mathrm{div}_{\mathbf{g}}\left(\nabla^{\mathbf{g}}\mathbf{u}\cdot\mathring{\boldsymbol{\sigma}}\right) + \mathring{\boldsymbol{\sigma}}\cdot\mathbf{Ric}_{\mathbf{g}}\cdot\mathbf{u} + \mathring{\rho}(\delta\mathbf{b} - \delta\mathbf{a})\right] = \mathbf{0},
\end{aligned} \tag{3.63}$$

i.e., the linearized balance of linear momentum is spatially covariant.

Linearized material balance of linear momentum reads $\mathrm{Div}(\mathbf{A}\!:\!\nabla_0^{\mathbf{g}}\mathbf{U}) + \mathring{\mathbf{F}}\mathring{\mathbf{P}}^{\star}\cdot\mathbf{Ric}_{\mathbf{g}}\circ\mathring{\varphi}\cdot\mathbf{U} + \rho_0(\delta\mathbf{B} - \delta\mathbf{A}) = \mathbf{0}$. In the new spatial frame this reads

$$\begin{aligned}
&\mathrm{Div}'(\mathbf{A}'\!:\!\nabla\mathbf{U}') + \mathring{\mathbf{F}}'\mathring{\mathbf{P}}'^{\star}\cdot\mathbf{Ric}'_{\mathbf{g}'}\circ\mathring{\varphi}'\cdot\mathbf{U}' + \rho_0'(\delta\mathbf{B}' - \delta\mathbf{A}') \\
&= \xi_*\left[\mathrm{Div}(\mathbf{A}\!:\!\nabla\mathbf{U}) + \mathring{\mathbf{F}}\mathring{\mathbf{P}}^{\star}\cdot\mathbf{Ric}_{\mathbf{g}}\circ\mathring{\varphi}\cdot\mathbf{U} + \rho_0(\delta\mathbf{B} - \delta\mathbf{A})\right] = \mathbf{0}.
\end{aligned} \tag{3.64}$$

The proofs of the spatial covariance of the balance of angular momentum and conservation of mass are straightforward.

## 3.2 Material Covariance of Linearized Elasticity

In this section, we find the transformations of the fields and governing equations of linearized elasticity under an arbitrary time-independent material diffeomorphism (change of coordinates in the reference configuration). Consider a referential change of frame (or a coordinate transformation in the reference configuration) $\Xi : \mathcal{B} \to \tilde{\mathcal{B}}$. Under this change of frame, the one-parameter family of deformations $\varphi_{t,\epsilon}$ is transformed to $\tilde{\varphi}_{t,\epsilon} = \varphi_{t,\epsilon}\circ\Xi^{-1} : \mathcal{B} \to \mathcal{S}$. The transformed variation field is defined as

$$\delta\tilde{\varphi}_t(\tilde{X}) = \tilde{\mathbf{U}}(\tilde{X},t) = \frac{d}{d\epsilon}\Big|_{\epsilon=0}\varphi_{t,\epsilon}(\Xi^{-1}(\tilde{X})) = \delta\varphi_t\circ\Xi^{-1}(\tilde{X}) = \mathbf{U}(\Xi^{-1}(\tilde{X}),t). \tag{3.65}$$

Hence, $\delta\tilde{\varphi}_t = \delta\varphi_t\circ\Xi^{-1} = \mathbf{U}\circ\Xi^{-1}$.

**Transformation of the linearized velocity, acceleration, and deformation gradient.** Linearization of the material velocity with respect to the new reference configuration is calculated as

$$\delta\tilde{\mathbf{V}}(\tilde{X},t) = \nabla_{\frac{\partial}{\partial\epsilon}}\frac{\partial\tilde{\varphi}_{t,\epsilon}(\tilde{X})}{\partial t}\Big|_{\epsilon=0} = \nabla_{\frac{\partial}{\partial\epsilon}}\frac{\partial\varphi_{t,\epsilon}(\Xi^{-1}(\tilde{X}))}{\partial t}\Big|_{\epsilon=0} = \delta\mathbf{V}(\Xi^{-1}(\tilde{X}),t). \tag{3.66}$$

Thus, $\delta\tilde{\mathbf{V}} = \delta\mathbf{V}\circ\Xi^{-1} = \dot{\mathbf{U}}\circ\Xi^{-1}$.

Linearization of the material acceleration with respect the new reference configuration is calculated as

$$\begin{aligned}
\delta\tilde{\mathbf{A}}(\tilde{X},t) &= \nabla_{\frac{\partial}{\partial\epsilon}}\nabla_{\frac{\partial}{\partial t}}\frac{\partial\tilde{\varphi}_{t,\epsilon}(\tilde{X})}{\partial t}\Big|_{\epsilon=0} \\
&= \nabla_{\frac{\partial}{\partial t}}\nabla_{\frac{\partial}{\partial\epsilon}}\frac{\partial\varphi_{t,\epsilon}(\Xi^{-1}(\tilde{X}))}{\partial t}\Big|_{\epsilon=0} + \nabla_{[\frac{\partial}{\partial\epsilon},\frac{\partial}{\partial t}]}\frac{\partial\varphi_{t,\epsilon}(\Xi^{-1}(\tilde{X}))}{\partial t}\Big|_{\epsilon=0} + \boldsymbol{\mathcal{R}}_{\mathbf{g}}\left(\frac{\partial}{\partial\epsilon},\frac{\partial}{\partial t}\right)\frac{\partial\varphi_{t,\epsilon}(\Xi^{-1}(\tilde{X}))}{\partial t}\Big|_{\epsilon=0} \\
&= \nabla_{\frac{\partial}{\partial t}}\nabla_{\frac{\partial}{\partial t}}\frac{\partial\varphi_{t,\epsilon}(\Xi^{-1}(\tilde{X}))}{\partial\epsilon}\Big|_{\epsilon=0} + \nabla_{[\tilde{\mathbf{U}},\tilde{\mathring{\mathbf{V}}}]}\tilde{\mathring{\mathbf{V}}} + \boldsymbol{\mathcal{R}}_{\mathbf{g}}\left(\tilde{\mathbf{U}},\tilde{\mathring{\mathbf{V}}},\tilde{\mathring{\mathbf{V}}}\right) \\
&= \delta\mathbf{A}(\Xi^{-1}(\tilde{X}),t),
\end{aligned} \tag{3.67}$$

i.e., $\delta\tilde{\mathbf{A}} = \delta\mathbf{A}\circ\Xi^{-1}$.

Deformation gradient for the motion $\tilde{\varphi}_{t,\epsilon}$ is written as $\tilde{\mathbf{F}}_{t,\epsilon} = T\tilde{\varphi}_{t,\epsilon} = T(\varphi_{t,\epsilon}\circ\Xi^{-1}) = T\varphi_{t,\epsilon}\circ(T\Xi)^{-1} = \Xi_{t*}T\varphi_{t,\epsilon}$. Therefore

$$\delta\tilde{\mathbf{F}} = \nabla_{\Xi_*\frac{\partial}{\partial\epsilon}}\Xi_{t*}T\varphi_{t,\epsilon}\Big|_{\epsilon=0} = \Xi_*\nabla_{\frac{\partial}{\partial\epsilon}}T\varphi_{t,\epsilon}\Big|_{\epsilon=0} = \Xi_*\delta\mathbf{F} = \Xi_*\nabla\mathbf{U}. \tag{3.68}$$

In components, $(\delta\tilde{F})^a{}_{\tilde{A}} = (\bar{\bar{F}}^{-1})^A{}_{\tilde{A}}\,U^a{}_{|A}$.

**Transformation of the linearized first Piola-Kirchhoff stress and the first elasticity tensor.** With respect to the new reference configuration (cf. (2.13))

$$\tilde{W}_\epsilon = W(\tilde{X},\tilde{\mathbf{F}}_\epsilon,\tilde{\mathbf{G}},\mathbf{g}\circ\tilde{\varphi}_\epsilon,\tilde{\boldsymbol{\zeta}}_1,\ldots,\tilde{\boldsymbol{\zeta}}_n) = W(X,\Xi_*\mathbf{F}_\epsilon,\Xi_*\mathbf{G},\mathbf{g}\circ\varphi_\epsilon\circ\Xi^{-1},\Xi_*\boldsymbol{\zeta}_1,\ldots,\Xi_*\boldsymbol{\zeta}_n) = \Xi_*W_\epsilon. \tag{3.69}$$

The above relation, in particular, implies that $\delta\tilde{W} = \Xi_*\delta W$. Note that $\delta W = \frac{\partial W}{\partial\mathbf{F}}\cdot\delta\mathbf{F}$. Similarly

$$\delta\tilde{W} = \frac{\partial\tilde{W}}{\partial\tilde{\mathbf{F}}}\cdot\delta\tilde{\mathbf{F}} = \frac{\partial\tilde{W}}{\partial\tilde{\mathbf{F}}}\cdot\Xi_*\delta\mathbf{F} = \Xi_*\left(\Xi^*\frac{\partial\tilde{W}}{\partial\tilde{\mathbf{F}}}\cdot\delta\mathbf{F}\right). \tag{3.70}$$

Hence

$$\Xi^*\frac{\partial\tilde{W}}{\partial\tilde{\mathbf{F}}}\cdot\nabla\mathbf{U} = \frac{\partial W}{\partial\mathbf{F}}\cdot\nabla\mathbf{U}, \quad \forall\mathbf{U}. \tag{3.71}$$

This implies that

$$\frac{\partial\tilde{W}}{\partial\tilde{\mathbf{F}}} = \Xi_*\frac{\partial W}{\partial\mathbf{F}}. \tag{3.72}$$

Therefore

$$\delta\tilde{\mathbf{P}} = \mathbf{g}^\sharp\nabla_{\frac{\partial}{\partial\epsilon}}\frac{\partial\tilde{W}_\epsilon}{\partial\tilde{\mathbf{F}}_\epsilon}\Bigg|_{\epsilon=0} = \mathbf{g}^\sharp\nabla_{\frac{\partial}{\partial\epsilon}}\Xi_*\frac{\partial W_\epsilon}{\partial\mathbf{F}_\epsilon}\Bigg|_{\epsilon=0} = \Xi_*\left(\mathbf{g}^\sharp\frac{\partial^2 W}{\partial\mathring{\mathbf{F}}\partial\mathring{\mathbf{F}}}\cdot\nabla\mathbf{U}\right). \tag{3.73}$$

Hence, $\delta\tilde{\mathbf{P}} = \tilde{\mathsf{A}}\cdot\nabla\tilde{\mathbf{U}} = \Xi_*\mathsf{A}\cdot\Xi_*(\nabla\mathbf{U})$. In particular, the first elasticity tensor is transformed as $\tilde{\mathsf{A}}^{a\tilde{A}}{}_b{}^{\tilde{B}} = \bar{\bar{F}}^{\tilde{A}}{}_A\bar{\bar{F}}^{\tilde{B}}{}_B\;\mathsf{A}^{aA}{}_b{}^B$.

**Transformation of the second Piola-Kirchhoff and the second elasticity tensor.** Note that the second elasticity tensor for hyperelastic solids (for which an energy function exists) is defined in components as $\mathsf{C}^{ABCD} = 4\frac{\partial^2 W}{\partial C_{AB}\partial C_{CD}}$. Under a referential change of coordinates $\mathsf{C}$ is transformed to $\tilde{\mathsf{C}} = \Xi_*\mathsf{C}$, which in components reads

$$\tilde{\mathsf{C}}^{\tilde{A}\tilde{B}\tilde{C}\tilde{D}} = \bar{\bar{F}}^{\tilde{A}}{}_A\bar{\bar{F}}^{\tilde{B}}{}_B\bar{\bar{F}}^{\tilde{C}}{}_C\bar{\bar{F}}^{\tilde{D}}{}_D\;\mathsf{C}^{ABCD}\circ\Xi^{-1}. \tag{3.74}$$

It then immediately follows that if $\mathsf{C}$ possesses the minor ($\mathsf{C}^{ABCD} = \mathsf{C}^{BACD} = \mathsf{C}^{ABDC}$) and major ($\mathsf{C}^{ABCD} = \mathsf{C}^{CDAB}$) symmetries, so does $\tilde{\mathsf{C}}$, i.e., the major and minor symmetries of the elasticity tensor are preserved under a material diffeomorphism. The symmetry group $\mathrm{Sym}_{\mathbf{G}}(\mathbf{T})$ of an $m$-order tensor field $\mathbf{T}$ is the subgroup of $\mathbf{G}$-orthogonal transformations defined as

$$\langle \mathbf{Q} \rangle_m \mathbf{T} = \mathbf{T}\,, \quad \forall\, \mathbf{Q} \in \mathrm{Sym}_{\mathbf{G}}(\mathbf{T}) \leqslant \mathrm{Orth}\,(\mathbf{G})\,. \tag{3.75}$$

The space of elasticity tensors $\mathbb{E}\mathrm{la}$ (consisting of all those fourth-order tensors satisfying the major and minor symmetries) may be endowed with an equivalence relation such that

$$\mathsf{C}_1 \sim \mathsf{C}_2 \iff \exists \mathbf{Q} \in \mathrm{Orth}\,(\mathbf{G}): \quad \langle \mathbf{Q} \rangle_4 \mathsf{C}_2 = \mathsf{C}_1\,. \tag{3.76}$$

Equivalently[15]

$$\mathsf{C}_1 \sim \mathsf{C}_2 \iff \exists \mathbf{Q} \in \mathrm{Orth}\,(\mathbf{G}): \quad \mathrm{Sym}_{\mathbf{G}}(\mathsf{C}_1) = \mathbf{Q}\, \mathrm{Sym}_{\mathbf{G}}(\mathsf{C}_2)\, \mathbf{Q}^{\mathsf{T}}\,. \tag{3.77}$$

That is, two elasticity tensors are equivalent if and only if their symmetry groups are conjugate subgroups of $SO(3)$. Thus, by an abuse of notation, one may write $\mathsf{C}_1 \sim \mathsf{C}_2$ if and only if $\mathrm{Sym}_{\mathbf{G}}(\mathsf{C}_1) \sim \mathrm{Sym}_{\mathbf{G}}(\mathsf{C}_2)$, where use was made of the fact that conjugacy is also an equivalence relation. Under this equivalence relation, $\mathbb{E}\mathrm{la}$ is divided into eight equivalence classes, known as *symmetry classes*, namely triclinic, monoclinic, trigonal, orthotropic, tetragonal, cubic, transversely isotropic, and isotropic (for proofs,[16] see [Forte and Vianello, 1996, Chadwick et al., 2001]). We note that under the material diffeomorphsim the elasticity tensor is transformed such that $\tilde{\mathring{\mathsf{C}}} = \Xi_* \mathring{\mathsf{C}}$, where $\mathring{\mathsf{C}}$ is the linearized elasticity tensor with respect to the reference motion $\mathring{\varphi}_t$. Therefore, from (3.74), it follows that $\tilde{\mathring{\mathsf{C}}} = \langle \bar{\bar{\mathbf{F}}} \rangle_4 (\mathring{\mathsf{C}} \circ \Xi^{-1})$. It is straightforward to verify that [Lu and Papadopoulos, 2000]

$$\mathbf{Q} \in \mathrm{Orth}(\mathbf{G}) \iff \Xi_* \mathbf{Q} = \bar{\bar{\mathbf{F}}} \mathbf{Q} \bar{\bar{\mathbf{F}}}^{-1} \in \mathrm{Orth}(\tilde{\mathbf{G}})\,, \tag{3.78}$$

where $\tilde{\mathbf{G}} = \Xi_* \mathbf{G}$. Using the properties of the Kronecker product (see, e.g., [Zheng and Spencer, 1993b]), one concludes that

$$\mathrm{Sym}_{\tilde{\mathbf{G}}}(\tilde{\mathring{\mathsf{C}}}) = \bar{\bar{\mathbf{F}}}\, \mathrm{Sym}_{\mathbf{G}}(\mathring{\mathsf{C}})\, \bar{\bar{\mathbf{F}}}^{-1}\,. \tag{3.79}$$

That is, the symmetry groups of $\tilde{\mathring{\mathsf{C}}}$ and $\mathring{\mathsf{C}}$ are conjugate, and hence, isomorphic.[17] It then follows from (3.77) and (3.79) that

$$\begin{aligned} \mathring{\mathsf{C}}_1 \sim \mathring{\mathsf{C}}_2 &\iff \exists \mathbf{Q} \in \mathrm{Orth}(\mathbf{G}) : \mathrm{Sym}_{\mathbf{G}}(\mathring{\mathsf{C}}_1) = \mathbf{Q}\, \mathrm{Sym}_{\mathbf{G}}(\mathring{\mathsf{C}}_2)\, \mathbf{Q}^{\mathsf{T}} \\ &\iff \exists \tilde{\mathbf{Q}} \in \mathrm{Orth}(\tilde{\mathbf{G}}) : \mathrm{Sym}_{\tilde{\mathbf{G}}}(\tilde{\mathring{\mathsf{C}}}_1) = \tilde{\mathbf{Q}}\, \mathrm{Sym}_{\tilde{\mathbf{G}}}(\tilde{\mathring{\mathsf{C}}}_2)\, \tilde{\mathbf{Q}}^{\mathsf{T}} \\ &\iff \tilde{\mathring{\mathsf{C}}}_1 \sim \tilde{\mathring{\mathsf{C}}}_2\,, \end{aligned} \tag{3.80}$$

where $\tilde{\mathbf{Q}} = \Xi_* \mathbf{Q}$. Therefore, $\mathring{\mathsf{C}}_1 \sim \mathring{\mathsf{C}}_2 \iff \tilde{\mathring{\mathsf{C}}}_1 \sim \tilde{\mathring{\mathsf{C}}}_2$, i.e., there is a one-to-one correspondence between the symmetry classes of the elasticity tensors in the initial and transformed reference configurations. It is important to notice that the symmetry of a tensor explicitly depends on the metric. In particular, a tensor, e.g., the elasticity tensor, may belong to a different type of symmetry class with respect to a different metric under consideration. Note that if $\bar{\bar{\mathbf{F}}}^{-\star} \mathbf{G} \bar{\bar{\mathbf{F}}}^{-1} = \mathbf{G}$, i.e., if $\bar{\bar{\mathbf{F}}}$ is $\mathbf{G}$-orthogonal ($\in \mathrm{Orth}\,(\mathbf{G})$), then the symmetry class is trivially unaffected under a change of reference configuration. Mazzucato and Rachele [2006] characterized the orbits of different symmetry classes of the elasticity tensor (i.e., $\psi^* \mathsf{C}$, where $\psi$ is a diffeomorphism in the material manifold) with respect to the Euclidean metric under a change of reference configuration that fixes the boundary to the first-order. In particular, they found that the orbit of isotropic materials consists of some orthotropic and some transversely isotropic materials, in addition to isotropic materials. As they describe the material symmetry groups of $\mathsf{C}$ and $\tilde{\mathsf{C}}$ with respect to the Euclidean metric for both configurations, they are implicitly characterizing the symmetry group of the transformed elasticity tensor $\tilde{\mathsf{C}}$ with respect to the untransformed metric $\mathbf{G}$, i.e., $\mathrm{Sym}_{\mathbf{G}}(\tilde{\mathsf{C}})$.

[15] Note that (3.76) and (3.77) imply that two elasticity tensors are equivalent (represent the same type of material anisotropy) if and only if there exists an orthogonal transformation such that under its action the two elasticity tensors (or their symmetry groups) coincide.

[16] Also, see [Olive and Auffray, 2013] for the determination of the symmetry classes of an even-order tensor space.

[17] The isomorphism is trivially given by the conjugacy relations as follows. Let $\phi : \mathrm{Sym}_{\mathbf{G}}(\mathring{\mathsf{C}}) \to \mathrm{Sym}_{\tilde{\mathbf{G}}}(\tilde{\mathring{\mathsf{C}}})$ and let $\mathbf{H} \in \mathrm{Sym}_{\mathbf{G}}(\mathring{\mathsf{C}})$. Then, $\phi(\mathbf{H}) = \bar{\bar{\mathbf{F}}} \mathbf{H} \bar{\bar{\mathbf{F}}}^{-1} \in \mathrm{Sym}_{\tilde{\mathbf{G}}}(\tilde{\mathring{\mathsf{C}}})$. It is straightforward to see that $\phi(\mathbf{H}_1\mathbf{H}_2) = \phi(\mathbf{H}_1)\phi(\mathbf{H}_2)$ for $\mathbf{H}_1, \mathbf{H}_2 \in \mathrm{Sym}_{\mathbf{G}}(\mathring{\mathsf{C}})$. Also, it is straightforward to see that $\phi$ is one-to-one, and thus, an isomorphism.

Next, using (3.79) and (3.80), we show that symmetry classes are preserved under a change of reference configuration. In other words, the elasticity tensor $\mathring{\mathsf{C}}$ belongs to a given symmetry class if and only if $\tilde{\mathring{\mathsf{C}}}$ belongs to the same type of the symmetry class in the transformed reference configuration. For instance, let $\mathring{\mathsf{C}}$ belong to the transversely isotropic symmetry class in the initial reference configuration at a material point $X$ with the unit vector $\mathbf{N} \in T_X\mathcal{B}$ identifying the material preferred direction. It follows that the symmetry group of $\mathring{\mathsf{C}}$ is given by

$$\mathrm{Sym}_{\mathbf{G}}(\mathring{\mathsf{C}}) = \left\{ \mathbf{Q} \in \mathrm{Orth}\,(\mathbf{G}) : \mathbf{Q}\mathbf{N} = \pm\mathbf{N} \right\}. \tag{3.81}$$

From (3.79)

$$\mathrm{Sym}_{\tilde{\mathbf{G}}}(\tilde{\mathring{\mathsf{C}}}) = \left\{ \tilde{\mathbf{Q}} \in \mathrm{Orth}(\tilde{\mathbf{G}}) : \tilde{\mathbf{Q}} = \bar{\bar{\mathbf{F}}}\mathbf{Q}\bar{\bar{\mathbf{F}}}^{-1},\ \mathbf{Q} \in \mathrm{Sym}_{\mathbf{G}}(\mathring{\mathsf{C}}) \right\}. \tag{3.82}$$

The material preferred direction is transformed to $\tilde{\mathbf{N}} = \bar{\bar{\mathbf{F}}}\mathbf{N}$. Clearly, employing (3.81) and (3.82), $\forall \tilde{\mathbf{Q}} \in \mathrm{Sym}_{\tilde{\mathbf{G}}}(\tilde{\mathring{\mathsf{C}}})$ one has $\tilde{\mathbf{Q}}\tilde{\mathbf{N}} = \pm\tilde{\mathbf{N}}$. Moreover, if $\tilde{\mathbf{Q}} \in \mathrm{Orth}(\tilde{\mathbf{G}})$ such that $\tilde{\mathbf{Q}}\tilde{\mathbf{N}} = \pm\tilde{\mathbf{N}}$, then, from (3.78), there exists $\mathbf{Q} = \bar{\bar{\mathbf{F}}}^{-1}\tilde{\mathbf{Q}}\bar{\bar{\mathbf{F}}} \in \mathrm{Orth}(\tilde{\mathbf{G}})$, for which one concludes that $\mathbf{Q}\mathbf{N} = \pm\mathbf{N}$, i.e., $\mathbf{Q} \in \mathrm{Sym}_{\mathbf{G}}(\mathring{\mathsf{C}})$. Thus

$$\mathrm{Sym}_{\tilde{\mathbf{G}}}(\tilde{\mathring{\mathsf{C}}}) = \left\{ \tilde{\mathbf{Q}} \in \mathrm{Orth}(\tilde{\mathbf{G}}) : \tilde{\mathbf{Q}}\tilde{\mathbf{N}} = \pm\tilde{\mathbf{N}} \right\}, \tag{3.83}$$

and hence, $\tilde{\mathring{\mathsf{C}}}$ belongs to the transversely isotropic symmetry class in the transformed reference configuration. The generalization of this proof to the orthotropic case is immediate. One only needs to consider three $\mathbf{G}$-orthonormal vectors $\mathbf{N}_1$, $\mathbf{N}_2$, and $\mathbf{N}_3$ specifying the orthotropic axes and define the symmetry group in the orthotropic case as $\mathbf{Q} \in \mathrm{Orth}\,(\mathbf{G})$ such that $\mathbf{Q}\mathbf{N}_i = \pm\mathbf{N}_i$, $i = 1, 2, 3$. One can similarly prove that other symmetry classes are preserved under material diffeomorphisms as well. For the isotropic case, it immediately follows from (3.78) and (3.79) that $\mathrm{Sym}_{\mathbf{G}}(\mathring{\mathsf{C}}) = \mathrm{Orth}\,(\mathbf{G}) \iff \mathrm{Sym}_{\tilde{\mathbf{G}}}(\tilde{\mathring{\mathsf{C}}}) = \mathrm{Orth}(\tilde{\mathbf{G}})$. Therefore, we have proved the following result.

**Proposition 3.2.** *The elasticity tensor $\mathring{\mathsf{C}}$ in the reference configuration $(\mathcal{B}, \mathbf{G})$ with respect to the reference motion $\mathring{\varphi} : \mathcal{B} \to \mathcal{S}$ belongs to a given symmetry class if and only if $\tilde{\mathring{\mathsf{C}}}$, which is the elasticity tensor in the transformed reference configuration $(\tilde{\mathcal{B}}, \Xi_*\mathbf{G})$ with respect to the reference motion $\tilde{\mathring{\varphi}} : \tilde{\mathcal{B}} \to \mathcal{S}$, belongs to the same type of the symmetry class.*

Using (3.12), one can write

$$\delta\tilde{\mathbf{S}} = 2\frac{\partial^2 \tilde{W}_\epsilon}{\partial \tilde{\mathbf{C}}^\flat_\epsilon \partial \tilde{\mathbf{C}}^\flat_\epsilon} : \frac{d}{d\epsilon}\tilde{\mathbf{C}}^\flat_\epsilon \bigg|_{\epsilon=0} = \Xi_* \left( 4\frac{\partial^2 \hat{W}}{\partial \mathring{\mathbf{C}}^\flat \partial \mathring{\mathbf{C}}^\flat} : \mathring{\varphi}^* \boldsymbol{\epsilon} \right) = \tilde{\mathsf{C}} : \tilde{\mathring{\varphi}}^* \boldsymbol{\epsilon}, \tag{3.84}$$

where $\tilde{\mathsf{C}} = \Xi_*\mathsf{C}$ and $\tilde{\mathring{\varphi}}^*\boldsymbol{\epsilon} = \Xi_* \circ \mathring{\varphi}^*\boldsymbol{\epsilon} = \Xi_*(\mathring{\varphi}^*\boldsymbol{\epsilon})$, and thus, $\delta\tilde{\mathbf{S}} = \Xi_*\mathsf{C} : \Xi_*(\mathring{\varphi}^*\boldsymbol{\epsilon})$.

**Transformation of the linearized balances of linear and angular momenta.** The linearized spatial balance of linear momentum with respect to the new reference configuration reads

$$\mathrm{div}_{\tilde{\mathbf{g}}}(\tilde{\mathbb{C}} : \tilde{\boldsymbol{\epsilon}}) + \mathrm{div}_{\tilde{\mathbf{g}}}(\nabla^{\mathbf{g}}\tilde{\mathbf{u}} \cdot \tilde{\mathring{\boldsymbol{\sigma}}}) + \tilde{\mathring{\boldsymbol{\sigma}}} \cdot \widetilde{\mathsf{Ric}}_{\mathbf{g}} \cdot \tilde{\mathbf{u}} + (\mathring{\rho}\nabla^{\mathbf{g}}_{\mathbf{u}}\mathring{\mathbf{b}}) \circ \Xi^{-1} = \mathring{\rho} \left[ \ddot{\mathbf{U}} \circ \mathring{\varphi}_t^{-1} + \nabla^{\mathbf{g}}_{[\mathbf{u}, \mathring{\mathbf{v}}]}\mathring{\mathbf{v}} + \boldsymbol{\mathcal{R}}_{\mathbf{g}}(\mathbf{u}, \mathring{\mathbf{v}}, \mathring{\mathbf{v}}) \right] \circ \Xi^{-1}. \tag{3.85}$$

Note that $\tilde{\mathbb{C}} = \frac{1}{\mathring{J}}\tilde{\mathring{\varphi}}_*\tilde{\mathsf{C}} = \frac{1}{\mathring{J}}(\mathring{\varphi} \circ \Xi^{-1})_*\Xi_*\mathsf{C} = \frac{1}{\mathring{J}}\mathring{\varphi}_*\mathsf{C} = \mathbb{C}$. Knowing that all the spatial tensors transform like a scalar, i.e., $\tilde{\boldsymbol{\epsilon}} = \boldsymbol{\epsilon} \circ \Xi^{-1}$, $\tilde{\mathbf{g}} = \mathbf{g} \circ \Xi^{-1}$, $\widetilde{\mathsf{Ric}}_{\mathbf{g}} = \mathsf{Ric}_{\mathbf{g}} \circ \Xi^{-1}$, $\tilde{\mathbf{u}} = \mathbf{u} \circ \Xi^{-1}$, and $\tilde{\mathring{\boldsymbol{\sigma}}} = \mathring{\boldsymbol{\sigma}} \circ \Xi^{-1}$, one concludes that the linearized spatial balance of linear momentum is materially covariant.

Similarly, the linearized material balance of linear momentum transforms like a scalar, i.e.,

$$\begin{aligned}&\widetilde{\mathrm{Div}}(\tilde{\mathbf{A}}:\nabla\tilde{\mathbf{U}}) + \tilde{\mathring{\mathbf{F}}}\tilde{\mathring{\mathbf{P}}}^{\star}\cdot\widetilde{\mathbf{Ric}}_{\mathbf{g}}\circ\tilde{\mathring{\varphi}}\cdot\tilde{\mathbf{U}} + \tilde{\rho}_0\nabla_{\tilde{\mathbf{U}}}\tilde{\mathring{\mathbf{B}}} - \tilde{\rho}_0\left[\ddot{\tilde{\mathbf{U}}} + \nabla_{[\tilde{\mathbf{U}},\tilde{\mathring{\mathbf{V}}}]}\tilde{\mathring{\mathbf{V}}} + \tilde{\boldsymbol{\mathcal{R}}}_{\mathbf{g}}(\tilde{\mathbf{U}},\tilde{\mathring{\mathbf{U}}},\tilde{\mathring{\mathbf{V}}})\right] \\ &= \left\{\mathrm{Div}(\mathbf{A}:\nabla\mathbf{U}) + \mathring{\mathbf{F}}\mathring{\mathbf{P}}^{\star}\cdot\mathbf{Ric}_{\mathbf{g}}\circ\mathring{\varphi}\cdot\mathbf{U} + \rho_0\nabla_{\mathbf{U}}\mathring{\mathbf{B}} - \rho_0\left[\ddot{\mathbf{U}} + \nabla_{[\mathbf{U},\mathring{\mathbf{V}}]}\mathring{\mathbf{V}} + \boldsymbol{\mathcal{R}}_{\mathbf{g}}(\mathbf{U},\mathring{\mathbf{U}},\mathring{\mathbf{V}})\right]\right\}\circ\Xi^{-1} = \mathbf{0}.\end{aligned} \tag{3.86}$$

The balance of angular momentum transforms like a scalar as well.

**Wave equation and its spatial and material covariance.** Let us assume that $\mathbf{u}(x,t) = \bar{\mathbf{u}}(x)e^{-\mathrm{i}\omega t}$, where $\omega$ is the frequency. For a Euclidean ambient space, the wave equation is written as $\mathrm{div}_{\mathbf{g}}(\mathbb{C}:\bar{\boldsymbol{\epsilon}}) + \mathrm{div}_{\mathbf{g}}(\nabla^{\mathbf{g}}\bar{\mathbf{u}}\cdot\mathring{\boldsymbol{\sigma}}) + \mathring{\rho}\nabla^{\mathbf{g}}_{\bar{\mathbf{u}}}\mathring{\mathbf{b}} = \mathring{\rho}\left[-\omega^2\bar{\mathbf{u}} + \nabla^{\mathbf{g}}_{[\bar{\mathbf{u}},\mathring{\mathbf{v}}]}\mathring{\mathbf{v}}\right]$, where $2\bar{\boldsymbol{\epsilon}} = \nabla^{\mathbf{g}}\bar{\mathbf{u}}^{\flat} + (\nabla^{\mathbf{g}}\bar{\mathbf{u}}^{\flat})^{\star}$. Similarly, in material form $\mathbf{U}(X,t) = \bar{\mathbf{U}}(X)e^{-\mathrm{i}\omega t}$, and the wave equation in material form reads

$$\mathrm{Div}(\mathbf{A}:\nabla\bar{\mathbf{U}}) + \rho_0\nabla_{\bar{\mathbf{U}}}\mathring{\mathbf{B}} = \rho_0\left[-\omega^2\bar{\mathbf{U}} + \nabla_{[\bar{\mathbf{U}},\mathring{\mathbf{V}}]}\mathring{\mathbf{V}}\right]. \tag{3.87}$$

If the reference motion is static, i.e., $\mathring{\mathbf{V}} = \mathbf{0}$ and body force is ignored, the spatial and material wave equations are simplified to read

$$\begin{aligned}&\mathrm{div}_{\mathbf{g}}(\mathbb{C}:\nabla^{\mathbf{g}}\bar{\mathbf{u}}^{\flat}) + \mathrm{div}_{\mathbf{g}}(\nabla^{\mathbf{g}}\bar{\mathbf{u}}\cdot\mathring{\boldsymbol{\sigma}}) + \mathring{\rho}\omega^2\bar{\mathbf{u}} = \mathbf{0}, \\ &\mathrm{Div}(\mathbf{A}:\nabla\bar{\mathbf{U}}) + \rho_0\omega^2\bar{\mathbf{U}} = \mathbf{0}.\end{aligned} \tag{3.88}$$

Note that as a consequence of the spatial and material covariance of the balance of linear momentum and its linearization, the wave equation is form-invariant under arbitrary spatial coordinate transformations and arbitrary time-independent referential coordinate transformations.

# 4 A Mathematical Formulation of the Problem of Cloaking a Cavity in Nonlinear and Linearized Elastodynamics

Suppose an object is to be hidden from elastic waves. This object lies in a cavity inside an elastic body. The hole needs to be reinforced by a cloak, which is a layer of specially designed inhomogeneous and anisotropic material that will deflect away any incoming elastic waves from the cloaked object. For an observer away from the hole the elastic waves are passing through the body as if there was no hole. Let us assume that the body is homogeneous and is made of an isotropic material. The idea of elastodynamics transformation cloaking is to first map the stress-free body in its reference configuration to a corresponding homogeneous and isotropic body in its stress-free reference configuration. We assume that the homogeneous and isotropic transformed body (virtual body) has a very small hole of radius $\epsilon$ ($\epsilon \to 0$). Consider a map that shrinks the hole to a very small hole and is the identity outside the cloak. Note that there are many such mappings. Next, the important requirement is that the physical body with the hole and the homogeneous and isotropic virtual body must have identical current configurations outside the cloak. In other words, to an elastic wave the two bodies are identical outside the cloaking region. Inside the cloak we will impose certain requirements. The last step is to check if these requirements are enough to specify the elastic properties of the cloak, its mass density, and the external loads and the boundary conditions in the virtual structure, or if they result in an overdetermined system with no solution. One should also check if all the balance laws are satisfied in both the virtual and physical bodies. To formulate this problem the understanding of the transformation properties of the governing equations of nonlinear and linearized elasticity that we established in the previous two sections is crucial.

## 4.1 Nonlinear Elastodynamics Transformation Cloaking

Let us consider a body $\mathcal{B}$ with a hole $\mathcal{H}$ (see Fig.4). An object is placed inside $\mathcal{H}$ and needs to be hidden from elastic waves. The hole is reinforced by a cloak $\mathcal{C}$, which we can assume is an annulus (or spherical shell). The elastic properties and mass density of $\mathcal{C}$ are, in general, inhomogeneous and anisotropic. For the sake of simplicity and without loss of generality, let us assume that in $\mathcal{B} \setminus \mathcal{C}$ the body is homogeneous and isotropic. This means that the mass density $\rho_0$ is a constant and the body has an energy function $W = W(I_1, I_2, I_3)$, where $I_i$ $(i = 1, 2, 3)$ are the principal invariants of the left (or right) Cauchy-Green strain. Motion of $\mathcal{B}$ is represented by a map $\varphi_t : \mathcal{B} \to \mathcal{S}$ in Fig.4. A spatial diffeomorphism leaves the governing equations form invariant. However, spatial diffeomorphisms (spatial coordinate transformations) would not be useful in designing a cloak because a change of spatial coordinates more or less corresponds to the same body as seen through a warped lens — but the material is the same. The cloaking transformation is assumed to be a time-independent map $\Xi : \mathcal{B} \to \tilde{\mathcal{B}}$ such that the annulus $\mathcal{C}$ is transformed to a disk with a very small hole (or a spherical ball with a very small hole in 3D). The mapping $\Xi$ is assumed to be the identity in $\mathcal{B} \setminus \mathcal{C}$. It is also assumed that the virtual body has the uniform and isotropic mechanical properties of $\mathcal{B} \setminus \mathcal{C}$.

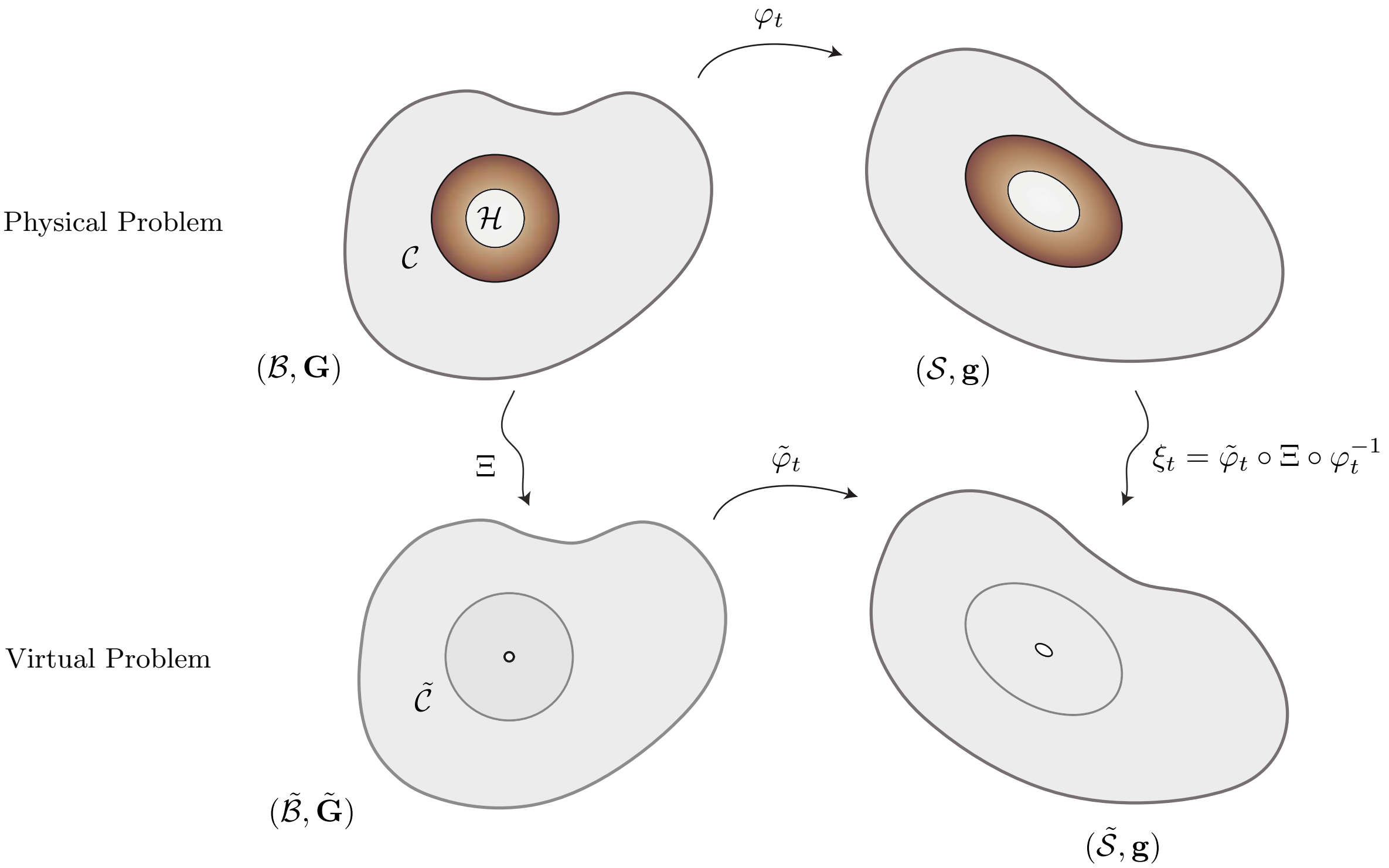

Figure 4: *A cloaking transformation $\Xi$ transforms a body with a hole $\mathcal{H}$ to another body with an infinitesimal hole that is homogeneous and isotropic. The cloaking transformation is defined to be the identity map outside the cloak $\mathcal{C}$. Note that $\Xi$ is not a referential change of coordinates and $\xi$ is not a spatial change of coordinates.*

One may be tempted to think that the cloak can be designed using a referential change of frame (coordinate transformation). We have shown that the governing equations of both nonlinear and linearized elasticity are spatially and referentially covariant. In other words, the governing equations of the same body are invariant under arbitrary time-dependent coordinate transformations in the current configuration and arbitrary time-independent coordinate transformations in the reference configuration. This, however, is not useful for cloaking applications, because a change of the material frame does not lead to a new elastic body. In other words, the two bodies $(\mathcal{B}, \mathbf{G})$ and $(\Xi(\mathcal{B}), \Xi_* \mathbf{G})$ are isometric and essentially the same elastic body with the same mechanical response. One instead needs two different elastic bodies that cannot be distinguished by an elastic wave in their current configurations outside the cloak. Motion of the virtual (homogeneous and isotropic) body is represented by $\tilde{\varphi}_t$ as is schematically shown in Fig.4. The spatial configurations of the two bodies with their corresponding deformations $\varphi_t$ and $\tilde{\varphi}_t$ are required to be identical outside the cloaking region, i.e., in $\mathcal{B} \setminus \mathcal{C}$. In particular, this implies that any elastic measurements made in the current configurations of the two bodies outside the cloak are identical, and hence, the two bodies cannot be distinguished by an observer located in $\mathcal{B} \setminus \mathcal{C}$. Moreover, in the virtual body $\tilde{\mathcal{B}}$, the influence of the hole is negligible as $\mathcal{H}$ in $\mathcal{B}$ is mapped to an infinitesimal hole in $\tilde{\mathcal{B}}$. It is important to notice that our approach does not impose any restrictions on the size of the hole, and thus, that of the concealment as illustrated in Fig.4.

Let us consider a body (physical body) that has a material manifold $(\mathcal{B}, \mathbf{G})$ and is in a time-dependent current configuration $\varphi_t(\mathcal{B})$. Suppose that the stress-free reference configuration of the body is in a one-to-one correspondence with that of another body $\tilde{\mathcal{B}}$ (virtual body) in its reference configuration $(\tilde{\mathcal{B}}, \tilde{\mathbf{G}})$ (see Fig.4). Let us denote the bijection between the two bodies by $\Xi$. Hence, $\Xi : \mathcal{B} \to \tilde{\mathcal{B}}$ is a diffeomorphism. We assume that the two stress-free bodies are embedded in the Euclidean space, and hence, $\mathbf{G}$ and $\tilde{\mathbf{G}}$ are their corresponding induced Euclidean metrics. This immediately implies that $\Xi$ is not a simple change of material frame (referential coordinate transformation) as $\tilde{\mathbf{G}} \neq \Xi_* \mathbf{G}$, in general. The boundary-value problems related to the motions $\varphi_t : \mathcal{B} \to \mathcal{S}$ and $\tilde{\varphi}_t : \tilde{\mathcal{B}} \to \mathcal{S}$ are called the physical and virtual problems, respectively. The deformation gradients corresponding to the physical and virtual problems are denoted by $\mathbf{F} = T\varphi_t$ and $\tilde{\mathbf{F}} = T\tilde{\varphi}_t$, respectively. The energy function of the physical problem in $\mathcal{B} \setminus \mathcal{C}$ is known, while it is not known a priori in $\mathcal{C}$. Energy function of the virtual body is known everywhere in $\tilde{\mathcal{B}}$. The applied loads (body force and traction boundary conditions) and the essential boundary conditions are given for the physical problem. They are, however, not known a priori for the virtual problem.

**Shifters in Euclidean ambient space.** We assume that the reference configurations of both the physical and virtual bodies are embedded in the Euclidean space. To relate vector fields in the physical problem to those in the virtual problem we would need to use shifters. We first consider the physical body $\mathcal{B} \subset \mathcal{S} = \mathbb{R}^n$ ($n = 2$ or 3). The map $\mathsf{s} : T\mathcal{S} \to T\mathcal{S}$, $\mathsf{s}(x, \mathbf{w}) = (\tilde{x}, \mathbf{w})$ is called the shifter map (see Fig.5). The restriction of $\mathsf{s}$ to $x \in \mathcal{S}$ is denoted by $\mathsf{s}_x = \mathsf{s}(x) : T_x\mathcal{S} \to T_{\tilde{x}}\mathcal{S}$, and shifts $\mathbf{w}$ based at $x \in \mathcal{S}$ to $\mathbf{w}$ based at $\tilde{x} \in \mathcal{S}$ (shifter maps in the reference configuration is defined similarly). For $\mathcal{S}$ we choose two global colinear Cartesian coordinates $\{\tilde{z}^{\tilde{i}}\}$ and $\{z^i\}$ for the virtual and physical deformed configurations, respectively. We also use curvilinear coordinates $\{\tilde{x}^{\tilde{a}}\}$ and $\{x^a\}$ for these configurations. Note that $\mathsf{s}^{\tilde{i}}{}_i = \delta^{\tilde{i}}_i$. One can show that [Marsden and Hughes, 1983]

$$\mathsf{s}^{\tilde{a}}{}_a(x) = \frac{\partial \tilde{x}^{\tilde{a}}}{\partial \tilde{z}^{\tilde{i}}}(\tilde{x}) \frac{\partial z^i}{\partial x^a}(x) \delta^{\tilde{i}}_i \,. \tag{4.1}$$

Note that $\mathsf{s}$ preserves inner products, and hence, $\mathsf{s}^{\mathsf{T}} = \mathsf{s}^{-1}$. In components, $(\mathsf{s}^{\mathsf{T}})^a{}_{\tilde{a}} = g^{ab}\mathsf{s}^{\tilde{b}}{}_b \tilde{g}_{\tilde{a}\tilde{b}}$. Note also that

$$\mathsf{s}^{\tilde{a}}{}_{a|\tilde{b}} = \frac{\partial \mathsf{s}^{\tilde{a}}{}_a}{\partial \tilde{x}^{\tilde{b}}} + \tilde{\gamma}^{\tilde{a}}{}_{\tilde{b}\tilde{c}} \mathsf{s}^{\tilde{c}}{}_a - \frac{\partial x^b}{\partial \tilde{x}^{\tilde{b}}} \gamma^c{}_{ab} \mathsf{s}^{\tilde{a}}{}_c \,, \tag{4.2}$$

where $\gamma^a{}_{bc} = \frac{\partial x^a}{\partial z^k} \frac{\partial^2 z^k}{\partial x^b \partial x^c}$ and $\tilde{\gamma}^{\tilde{a}}{}_{\tilde{b}\tilde{c}} = \frac{\partial \tilde{x}^{\tilde{a}}}{\partial \tilde{z}^{\tilde{k}}} \frac{\partial^2 \tilde{z}^{\tilde{k}}}{\partial \tilde{x}^{\tilde{b}} \partial \tilde{x}^{\tilde{c}}}$. It is easily shown that $\mathsf{s}^{\tilde{a}}{}_{a|\tilde{b}} = 0$, i.e., the shifter is covariantly constant.

**Example 4.1.** Consider cylindrical coordinates $(r, \theta, z)$ and $(\tilde{r}, \tilde{\theta}, \tilde{z})$ at $x \in \mathbb{R}^3$ and $\tilde{x} \in \mathbb{R}^3$, respectively. The shifter map has the following matrix representation with respect to these coordinates

$$\mathsf{s} = \begin{bmatrix} \cos(\tilde{\theta} - \theta) & r\sin(\tilde{\theta} - \theta) & 0 \\ -\sin(\tilde{\theta} - \theta)/\tilde{r} & r\cos(\tilde{\theta} - \theta)/\tilde{r} & 0 \\ 0 & 0 & 1 \end{bmatrix} . \tag{4.3}$$

**Example 4.2.** Consider spherical coordinates $(r, \theta, \phi)$ and $(\tilde{r}, \tilde{\theta}, \tilde{\phi})$ at $x \in \mathbb{R}^3$ and $\tilde{x} \in \mathbb{R}^3$, respectively. The shifter map has the following matrix representation with respect to these coordinates

$$\mathsf{s} = \begin{bmatrix} \cos(\tilde{\phi} - \phi)\sin\tilde{\theta}\sin\theta + \cos\tilde{\theta}\cos\theta & r[\cos(\tilde{\phi} - \phi)\sin\tilde{\theta}\cos\theta - \cos\tilde{\theta}\sin\theta] & r\sin(\tilde{\phi} - \phi)\sin\tilde{\theta}\sin\theta \\ [\cos(\tilde{\phi} - \phi)\cos\tilde{\theta}\sin\theta - \sin\tilde{\theta}\cos\theta]/\tilde{r} & r[\cos(\tilde{\phi} - \phi)\cos\tilde{\theta}\cos\theta + \sin\tilde{\theta}\sin\theta]/\tilde{r} & r\sin(\tilde{\phi} - \phi)\cos\tilde{\theta}\sin\theta/\tilde{r} \\ -\sin(\tilde{\phi} - \phi)\sin\theta/(\tilde{r}\sin\tilde{\theta}) & -r\sin(\tilde{\phi} - \phi)\cos\theta/(\tilde{r}\sin\tilde{\theta}) & r\cos(\tilde{\phi} - \phi)\sin\theta/(\tilde{r}\sin\tilde{\theta}) \end{bmatrix} . \tag{4.4}$$

**Balance of linear momentum in the physical and virtual bodies.** The balance of linear momentum for the physical body reads: $\operatorname{Div}\mathbf{P} + \rho_0\mathbf{B} = \rho_0\mathbf{A}$. We use the Piola identity and write the divergence term with respect to the reference configuration of the virtual body. The Piola identity [Marsden and Hughes, 1983] tells us that for a vector field $\mathbf{W}$ on $\mathcal{B}$, one can write (see the appendix)

$$\operatorname{Div}\mathbf{W} = J_\Xi\,\widetilde{\operatorname{Div}}\tilde{\mathbf{W}}, \quad \tilde{\mathbf{W}} = J_\Xi^{-1}\overset{\equiv}{\mathbf{F}}\mathbf{W}\,. \tag{4.5}$$

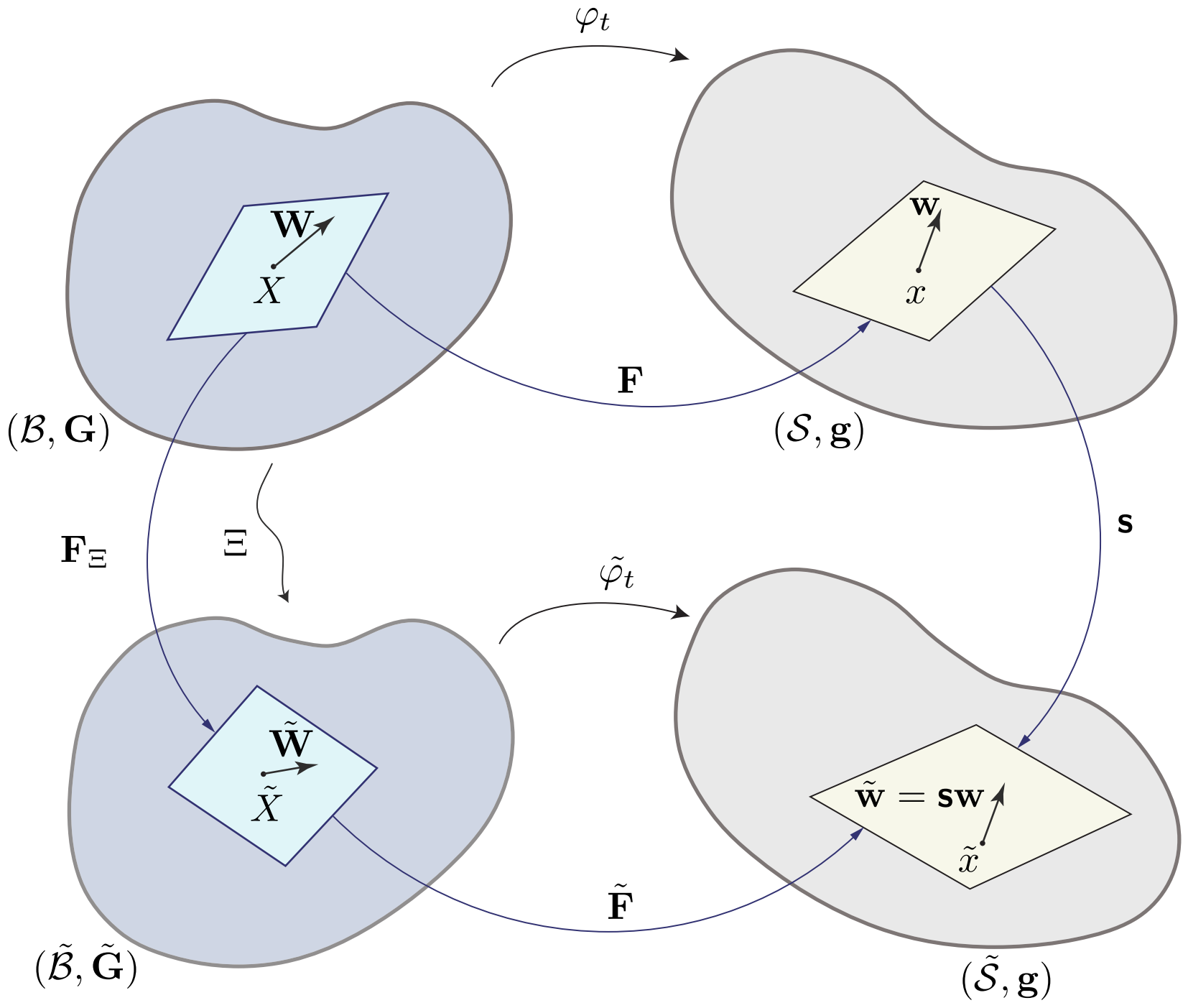

Figure 5: *The mapping* $\mathbf{s}$ *is the shifter map, which parallel transports vectors in* $T_x\mathcal{S}$ *to vectors in* $T_{\tilde{x}}\tilde{\mathcal{S}}$.

In components, this reads

$$W^A{}_{|A} = J_\Xi\left[J_\Xi^{-1}\overset{\equiv}{F}{}^{\tilde{A}}{}_A W^A\right]_{|\tilde{A}}\,. \tag{4.6}$$

$\tilde{\mathbf{W}}$ is called the Piola transform of $\mathbf{W}$. We use the local coordinate charts $\{X^A\}$ and $\{\tilde{X}^{\tilde{A}}\}$ for $\mathcal{B}$ and $\tilde{\mathcal{B}}$, respectively. Let us start with $\operatorname{Div}\mathbf{P}$ and rewrite it with respect to the reference configuration of the virtual body. In coordinates and using the Piola identity one can write[18]

$$P^{aA}{}_{|A} = J_\Xi\left(J_\Xi^{-1}\overset{\equiv}{F}{}^{\tilde{A}}{}_A P^{aA}\right)_{|\tilde{A}} = J_\Xi(\tilde{P}^{a\tilde{A}}{}_{|\tilde{A}})\circ\Xi\,, \tag{4.8}$$

where

$$\tilde{P}^{a\tilde{A}}\circ\Xi = J_\Xi^{-1}\overset{\equiv}{F}{}^{\tilde{A}}{}_A P^{aA}, \tag{4.9}$$

[18] Note that the Piola identity can be written as

$$\left[J_\Xi^{-1}\overset{\equiv}{F}{}^{\tilde{A}}{}_A\right]_{|\tilde{A}} = 0\,. \tag{4.7}$$

Thus

$$\left[J_\Xi^{-1}\overset{\equiv}{F}{}^{\tilde{A}}{}_A P^{aA}\right]_{|\tilde{A}} = \left[J_\Xi^{-1}\overset{\equiv}{F}{}^{\tilde{A}}{}_A\right]_{|\tilde{A}} P^{aA} + J_\Xi^{-1}\overset{\equiv}{F}{}^{\tilde{A}}{}_A P^{aA}{}_{|\tilde{A}}\,.$$

Therefore, using (4.7) one can write

$$J_\Xi\left[J_\Xi^{-1}\overset{\equiv}{F}{}^{\tilde{A}}{}_A P^{aA}\right]_{|\tilde{A}} = \overset{\equiv}{F}{}^{\tilde{A}}{}_A P^{aA}{}_{|\tilde{A}} = P^{aA}{}_{|A}\,.$$

and $J_\Xi$ is the Jacobian of the *cloaking transformation* $\Xi$, which is written as

$$J_\Xi = \sqrt{\frac{\det \tilde{\mathbf{G}} \circ \Xi}{\det \mathbf{G}}} \det \overset{\Xi}{\mathbf{F}} \,. \tag{4.10}$$

However, this is a vector in $T_x\mathcal{S}$. The corresponding vector in $T_{\tilde{x}}\mathcal{S}$ is obtained using the shifter $\mathsf{s}(x)$ as

$$\mathsf{s}^{\tilde{a}}{}_a \tilde{P}^{a\tilde{A}} \circ \Xi = J_\Xi^{-1} \mathsf{s}^{\tilde{a}}{}_a P^{aA} \overset{\Xi}{F}{}^{\tilde{A}}{}_A. \tag{4.11}$$

Note that $\mathsf{s}^{\tilde{a}}{}_a P^{aA}{}_{|A} = (\mathsf{s}^{\tilde{a}}{}_a P^{aA})_{|A}$. Therefore

$$\mathsf{s} \circ \varphi \operatorname{Div} \mathbf{P} = J_\Xi (\widetilde{\operatorname{Div}} \tilde{\mathbf{P}}) \circ \Xi \,, \tag{4.12}$$

where

$$\tilde{\mathbf{P}} = J_\Xi^{-1} \mathsf{s} \circ \varphi \mathbf{P} \overset{\Xi}{\mathbf{F}}{}^\star, \text{ or } \quad \tilde{P}^{\tilde{a}\tilde{A}} \circ \Xi = J_\Xi^{-1} \overset{\Xi}{F}{}^{\tilde{A}}{}_A \mathsf{s}^{\tilde{a}}{}_a P^{aA}. \tag{4.13}$$

Equivalently, the transformed second Piola-Kirchhoff stress reads

$$\tilde{\mathbf{S}} \circ \Xi = J_\Xi^{-1} \tilde{\mathbf{F}}^{-1} \circ \tilde{\varphi} \, \mathsf{s} \circ \varphi \, \mathbf{F} \mathbf{S} \overset{\Xi}{\mathbf{F}}{}^\star. \tag{4.14}$$

Note that force on an infinitesimal area $dA$ is calculated as

$$t^a dA = P^{aA} N_A dA = P^{aA} J_\Xi^{-1} \overset{\Xi}{F}{}^{\tilde{A}}{}_A \tilde{N}_{\tilde{A}} d\tilde{A} = \tilde{P}^{a\tilde{A}} \tilde{N}_{\tilde{A}} d\tilde{A} = \tilde{t}^a d\tilde{A} \,, \tag{4.15}$$

and hence $\mathsf{s}^{\tilde{a}}{}_a t^a dA = \tilde{t}^{\tilde{a}} d\tilde{A}$, i.e., this force is the same in the physical and the virtual bodies if we assume that the first Piola-Kirchhoff stress in the virtual body is given by (4.13). Now in the virtual body $\tilde{\mathbf{P}} = J_\Xi^{-1} \mathsf{s} \circ \varphi \mathbf{P} \overset{\Xi}{\mathbf{F}}{}^\star$ is the shifted Piola transform of $\mathbf{P}$ and

$$\mathbf{P} = \mathbf{g}^\sharp \frac{\partial W}{\partial \mathbf{F}}, \qquad \tilde{\mathbf{P}} = \mathbf{g}^\sharp \frac{\partial \tilde{W}}{\partial \tilde{\mathbf{F}}}, \tag{4.16}$$

where $\tilde{W} = \tilde{W}(\tilde{\mathbf{G}}, \mathbf{g} \circ \tilde{\varphi}, \tilde{\mathbf{F}})$ as we assume that the virtual body is isotropic and homogeneous. To ensure that the relation $\tilde{\mathbf{P}} = J_\Xi^{-1} \mathsf{s} \circ \varphi \mathbf{P} \overset{\Xi}{\mathbf{F}}{}^\star$ holds one needs to define $W$ in $\mathcal{B} \setminus \mathcal{C}$ as $W|_{\mathcal{B}\setminus\mathcal{C}} = \tilde{W} \circ \Xi$, $\forall X \in \mathcal{B} \setminus \mathcal{C}$. This is because $\Xi$ is defined to be the identity map in $\mathcal{B} \setminus \mathcal{C}$, and hence, the material outside the cloak will be isotropic and identical to that of the virtual body. The cloaking region $\mathcal{C}$, nevertheless, will be anisotropic, in general, i.e., $W_\mathcal{C} = W_\mathcal{C}(X, \mathbf{G}, \mathbf{g} \circ \varphi, \mathbf{F}, \boldsymbol{\zeta}_1, \ldots, \boldsymbol{\zeta}_m)$, $\forall X \in \mathcal{C}$, where $\boldsymbol{\zeta}_1, \ldots, \boldsymbol{\zeta}_m$ are the structural tensors that characterize the symmetry group of the material in the cloaking region.[19] Note that $W_\mathcal{C}$ must satisfy the following relation

$$(\mathbf{g}^\sharp \circ \varphi_t) \frac{\partial W_\mathcal{C}}{\partial \mathbf{F}} = J_\Xi (\mathsf{s}^{-1} \circ \tilde{\varphi}) \, (\mathbf{g}^\sharp \circ \tilde{\varphi}_t \circ \Xi) \left( \frac{\partial \tilde{W}}{\partial \tilde{\mathbf{F}}} \circ \Xi \right) \overset{\Xi}{\mathbf{F}}{}^{-\star} \,. \tag{4.17}$$

One can now rewrite the balance of linear momentum $\operatorname{Div} \mathbf{P} + \rho_0 \mathbf{B} = \rho_0 \mathbf{A}$ as

$$\widetilde{\operatorname{Div}} \tilde{\mathbf{P}} + \tilde{\rho}_0 \tilde{\mathbf{B}} = \tilde{\rho}_0 \tilde{\mathbf{A}} \,, \tag{4.18}$$

where $\tilde{\rho}_0 = J_\Xi^{-1} \rho_0 \circ \Xi^{-1}$, $\tilde{\mathbf{B}} = (\mathsf{s} \circ \varphi) \mathbf{B} \circ \Xi^{-1}$, and $\tilde{\mathbf{A}} = (\mathsf{s} \circ \varphi) \mathbf{A} \circ \Xi^{-1}$. In other words, assuming that in the virtual body the first Piola-Kirchhoff stress is given by (4.13), the virtual body has the mass density $\tilde{\rho}_0$, is under the body force $\tilde{\mathbf{B}}$, and has the material acceleration $\tilde{\mathbf{A}}$, the balance of linear momentum is satisfied. Conversely, if the balance of linear momentum is satisfied for the virtual body with the above transformed fields, it is satisfied for the physical body as well.

[19] Note that we do not know a priori how many structural tensors are needed and what they are. In other words, we cannot explicitly determine how the symmetry group transforms under arbitrary diffeomorphisms.

**Balance of angular momentum in the physical and virtual bodies.** In the physical body the balance of angular momentum is equivalent to $S^{AB} = S^{BA}$, or $\mathbf{S}^\star = \mathbf{S}$. The balance of angular momentum must hold in the virtual body as well, i.e., $\tilde{\mathbf{S}}^\star = \tilde{\mathbf{S}}$, which holds if and only if

$$(\mathbf{RS})^\star = \mathbf{RS}\,, \tag{4.19}$$

where $\mathbf{R} = (\bar{\bar{\mathbf{F}}}^{-1} \circ \Xi)(\tilde{\mathbf{F}}^{-1} \circ \tilde{\varphi} \circ \Xi)(\mathbf{s} \circ \varphi)\mathbf{F}$. In terms of $\mathbf{R}$, the second Piola-Kirchhoff stress in the physical and virtual bodies are related as

$$\tilde{\mathbf{S}} = J_\Xi^{-1}\bar{\bar{\mathbf{F}}}\mathbf{RS}\bar{\bar{\mathbf{F}}}^\star, \quad \text{or} \quad \mathbf{S} = J_\Xi \mathbf{R}^{-1}\bar{\bar{\mathbf{F}}}^{-1}\tilde{\mathbf{S}}\bar{\bar{\mathbf{F}}}^{-\star}\,. \tag{4.20}$$

Note that assuming that balance of angular momentum is satisfied in one configuration it may or may not be satisfied in the other configuration. We will see in §4.3 that the balance of angular momentum is the obstruction to transformation cloaking in classical linear elastodynamics.

**Incompressible solids.** In the case of incompressible solids, $\mathbf{P} = -pJ\mathbf{g}^\sharp\mathbf{F}^{-\star} + \bar{\mathbf{P}}$, or in components $P^{aA} = -pJ(F^{-1})^A{}_b g^{ab} + \bar{P}^{aA}$, where $\bar{\mathbf{P}}$ is the constitutive part of the stress. We assume that if the virtual body is incompressible at a point $\tilde{X} \in \tilde{\mathcal{B}}$ the corresponding point $X \in \mathcal{B}$ in the physical body is incompressible as well and vice versa. We also assume that $\tilde{\bar{\mathbf{P}}} = J_\Xi^{-1}\mathbf{s}\bar{\mathbf{P}}\bar{\bar{\mathbf{F}}}^\star$, i.e., (4.17) still holds.

**Remark 4.3.** Note that although we use the same metric for the ambient space for the two bodies, after deformation, a physical point and its corresponding virtual point will be mapped to two different points in the ambient space, in general. For example, in cylindrical coordinates, $r(R,\Theta,\Phi) \neq \tilde{r}(\tilde{R},\tilde{\Theta},\tilde{\Phi})$, in general, where $\Xi(R,\Theta,\Phi) = (\tilde{R},\tilde{\Theta},\tilde{\Phi})$. Therefore, when calculating $J$ and $\tilde{J}$, $\det \mathbf{g}$ at two different points is used.

**Remark 4.4.** We showed that if $\tilde{\mathbf{G}} = \Xi_*\mathbf{G}$, then the symmetry class of the material is preserved under $\Xi$ (cf. (2.29) in the nonlinear case and Prop. 3.2 in the linear case). However, for general $\Xi$ and $\tilde{\mathbf{G}}$, the type of the symmetry class of the body $\tilde{\mathcal{B}}$ with respect to the metric $\tilde{\mathbf{G}}$ can be quite non-trivial and different from that of the body $\mathcal{B}$ with respect to the metric $\mathbf{G}$. This is a geometric interpretation of the expected anisotropy of an elastic cloak.

**Remark 4.5.** The relation $\tilde{\rho}_0 = J_\Xi^{-1}\rho_0\circ\Xi^{-1}$ implies that the total mass of the cloak is equal to the mass of the corresponding transformed region in the virtual structure with uniform density $\rho_0$, i.e., $m(\mathcal{C}) = \rho_0 \operatorname{vol}_{\tilde{\mathbf{G}}}(\tilde{\mathcal{C}})$. For example, the mass of a cylindrical cloak, which is an annulus with inner radius $R_i$ and outer radius $R_o$ in the physical structure is equal to the mass of the annulus in the virtual structure (filled with a homogeneous solid) with inner radius $f(R_i) = \epsilon$ and outer radius $f(R_o) = R_o$. Therefore, as the virtual and physical structures are identical outside the cloak, one concludes that the two bodies have the same total mass.

Even if the balance of angular momentum is satisfied in both configurations, it turns out that hiding a hole in a nonlinear elastic medium from arbitrary finite-amplitude excitations is not possible.

**Proposition 4.6.** *Nonlinear elastodynamics transformation cloaking is not possible. In other words, it is not possible to design a cloak that would hide a cavity from any (finite) time-dependent elastic disturbance (or wave).*

*Proof.* The idea of nonlinear transformation elastodynamics cloaking can be summarized as follows. Starting from the balance of linear momentum we assume the relation (4.13) between the first Piola-Kirchhoff stresses in the virtual and physical bodies. The energy function of the virtual body is given while that of the cloak in the physical body is not known. This implies that to be able to find the energy function of the cloak one would need to relate the kinematics of the physical and virtual bodies. This kinematics constraint is the relation between accelerations: $\tilde{\mathbf{A}} \circ \Xi = (\mathbf{s} \circ \varphi)\mathbf{A}$.[20] All these will have to be consistent with the balance of angular momentum (4.19). We start from the relation $\tilde{\mathbf{A}} \circ \Xi = (\mathbf{s}\circ\varphi)\mathbf{A}$ and write it in Cartesian coordinates: $\tilde{A}^{\tilde{i}} = \mathsf{s}^{\tilde{i}}{}_i A^i = \delta^{\tilde{i}}_i A^i$. Thus, $\tilde{V}^{\tilde{i}}(\Xi(X),t) = \delta^{\tilde{i}}_i V^i(X,t) + a_1$, where $a_1$ is a constant. Assuming that

[20]The kinematics relationship in linear elastodynamics cloaking turns out to be $\tilde{\mathbf{U}} \circ \Xi = \mathbf{U}$.

$\tilde{V}^{\tilde{i}}(\Xi(X),0) = 0$ and $V^i(X,0) = 0$, $a_1 = 0$, and hence, $\tilde{\varphi}^{\tilde{i}}(\Xi(X),t) = \delta^{\tilde{i}}_i \varphi^i(X,t) + a_0$, where $a_0$ is a constant. Knowing that $\tilde{\varphi}$ and $\varphi$ satisfy the same essential boundary conditions on $\partial_d\tilde{\mathcal{B}} = \partial_d\mathcal{B}$, $a_0 = 0$. Therefore, we have concluded that

$$\tilde{\varphi}^{\tilde{i}}(\Xi(X),t) = \delta^{\tilde{i}}_i \varphi^i(X,t)\,. \tag{4.21}$$

In the virtual body there is a circular (spherical) hole of radius $\epsilon \to 0$. (4.21) is telling us that the deformed configuration of $\partial\mathcal{H}$ is identical to that of the infinitesimal cavity in the virtual body. This is true for any loading of the physical structure, and even in the limit of vanishing loads. This implies that the hole surrounded by a cloak in the physical body would collapse to a cavity of radius $\epsilon \to 0$ (anti-cavitation), i.e., the physical body is unstable. Therefore, nonlinear elastodynamics cloaking is not possible. □

However, nonlinear elastostatics cloaking may be possible as discussed next.

## 4.2 Nonlinear Elastostatics Cloaking

We ignore inertial forces and explore the possibility of static cloaking in nonlinear elasticity. Note that the virtual structure is assumed to be isotropic, and therefore, its constitutive equation is given by

$$\tilde{\mathbf{S}} = 2\frac{\partial \tilde{W}}{\partial \tilde{\mathbf{C}}^\flat} = 2\tilde{W}_{\tilde{I}_1}\tilde{\mathbf{G}}^\sharp + 2\tilde{W}_{\tilde{I}_2}(\tilde{I}_2\tilde{\mathbf{C}}^{-1} - \tilde{I}_3\tilde{\mathbf{C}}^{-2}) + 2\tilde{W}_{\tilde{I}_3}\tilde{I}_3\tilde{\mathbf{C}}^{-1}\,, \tag{4.22}$$

where $\tilde{\mathbf{S}}$ is the second Piola-Kirchhoff stress and

$$\tilde{I}_1 = \operatorname{tr}\tilde{\mathbf{C}}\,,\quad \tilde{I}_2 = \det\tilde{\mathbf{C}}\operatorname{tr}\tilde{\mathbf{C}}^{-1}\,,\quad \tilde{I}_3 = \det\tilde{\mathbf{C}}\,,\quad \text{and}\quad \tilde{W}_{\tilde{I}_n} := \frac{\partial\tilde{W}}{\partial\tilde{I}_n}\,,\quad n = 1,2,3. \tag{4.23}$$

Note that

$$\tilde{\mathbf{C}}^{-1} = \tilde{\mathbf{F}}^{-1}\tilde{\mathbf{g}}^\sharp\tilde{\mathbf{F}}^{-\star},\quad \tilde{\mathbf{C}}^{-2} = \tilde{\mathbf{C}}^{-1}\tilde{\mathbf{G}}\tilde{\mathbf{C}}^{-1},\quad \boldsymbol{s}^{-1}\tilde{\mathbf{g}}^\sharp = \mathbf{g}^\sharp\boldsymbol{s}^\star,\quad \mathbf{C}^{-1} = \mathbf{F}^{-1}\mathbf{g}^\sharp\mathbf{F}^{-\star}. \tag{4.24}$$

Hence, from (4.20), (4.22), and (4.24) one can write

$$J_\Xi^{-1}\mathbf{S}_\mathcal{C} = \alpha\mathbf{R}^{-1}\Xi^*\tilde{\mathbf{G}}^\sharp + (\tilde{I}_2\beta + \tilde{I}_3\gamma)\mathbf{C}^{-1}\mathbf{R}^\star - \tilde{I}_3\beta\mathbf{C}^{-1}\mathbf{R}^\star(\Xi^*\tilde{\mathbf{G}})\mathbf{R}\mathbf{C}^{-1}\mathbf{R}^\star\,, \tag{4.25}$$

where $\alpha = 2\tilde{W}_{\tilde{I}_1}$, $\beta = 2\tilde{W}_{\tilde{I}_2}$, and $\gamma = 2\tilde{W}_{\tilde{I}_3}$. In components

$$\begin{aligned} J_\Xi^{-1}S_\mathcal{C}^{AB} &= \alpha(R^{-1})^A{}_M(\Xi^*\tilde{G})^{MB} + (\tilde{I}_2\beta + \tilde{I}_3\gamma)(C^{-1})^{AM}R^B{}_M \\ &\quad - \tilde{I}_3\beta(C^{-1})^{AM}R^K{}_M(\Xi^*\tilde{G})_{KL}R^L{}_P(C^{-1})^{PQ}R^B{}_Q\,, \end{aligned} \tag{4.26}$$

where $(\Xi^*\tilde{G})^{MB} = (\bar{\bar{\tilde{F}}}^{-1})^M{}_{\tilde{M}}(\bar{\bar{\tilde{F}}}^{-1})^B{}_{\tilde{N}}\tilde{G}^{\tilde{M}\tilde{N}}$ and $(\Xi^*\tilde{G})_{KL} = \bar{\bar{\tilde{F}}}^{\tilde{K}}{}_K\bar{\bar{\tilde{F}}}^{\tilde{L}}{}_L\tilde{G}_{\tilde{K}\tilde{L}}$. For an incompressible solid ($\tilde{I}_3 = 1$)

$$\tilde{\mathbf{S}} = -\tilde{p}\tilde{\mathbf{C}}^{-1} + 2\tilde{W}_{\tilde{I}_1}\tilde{\mathbf{G}}^\sharp - 2\tilde{W}_{\tilde{I}_2}\tilde{\mathbf{C}}^{-2}\,, \tag{4.27}$$

where $\tilde{p}$ is a pressure-like field associated with the incompressibility constraint $\tilde{J} = 1$. We assume that the cloak is made of an incompressible solid as well and hence

$$J_\Xi^{-1}\mathbf{S}_\mathcal{C} = -p\mathbf{C}^{-1}\mathbf{R}^\star + \alpha\mathbf{R}^{-1}\Xi^*\tilde{\mathbf{G}}^\sharp - \beta\mathbf{C}^{-1}\mathbf{R}^\star(\Xi^*\tilde{\mathbf{G}})\mathbf{R}\mathbf{C}^{-1}\mathbf{R}^\star\,, \tag{4.28}$$

where $p$ is a pressure-like variable associated with the incompressibility constraint $J = 1$. Note that

$$\tilde{J} = \sqrt{\frac{\det(\mathbf{g}\circ\tilde{\varphi}_t)}{\det\tilde{\mathbf{G}}}}\det\tilde{\mathbf{F}} = 1,\quad J = \sqrt{\frac{\det(\mathbf{g}\circ\varphi_t)}{\det\mathbf{G}}}\det\mathbf{F}\,. \tag{4.29}$$

Next we consider an infinite body with a cylindrical hole and an infinite body with a spherical cavity. In both examples we assume radial deformations. Under this assumption we will find the constitutive equations of the corresponding cloaks. It should be emphasized that these will be static cloaks. The following examples are nonlinear analogues of Mansfield [1953]'s neutral holes.

**Example 1: A nonlinear static cylindrical cloak.** Let us consider radial deformations of an infinitely-long solid cylinder with a cylindrical hole of radius $R_i$. For the physical body in the cylindrical coordinates $(R, \Theta, Z)$ and $(r, \theta, z)$ for the reference configuration and the ambient space, respectively, we consider deformations of the form $(r, \theta, z) = (r(R), \Theta, Z)$. Similarly, for the virtual body in the cylindrical coordinates $(\tilde{R}, \tilde{\Theta}, \tilde{Z})$ and $(\tilde{r}, \tilde{\theta}, \tilde{z})$ we have $(\tilde{r}, \tilde{\theta}, \tilde{z}) = (\tilde{r}(\tilde{R}), \tilde{\Theta}, \tilde{Z})$. Note that $(\tilde{R}, \tilde{\Theta}, \tilde{Z}) = \Xi(R, \Theta, Z) = (f(R), \Theta, Z)$. We assume the following relation between the kinematics of deformations in the physical and virtual bodies[21]

$$\frac{\tilde{r}(\tilde{R})}{\tilde{R}} = \frac{r(R)}{R}. \tag{4.30}$$

Therefore

$$(\tilde{r}, \tilde{\theta}, \tilde{z}) = (\tilde{r}(\tilde{R}), \tilde{\Theta}, \tilde{Z}) = \left( \frac{f(R)}{R} r(R), \Theta, Z \right). \tag{4.31}$$

Note that (in particular, $\tilde{r}'(\tilde{R}_o) = r'(R_o)$)

$$\tilde{r}'(\tilde{R}) = \frac{1}{f'(R)} \frac{d}{dR} \left[ \frac{f(R)}{R} r(R) \right] = \frac{f(R)}{Rf'(R)} r'(R) + \left[ 1 - \frac{f(R)}{Rf'(R)} \right] \frac{r(R)}{R}. \tag{4.32}$$

We will design a nonlinear elastic cloak of inner and outer radii $R_i$ and $R_o$ in the physical body such that the static response of the body with a hole in radial deformations outside the cloak is identical to that of an isotropic and homogeneous elastic body with an infinitesimal hole. We assume that $f(R_i) = \epsilon$, and $f(R_o) = R_o$. As we will see in §5.2, the function $f$ must satisfy the extra condition: $f'(R_o) = 1$. Note that the spatial shifter map has the following coordinate representation

$$\mathsf{s} = \begin{bmatrix} 1 & 0 & 0 \\ 0 & r(R)/\tilde{r}(\tilde{R}) & 0 \\ 0 & 0 & 1 \end{bmatrix} = \begin{bmatrix} 1 & 0 & 0 \\ 0 & R/f(R) & 0 \\ 0 & 0 & 1 \end{bmatrix}. \tag{4.33}$$

Therefore

$$\mathbf{R} = \begin{bmatrix} r'(R)/[f'(R)\tilde{r}'(\tilde{R})] & 0 & 0 \\ 0 & R/f(R) & 0 \\ 0 & 0 & 1 \end{bmatrix}, \quad \mathbf{C}^{-1} = \begin{bmatrix} 1/r'(R)^2 & 0 & 0 \\ 0 & 1/r(R)^2 & 0 \\ 0 & 0 & 1 \end{bmatrix}. \tag{4.34}$$

Using (4.25) the constitutive equations of the cloak read

$$\begin{aligned} \mathbf{S} =& \alpha \begin{bmatrix} f\tilde{r}'/(Rr') & 0 & 0 \\ 0 & f'/R^2 & 0 \\ 0 & 0 & ff'/R \end{bmatrix} + (\tilde{I}_2\beta + \tilde{I}_3\gamma) \begin{bmatrix} f/(Rr'\tilde{r}') & 0 & 0 \\ 0 & f'/r^2 & 0 \\ 0 & 0 & ff'/R \end{bmatrix} \\ & - \tilde{I}_3\beta \begin{bmatrix} f/(Rr\tilde{r}'^3) & 0 & 0 \\ 0 & fR^2/r^4 & 0 \\ 0 & 0 & ff'/R \end{bmatrix}, \end{aligned} \tag{4.35}$$

where

$$\tilde{I}_1 = 1 + \tilde{r}'^2(f(R)) + \frac{r^2(R)}{R^2}, \quad \tilde{I}_2 = \frac{1}{2}\tilde{r}'^2(1 - \tilde{r}'^2) + \frac{r^2(R)}{2R^2} \left[ 1 - \frac{r^2(R)}{R^2} \right], \quad \tilde{I}_3 = \tilde{J}^2 = \left[ \frac{r(R)}{R} \tilde{r}'(f(R)) \right]^2. \tag{4.36}$$

[21] In the case of elastodynamics cloaking the kinematic relation between the physical and virtual problems was the equality (up to a shift) of acceleration vectors. In elastostatics there is no such constraint and one has freedom in choosing a kinematic relation between the two problems. Note also that (4.30) is just one choice. One may assume

$$\frac{\tilde{r}(\tilde{R})}{\tilde{R}} = h\left( \frac{r(R)}{R} \right),$$

for any positive and strictly increasing function $h$ such that $h(1) = 1$. Note that (4.30) is a nonlinear analogue of Olsson and Wall [2011]'s kinematic assumption.

It is seen that $\mathbf{S}$ is symmetric, and hence, the balance of angular momentum is satisfied. The surface of the cavity is traction free, and hence, we have the boundary condition $S^{RR}(R_i) = 0$. Note that radial deformations cannot be maintained in an arbitrary compressible isotropic solid only by applying boundary tractions. This is a result of Ericksen's theorem, which states that the only universal deformations in compressible isotropic solids are homogeneous deformations [Ericksen, 1955] (see [Yavari and Goriely, 2016] for an extension of this result to compressible solids with finite eigenstrains). Note, however, that if the radial stretch is uniform, i.e., $\tilde{r}(\tilde{R})/\tilde{R} = r(R)/R = \lambda$, then the deformation will be homogeneous, and hence, universal, i.e., it can be maintained solely by applying boundary tractions in any compressible isotropic solid.

Suppose the virtual body is incompressible, i.e., $\tilde{J} = \tilde{r}\tilde{r}'/\tilde{R} = 1$. Hence, $\tilde{r}(\tilde{R}) = \sqrt{\tilde{R}^2 + C}$, for some constant $C$. Note that $\tilde{R}_i = \epsilon$, and thus $\tilde{r}(\tilde{R}_i) = \sqrt{\epsilon^2 + C}$. In the limit when the virtual hole becomes smaller and smaller ($\epsilon \to 0$), we expect that $\tilde{r}(\tilde{R}_i) \to 0$. Therefore, $C = 0$, i.e., $\tilde{r}(\tilde{R}) = \tilde{R}$. Now (4.30) gives us $r(R) = R$, and hence $J = rr'/R = 1$, i.e., the physical body is incompressible as well. For incompressible solids radial deformations are universal [Ericksen, 1954], i.e., such deformations can be maintained in any isotropic incompressible solid by applying only boundary tractions. From (4.28) the constitutive equations of the incompressible cloak are written as

$$\begin{aligned}\mathbf{S} &= -p\begin{bmatrix}1/(f'r'\tilde{r}') & 0 & 0\\ 0 & R/(fr^2) & 0\\ 0 & 0 & 1\end{bmatrix} + \alpha\begin{bmatrix}f\tilde{r}'/(Rr') & 0 & 0\\ 0 & f'/R^2 & 0\\ 0 & 0 & ff'/R\end{bmatrix} - \beta\begin{bmatrix}f/(Rr'\tilde{r}'^{\,3}) & 0 & 0\\ 0 & R^2f'/r^4 & 0\\ 0 & 0 & ff'/R\end{bmatrix},\\ &= -p\begin{bmatrix}1/f'(R) & 0 & 0\\ 0 & 1/(Rf(R)) & 0\\ 0 & 0 & 1\end{bmatrix} + \alpha\begin{bmatrix}f(R)/R & 0 & 0\\ 0 & f'(R)/R^2 & 0\\ 0 & 0 & f(R)f'(R)/R\end{bmatrix} - \beta\begin{bmatrix}f(R)/R & 0 & 0\\ 0 & f'(R)/R^2 & 0\\ 0 & 0 & f(R)f'(R)/R\end{bmatrix},\\ &= -p\begin{bmatrix}1/f'(R) & 0 & 0\\ 0 & 1/(Rf(R)) & 0\\ 0 & 0 & 1\end{bmatrix} + (\alpha-\beta)\begin{bmatrix}f(R)/R & 0 & 0\\ 0 & f'(R)R^2 & 0\\ 0 & 0 & f(R)f'(R)/R\end{bmatrix}.\end{aligned} \tag{4.37}$$

It is seen that in this case too $\mathbf{S}$ is symmetric, i.e., the balance of angular momentum is satisfied.

**Example 2: A nonlinear static spherical cloak.** Let us consider radial deformations of a finite solid sphere with a spherical cavity of radius $R_i$. For the physical body in the spherical coordinates $(R, \Theta, \Phi)$ and $(r, \theta, \phi)$ for the reference configuration and the ambient space, respectively, we consider deformations of the form $(r, \theta, \phi) = (r(R), \Theta, \Phi)$. Similarly, for the virtual body in the cylindrical coordinates $(\tilde{R}, \tilde{\Theta}, \tilde{\Phi})$ and $(\tilde{r}, \tilde{\theta}, \tilde{\phi})$ we have $(\tilde{r}, \tilde{\theta}, \tilde{\phi}) = (\tilde{r}(\tilde{R}), \tilde{\Theta}, \tilde{\Phi})$. Notice that $(\tilde{R}, \tilde{\Theta}, \tilde{\Phi}) = \Xi(R, \Theta, Z) = (f(R), \Theta, \Phi)$. We again assume the following relation between the kinematics of deformations in the physical and virtual bodies $\tilde{r}(\tilde{R})/\tilde{R} = r(R)/R$. Therefore

$$(\tilde{r}, \tilde{\theta}, \tilde{\phi}) = (\tilde{r}(\tilde{R}), \tilde{\Theta}, \tilde{\Phi}) = \left(\frac{f(R)}{R} r(R), \Theta, \Phi\right). \tag{4.38}$$

Note that $\tilde{r}'(\tilde{R})$ is given in (4.32). We will design a nonlinear elastic cloak of inner and outer radii $R_i$ and $R_o$ in the physical body such that the static response of the body with a cavity in radial deformations outside the cloak be identical to that of an isotropic and homogeneous elastic body with an infinitesimal cavity. The function $f$ satisfies the following conditions: $f(R_i) = \epsilon$, $f(R_o) = R_o$, and $f'(R_o) = 1$. Note that the spatial shifter map has the following coordinate representation

$$\mathsf{s} = \begin{bmatrix}1 & 0 & 0\\ 0 & r(R)/\tilde{r}(\tilde{R}) & 0\\ 0 & 0 & r(R)/\tilde{r}(\tilde{R})\end{bmatrix} = \begin{bmatrix}1 & 0 & 0\\ 0 & R/f(R) & 0\\ 0 & 0 & R/f(R)\end{bmatrix}. \tag{4.39}$$

Therefore

$$\mathbf{R} = \begin{bmatrix}r'(R)/(f'(R)\tilde{r}'(\tilde{R})) & 0 & 0\\ 0 & R/f(R) & 0\\ 0 & 0 & R/f(R)\end{bmatrix}, \quad \mathbf{C}^{-1} = \begin{bmatrix}1/r'(R)^2 & 0 & 0\\ 0 & 1/r(R)^2 & 0\\ 0 & 0 & 1/(r(R)^2\sin^2\Theta)\end{bmatrix}. \tag{4.40}$$

The constitutive equations of the cloak read

$$\begin{aligned}\mathbf{S} =&\alpha \begin{bmatrix} f^2\tilde{r}'/(R^2r') & 0 & 0 \\ 0 & ff'/R^3 & 0 \\ 0 & 0 & ff'/(R^3\sin^2\Theta) \end{bmatrix} + (\tilde{I}_2\beta + \tilde{I}_3\gamma) \begin{bmatrix} f^2/(R^2r'\tilde{r}') & 0 & 0 \\ 0 & ff'/(Rr^2) & 0 \\ 0 & 0 & ff'/(Rr^2\sin^2\Theta) \end{bmatrix} \\ &- \tilde{I}_3\beta \begin{bmatrix} f^2/(R^2r'\tilde{r}'^{\,3}) & 0 & 0 \\ 0 & Rff'/r^4 & 0 \\ 0 & 0 & Rff'/(r^4\sin^2\Theta) \end{bmatrix}, \end{aligned} \tag{4.41}$$

where

$$\tilde{I}_1 = \tilde{r}'^{\,2}(f(R)) + \frac{2r^2(R)}{R^2}, \quad \tilde{I}_2 = \frac{1}{2}\tilde{r}'^{\,2}(1-\tilde{r}'^{\,2}) + \frac{r^2(R)}{R^2}\left[1 - \frac{r^2(R)}{R^2}\right], \quad \tilde{I}_3 = \tilde{J}^2 = \left[\frac{r(R)}{R}\tilde{r}'(f(R))\right]^4. \tag{4.42}$$

It is seen that $\mathbf{S}$ is symmetric, and hence, the balance of angular momentum is satisfied. The surface of the cavity is traction free, and hence, we have the boundary condition $S^{RR}(R_i) = 0$.

If the virtual body is incompressible, i.e., $\tilde{J} = \tilde{r}^2\tilde{r}'/\tilde{R}^2 = 1$, and hence $\tilde{r}(\tilde{R}) = (\tilde{R}^3 + C)^{\frac{1}{3}}$, for some constant $C$. Similar to the cylindrical cloak $\lim_{\epsilon\to 0}\tilde{r}(\tilde{R}_i) = 0$, and hence $C = 0$. Therefore, $\tilde{r}(\tilde{R}) = \tilde{R}$ and consequently, from (4.30), $r(R) = R$, and $J = rr'/R = 1$, i.e., the cloak is made of an incompressible solid as well. From (4.28) the constitutive equations of the incompressible spherical cloak read

$$\begin{aligned}\mathbf{S} =& - p\begin{bmatrix} 1/(f'r'\tilde{r}') & 0 & 0 \\ 0 & R/(fr^2) & 0 \\ 0 & 0 & R/(fr^2\sin^2\Theta) \end{bmatrix} + \alpha \begin{bmatrix} f^2\tilde{r}'/(R^2r') & 0 & 0 \\ 0 & ff'/R^3 & 0 \\ 0 & 0 & ff'/(R^3\sin^2\Theta) \end{bmatrix} \\ &- \beta \begin{bmatrix} f^2/(R^2r'\tilde{r}'^{\,3}) & 0 & 0 \\ 0 & Rff'/r^4 & 0 \\ 0 & 0 & Rff'/(r^4\sin^2\Theta) \end{bmatrix}, \\ =& -p\begin{bmatrix} 1/f'(R) & 0 & 0 \\ 0 & 1/(Rf(R)) & 0 \\ 0 & 0 & 1/(Rf(R)\sin^2\Theta) \end{bmatrix} + \alpha \begin{bmatrix} f^2/R^2 & 0 & 0 \\ 0 & ff'/R^3 & 0 \\ 0 & 0 & ff'/(R^3\sin^2\Theta) \end{bmatrix} \\ &- \beta \begin{bmatrix} f^2/R^2 & 0 & 0 \\ 0 & ff'/R^3 & 0 \\ 0 & 0 & ff'/(R^3\sin^2\Theta) \end{bmatrix}, \\ =& -p\begin{bmatrix} 1/f'(R) & 0 & 0 \\ 0 & 1/(Rf(R)) & 0 \\ 0 & 0 & 1/(Rf(R)\sin^2\Theta) \end{bmatrix} + (\alpha - \beta)\begin{bmatrix} f^2/R^2 & 0 & 0 \\ 0 & ff'/R^3 & 0 \\ 0 & 0 & ff'/(R^3\sin^2\Theta) \end{bmatrix}. \end{aligned} \tag{4.43}$$

Note that in this case too $\mathbf{S}$ is symmetric, i.e., the balance of angular momentum is satisfied.

## 4.3 Linear Elastodynamics Transformation Cloaking

In this section we explore the possibility of transformation cloaking in classical linear elasticity. Let us assume that the reference motion is static, i.e., $\mathring{\mathbf{V}}(X,t) = \mathbf{0}$, and the ambient space is Euclidean, that is, $\mathsf{Ric}_{\mathbf{g}} = \mathbf{0}$. In this special case, the linearized material balance of linear momentum is simplified to read

$$\operatorname{Div}(\mathbf{A}\!:\!\nabla_0^{\mathbf{g}}\mathbf{U}) + \rho_0\nabla_{\mathbf{U}}^{\mathbf{g}}\mathring{\mathbf{B}} = \rho_0\ddot{\mathbf{U}}\,. \tag{4.44}$$

Now suppose that the reference configuration of the physical body $\mathcal{B}$ is mapped to the reference configuration of the virtual body $\tilde{\mathcal{B}}$, where $\mathcal{B}$ and $\tilde{\mathcal{B}}$ are endowed with the Euclidean metrics $\mathbf{G}(X)$ and $\tilde{\mathbf{G}}(\tilde{X})$, respectively. Let us denote the map between the two stress-free reference configurations by $\Xi : \mathcal{B} \to \tilde{\mathcal{B}}$. The Jacobian of this cloaking map is calculated as

$$J_\Xi = \sqrt{\frac{\det \tilde{\mathbf{G}}(\tilde{X}(X))}{\det \mathbf{G}(X)}}\,\det \overset{\Xi}{\mathbf{F}}\,. \tag{4.45}$$

**Linearized balance of linear momentum in the physical and virtual bodies.** Using the Piola identity, the divergence term in (4.44) can be rewritten as

$$\left(\mathsf{A}^{aA}{}_b{}^B U^b{}_{|B}\right)_{|A} = J_\Xi \left[ J_\Xi^{-1} \overset{\equiv}{F}{}^{\tilde{A}}{}_A \mathsf{A}^{aA}{}_b{}^B U^b{}_{|B} \right]_{|\tilde{A}} = J_\Xi \left( \tilde{\mathsf{A}}^{a\tilde{A}}{}_b{}^{\tilde{B}} U^b{}_{|\tilde{B}} \right)_{|\tilde{A}} , \tag{4.46}$$

where

$$U^b{}_{|\tilde{B}} = (\overset{\equiv}{F}{}^{-1})^B{}_{\tilde{B}} U^b{}_{|B}, \quad (\tilde{\mathsf{A}}^{a\tilde{A}}{}_b{}^{\tilde{B}}) \circ \Xi = J_\Xi^{-1} \overset{\equiv}{F}{}^{\tilde{A}}{}_A \overset{\equiv}{F}{}^{\tilde{B}}{}_B \mathsf{A}^{aA}{}_b{}^B . \tag{4.47}$$

Hence, using the shifter map and knowing that it is covariantly constant one can write

$$\mathsf{s}^{\tilde{a}}{}_a \left(\mathsf{A}^{aA}{}_b{}^B U^b{}_{|B}\right)_{|A} = J_\Xi \left( \tilde{\mathsf{A}}^{\tilde{a}\tilde{A}}{}_{\tilde{b}}{}^{\tilde{B}} \tilde{U}^{\tilde{b}}{}_{|\tilde{B}} \right)_{|\tilde{A}} , \tag{4.48}$$

where

$$\tilde{\mathsf{A}}^{\tilde{a}\tilde{A}}{}_{\tilde{b}}{}^{\tilde{B}} \circ \Xi = J_\Xi^{-1} \overset{\equiv}{F}{}^{\tilde{A}}{}_A \overset{\equiv}{F}{}^{\tilde{B}}{}_B \, \mathsf{s}^{\tilde{a}}{}_a (\mathsf{s}^{-1})^b{}_{\tilde{b}} \, \mathsf{A}^{aA}{}_b{}^B, \quad \tilde{U}^{\tilde{a}} \circ \Xi = \mathsf{s}^{\tilde{a}}{}_a \circ \varphi \, U^a . \tag{4.49}$$

Therefore

$$\mathbf{s} \circ \varphi \, \mathrm{Div}(\mathbf{A} \!:\! \nabla_0^{\mathbf{g}} \mathbf{U}) = J_\Xi \, \widetilde{\mathrm{Div}}(\tilde{\mathbf{A}} \!:\! \tilde{\nabla}_0^{\mathbf{g}} \tilde{\mathbf{U}}) \circ \Xi , \tag{4.50}$$

where $\tilde{\mathbf{U}} \circ \Xi = \mathbf{s} \circ \varphi \mathbf{U}$. Again, knowing that the shifter is covariantly constant, one obtains $\dot{\tilde{\mathbf{U}}} = (\mathbf{s} \circ \varphi) \dot{\mathbf{U}} \circ \Xi^{-1}$. Similarly, , $\ddot{\tilde{\mathbf{U}}} = (\mathbf{s} \circ \mathring{\varphi}) \ddot{\mathbf{U}} \circ \Xi^{-1}$. Substituting (4.50) into $(4.44)_1$, one can write the linearized balance of linear momentum in the virtual body as

$$\widetilde{\mathrm{Div}}(\tilde{\mathbf{A}} \!:\! \tilde{\nabla}_0^{\mathbf{g}} \tilde{\mathbf{U}}) + \tilde{\rho}_0 \tilde{\nabla}_{\tilde{\mathbf{U}}}^{\mathbf{g}} \mathring{\tilde{\mathbf{B}}} = \tilde{\rho}_0 \ddot{\tilde{\mathbf{U}}} , \tag{4.51}$$

where $\tilde{\rho}_0 = J_\Xi^{-1} \rho_0 \circ \Xi^{-1}$ and $\mathring{\tilde{\mathbf{B}}} = \mathbf{s} \circ \mathring{\varphi} \mathring{\mathbf{B}} \circ \Xi^{-1}$. In summary, the linearized balance of linear momentum is form-invariant under the following *cloaking transformations*:

$$\begin{aligned} &\tilde{X} = \Xi(X), \quad \tilde{\mathbf{U}} = \mathbf{s} \circ \mathring{\varphi} \; \mathbf{U} \circ \Xi^{-1}, \quad \mathring{\tilde{\mathbf{B}}} = \mathbf{s} \circ \mathring{\varphi} \; \mathring{\mathbf{B}} \circ \Xi^{-1}, \\ &\tilde{\rho}_0 = J_\Xi^{-1} \rho_0 \circ \Xi^{-1}, \quad \tilde{\mathbf{A}} = (J_\Xi^{-1} \mathbf{s}^{-1} \circ \mathring{\varphi} \; \overset{\equiv}{\mathbf{F}} \; \mathbf{s} \circ \mathring{\varphi} \; \mathbf{A} \; \overset{\equiv}{\mathbf{F}}{}^{\star}) \circ \Xi^{-1} . \end{aligned} \tag{4.52}$$

**Linearized balance of angular momentum in the physical and virtual bodies.** Suppose that the reference motions of both the physical and virtual bodies are their corresponding undeformed configurations, i.e., both $\mathring{\mathbf{F}}$ and $\mathring{\tilde{\mathbf{F}}}$ are identity maps (there are no initial stresses). Therefore, for any $\mathbf{W} \in T_X\mathcal{B}$, one has $\mathring{\tilde{\mathbf{F}}} \mathbf{S} \mathbf{W} = \mathbf{s} \mathring{\mathbf{F}} \mathbf{W}$, and hence, $\mathbf{S} = \mathring{\tilde{\mathbf{F}}}{}^{-1} \mathbf{s} \mathring{\mathbf{F}}$ (in components, $\mathsf{S}^{\tilde{A}}{}_A = \delta^{\tilde{A}}_{\tilde{a}} \, \mathsf{s}^{\tilde{a}}{}_a \delta^a_A$). Let us next linearize (4.19) with respect to $\mathring{\varphi}$ and $\mathring{\tilde{\varphi}}$, which reads $(\mathring{\mathbf{R}} \delta \mathbf{S})^\star = \mathring{\mathbf{R}} \delta \mathbf{S}$, where $\mathring{\mathbf{R}} = (\overset{\equiv}{\mathbf{F}}{}^{-1} \circ \Xi) \mathbf{S}$. In components

$$(\overset{\equiv}{F}{}^{-1})^A{}_{\tilde{A}} \mathsf{S}^{\tilde{A}}{}_M \mathsf{C}^{MBKL} = (\overset{\equiv}{F}{}^{-1})^B{}_{\tilde{A}} \mathsf{S}^{\tilde{A}}{}_M \mathsf{C}^{MAKL} . \tag{4.53}$$

Clearly, the balance of angular momentum may not be satisfied for a given cloaking map $\Xi$. In terms of the first Piola-Kirchhoff stress, the balance of angular momentum reads $P^{[aA} F^{c]}{}_A = 0$, where $2P^{[aA} F^{c]}{}_A = P^{aA} F^c{}_A - P^{cA} F^a{}_A$. Assuming that there is no initial stress, linearized balance of angular momentum reads $\delta P^{[aA} F^{c]}{}_A = 0$, or $\mathsf{A}^{[aAbB} \mathring{F}^{c]}{}_A = 0$. Assuming that balance of angular momentum in the virtual body is satisfied, i.e., $\tilde{\mathsf{A}}^{[\tilde{a}\tilde{A}\tilde{b}\tilde{B}} \mathring{\tilde{F}}^{\tilde{c}]}{}_{\tilde{A}} = 0$, the balance of angular momentum in the physical body requires that

$$(\mathsf{s}^{-1})^{[a}{}_{\tilde{a}} \mathring{F}^{c]}{}_A (\mathsf{s}^{-1})^b{}_{\tilde{b}} (\overset{\equiv}{F}{}^{-1})^A{}_{\tilde{A}} (\overset{\equiv}{F}{}^{-1})^B{}_{\tilde{B}} \tilde{\mathsf{A}}^{\tilde{a}\tilde{A}\tilde{b}\tilde{B}} = 0 . \tag{4.54}$$

Note that (4.54) is not necessarily satisfied.

**Transformed second elasticity tensor.** Using (3.15) and the relation $\mathbf{S} = \mathring{\tilde{\mathbf{F}}}^{-1}\mathbf{s}\mathring{\mathbf{F}}$, it is straightforward to show that $\tilde{\mathbf{C}} = J_\Xi^{-1}\bar{\bar{\mathbf{F}}}\mathbf{S}\bar{\bar{\mathbf{F}}}\mathbf{S}\mathbf{C}\circ\Xi$. In components one has

$$\tilde{\mathsf{C}}^{\tilde{A}\tilde{B}\tilde{C}\tilde{D}} = J_\Xi^{-1}\bar{\bar{F}}^{\tilde{A}}{}_A\,\mathsf{S}^{\tilde{B}}{}_B\,\bar{\bar{F}}^{\tilde{C}}{}_C\,\mathsf{S}^{\tilde{D}}{}_D\,\mathsf{C}^{ABCD}\,. \tag{4.55}$$

Or[22]

$$\mathsf{C}^{ABCD} = J_\Xi(\bar{\bar{F}}^{-1})^A{}_{\tilde{A}}\,(\mathsf{S}^{-1})^B{}_{\tilde{B}}\,(\bar{\bar{F}}^{-1})^C{}_{\tilde{C}}\,(\mathsf{S}^{-1})^D{}_{\tilde{D}}\,\tilde{\mathsf{C}}^{\tilde{A}\tilde{B}\tilde{C}\tilde{D}}\,. \tag{4.56}$$

We assume that the virtual structure is isotropic, and hence, $\tilde{\mathsf{C}}^{\tilde{A}\tilde{B}\tilde{C}\tilde{D}} = \lambda\tilde{G}^{\tilde{A}\tilde{B}}\tilde{G}^{\tilde{C}\tilde{D}}+\mu(\tilde{G}^{\tilde{A}\tilde{C}}\tilde{G}^{\tilde{B}\tilde{D}}+\tilde{G}^{\tilde{A}\tilde{D}}\tilde{G}^{\tilde{C}\tilde{B}})$ are given. Note that the transformed second elasticity tensor $\mathbf{C}$ possesses the major symmetries, i.e., $\mathsf{C}^{ABCD} = \mathsf{C}^{CDAB}$. However, the minor symmetries are not satisfied, in general. This is in agreement with the previous works in the literature.

**Remark 4.7.** If the initial stress is not zero, i.e., linearized elasticity with respect to a stressed configuration, one may expect that the elasticity constants in the physical body may become fully symmetric. However, initial stress in the cloak must vanish because from (4.20) one can write

$$\mathring{\tilde{S}}^{\tilde{A}\tilde{B}} = J_\Xi^{-1}\bar{\bar{F}}^{\tilde{A}}{}_C\mathring{R}^C{}_A\mathring{S}^{AB}\bar{\bar{F}}_B{}^{\tilde{B}}\,, \tag{4.57}$$

where $\mathring{\mathbf{R}} = \bar{\bar{\mathbf{F}}}^{-1}\mathring{\tilde{\mathbf{F}}}^{-1}\mathbf{s}\mathring{\mathbf{F}}$. Therefore, if the physical body is initially stressed, so is the virtual body. Nevertheless, note that if the virtual body is initially stressed, the wave patterns would be affected due to the anisotropy induced by the finite deformation the virtual body has undergone unless the initial stress is hydrostatic. But then a hydrostatic state of stress would violate the traction-free boundary condition on the inner boundary of the cloak, i.e., the hole. Therefore, the initial stress in the virtual body must be zero.

**Proposition 4.8.** *Classical linear elasticity is not flexible enough to allow for elastodynamics transformation cloaking regardless of the shape of the hole and the cloak. In other words, elastodynamics transformation cloaking is not possible if the cloak is required to be made of a classical linear elastic solid.*

*Proof.* The balance of angular momentum in the physical body (4.54) is expanded to read

$$(\mathsf{s}^{-1})^a{}_{\tilde{a}}\mathring{F}^c{}_A(\mathsf{s}^{-1})^b{}_{\tilde{b}}(\bar{\bar{F}}^{-1})^A{}_{\tilde{A}}(\bar{\bar{F}}^{-1})^B{}_{\tilde{B}}\tilde{\mathsf{A}}^{\tilde{a}\tilde{A}\tilde{b}\tilde{B}} = (\mathsf{s}^{-1})^c{}_{\tilde{a}}\mathring{F}^a{}_A(\mathsf{s}^{-1})^b{}_{\tilde{b}}(\bar{\bar{F}}^{-1})^A{}_{\tilde{A}}(\bar{\bar{F}}^{-1})^B{}_{\tilde{B}}\tilde{\mathsf{A}}^{\tilde{a}\tilde{A}\tilde{b}\tilde{B}}\,. \tag{4.58}$$

Thus, knowing that $\mathring{F}^a{}_A = \delta^a_A$, one obtains

$$\begin{aligned}&(\mathsf{s}^{-1})^a{}_{\tilde{a}}(\mathsf{s}^{-1})^b{}_{\tilde{b}}(\bar{\bar{F}}^{-1})^c{}_{\tilde{A}}(\bar{\bar{F}}^{-1})^B{}_{\tilde{B}}\tilde{\mathsf{A}}^{\tilde{a}\tilde{A}\tilde{b}\tilde{B}}\\ &\quad= (\mathsf{s}^{-1})^c{}_{\tilde{a}}(\mathsf{s}^{-1})^b{}_{\tilde{b}}(\bar{\bar{F}}^{-1})^a{}_{\tilde{A}}(\bar{\bar{F}}^{-1})^B{}_{\tilde{B}}\tilde{\mathsf{A}}^{\tilde{a}\tilde{A}\tilde{b}\tilde{B}}\,,\quad \forall\, a,b,c,B\in\{1,2,3\}.\end{aligned} \tag{4.59}$$

Without loss of generality, we may represent (4.59) in the Cartesian coordinates in which the shifter is the identity, i.e., $\mathsf{s}^{\tilde{a}}{}_a = \delta^{\tilde{a}}{}_a$. Therefore

$$(\bar{\bar{F}}^{-1})^B{}_{\tilde{B}}\left[(\bar{\bar{F}}^{-1})^c{}_{\tilde{A}}\tilde{\mathsf{A}}^{a\tilde{A}b\tilde{B}} - (\bar{\bar{F}}^{-1})^a{}_{\tilde{A}}\tilde{\mathsf{A}}^{c\tilde{A}b\tilde{B}}\right] = 0\,,\quad \forall\, a,b,c,B\in\{1,2,3\}. \tag{4.60}$$

Knowing that in the Cartesian coordinates $\tilde{\mathsf{A}}^{\tilde{a}\tilde{A}\tilde{b}\tilde{B}} = \lambda\delta^{\tilde{a}\tilde{A}}\delta^{\tilde{b}\tilde{B}}+\mu(\delta^{\tilde{a}\tilde{b}}\delta^{\tilde{A}\tilde{B}}+\delta^{\tilde{a}\tilde{B}}\delta^{\tilde{b}\tilde{A}})$, for an arbitrary cloaking transformation with components $(\bar{\bar{F}}^{-1})^i{}_j = \mathsf{F}_{ij}$, $i,j\in\{1,2,3\}$, (4.60) is simplified for $i\neq j\neq k\in\{1,2,3\}$ to read

$$(\lambda+\mu)\mathsf{F}_{ii}(\mathsf{F}_{ij}-\mathsf{F}_{ji}) - \mu(\mathsf{F}_{ij}\mathsf{F}_{jj}+\mathsf{F}_{ik}\mathsf{F}_{jk}+\mathsf{F}_{ii}\mathsf{F}_{ji}) = 0\,, \tag{4.61}$$

$$\mu(\mathsf{F}_{ij}\mathsf{F}_{ik}-\mathsf{F}_{ii}\mathsf{F}_{kj}) + \lambda\mathsf{F}_{ij}(\mathsf{F}_{ik}-\mathsf{F}_{ki}) = 0\,, \tag{4.62}$$

$$\mu(\mathsf{F}_{ii}^2+\mathsf{F}_{ik}^2) + (\lambda+2\mu)\mathsf{F}_{ij}^2 - \mu\mathsf{F}_{ii}\mathsf{F}_{jj} - \lambda\mathsf{F}_{ij}\mathsf{F}_{ji} = 0\,. \tag{4.63}$$

[22]When Cartesian coordinates are used in both the physical and virtual bodies one writes $\tilde{\mathsf{C}}^{\tilde{A}\tilde{B}\tilde{C}\tilde{D}} = J_\Xi^{-1}\bar{\bar{F}}^{\tilde{A}}{}_A\,\bar{\bar{F}}^{\tilde{C}}{}_C\,\delta^{\tilde{B}}_B\delta^{\tilde{D}}_D\,\mathsf{C}^{ABCD}$. This is identical to Norris and Parnell [2012]'s Eq. (2.6).

(4.61) can be rewritten as ($i \leftrightarrow j$)

$$(\lambda+\mu)\mathsf{F}_{jj}(\mathsf{F}_{ji}-\mathsf{F}_{ij})-\mu(\mathsf{F}_{ji}\mathsf{F}_{ii}+\mathsf{F}_{jk}\mathsf{F}_{ik}+\mathsf{F}_{jj}\mathsf{F}_{ij})=0\,. \tag{4.64}$$

Using (4.61) and (4.64), one obtains

$$(\lambda+\mu)(\mathsf{F}_{ii}+\mathsf{F}_{jj})(\mathsf{F}_{ij}-\mathsf{F}_{ji})=0. \tag{4.65}$$

Note that $\lambda+\mu=\frac{1}{3}(3\lambda+2\mu)+\frac{1}{3}\mu>0$. Thus, either $\mathsf{F}_{ij}=\mathsf{F}_{ji}$, or $\mathsf{F}_{ii}=-\mathsf{F}_{jj}$. Note that $\mathsf{F}_{ii}=-\mathsf{F}_{jj}$ immediately implies that $\mathsf{F}_{ii}=0$, and from (4.61), $\mathsf{F}_{jk}=0$, resulting in $\bar{\bar{\bar{\mathbf{F}}}}^{-1}=\mathbf{0}$, which is not possible as the tangent map of a cloaking transformation cannot be singular. Hence, $\mathsf{F}_{ij}=0$ for $i\neq j\in\{1,2,3\}$. It is then straightforward to see that $\bar{\bar{\bar{\mathbf{F}}}}=\beta\mathbf{I}$, where $\mathbf{I}$ denotes the identity and $\beta$ is a scalar. As the cloaking transformation $\Xi$ must be the identity on the boundary of the cloak $\partial_o\mathcal{C}$ one concludes that $\beta=1$, and thus, $\Xi=id$. Therefore, transformation cloaking is not possible in classical linear elasticity. □

**Remark 4.9.** We observe that the balance of angular momentum is the obstruction to cloaking, and not the acceleration term. This implies that transformation cloaking is not possible in classical linear elasticity at fixed frequency or classical linear elastostatics either.

**Remark 4.10.** Note that when $\mu=0$ (a pentamode material [Milton and Cherkaev, 1995]) transformation cloaking may be possible. When $\mu=0$, the only constraint imposed by the balance of angular momentum for the physical body (cf. (4.61), (4.62), and (4.63)) is that $\bar{\bar{\bar{\mathbf{F}}}}$ be symmetric. One should note that $\bar{\bar{\bar{\mathbf{F}}}}$ is a two-point tensor. However, as the reference configuration of both the virtual and the physical bodies are embedded in the Euclidean space, symmetry of $\bar{\bar{\bar{\mathbf{F}}}}$ makes sense. For an arbitrary hole surrounded by a cloak (with an arbitrary shape), if the derivative of a cloaking map is symmetric, the elastic constants of the cloak are fully symmetric. This happens to be the case for cylindrical and spherical cloaks. As we will see in §5.3, for the virtual and physical boundary-value problems to be equivalent outside the cloak, $\bar{\bar{\bar{\mathbf{F}}}}$ must be the identity map on the outer boundary of the cloak, i.e., $\bar{\bar{\bar{\mathbf{F}}}}|_{\partial_o\mathcal{C}}=id$.

Next we critically examine a cloaking geometry that has been suggested in several previous works in the literature. Consider an infinitely-long hollow solid cylinder that in its stress-free reference configuration has inner and outer radii $R_i$ and $R_o$, respectively. Let us transform the reference configuration to the reference configuration of another body (virtual body) that is a hollow cylinder with inner and outer radii $\epsilon$ and $R_o$, respectively, using a cloaking map $\Xi(R,\Theta,Z)=(f(R),\Theta,Z)$ such that $f(R_o)=R_o$. For such a map we have

$$\bar{\bar{\bar{\mathbf{F}}}}=\begin{bmatrix} f'(R) & 0 & 0\\ 0 & 1 & 0\\ 0 & 0 & 1\end{bmatrix}. \tag{4.66}$$

Let us consider an infinitely extended isotropic homogeneous elastic medium with shear modulus $\mu$, Lamé constant $\lambda$, and mass density $\rho_0$ containing an infinitely-long cylindrical cavity $\mathcal{H}$ with radius $R_i$ in its stress-free reference configuration. Let $(R,\Theta,Z)$ be the cylindrical coordinates such that $R=0$ corresponds to the centerline of the cavity. The cloaking device $\mathcal{C}$ is an infinitely-long hollow solid cylinder with inner radius $R_i$ (radius of the hole) and outer radius $R_o$, where $R_o<R_s$ ($R_s$ is the distance of a line source from the origin) surrounding the hole. The elastic properties of the cloaking device are to be determined. In doing so, the reference configuration of the physical body $\mathcal{B}$ is mapped to that of the virtual body $\tilde{\mathcal{B}}$ via a mapping $\Xi:\mathcal{B}\to\tilde{\mathcal{B}}$, where for $R_i\leq R\leq R_o$, it is defined as, $(\tilde{R},\tilde{\Theta},\tilde{Z})=\Xi(R,\Theta,Z)=(f(R),\Theta,Z)$ such that $f(R_i)=\epsilon$ and $f(R_o)=R_o$, and for $R\geq R_o$ it is the identity map. We assume that the transformed reference configuration is also isotropic and homogeneous with the same elastic properties as the medium in the physical reference configuration. Let the reference configurations of $\mathcal{B}$ and $\tilde{\mathcal{B}}$ be endowed with the flat metric in the cylindrical coordinates, i.e., $\mathbf{G}=\operatorname{diag}(1,R^2,1)$ and $\tilde{\mathbf{G}}=\operatorname{diag}(1,\tilde{R}^2,1)$ in the cylindrical coordinates $(R,\Theta,Z)$ and $(\tilde{R},\tilde{\Theta},\tilde{Z})$, respectively. Also, let the ambient space be endowed with the Euclidean metric $\mathbf{g}=\operatorname{diag}(1,r^2,1)$ in the coordinates $(r,\theta,z)$. Therefore, the elasticity tensor for the material in the virtual body reads

$$\tilde{\mathsf{C}}^{\tilde{A}\tilde{B}\tilde{C}\tilde{D}}=\lambda\tilde{G}^{\tilde{A}\tilde{B}}\tilde{G}^{\tilde{C}\tilde{D}}+\mu(\tilde{G}^{\tilde{A}\tilde{C}}\tilde{G}^{\tilde{B}\tilde{D}}+\tilde{G}^{\tilde{A}\tilde{D}}\tilde{G}^{\tilde{C}\tilde{B}})\,. \tag{4.67}$$

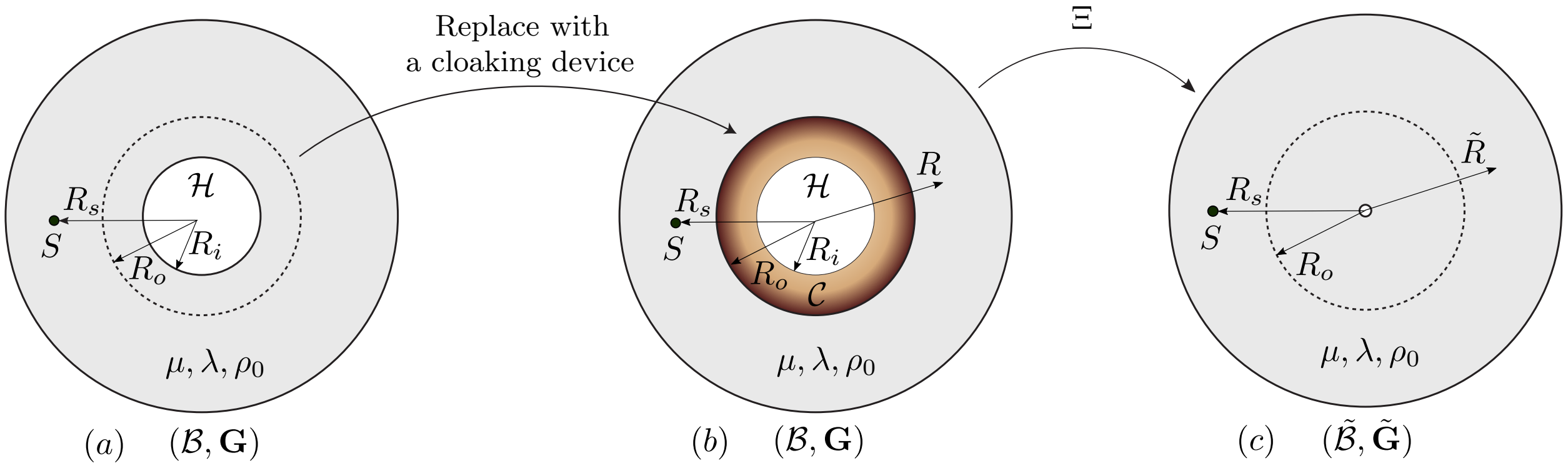

Figure 6: *Cloaking an object inside a hole $\mathcal{H}$ from elastic waves generated by a line source located at a distance $R_s$ from the center of the hole. The system (a) is an isotropic and homogeneous medium containing a finite hole $\mathcal{H}$. The system (c) is isotropic and homogeneous with the same elastic properties as the medium in the reference configuration (b) outside the cloaking region $\mathcal{C}$. The configuration (b) is mapped to the new reference configuration (c) such that the hole $\mathcal{H}$ is mapped to an infinitesimal hole with negligible effects on the elastic waves. The cloaking transformation is the identity mapping in $\mathcal{B} \setminus \mathcal{C}$.*

In Voigt representation $\tilde{\mathsf{C}}$ reads[23]

$$\tilde{\mathsf{C}} = \begin{bmatrix} \lambda + 2\mu & \lambda/\tilde{R}^2 & \lambda & 0 & 0 & 0 \\ \lambda/\tilde{R}^2 & (\lambda+2\mu)/\tilde{R}^4 & \lambda/\tilde{R}^2 & 0 & 0 & 0 \\ \lambda & \lambda/\tilde{R}^2 & \lambda+2\mu & 0 & 0 & 0 \\ 0 & 0 & 0 & \mu/\tilde{R}^2 & 0 & 0 \\ 0 & 0 & 0 & 0 & \mu & 0 \\ 0 & 0 & 0 & 0 & 0 & \mu/\tilde{R}^2 \end{bmatrix} . \tag{4.68}$$

As $\Xi$ is the identity mapping for $R \geq R_o$, we note that $\mathsf{C} = \tilde{\mathsf{C}}$ in $\mathcal{B} \setminus \mathcal{C}$, for which $R \geq R_o$. For $R_i \leq R \leq R_o$ (the cloaking device region $\mathcal{C}$), we have

$$\mathsf{C}_{\mathcal{C}}^{ABCD} = J_{\Xi}^{\mathcal{C}} (\bar{\bar{F}}^{-1})^A{}_{\tilde{A}} (\mathsf{S}^{-1})^B{}_{\tilde{B}} (\bar{\bar{F}}^{-1})^C{}_{\tilde{C}} (\mathsf{S}^{-1})^D{}_{\tilde{D}} \tilde{\mathsf{C}}^{\tilde{A}\tilde{B}\tilde{C}\tilde{D}} \circ \Xi , \tag{4.69}$$

where $\bar{\bar{\mathbf{F}}}^{-1} = \operatorname{diag}(f'(R)^{-1}, 1, 1)$, and hence

$$J_{\Xi}^{\mathcal{C}} = \sqrt{\frac{\det \tilde{\mathbf{G}}(\tilde{X}(X))}{\det \mathbf{G}(X)}} \det \bar{\bar{\mathbf{F}}} = \frac{f(R) f'(R)}{R} , \quad R_i \leq R \leq R_o . \tag{4.70}$$

Note that

$$\mathsf{S} = \begin{bmatrix} 1 & 0 & 0 \\ 0 & R/f(R) & 0 \\ 0 & 0 & 1 \end{bmatrix} . \tag{4.71}$$

Therefore, in the cloaking device ($R_i \leq R \leq R_o$), the elasticity tensor has the following non-zero physical

[23] Note that in the Voigt notation one has the following bijection between indices $\{11, 22, 33, 23, 13, 12\} \to \{1, 2, 3, 4, 5, 6\}$.

components[24]

$$
\begin{aligned}
&\hat{\mathsf{C}}^{RRRR} = \frac{(\lambda+2\mu)f(R)}{Rf'(R)}, && \hat{\mathsf{C}}^{\Theta\Theta\Theta\Theta} = \frac{(\lambda+2\mu)Rf'(R)}{f(R)}, && \hat{\mathsf{C}}^{RR\Theta\Theta} = \hat{\mathsf{C}}^{\Theta\Theta RR} = \lambda, \\
&\hat{\mathsf{C}}^{R\Theta R\Theta} = \frac{\mu f(R)}{Rf'(R)}, && \hat{\mathsf{C}}^{\Theta R\Theta R} = \frac{\mu Rf'(R)}{f(R)}, && \hat{\mathsf{C}}^{\Theta RR\Theta} = \hat{\mathsf{C}}^{R\Theta\Theta R} = \mu\,, \\
&\hat{\mathsf{C}}^{RRZZ} = \hat{\mathsf{C}}^{ZZRR} = \frac{\lambda f(R)}{R}, && \hat{\mathsf{C}}^{\Theta\Theta ZZ} = \hat{\mathsf{C}}^{ZZ\Theta\Theta} = \lambda f'(R), && \hat{\mathsf{C}}^{ZZZZ} = \frac{(\lambda+2\mu)f(R)f'(R)}{R}\,, \\
&\hat{\mathsf{C}}^{RZRZ} = \frac{\mu f(R)}{Rf'(R)}\,, && \hat{\mathsf{C}}^{ZRZR} = \frac{\mu f(R)f'(R)}{R}\,, && \hat{\mathsf{C}}^{ZRRZ} = \hat{\mathsf{C}}^{RZZR} = \frac{\mu f(R)}{R}\,, \\
&\hat{\mathsf{C}}^{Z\Theta Z\Theta} = \frac{\mu f(R)f'(R)}{R}\,, && \hat{\mathsf{C}}^{\Theta Z\Theta Z} = \frac{\mu Rf'(R)}{f(R)}\,, && \hat{\mathsf{C}}^{\Theta ZZ\Theta} = \hat{\mathsf{C}}^{Z\Theta\Theta Z} = \mu f'(R)\,.
\end{aligned}
\tag{4.72}
$$

**Remark 4.11.** When $\mu = 0$ (energy density is not positive-definite anymore), the elastic constants of the cloak can be written as

$$
\begin{aligned}
&\hat{\mathsf{C}}^{RRRR} = \frac{\lambda f(R)}{Rf'(R)}, \quad \hat{\mathsf{C}}^{\Theta\Theta\Theta\Theta} = \frac{\lambda Rf'(R)}{f(R)}, \quad \hat{\mathsf{C}}^{RR\Theta\Theta} = \hat{\mathsf{C}}^{\Theta\Theta RR} = \lambda, \\
&\hat{\mathsf{C}}^{R\Theta R\Theta} = \hat{\mathsf{C}}^{\Theta R\Theta R} = \hat{\mathsf{C}}^{\Theta RR\Theta} = \hat{\mathsf{C}}^{R\Theta\Theta R} = 0\,, \\
&\hat{\mathsf{C}}^{RRZZ} = \hat{\mathsf{C}}^{ZZRR} = \frac{\lambda f(R)}{R}, \quad \hat{\mathsf{C}}^{\Theta\Theta ZZ} = \hat{\mathsf{C}}^{ZZ\Theta\Theta} = \lambda f'(R), \quad \hat{\mathsf{C}}^{ZZZZ} = \frac{\lambda f(R)f'(R)}{R}\,, \\
&\hat{\mathsf{C}}^{RZRZ} = \hat{\mathsf{C}}^{ZRZR} = \hat{\mathsf{C}}^{ZRRZ} = \hat{\mathsf{C}}^{RZZR} = 0\,, \\
&\hat{\mathsf{C}}^{Z\Theta Z\Theta} = \hat{\mathsf{C}}^{\Theta Z\Theta Z} = \hat{\mathsf{C}}^{\Theta ZZ\Theta} = \hat{\mathsf{C}}^{Z\Theta\Theta Z} = 0\,.
\end{aligned}
\tag{4.73}
$$

It is seen that in this special case the elastic constants of the cloak are fully symmetric.

Note that $\rho = J_\Xi\, \tilde{\rho} \circ \Xi$, and therefore, the mass density in $\mathcal{B} \setminus \mathcal{C}$ is homogeneous and is equal to $\rho_0$. The mass density in the cloaking device is inhomogeneous and is given by

$$
\rho_{\mathcal{C}}(R) = \frac{f(R)f'(R)}{R}\rho_0\,, \qquad R_i \leq R \leq R_o\,. \tag{4.74}
$$

In the case of anti-plane waves, the only non-trivial equilibrium equation is the one in the $z$-direction, which using (3.15) for the physical body in the cloaking region is written as [25]

$$
\frac{1}{R}\frac{\partial}{\partial R}\left(R\hat{\mathsf{C}}^{RZRZ}(R)\frac{\partial W}{\partial R}\right) + \frac{1}{R^2}\hat{\mathsf{C}}^{\Theta Z\Theta Z}(R)\frac{\partial^2 W}{\partial \Theta^2} + \rho_{\mathcal{C}}(R)\omega^2 W = \frac{a_m}{R_s}\delta(R-R_s)\delta(\Theta-\Theta_s)\,, \tag{4.75}
$$

where $a_m$ is the amplitude of a time-harmonic line source with frequency $\omega$ located at $(R_s, \Theta_s)$ such that the induced displacement is given as $\delta\varphi_t(R,\Theta) = \mathbf{U}(R,\Theta,t) = \big(0, 0, \Re\big[W(R,\Theta)e^{-\mathrm{i}\omega t}\big]\big)$. The only non-zero stress components in the cloak read

$$
\delta S^{RZ} = \mathsf{C}^{ZRRZ} W_{|R}\,, \quad \delta S^{ZR} = \mathsf{C}^{RZRZ} W_{|R}\,, \quad \delta S^{\Theta Z} = \mathsf{C}^{Z\Theta\Theta Z} W_{|\Theta}\,, \quad \delta S^{Z\Theta} = \mathsf{C}^{\Theta Z\Theta Z} W_{|\Theta}\,. \tag{4.76}
$$

[24] Note that the physical components of the elasticity tensor $\hat{C}^{ABCD}$ are related to the components of the elasticity tensor as $\hat{C}^{ABCD} = \sqrt{G_{AA}}\sqrt{G_{BB}}\sqrt{G_{CC}}\sqrt{G_{DD}}C^{ABCD}$ (no summation) [Truesdell, 1953].

[25] Using (3.15), one obtains

$$
\left(\mathsf{A}^{aA}{}_z{}^C W^z{}_{|C}\right)_{|A} = \left(\mathsf{C}^{ABCN} \mathring{F}^a{}_B \mathring{F}^n{}_N g_{zn} W^z{}_{|C}\right)_{|A} = \left(\mathsf{C}^{ABCZ} \delta^a{}_B \delta^z{}_Z g_{zz} W^z{}_{|C}\right)_{|A}\,.
$$

Note that

$$
\begin{aligned}
\left(\mathsf{C}^{ABCZ} W_{|C}\right)_{|A} &= \left(\mathsf{C}^{RBRZ} W_{|R} + \mathsf{C}^{RB\Theta Z} W_{|\Theta}\right)_{|R} + \left(\mathsf{C}^{\Theta BRZ} W_{|R} + \mathsf{C}^{\Theta B\Theta Z} W_{|\Theta}\right)_{|\Theta} \\
&= \left(\mathsf{C}^{RZRZ} W_{|R}\right)_{|R} + \left(\mathsf{C}^{\Theta Z\Theta Z} W_{|\Theta}\right)_{|\Theta}\,.
\end{aligned}
$$

As $\hat{\mathsf{C}}^{ZRRZ} \neq \hat{\mathsf{C}}^{RZRZ}$ and $\hat{\mathsf{C}}^{Z\Theta\Theta Z} \neq \hat{\mathsf{C}}^{\Theta Z\Theta Z}$ (cf. (4.72)), the balance of angular momentum is not satisfied in the simple case of antiplane waves. One should note that the imbalance of angular momentum is actually significant as $\hat{\mathsf{C}}^{Z\Theta\Theta Z} - \hat{\mathsf{C}}^{\Theta Z\Theta Z} = \mu f'(R)\left[1 - \frac{R}{f(R)}\right]$ will be unbounded at the inner boundary of the cloak ($R = R_i$) in the limit of $f(R_i) = \epsilon \to 0$. Thus, the loss of the minor symmetry of the elasticity tensor cannot be ignored. For a linear mapping[26] one has

$$\hat{\mathsf{C}}^{RZRZ}(R) = \mu\frac{R - R_i}{R}\,, \quad \hat{\mathsf{C}}^{\Theta Z\Theta Z}(R) = \mu\frac{R}{R - R_i}\,, \quad \rho_{\mathcal{C}}(R) = \frac{R_o^2}{(R_o - R_i)^2}\left(1 - \frac{R_i}{R}\right)\rho_0\,. \tag{4.77}$$

These are identical to Parnell [2012]'s Eq.(2.5). Brun et al. [2009] and many other researchers have used a linear function for $f(R)$, which has been borrowed from the similar calculations in electromagnetism. However, a linear $f(R)$ is not acceptable for elasticity as will be explained in §5.2. Let us ignore this condition and use a linear $f(R)$, i.e.,

$$f(R) = \frac{R_o(R - R_i)}{R_o - R_i}\,. \tag{4.78}$$

For this (inappropriate) choice of $f(R)$ the mass density in the cloak reads

$$\rho_{\mathcal{C}}(R) = \frac{R_o^2}{(R_o - R_i)^2}\left(1 - \frac{R_i}{R}\right)\rho_0\,, \tag{4.79}$$

which is identical to Brun et al. [2009]'s mass density. Our elastic constants in the cloak read

$$\begin{aligned}
&\hat{\mathsf{C}}^{RRRR} = (\lambda + 2\mu)\frac{R - R_i}{R}, \quad &&\hat{\mathsf{C}}^{\Theta\Theta\Theta\Theta} = (\lambda + 2\mu)\frac{R}{R - R_i}, \quad &&\hat{\mathsf{C}}^{RR\Theta\Theta} = \hat{\mathsf{C}}^{\Theta\Theta RR} = \lambda, \\
&\hat{\mathsf{C}}^{\Theta RR\Theta} = \hat{\mathsf{C}}^{R\Theta\Theta R} = \mu, &&\hat{\mathsf{C}}^{R\Theta R\Theta} = \frac{R - R_i}{R}\mu, &&\hat{\mathsf{C}}^{\Theta R\Theta R} = \frac{R}{R - R_i}\mu\,,
\end{aligned} \tag{4.80}$$

which are identical to Brun et al. [2009]'s as well. Brun et al. [2009] and many other authors (e.g., Norris and Shuvalov [2011]) have claimed that the lack of minor symmetry in the elastic constants of the cloak implies that a cloak is made of a Cosserat solid. This claim is, unfortunately, incorrect. We will show in §5.2 that transformation cloaking is not possible in (generalized) Cosserat solids.

# 5 Elastodynamics Transformation Cloaking in Solids with Microstructure

Having established that classical linear elasticity is not flexible enough to allow for transformation cloaking a possible solution would be to see if transformation cloaking may be achieved in solids with microstructure. In such continua the Cauchy stress does not need to be symmetric. This is the reason that in the literature of elastodynamics cloaking it has been suggested that a cloak should be made of a Cosserat solid. The first systematic formulation of generalized continua goes back to the seminal work of Cosserat brothers [Cosserat and Cosserat, 1909], which remained unnoticed until the interest in continua with microstructure was revived in the 1950s, 1960s, and 1970s [Ericksen and Truesdell, 1957, Toupin, 1962, 1964, Mindlin and Tiersten, 1962, Mindlin, 1964, Eringen, 2012] and now there is a vast literature on generalized continua.

In formulating the cloaking problem, we assume that both the virtual and physical bodies are made of solids with microstructure. We can follow two paths: i) Assume that energy depends on $\nabla\mathbf{F}$ with components $F^a{}_{A|B}$, i.e., gradient elasticity. This seems to be more natural for our purposes because first there is no ambiguity in the physical meaning of microstructure, and second there are well-established connections between strain gradient elasticity and atomistic calculations (see [Maranganti and Sharma, 2007] and references therein). ii) In addition to the deformation mapping $\varphi$ assume a set of director fields that have their own independent kinematics, i.e., (generalized) Cosserat elasticity. Energy depends on these extra fields as well. One should note that there are many different choices for describing microstcuture. This has been discussed in some detail in the monograph [Capriz, 2013]. See also [Yavari and Marsden, 2009a]. We discuss elastodynamics transformation cloaking in both gradient and generalized Cosserat solids.

[26] This map does not satisfy the required traction continuity condition, i.e., $f'(R_o) = 1$, but nevertheless has been extensively used in the literature (see §5.2).

## 5.1 Elastodynamics Transformation Cloaking in Gradient Elastic Solids

In gradient elasticity (or strain-gradient elasticity) energy function depends on the (covariant) derivative of the deformation gradient as well, i.e., [Toupin, 1964]

$$W = W(X, \mathbf{F}, \nabla\mathbf{F}, \mathbf{G}, \mathbf{g}\circ\varphi)\,. \tag{5.1}$$

Note that (bulk) compatibility equations are written as ${F^a}_{A|B} = {F^a}_{B|A}$ [Yavari, 2013]. Objectivity of an energy function means invariance under (rigid-body) rotations in the ambient space. One can think of $\mathbf{F}$ as a vector-valued 1-form [Yavari, 2008]. The 1-form part would not be affected by changes of coordinates in the ambient space. Now the question is how should $W$ depend on $\mathbf{F}$ in order to be isotropic? Suppose $f = f(\mathbf{u})$, where $\mathbf{u}$ is a vector. We know that for $f$ to be isotropic it should have the form $f = \hat{f}(\mathbf{u}\cdot\mathbf{u})$, where $\mathbf{u}\cdot\mathbf{u} = u^a u^b g_{ab}$ [Spencer, 1971]. This new variable in the case of deformation gradient is ${F^a}_A {F^b}_A g_{ab} = C_{AB}$. In gradient elasticity one has an extra independent variable. Let us first consider a scalar function of two vectors $f = f(\mathbf{u},\mathbf{v})$. For $f$ to be isotropic, one must have $f = \hat{f}(\mathbf{u}\cdot\mathbf{u}, \mathbf{u}\cdot\mathbf{v}, \mathbf{v}\cdot\mathbf{v})$. In the case of energy function these three variables are

$${F^a}_A {F^b}_B g_{ab} = C_{AB}\,, \quad {F^a}_A {F^b}_{B|C} g_{ab} =: D_{ABC}\,, \quad {F^a}_{A|B} {F^b}_{C|D} g_{ab} =: E_{ABCD}\,. \tag{5.2}$$

Note the symmetries of the new measures of strain: $D_{ABC} = D_{ACB}$, $E_{ABCD} = E_{CDAB} = E_{BACD} = E_{ABDC} = E_{BADC}$. The strain measure $\mathbf{E}$ is, however, functionally dependent on $\mathbf{C}^\flat$ and $\mathbf{D}$ as noticed by Toupin [1964]. More specifically, given $\mathbf{D}$, one can write ${F^b}_{B|C} = g^{ab}(F^{-1})^A{}_a D_{ABC}$. Hence, $E_{ABCD} = C^{-MN} D_{MAB} D_{NCD}$. Therefore, $W = \hat{W}(X, C_{AB}, D_{ABC}, G_{AB})$. Note that $C_{AB|C} = {F^a}_{A|C} {F^b}_B g_{ab} + {F^a}_A {F^b}_{B|C} g_{ab} = D_{BAC} + D_{ABC}$. Thus, $D_{ABC} = \frac{1}{2}\left(C_{AB|C} + C_{AC|B} - C_{BC|A}\right)$. Therefore, one can assume that energy density depends on $\mathbf{C}^\flat$ and $\nabla^{\mathbf{G}}\mathbf{C}^\flat$, i.e., $W = \hat{W}(X, C_{AB}, C_{AB|C}, G_{AB})$, which is exactly the way Toupin expressed the energy density.

**Balance of linear momentum.** We next derive the governing equations of gradient elasticity using Hamilton's principle of least action. Assuming the Lagrangian density $\mathcal{L} = W - T$, where $T = \frac{1}{2}\rho_0 \langle\!\langle \mathbf{V}, \mathbf{V} \rangle\!\rangle_{\mathbf{g}} = \frac{1}{2}\rho_0 V^a V^b g_{ab}$ is the classical kinetic energy density, the action is defined as

$$S = \int_{t_1}^{t_2} \int_{\mathcal{B}} (W - T)\, dV dt\,, \tag{5.3}$$

where $t_1 < t_2$ are arbitrary time instances, and $dV$ is the Riemannian volume element of $(\mathcal{B}, \mathbf{G})$. Hamilton's principle is written as

$$\delta S = \int_{t_1}^{t_2} \int_{\mathcal{B}} (\delta W - \delta T)\, dV dt = \int_{\partial_t \mathcal{B}} T^a \delta\varphi^b g_{ab}\, dA\,, \tag{5.4}$$

where $T^a$ is traction, and $\partial_t\mathcal{B} \subset \partial\mathcal{B}$ is part of the boundary on which tractions are specified. The only term that is different from that of classical nonlinear elasticity is $\delta W$, which is calculated as

$$\delta W = \nabla^{\varphi_\epsilon}_{\frac{\partial}{\partial\epsilon}} W \Big|_{\epsilon=0} = \frac{\partial W}{\partial \mathbf{F}} : \delta\mathbf{F} + \frac{\partial W}{\partial \nabla\mathbf{F}} : \delta\nabla^{\varphi}\mathbf{F} = \frac{\partial W}{\partial {F^a}_A} \delta {F^a}_A + \frac{\partial W}{\partial {F^a}_{A|B}} \delta {F^a}_{A|B}\,, \tag{5.5}$$

where we used the fact that $\nabla\mathbf{g} = \mathbf{0}$. The covariant derivative of the deformation gradient is linearized as follows.

$$\delta {F^a}_{A|B} = \nabla_{\frac{\partial}{\partial\epsilon}} \nabla_{\frac{\partial}{\partial X^B}} \frac{\partial\varphi^a_\epsilon}{\partial X^A}\Bigg|_{\epsilon=0} = \nabla_{\frac{\partial}{\partial X^B}} \nabla_{\frac{\partial}{\partial\epsilon}} \frac{\partial\varphi^a_\epsilon}{\partial X^A}\Bigg|_{\epsilon=0} = \nabla_{\frac{\partial}{\partial X^B}} \delta {F^a}_A = {U^a}_{|A|B}\,, \tag{5.6}$$

where use was made of the fact that $\nabla$ is a flat connection, and hence, order of covariant differentiation can be interchanged (see the appendix). Therefore

$$\delta W = \frac{\partial W}{\partial F^a{}_A}\delta\varphi^a{}_{|A} + \frac{\partial W}{\partial F^a{}_{A|B}}\delta\varphi^a{}_{|A|B}\,. \tag{5.7}$$

Note that $\frac{\partial W}{\partial F^a{}_A}\delta\varphi^a{}_{|A} = \left(\frac{\partial W}{\partial F^a{}_A}\delta\varphi^a\right)_{|A} - \left(\frac{\partial W}{\partial F^a{}_A}\right)_{|A}\delta\varphi^a$. Thus

$$\int_{\mathcal{B}} \frac{\partial W}{\partial F^a{}_A}\delta\varphi^a{}_{|A}dV = -\int_{\mathcal{B}}\left(\frac{\partial W}{\partial F^a{}_A}\right)_{|A}\delta\varphi^a dV + \int_{\partial\mathcal{B}}\frac{\partial W}{\partial F^a{}_A}N_A\delta\varphi^a dA\,, \tag{5.8}$$

where $\mathbf{N}$ is the unit normal vector to $\partial\mathcal{B}$, and $N_A$ are components of the corresponding 1-form $\mathbf{N}^\flat$. Similarly, for the second term one can write

$$\begin{aligned}\int_{\mathcal{B}} \frac{\partial W}{\partial F^a{}_{A|B}}\delta\varphi^a{}_{|A|B}dV &= -\int_{\mathcal{B}}\left(\frac{\partial W}{\partial F^a{}_{A|B}}\right)_{|B}\delta\varphi^a{}_{|A}dV + \int_{\partial\mathcal{B}}\frac{\partial W}{\partial F^a{}_{A|B}}N_B\delta\varphi^a{}_{|A}dA \\ &= \int_{\mathcal{B}}\left(\frac{\partial W}{\partial F^a{}_{A|B}}\right)_{|B|A}\delta\varphi^a dV - \int_{\partial\mathcal{B}}\left[\left(\frac{\partial W}{\partial F^a{}_{A|B}}N_B\right)_{|A} + \left(\frac{\partial W}{\partial F^a{}_{A|B}}\right)_{|B}N_A\right]\delta\varphi^a dA\,.\end{aligned} \tag{5.9}$$

In deriving the above relation we used the topological fact that boundary of a boundary is empty, i.e., $\partial\partial\mathcal{B} = \varnothing$. Therefore

$$\begin{aligned}\delta W = &-\int_{\mathcal{B}}\left[\frac{\partial W}{\partial F^a{}_A} - \left(\frac{\partial W}{\partial F^a{}_{A|B}}\right)_{|B}\right]_{|A}\delta\varphi^a dV + \int_{\partial\mathcal{B}}\left[\frac{\partial W}{\partial F^a{}_A} - \left(\frac{\partial W}{\partial F^a{}_{A|B}}\right)_{|B}\right]N_A\delta\varphi^a dA \\ &-\int_{\partial\mathcal{B}}\left(\frac{\partial W}{\partial F^a{}_{A|B}}N_B\right)_{|A}\delta\varphi^a dA\,.\end{aligned} \tag{5.10}$$

The first term in (5.10) implies that the first Piola-Kirchhoff stress in gradient elasticity has the following representation

$$P^{aA} = g^{ab}\left[\frac{\partial W}{\partial F^b{}_A} - \left(\frac{\partial W}{\partial F^b{}_{A|B}}\right)_{|B}\right]. \tag{5.11}$$

Following Toupin [1964] we define a *hyper-stress* $H_a{}^{AB} = H_a{}^{BA} = \frac{\partial W}{\partial F^a{}_{A|B}}$. The integrand of the third term in (5.10) is simplified as

$$(H_a{}^{AB}N_B)_{|A} = H_a{}^{AB}{}_{|A}N_B - \mathfrak{B}_{AB}H_a{}^{AB}, \tag{5.12}$$

where $\mathfrak{B}_{AB} = \mathfrak{B}_{BA} = -N_{A|B}$ is the second fundamental form of the surface $\partial\mathcal{B}$ embedded in the Euclidean space (we assume that the undeformed body is embedded in a Euclidean ambient space). Therefore, traction in gradient elasticity is written as

$$T^a = P^{aA}N_A - H^{aAB}{}_{|B}N_A + H^{aAB}\mathfrak{B}_{AB}. \tag{5.13}$$

Similar to classical nonlinear elasticity, $\delta T = \rho_0\langle\!\langle \mathbf{V}, D_t^{\mathbf{g}}\delta\varphi\rangle\!\rangle_{\mathbf{g}} = \frac{d}{dt}\rho_0\langle\!\langle \mathbf{V}, \delta\varphi\rangle\!\rangle_{\mathbf{g}} - \rho_0\langle\!\langle \mathbf{A}, \delta\varphi\rangle\!\rangle_{\mathbf{g}}$. Assuming that $\delta\varphi(X,t_1) = \delta\varphi(X,t_2) = 0$, one obtains

$$\delta\int_{t_1}^{t_2}\int_{\mathcal{B}} -T dV dt = \int_{t_1}^{t_2}\int_{\mathcal{B}}\rho_0\langle\!\langle \mathbf{A}, \delta\varphi\rangle\!\rangle_{\mathbf{g}}dV dt. \tag{5.14}$$

Therefore, the balance of linear momentum (the Euler-Lagrange equations) reads $P^{aA}{}_{|A} = \rho_0 A^a$ (inclusion of body forces would be straightforward using Lagrange-D'Alembert principle) and traction vector is given in (5.13).

The Cauchy stress $\sigma^{ab} = J^{-1}F^a{}_A P^{bA}$ in gradient elasticity has the following representation (note the typo in Toupin [1964]'s Eq.(10.18))

$$\sigma^{ab} = J^{-1}g^{bc}\left[F^a{}_A\frac{\partial W}{\partial F^c{}_A} + F^a{}_{A|B}\frac{\partial W}{\partial F^c{}_{A|B}}\right] - h^{bac}{}_{|c}, \tag{5.15}$$

where $h^{bac} = J^{-1}F^a{}_A F^c{}_B H^{bAB}$.

**Remark 5.1.** When a gradient elastic solid is in a stress-free state the traction vector at every point on any surface vanishes. From (5.13) this implies that $P^{aA} - H^{aAB}{}_{|B} = 0$, and $H^{aAB} = 0$. Therefore, in a stress-free state both the (total) first Piola-Kirchhoff stress and hyper-stress vanish.

**Balance of angular momentum.** In the setting of Lagrangian mechanics of continua the balance of angular momentum is derived using Noether's theorem. Consider a flow $\psi_s : \mathcal{S} \to \mathcal{S}$ on the (Euclidean) ambient space. According to Noether's theorem any symmetry of the Lagrangian density corresponds to a conserved quantity. In particular, invariance of a Lagrangian density under rotations of the ambient space corresponds to the balance of angular momentum. For the balance of angular momentum in a Euclidean ambient space $\psi_s(x) = x + s\mathbf{\Omega}x$, where $\mathbf{\Omega}$ is an anti-symmetric matrix. As the kinetic energy density is invariant under rotations one only needs to require invariance of the energy function under flows of rotations of the ambient space, i.e.,

$$W(X, \mathbf{F}, \nabla\mathbf{F}, \mathbf{G}, \mathbf{g}) = W(X, \psi_{s*}\mathbf{F}, \psi_{s*}\nabla\mathbf{F}, \mathbf{G}, \psi_{s*}\mathbf{g}). \tag{5.16}$$

Taking derivative with respect to $s$ of both sides and evaluating at $s = 0$, one obtains

$$g^{ac}\left[F^b{}_A\frac{\partial W}{\partial F^c{}_A} + F^b{}_{A|B}\frac{\partial W}{\partial F^c{}_{A|B}}\right]\Omega_{ab} = 0\,. \tag{5.17}$$

As $\Omega_{ba} = -\Omega_{ab}$, one concludes that $\Pi^{ab} = \Pi^{ba}$, or $\Pi^{[ab]} = 0$,[27] where

$$\Pi^{ab} = g^{ac}\left[F^b{}_A\frac{\partial W}{\partial F^c{}_A} + F^b{}_{A|B}\frac{\partial W}{\partial F^c{}_{A|B}}\right] = g^{ac}F^b{}_A\frac{\partial W}{\partial F^c{}_A} + H^{aAB}F^b{}_{A|B}\,. \tag{5.18}$$

In terms of the first Piola-Kirchhoff stress balance of angular momentum reads

$$P^{[aA}F^{b]}{}_A + \left(H^{[aAB}F^{b]}{}_A\right)_{|B} = 0\,. \tag{5.19}$$

In terms of the Cauchy stress, $\sigma^{[ab]} + m^{bac}{}_{|c} = 0$, where $m^{bac} = h^{[ba]c}$ is Toupin's *couple-stress*.

**Linearized balance of linear momentum.** Linearizing the balance of linear momentum about a motion $\mathring{\varphi}$ one obtains $(\delta P^{aA})_{|A} + \rho_0\delta B^a = \rho_0\ddot{U}^a$. Note that $\delta(P^{aA}{}_{|A}) = \delta P^{aA}{}_{|A}$, where

$$\delta P^{aA} = \frac{\partial P^{aA}}{\partial F^b{}_B}\delta F^b{}_B + \frac{\partial P^{aA}}{\partial F^b{}_{B|C}}\delta F^b{}_{B|C} = \mathsf{A}^{aA}{}_b{}^B\, U^b{}_{|B} + \mathsf{B}^{aA}{}_b{}^{BC}\, U^b{}_{|B|C}\,, \tag{5.20}$$

and

$$\mathsf{A}^{aA}{}_b{}^B = \frac{\partial P^{aA}}{\partial F^b{}_B}, \quad \mathsf{B}^{aA}{}_b{}^{BC} = \frac{\partial P^{aA}}{\partial F^b{}_{B|C}}\,. \tag{5.21}$$

Following DiVincenzo [1986] we call $\mathsf{\mathbf{A}}$ and $\mathsf{\mathbf{B}}$ *dynamic elastic constants*. Notice that $\mathsf{B}^{aAbBC} = \mathsf{B}^{aAbCB}$. Noting that $P^{aA} = g^{am}\frac{\partial W}{\partial F^m{}_A} - H^{aAM}{}_{|M}$, one can write

$$\begin{aligned}\delta P^{aA} &= g^{am}\frac{\partial^2 W}{\partial F^m{}_A\partial F^n{}_N}\delta F^n{}_N + g^{am}\frac{\partial^2 W}{\partial F^m{}_A\partial F^n{}_{N|M}}\delta F^n{}_{N|M} - \delta(H^{aAM}{}_{|M}) \\ &= g^{am}\frac{\partial^2 W}{\partial F^m{}_A\partial F^n{}_N}U^n{}_{|N} + g^{am}\frac{\partial^2 W}{\partial F^m{}_A\partial F^n{}_{N|M}}U^n{}_{|N|M} - (\delta H^{aAM})_{|M}\,.\end{aligned} \tag{5.22}$$

[27] We use the standard notation $\Pi^{[ab]} = \frac{1}{2}\left(\Pi^{ab} - \Pi^{ba}\right)$.

But

$$\begin{aligned}\delta H^{aAM} &= \frac{\partial H^{aAM}}{\partial F^c{}_C}\delta F^c{}_C + \frac{\partial H^{aAM}}{\partial F^c{}_{C|D}}\delta F^c{}_{C|D} \\ &= g^{am}\frac{\partial^2 W}{\partial F^c{}_C \partial F^m{}_{A|M}} U^c{}_{|C} + g^{am}\frac{\partial^2 W}{\partial F^c{}_{C|D}\partial F^m{}_{A|M}} U^c{}_{|C|D} \,.\end{aligned} \tag{5.23}$$

Let us define the following three *static elastic constants* [DiVincenzo, 1986]

$$\mathbb{A}_a{}^A{}_b{}^B = \frac{\partial^2 W}{\partial F^a{}_A \partial F^b{}_B}, \quad \mathbb{B}_a{}^A{}_b{}^{BC} = \frac{\partial^2 W}{\partial F^a{}_A \partial F^b{}_{B|C}}, \quad \mathbb{C}_a{}^{AB}{}_b{}^{CD} = \frac{\partial^2 W}{\partial F^a{}_{A|B}\partial F^b{}_{C|D}} \,. \tag{5.24}$$

Therefore, the static elastic constants satisfy the following symmetries: $\mathbb{A}_a{}^A{}_b{}^B = \mathbb{A}_b{}^B{}_a{}^A$, $\mathbb{B}_a{}^A{}_b{}^{BC} = \mathbb{B}_a{}^A{}_b{}^{CB}$, and $\mathbb{C}_a{}^{AB}{}_b{}^{CD} = \mathbb{C}_a{}^{BA}{}_b{}^{CD} = \mathbb{C}_a{}^{BA}{}_b{}^{DC} = \mathbb{C}_b{}^{DC}{}_a{}^{BA}$ (note that $\mathbb{C}$ has 21 and 171 independent components in 2D and 3D, respectively). Thus, $\delta H^{aAM} = \mathbb{B}_b{}^{BaAM} U^b{}_{|B} + \mathbb{C}_b{}^{BCaAM} U^b{}_{|B|C}$, and hence

$$(\delta H^{aAM})_{|M} = (\mathbb{B}_b{}^{BaAM} U^b{}_{|B} + \mathbb{C}_b{}^{BCaAM} U^b{}_{|B|C})_{|M} \,. \tag{5.25}$$

Therefore

$$\begin{aligned}\mathsf{A}^{aA}{}_b{}^B &= \mathbb{A}^{aA}{}_b{}^B - \mathbb{B}_b{}^{BaAM}{}_{|M}, \\ \mathsf{B}^{aA}{}_b{}^{BC} &= \mathbb{B}^{aA}{}_b{}^{BC} - \mathbb{B}_b{}^{BaAC} - \mathbb{C}^{aAM}{}_b{}^{BC}{}_{|M} \,.\end{aligned} \tag{5.26}$$

Or equivalently

$$\begin{aligned}\mathsf{A}^{aAbB} &= \mathbb{A}^{aAbB} - \mathbb{B}^{bBaAM}{}_{|M}, \\ \mathsf{B}^{aAbBC} &= \mathbb{B}^{aAbBC} - \mathbb{B}^{bBaAC} - \mathbb{C}^{aAMbBC}{}_{|M} \,.\end{aligned} \tag{5.27}$$

In deriving the second relation we ignored the term $U^b{}_{|B|C|M}$ in $\delta H^{aAM}{}_{|M}$ as we are assuming a second-gradient elasticity in which displacement derivatives of orders three or higher are neglected. Note that in [DiVincenzo, 1986] homogeneous solids were considered. Here, for cloaking purposes, we must consider inhomogeneous solids and that is why derivatives of the static elastic constants appear in (5.27).

Note that $(5.27)_2$ requires that $\mathbb{B}^{bBaAC} = \mathbb{B}^{bCaAB} = \mathbb{B}^{bAaBC}$, reducing the number of independent components of $\mathbb{B}$ to 16 and 90 in 2D and 3D, respectively. After some manipulations and using the major symmetry of $\mathbb{C}$, one also finds that

$$\mathsf{B}^{aAbBC} + \mathsf{B}^{bBaAC} = -\left(\mathbb{C}^{bBMaAC} + \mathbb{C}^{aAMbBC}\right)_{|M} = -\left(\mathbb{C}^{aACbBM} + \mathbb{C}^{bBCaAM}\right)_{|M} \,. \tag{5.28}$$

Similarly, from $(5.27)_1$ and knowing that $\mathbb{A}$ has the major symmetry one obtains

$$\mathsf{A}^{aAbB} - \mathsf{A}^{bBaA} = \left(\mathbb{B}^{aAbBM} - \mathbb{B}^{bBaAM}\right)_{|M} = \left(\mathbb{B}^{aBbAM} - \mathbb{B}^{bBaAM}\right)_{|M} \,. \tag{5.29}$$

**Linearized balance of angular momentum.** The balance of angular momentum is equivalent to $\Pi^{ab} = \Pi^{ba}$, or $\Pi^{[ab]} = 0$. Suppose the reference motion is an isometric embedding of an initially stress-free body into the Euclidean space, i.e., $\mathring{F}^a{}_A = \delta^a_A$, which implies that $\mathring{F}^a{}_{A|B} = 0$. Assuming that $\mathring{P}^{aA} = 0$, one concludes that

$$\mathring{H}^{aAB}{}_{|B} = g^{ab}\frac{\partial W}{\partial F^a{}_A}\bigg|_{\mathbf{F}=\mathring{\mathbf{F}}} =: \mathring{P}^{aA}_{\text{cl.}} \,. \tag{5.30}$$

Therefore

$$\begin{aligned}\delta\Pi^{ab} &= \mathring{P}^{aA}_{\text{cl.}} U^b{}_{|A} + \mathring{H}^{aAB} U^b{}_{|A|B} + \mathbb{A}^{aA}{}_m{}^M \mathring{F}^b{}_A U^m{}_{|M} + \mathbb{B}^{aA}{}_m{}^{MN} \mathring{F}^b{}_A U^m{}_{|M|N} \\ &= \left(\mathbb{A}^{aM}{}_m{}^A \mathring{F}^b{}_A + \mathring{H}^{aAB}{}_{|B}\delta^b_m\right) U^m{}_{|A} + \left(\mathbb{B}^{aM}{}_m{}^{AB}\mathring{F}^b{}_M + \mathring{H}^{aAB}\delta^b_m\right) U^m{}_{|A|B} \,.\end{aligned} \tag{5.31}$$

Knowing that $\delta\Pi^{[ab]} = 0$, and that the first and the second covariant derivatives of the displacement field are independent, one concludes that

$$\begin{aligned}\mathbb{A}^{[aM}{}_m{}^A \mathring{F}^{b]}{}_M + \mathring{H}^{[aAB}{}_{|B}\delta^{b]}_m &= 0, \\ \mathbb{B}^{[aM}{}_m{}^{AB}\mathring{F}^{b]}{}_M + \mathring{H}^{[aAB}\delta^{b]}_m &= 0.\end{aligned} \tag{5.32}$$

Note that the issue with the acceleration term that was discussed in classical nonlinear elasticity in §4.1 persists even in nonlinear gradient elasticity, and hence, we do not discuss transformation cloaking in nonlinear gradient elastodynamics.

**Transformation cloaking in linearized gradient elastodynamics.** We start from a virtual body that is made of a homogeneous, isotropic, and centro-symmetric gradient elastic solid. We then consider a cloaking transformation and try to find the elastic constants of the physical body induced from the cloaking transformation such that the balance of linear and angular momenta are respected in both the virtual and physical bodies. Let us start from the balance of linear momentum in the physical body, i.e., $\mathrm{Div}\,\mathbf{P}+\rho_0\mathbf{B}=\rho_0\mathbf{A}$. Its linearization reads $\delta\left(\mathrm{Div}\,\mathbf{P}\right)+\rho_0\delta\mathbf{B}=\rho_0\mathbf{A}$, where

$$\delta\left(\mathrm{Div}\,\mathbf{P}\right)=\mathrm{Div}\,\delta\mathbf{P}=\mathrm{Div}\left(\boldsymbol{\mathsf{A}}:\nabla\mathbf{U}+\boldsymbol{\mathsf{B}}:\nabla\nabla\mathbf{U}\right)=\left(\mathsf{A}^{aA}{}_b{}^B\;U^b{}_{|B}+\mathsf{B}^{aA}{}_b{}^{BC}\;U^b{}_{|B|C}\right)_{|A}\frac{\partial}{\partial x^a}. \tag{5.33}$$

Under a cloaking transformation $\Xi:\mathcal{B}\to\tilde{\mathcal{B}}$ and using the shifter map, $\mathrm{Div}\,\delta\mathbf{P}$ is transformed to

$$J_\Xi\left(\tilde{\mathsf{A}}^{\tilde{a}\tilde{A}}{}_{\tilde{b}}{}^{\tilde{B}}\;\tilde{U}^{\tilde{b}}{}_{|\tilde{B}}+\tilde{\mathsf{B}}^{\tilde{a}\tilde{A}}{}_{\tilde{b}}{}^{\tilde{B}\tilde{C}}\;\tilde{U}^{\tilde{b}}{}_{|\tilde{B}|\tilde{C}}\right)_{|\tilde{A}}\frac{\partial}{\partial\tilde{x}^{\tilde{a}}}, \tag{5.34}$$

where

$$\begin{aligned}
\tilde{U}^{\tilde{a}}&=\mathsf{s}^{\tilde{a}}{}_aU^a\,,\\
\tilde{\mathsf{A}}^{\tilde{a}\tilde{A}}{}_{\tilde{b}}{}^{\tilde{B}}&=J_\Xi^{-1}\mathsf{s}^{\tilde{a}}{}_a\overset{\Xi}{F}{}^{\tilde{A}}{}_A(\mathsf{s}^{-1})^b{}_{\tilde{b}}\overset{\Xi}{F}{}^{\tilde{B}}{}_B\;\mathsf{A}^{aA}{}_b{}^B+J_\Xi^{-1}\mathsf{s}^{\tilde{a}}{}_a\overset{\Xi}{F}{}^{\tilde{A}}{}_A(\mathsf{s}^{-1})^b{}_{\tilde{b}}\overset{\Xi}{F}{}^{\tilde{B}}{}_{B|C}\;\mathsf{B}^{aA}{}_b{}^{BC}\,,\\
\tilde{\mathsf{B}}^{\tilde{a}\tilde{A}}{}_{\tilde{b}}{}^{\tilde{B}\tilde{C}}&=J_\Xi^{-1}\mathsf{s}^{\tilde{a}}{}_a\overset{\Xi}{F}{}^{\tilde{A}}{}_A(\mathsf{s}^{-1})^b{}_{\tilde{b}}\overset{\Xi}{F}{}^{\tilde{B}}{}_B\overset{\Xi}{F}{}^{\tilde{C}}{}_C\;\mathsf{B}^{aA}{}_b{}^{BC}\,.
\end{aligned} \tag{5.35}$$

Or, equivalently

$$\begin{aligned}
\mathsf{A}^{aA}{}_b{}^B&=J_\Xi(\mathsf{s}^{-1})^a{}_{\tilde{a}}(\overset{\Xi}{F}{}^{-1})^A{}_{\tilde{A}}\mathsf{s}^{\tilde{b}}{}_b(\overset{\Xi}{F}{}^{-1})^B{}_{\tilde{B}}\;\tilde{\mathsf{A}}^{\tilde{a}\tilde{A}}{}_{\tilde{b}}{}^{\tilde{B}}+J_\Xi(\mathsf{s}^{-1})^a{}_{\tilde{a}}(\overset{\Xi}{F}{}^{-1})^A{}_{\tilde{A}}\mathsf{s}^{\tilde{b}}{}_b(\overset{\Xi}{F}{}^{-1})^B{}_{\tilde{B}|\tilde{C}}\;\tilde{\mathsf{B}}^{\tilde{a}\tilde{A}}{}_{\tilde{b}}{}^{\tilde{B}\tilde{C}}\,,\\
\mathsf{B}^{aA}{}_b{}^{BC}&=J_\Xi(\mathsf{s}^{-1})^a{}_{\tilde{a}}(\overset{\Xi}{F}{}^{-1})^A{}_{\tilde{A}}\mathsf{s}^{\tilde{b}}{}_b(\overset{\Xi}{F}{}^{-1})^B{}_{\tilde{B}}(\overset{\Xi}{F}{}^{-1})^C{}_{\tilde{C}}\;\tilde{\mathsf{B}}^{\tilde{a}\tilde{A}}{}_{\tilde{b}}{}^{\tilde{B}\tilde{C}}\,.
\end{aligned} \tag{5.36}$$

To see this note that

$$\begin{aligned}
&\left(\mathsf{A}^{aA}{}_b{}^B\;U^b{}_{|B}+\mathsf{B}^{aA}{}_b{}^{BC}\;U^b{}_{|B|C}\right)_{|A}=J_\Xi\left[J_\Xi^{-1}\overset{\Xi}{F}{}^{\tilde{A}}{}_A\mathsf{A}^{aA}{}_b{}^B\;U^b{}_{|B}+J_\Xi^{-1}\overset{\Xi}{F}{}^{\tilde{A}}{}_A\mathsf{B}^{aA}{}_b{}^{BC}\;U^b{}_{|B|C}\right]_{|\tilde{A}}\\
&=J_\Xi\left[J_\Xi^{-1}\overset{\Xi}{F}{}^{\tilde{A}}{}_A\mathsf{A}^{aA}{}_b{}^B\overset{\Xi}{F}{}^{\tilde{B}}{}_BU^b{}_{|\tilde{B}}+J_\Xi^{-1}\overset{\Xi}{F}{}^{\tilde{A}}{}_A\mathsf{B}^{aA}{}_b{}^{BC}\overset{\Xi}{F}{}^{\tilde{B}}{}_B\overset{\Xi}{F}{}^{\tilde{C}}{}_CU^b{}_{|\tilde{B}|\tilde{C}}+J_\Xi^{-1}\overset{\Xi}{F}{}^{\tilde{A}}{}_A\mathsf{B}^{aA}{}_b{}^{BC}\overset{\Xi}{F}{}^{\tilde{B}}{}_{B|C}U^b{}_{|\tilde{B}}\right]_{|\tilde{A}}\\
&=(\mathsf{s}^{-1})^a{}_{\tilde{a}}J_\Xi\left(\tilde{\mathsf{A}}^{\tilde{a}\tilde{A}}{}_{\tilde{b}}{}^{\tilde{B}}\;\tilde{U}^{\tilde{b}}{}_{|\tilde{B}}+\tilde{\mathsf{B}}^{\tilde{a}\tilde{A}}{}_{\tilde{b}}{}^{\tilde{B}\tilde{C}}\;\tilde{U}^{\tilde{b}}{}_{|\tilde{B}|\tilde{C}}\right)_{|\tilde{A}}\,,
\end{aligned} \tag{5.37}$$

where the following relations were used.

$$\begin{aligned}
U^b{}_{|B}&=(\mathsf{s}^{-1})^b{}_{\tilde{b}}\overset{\Xi}{F}{}^{\tilde{B}}{}_B\;\tilde{U}^{\tilde{b}}{}_{|\tilde{B}},\\
U^b{}_{|B|C}&=(\mathsf{s}^{-1})^b{}_{\tilde{b}}\left[\overset{\Xi}{F}{}^{\tilde{B}}{}_{B|C}\;\tilde{U}^{\tilde{b}}{}_{|\tilde{B}}+\overset{\Xi}{F}{}^{\tilde{B}}{}_B\overset{\Xi}{F}{}^{\tilde{C}}{}_C\;\tilde{U}^{\tilde{b}}{}_{|\tilde{B}|\tilde{C}}\right].
\end{aligned} \tag{5.38}$$

We assume that the reference motion for both the physical and virtual bodies are isometric embeddings in the Euclidean ambient space, i.e., $\mathring{F}^a{}_A=\delta^a_A$ and $\mathring{\tilde{F}}{}^{\tilde{a}}{}_{\tilde{A}}=\delta^{\tilde{a}}_{\tilde{A}}$. This implies that $\mathring{F}^a{}_{A|B}=0$, and $\mathring{\tilde{F}}{}^{\tilde{a}}{}_{\tilde{A}|\tilde{B}}=0$. It is also assumed that there is no initial stress in either configuration, i.e., $\mathring{P}^a{}_A=0$, $\mathring{H}^{aAB}=0$, and $\mathring{\tilde{P}}{}^{\tilde{a}}{}_{\tilde{A}}=0$, $\mathring{\tilde{H}}{}^{\tilde{a}\tilde{A}\tilde{B}}=0$. Therefore, from (5.32) the balance of angular momentum in the physical and virtual bodies read (minor symmetries of $\mathbb{A}$, $\mathbb{B}$, $\tilde{\mathbb{A}}$, and $\tilde{\mathbb{B}}$)

$$\mathbb{A}^{[aM}{}_m{}^A\mathring{F}^{b]}{}_M=0,\quad\mathbb{B}^{[aM}{}_m{}^{AB}\mathring{F}^{b]}{}_M=0, \tag{5.39}$$

$$\tilde{\mathbb{A}}^{[\tilde{a}\tilde{M}}{}_{\tilde{m}}{}^{\tilde{A}}\mathring{\tilde{F}}{}^{\tilde{b}]}{}_{\tilde{M}}=0,\quad\tilde{\mathbb{B}}^{[\tilde{a}\tilde{M}}{}_{\tilde{m}}{}^{\tilde{A}\tilde{B}}\mathring{\tilde{F}}{}^{\tilde{b}]}{}_{\tilde{M}}=0. \tag{5.40}$$

The virtual body being uniform its elastic constants are (covariantly) constant, and hence, from (5.26) one concludes that

$$\tilde{\mathsf{A}}^{\tilde{a}\tilde{A}}{}_{\tilde{b}}{}^{\tilde{B}} = \tilde{\mathbb{A}}^{\tilde{a}\tilde{A}}{}_{\tilde{b}}{}^{\tilde{B}}, \qquad \tilde{\mathsf{B}}^{\tilde{a}\tilde{A}}{}_{\tilde{b}}{}^{\tilde{B}\tilde{C}} = \tilde{\mathbb{B}}^{\tilde{a}\tilde{A}}{}_{\tilde{b}}{}^{\tilde{B}\tilde{C}} - \tilde{\mathbb{B}}_{\tilde{b}}{}^{\tilde{B}\tilde{a}\tilde{A}\tilde{C}}. \tag{5.41}$$

Therefore, from $(5.40)_1$ one obtains

$$\tilde{\mathsf{A}}^{[\tilde{a}\tilde{M}}{}_{\tilde{m}}{}^{\tilde{A}}\mathring{\tilde{F}}^{\tilde{b}]}{}_{\tilde{M}} = 0. \tag{5.42}$$

The virtual body is assumed to be isotropic and non-chiral (centro-symmetric).[28] Knowing that an odd-order tensor cannot be isotropic and centro-symmetric [Mindlin, 1964, Auffray et al., 2013] one concludes that $\tilde{\mathbb{B}}^{\tilde{a}\tilde{A}}{}_{\tilde{b}}{}^{\tilde{B}\tilde{C}} = 0$, and hence $\tilde{\mathsf{B}}^{\tilde{a}\tilde{A}}{}_{\tilde{b}}{}^{\tilde{B}\tilde{C}} = 0$. In particular, $(5.40)_2$ is trivially satisfied. From $(5.36)_2$ one obtains $\mathsf{B}^{aA}{}_{b}{}^{BC} = 0$, and hence from $(5.27)_2$

$$\begin{aligned}
&\mathbb{B}^{aA}{}_{b}{}^{BC} - \mathbb{B}_{b}{}^{BaAC} = \mathbb{C}^{aAM}{}_{b}{}^{BC}{}_{|M}, \\
&\mathsf{A}^{aA}{}_{b}{}^{B} = J_{\Xi}(\mathsf{s}^{-1})^{a}{}_{\tilde{a}}(\bar{\bar{F}}^{-1})^{A}{}_{\tilde{A}}\mathsf{s}^{\tilde{b}}{}_{b}(\bar{\bar{F}}^{-1})^{B}{}_{\tilde{B}}\ \tilde{\mathbb{A}}^{\tilde{a}\tilde{A}}{}_{\tilde{b}}{}^{\tilde{B}}.
\end{aligned} \tag{5.43}$$

Notice that the dynamic elastic constants $\mathbf{A}$ of the cloak possess the major symmetries, i.e., $\mathsf{A}^{aAbB} = \mathsf{A}^{bBaA} = J_{\Xi}(\mathsf{s}^{-1})^{a}{}_{\tilde{a}}(\bar{\bar{F}}^{-1})^{A}{}_{\tilde{A}}(\mathsf{s}^{-1})^{b}{}_{\tilde{b}}(\bar{\bar{F}}^{-1})^{B}{}_{\tilde{B}}\ \tilde{\mathbb{A}}^{\tilde{a}\tilde{A}\tilde{b}\tilde{B}}$. Thus, using (5.29), one obtains $\mathbb{B}^{[aBb]AM}{}_{|M} = 0$, i.e.,

$$\mathbb{B}^{aBbAM}{}_{|M} = \mathbb{B}^{bBaAM}{}_{|M}\,. \tag{5.44}$$

From $(5.43)_1$ and $(5.39)_2$, one obtains

$$\mathbb{C}^{[aAMbBC}{}_{|M}\mathring{F}^{c]}{}_{A} = -\mathbb{B}^{bB[aAC}\mathring{F}^{c]}{}_{A}\,. \tag{5.45}$$

Also, from (5.28) and knowing that $\mathbf{B}$ vanishes identically, one obtains

$$\left(\mathbb{C}^{aACbBM} + \mathbb{C}^{bBCaAM}\right)_{|M} = 0\,. \tag{5.46}$$

**Remark 5.2.** From $(5.43)_1$ and $(5.39)_2$, one can write

$$\mathbb{B}^{[aAbBC}{}_{|C}\mathring{F}^{c]}{}_{A} - \mathbb{B}^{bB[aAC}{}_{|C}\mathring{F}^{c]}{}_{A} = -\mathbb{B}^{bB[aAC}{}_{|C}\mathring{F}^{c]}{}_{A} = \mathbb{C}^{[aAMbBC}{}_{|M|C}\mathring{F}^{c]}{}_{A}. \tag{5.47}$$

Therefore

$$\mathbb{B}^{bB[aAC}{}_{|C}\mathring{F}^{c]}{}_{A} = -\mathbb{C}^{[aAMbBC}{}_{|M|C}\mathring{F}^{c]}{}_{A}. \tag{5.48}$$

Note that from $(5.27)_1$ and $(5.43)_1$ we have

$$\begin{aligned}
\mathbb{A}^{aAbB} &= \mathbb{B}^{bBaAM}{}_{|M} + J_{\Xi}(\mathsf{s}^{-1})^{a}{}_{\tilde{a}}(\bar{\bar{F}}^{-1})^{A}{}_{\tilde{A}}(\mathsf{s}^{-1})^{b}{}_{\tilde{b}}(\bar{\bar{F}}^{-1})^{B}{}_{\tilde{B}}\ \tilde{\mathbb{A}}^{\tilde{a}\tilde{A}\tilde{b}\tilde{B}} \\
&= \mathbb{B}^{aAbBM}{}_{|M} - \mathbb{C}^{aAMbBN}{}_{|M|N} + J_{\Xi}(\mathsf{s}^{-1})^{a}{}_{\tilde{a}}(\bar{\bar{F}}^{-1})^{A}{}_{\tilde{A}}(\mathsf{s}^{-1})^{b}{}_{\tilde{b}}(\bar{\bar{F}}^{-1})^{B}{}_{\tilde{B}}\ \tilde{\mathbb{A}}^{\tilde{a}\tilde{A}\tilde{b}\tilde{B}}.
\end{aligned} \tag{5.49}$$

From the above relation and (5.39) one obtains

$$0 = \mathbb{A}^{[aAbB}\mathring{F}^{c]}{}_{A} = -\mathbb{C}^{[aAMbBN}{}_{|M|N}\mathring{F}^{c]}{}_{A} + J_{\Xi}(\mathsf{s}^{-1})^{[a}{}_{\tilde{a}}(\bar{\bar{F}}^{-1})^{A}{}_{\tilde{A}}(\mathsf{s}^{-1})^{b}{}_{\tilde{b}}(\bar{\bar{F}}^{-1})^{B}{}_{\tilde{B}}\ \tilde{\mathbb{A}}^{\tilde{a}\tilde{A}\tilde{b}\tilde{B}}\mathring{F}^{c]}{}_{A}. \tag{5.50}$$

Therefore

$$\mathbb{C}^{[aAMbBN}{}_{|M|N}\mathring{F}^{c]}{}_{A} = J_{\Xi}(\mathsf{s}^{-1})^{[a}{}_{\tilde{a}}(\bar{\bar{F}}^{-1})^{A}{}_{\tilde{A}}(\mathsf{s}^{-1})^{b}{}_{\tilde{b}}(\bar{\bar{F}}^{-1})^{B}{}_{\tilde{B}}\ \tilde{\mathbb{A}}^{\tilde{a}\tilde{A}\tilde{b}\tilde{B}}\mathring{F}^{c]}{}_{A}. \tag{5.51}$$

Note that in classical linearized elasticity the right-hand side had to vanish, e.g., Eq. (4.54), which was an obstruction to transformation cloaking.

[28] Note that only the virtual body is assumed to be centro-symmetric. There is no such constraint on the physical body; it can be both non-centro-symmetric and anisotropic.

From (5.51) and (5.48), the system of first-order PDEs for $\mathbb{B}$ read

$$\mathbb{B}^{bB[aAC}{}_{|C}\mathring{F}^{c]}{}_{A} = -J_{\Xi}(\mathsf{s}^{-1})^{[a}{}_{\tilde{a}}(\bar{\bar{F}}^{-1})^{A}{}_{\tilde{A}}(\mathsf{s}^{-1})^{b}{}_{\tilde{b}}(\bar{\bar{F}}^{-1})^{B}{}_{\tilde{B}}\ \tilde{\mathbb{A}}^{\tilde{a}\tilde{A}\tilde{b}\tilde{B}}\mathring{F}^{c]}{}_{A}\,. \tag{5.52}$$

In summary, one is given the homogeneous static elastic constants of the virtual body. Because of isotropy and centro-symmetry only $\tilde{\mathbb{A}}$ and $\tilde{\mathbb{C}}$ are nonzero. These are both (covariantly) constant tensors. Balance of angular momentum implies that $\tilde{\mathbb{A}}$ has the classical minor symmetries. So, it will have the same form as $\tilde{\mathbb{A}}$ in classical linear elasticity. There is no relation between $\tilde{\mathbb{C}}$ and $\mathbb{C}$, and there are no constraints on $\tilde{\mathbb{C}}$ other than being positive-definite. Balance of linear momentum relates the dynamic elastic constants of the two problems. From $\tilde{\mathbf{B}} = \mathbf{0}$ in the virtual body one concludes that $\mathbf{B} = \mathbf{0}$ in the physical body. This gives a relation between $\mathbb{B}$ and $\mathbb{C}$ in the form of a system of PDEs. Balance of angular momentum in the physical body is written as a set of constraints in the form of a system of second-order PDEs for $\mathbb{C}$.

**The impossibility of transformation cloaking in gradient elasticity.** Next we show that transformation cloaking is not possible in gradient elasticity. We first show this for cylindrical and spherical cloaks. From $(5.39)_2$ and knowing that $\mathring{F}^{b}{}_{M} = \delta^{b}{}_{M}$, with an abuse of notation, one writes

$$\mathbb{B}^{abmAB} = \mathbb{B}^{bamAB}\,, \tag{5.53}$$

i.e., $\mathbb{B}$ must be symmetric with respect to the first two indices. This symmetry and (5.52), imply that the right-hand side of (5.52) is symmetric with respect to indices $b$ and $B$ as well. Thus

$$(\mathsf{s}^{-1})^{[a}{}_{\tilde{a}}(\bar{\bar{F}}^{-1})^{A}{}_{\tilde{A}}(\mathsf{s}^{-1})^{b}{}_{\tilde{b}}(\bar{\bar{F}}^{-1})^{B}{}_{\tilde{B}}\ \tilde{\mathbb{A}}^{\tilde{a}\tilde{A}\tilde{b}\tilde{B}}\mathring{F}^{c]}{}_{A} = (\mathsf{s}^{-1})^{[a}{}_{\tilde{a}}(\bar{\bar{F}}^{-1})^{A}{}_{\tilde{A}}(\mathsf{s}^{-1})^{B}{}_{\tilde{b}}(\bar{\bar{F}}^{-1})^{b}{}_{\tilde{B}}\ \tilde{\mathbb{A}}^{\tilde{a}\tilde{A}\tilde{b}\tilde{B}}\mathring{F}^{c]}{}_{A}\,. \tag{5.54}$$

This is simplified and rearranged to read

$$\begin{aligned}&(\mathsf{s}^{-1})^{b}{}_{\tilde{b}}(\bar{\bar{F}}^{-1})^{B}{}_{\tilde{B}}\left[(\mathsf{s}^{-1})^{c}{}_{\tilde{a}}(\bar{\bar{F}}^{-1})^{a}{}_{\tilde{A}} - (\mathsf{s}^{-1})^{a}{}_{\tilde{a}}(\bar{\bar{F}}^{-1})^{c}{}_{\tilde{A}}\right]\tilde{\mathbb{A}}^{\tilde{a}\tilde{A}\tilde{b}\tilde{B}}\\ &= (\mathsf{s}^{-1})^{B}{}_{\tilde{b}}(\bar{\bar{F}}^{-1})^{b}{}_{\tilde{B}}\left[(\mathsf{s}^{-1})^{c}{}_{\tilde{a}}(\bar{\bar{F}}^{-1})^{a}{}_{\tilde{A}} - (\mathsf{s}^{-1})^{a}{}_{\tilde{a}}(\bar{\bar{F}}^{-1})^{c}{}_{\tilde{A}}\right]\tilde{\mathbb{A}}^{\tilde{a}\tilde{A}\tilde{b}\tilde{B}}\,,\quad \forall\, b, B, c, a \in \{1,2,3\}\,.\end{aligned} \tag{5.55}$$

In particular, it is straightforward to verify that this condition cannot be satisfied for either a cylindrical or a spherical cloak. To see this, let us expand (5.55) for $b = 1$, $B = 3$, $a = 1$, and $c = 3$:

$$\begin{aligned}&(\mathsf{s}^{-1})^{1}{}_{\tilde{b}}(\bar{\bar{F}}^{-1})^{3}{}_{\tilde{B}}\left[(\mathsf{s}^{-1})^{3}{}_{\tilde{a}}(\bar{\bar{F}}^{-1})^{1}{}_{\tilde{A}} - (\mathsf{s}^{-1})^{1}{}_{\tilde{a}}(\bar{\bar{F}}^{-1})^{3}{}_{\tilde{A}}\right]\tilde{\mathbb{A}}^{\tilde{a}\tilde{A}\tilde{b}\tilde{B}}\\ &= (\mathsf{s}^{-1})^{3}{}_{\tilde{b}}(\bar{\bar{F}}^{-1})^{1}{}_{\tilde{B}}\left[(\mathsf{s}^{-1})^{3}{}_{\tilde{a}}(\bar{\bar{F}}^{-1})^{1}{}_{\tilde{A}} - (\mathsf{s}^{-1})^{1}{}_{\tilde{a}}(\bar{\bar{F}}^{-1})^{3}{}_{\tilde{A}}\right]\tilde{\mathbb{A}}^{\tilde{a}\tilde{A}\tilde{b}\tilde{B}}\,.\end{aligned} \tag{5.56}$$

Knowing that in the spherical (or cylindrical) coordinates and for a radial cloak $\mathbf{s}^{-1}$ and $\bar{\bar{\mathbf{F}}}^{-1}$ have diagonal representations, this is further simplified and reads

$$\begin{aligned}&(\mathsf{s}^{-1})^{1}{}_{1}(\bar{\bar{F}}^{-1})^{3}{}_{3}\left[(\mathsf{s}^{-1})^{3}{}_{3}(\bar{\bar{F}}^{-1})^{1}{}_{1}\tilde{\mathbb{A}}^{3113} - (\mathsf{s}^{-1})^{1}{}_{1}(\bar{\bar{F}}^{-1})^{3}{}_{3}\tilde{\mathbb{A}}^{1313}\right]\\ &= (\mathsf{s}^{-1})^{3}{}_{3}(\bar{\bar{F}}^{-1})^{1}{}_{1}\left[(\mathsf{s}^{-1})^{3}{}_{3}(\bar{\bar{F}}^{-1})^{1}{}_{1}\tilde{\mathbb{A}}^{3131} - (\mathsf{s}^{-1})^{1}{}_{1}(\bar{\bar{F}}^{-1})^{3}{}_{3}\tilde{\mathbb{A}}^{1331}\right]\,.\end{aligned} \tag{5.57}$$

But note that $\tilde{\mathbb{A}}$ (the elastic constants of the virtual body) possesses both the minor and major symmetries. Moreover, $\tilde{\mathbb{A}}^{1331} = \mu > 0$. Hence, one obtains

$$\begin{aligned}&(\mathsf{s}^{-1})^{1}{}_{1}(\bar{\bar{F}}^{-1})^{3}{}_{3}\left[(\mathsf{s}^{-1})^{3}{}_{3}(\bar{\bar{F}}^{-1})^{1}{}_{1} - (\mathsf{s}^{-1})^{1}{}_{1}(\bar{\bar{F}}^{-1})^{3}{}_{3}\right]\\ &= (\mathsf{s}^{-1})^{3}{}_{3}(\bar{\bar{F}}^{-1})^{1}{}_{1}\left[(\mathsf{s}^{-1})^{3}{}_{3}(\bar{\bar{F}}^{-1})^{1}{}_{1} - (\mathsf{s}^{-1})^{1}{}_{1}(\bar{\bar{F}}^{-1})^{3}{}_{3}\right]\,.\end{aligned} \tag{5.58}$$

Therefore, $(\bar{\bar{F}}^{-1})^{3}{}_{3}(\mathsf{s}^{-1})^{1}{}_{1} = (\mathsf{s}^{-1})^{3}{}_{3}(\bar{\bar{F}}^{-1})^{1}{}_{1}$. Recalling that $\bar{\bar{\mathbf{F}}} = \mathrm{diag}(f'(R), 1, 1)$ and $\mathbf{s} = \mathrm{diag}\,(1, R/f(R), 1)$ and $\mathbf{s} = \mathrm{diag}\,(1, R/f(R), R/f(R))$, in the cylindrical and spherical coordinates, respectively, one must have $f(R) = R$, i.e., $\Xi = id$, which is not acceptable for a cloaking map.

One may now ask whether choosing a less symmetric cloaking map may make transformation cloaking possible. We next show that this is not the case.

**Proposition 5.3.** *Assuming that the virtual body is isotropic and centro-symmetric, elastodynamics transformation cloaking is not possible for gradient elastic solids in either 2D or 3D for a hole (cavity) of any shape.*

*Proof.* Without loss of generality, we may write (5.55) in Cartesian coordinates for which the shifter has a trivial representation, i.e., $\mathsf{s}^{\tilde{a}}{}_{a} = \delta^{\tilde{a}}_{a}$. In 2D, one has

$$\begin{aligned}&(\bar{\bar{F}}^{-1})^{B}{}_{\tilde{B}}\left[(\bar{\bar{F}}^{-1})^{a}{}_{\tilde{A}}\tilde{\mathbb{A}}^{c\tilde{A}b\tilde{B}} - (\bar{\bar{F}}^{-1})^{c}{}_{\tilde{A}}\tilde{\mathbb{A}}^{a\tilde{A}b\tilde{B}}\right]\\&= (\bar{\bar{F}}^{-1})^{b}{}_{\tilde{B}}\left[(\bar{\bar{F}}^{-1})^{a}{}_{\tilde{A}}\tilde{\mathbb{A}}^{c\tilde{A}B\tilde{B}} - (\bar{\bar{F}}^{-1})^{c}{}_{\tilde{A}}\tilde{\mathbb{A}}^{a\tilde{A}B\tilde{B}}\right], \quad \forall\, b,B,c,a\in\{1,2\}.\end{aligned} \tag{5.59}$$

Let us consider an arbitrary cloaking transformation with the following components

$$\bar{\bar{\mathbf{F}}}^{-1} = \begin{bmatrix}\mathsf{F}_{11} & \mathsf{F}_{12}\\ \mathsf{F}_{21} & \mathsf{F}_{22}\end{bmatrix}. \tag{5.60}$$

Equation (5.59) is expanded to read

$$\lambda\left(\mathsf{F}_{12}-\mathsf{F}_{21}\right)^2 + \mu\left[\left(\mathsf{F}_{11}-\mathsf{F}_{22}\right)^2 + 2\left(\mathsf{F}_{12}^2+\mathsf{F}_{21}^2\right)\right] = 0\,. \tag{5.61}$$

We know that $\mu>0$, and $2\mu+3\lambda>0$ for the energy function to be positive-definite.[29] Thus, multiplying the above identity by 3, we have $3\lambda\left(\mathsf{F}_{12}-\mathsf{F}_{21}\right)^2+3\mu[(\mathsf{F}_{11}-\mathsf{F}_{22})^2+2(\mathsf{F}_{12}^2+\mathsf{F}_{21}^2)]=0$, which can be rewritten as

$$(3\lambda+2\mu)\left(\mathsf{F}_{12}-\mathsf{F}_{21}\right)^2 + 3\mu\left(\mathsf{F}_{11}-\mathsf{F}_{22}\right)^2 + \mu\left[3\left(\mathsf{F}_{12}+\mathsf{F}_{21}\right)^2 + \left(\mathsf{F}_{12}-\mathsf{F}_{21}\right)^2\right] = 0\,. \tag{5.62}$$

Note that the coefficient of each term is positive. Therefore, $\mathsf{F}_{12}=\mathsf{F}_{21}=0$, and $\mathsf{F}_{11}=\mathsf{F}_{22}$, i.e., $\bar{\bar{\mathbf{F}}}=\alpha\mathbf{I}$, where $\mathbf{I}$ is the identity matrix and $\alpha>0$ is a scalar.

In 3D, the balance of angular momentum in the physical body implies

$$\begin{aligned}&(\bar{\bar{F}}^{-1})^{B}{}_{\tilde{B}}\left[(\bar{\bar{F}}^{-1})^{a}{}_{\tilde{A}}\tilde{\mathbb{A}}^{c\tilde{A}b\tilde{B}} - (\bar{\bar{F}}^{-1})^{c}{}_{\tilde{A}}\tilde{\mathbb{A}}^{a\tilde{A}b\tilde{B}}\right]\\&= (\bar{\bar{F}}^{-1})^{b}{}_{\tilde{B}}\left[(\bar{\bar{F}}^{-1})^{a}{}_{\tilde{A}}\tilde{\mathbb{A}}^{c\tilde{A}B\tilde{B}} - (\bar{\bar{F}}^{-1})^{c}{}_{\tilde{A}}\tilde{\mathbb{A}}^{a\tilde{A}B\tilde{B}}\right], \quad \forall\, b,B,c,a\in\{1,2,3\}.\end{aligned} \tag{5.63}$$

This relation for $\{b\neq B\}$ and $\{c\neq a\}$ is nontrivial and gives six linearly independent algebraic equations. Let us consider an arbitrary cloaking transformation with the following components

$$\bar{\bar{\mathbf{F}}}^{-1} = \begin{bmatrix}\mathsf{F}_{11} & \mathsf{F}_{12} & \mathsf{F}_{13}\\ \mathsf{F}_{21} & \mathsf{F}_{22} & \mathsf{F}_{23}\\ \mathsf{F}_{31} & \mathsf{F}_{32} & \mathsf{F}_{33}\end{bmatrix}. \tag{5.64}$$

Equation (5.63) is expanded for $i\neq j\neq k\in\{1,2,3\}$ and reads

$$\lambda\left(\mathsf{F}_{ij}-\mathsf{F}_{ji}\right)^2 + \mu\left[2\left(\mathsf{F}_{ij}^2+\mathsf{F}_{ji}^2\right)+\mathsf{F}_{ik}^2+\mathsf{F}_{jk}^2+\left(\mathsf{F}_{ii}-\mathsf{F}_{jj}\right)^2\right] = 0\,, \tag{5.65}$$

$$\lambda\left(\mathsf{F}_{ij}-\mathsf{F}_{ji}\right)\left(\mathsf{F}_{ik}-\mathsf{F}_{ki}\right) + \mu\left[\mathsf{F}_{ij}\mathsf{F}_{ik}+\mathsf{F}_{jj}\mathsf{F}_{kj}+\mathsf{F}_{kk}\mathsf{F}_{jk}+2\mathsf{F}_{ji}\mathsf{F}_{ki}-\mathsf{F}_{ii}\left(\mathsf{F}_{jk}+\mathsf{F}_{kj}\right)\right] = 0\,. \tag{5.66}$$

[29] Note that

$$\begin{aligned}\delta W &= \frac{1}{2}\frac{\partial W}{\partial F^{a}{}_{A}\partial F^{b}{}_{B}}U^{a}{}_{|A}U^{b}{}_{|B} + \frac{\partial W}{\partial F^{a}{}_{A}\partial F^{b}{}_{B|C}}U^{a}{}_{|A}U^{b}{}_{|B|C} + \frac{1}{2}\frac{\partial W}{\partial F^{a}{}_{A|B}\partial F^{b}{}_{C|D}}U^{a}{}_{|A|B}U^{b}{}_{|C|D}\\&= \frac{1}{2}\mathbb{A}^{aAbB}U_{a|A}U_{b|B} + \mathbb{B}^{aAbBC}U_{a|A}U_{b|B|C} + \frac{1}{2}\mathbb{C}^{aABbCD}U_{a|A|B}U_{b|C|D}.\end{aligned}$$

Positive-definiteness of energy requires that $\delta W>0$ for any pair $(U_{a|A},U_{a|A|B})\neq(0,0)$. In particular, when $U_{a|A}\neq 0$, and $U_{a|A|B}=0$, $\mathbb{A}^{aAbB}U_{a|A}U_{b|B}>0$, which implies that $\mathbb{A}$ must be positive-definite. In the case of isotropic solids this is equivalent to $\mu>0$, and $3\lambda+2\mu>0$.

After some algebraic manipulations (5.65) is rewritten as

$$(3\lambda+2\mu)\left(\mathsf{F}_{ij}-\mathsf{F}_{ji}\right)^2+\mu\left[3\left(\mathsf{F}_{ij}+\mathsf{F}_{ji}\right)^2+\left(\mathsf{F}_{ij}-\mathsf{F}_{ji}\right)^2\right]+3\mu\left[\mathsf{F}_{ik}^2+\mathsf{F}_{jk}^2+\left(\mathsf{F}_{ii}-\mathsf{F}_{jj}\right)^2\right]=0\,. \tag{5.67}$$

Note that the coefficient of each term is positive. Thus, $\mathsf{F}_{ij}=0$ for $i\neq j$, and $\mathsf{F}_{11}=\mathsf{F}_{22}=\mathsf{F}_{33}$. These trivially satisfy (5.66). One concludes that $\stackrel{\Xi}{\mathbf{F}}=\alpha\mathbf{I}$, where $\mathbf{I}$ is the identity matrix and $\alpha>0$ is a scalar. Knowing that the cloaking map restricted to the outer boundary of the cloak is the identity map, one concludes that $\alpha=1$, and hence, $\Xi=id$, which is clearly not an acceptable cloaking transformation. □

## 5.2 Elastodynamics Transformation Cloaking in Generalized Cosserat Solids

We assume that the microstructure in the deformed configuration is described by three (or two in 2D) linearly independent vectors $\{\underset{\mathfrak{a}}{\mathbf{d}}(x,t),\ \mathfrak{a}=1,2,3\}$, which are called director fields or directors [Ericksen and Truesdell, 1957]. The directors in the reference configuration are denoted by $\left\{\underset{\mathfrak{a}}{\mathbf{D}}(X),\ \mathfrak{a}=1,2,3\right\}$. These are not material vectors in the sense that $\underset{\mathfrak{a}}{\mathbf{d}}(x,t)\neq(\varphi_*\underset{\mathfrak{a}}{\mathbf{D}})(x,t)$. The kinematics of an oriented body is described by the pair $\left(\varphi(X,t),\underset{\mathfrak{a}}{\mathbf{d}}(X,t)\right)$, where $\underset{\mathfrak{a}}{\mathbf{d}}(X,t)$ is short for $\underset{\mathfrak{a}}{\mathbf{d}}(\varphi(X,t),\underset{\mathfrak{a}}{\mathbf{D}}(X))$ [Toupin, 1964, Stojanović, 1970]. An oriented body with (deformable) directors is called a generalized Cosserat solid. If the directors are rigid, i.e.,

$$\underset{\mathfrak{a}}{\mathsf{d}}^a(X,t)\ \underset{\mathfrak{b}}{\mathsf{d}}^b(X,t)g_{ab}(\varphi(X,t))=\underset{\mathfrak{a}\mathfrak{b}}{\mathfrak{g}}(X)\,, \tag{5.68}$$

for some symmetric, positive-definite, and time-independent matrix $\underset{\mathfrak{a}\mathfrak{b}}{\mathfrak{g}}$, the oriented body is referred to as a Cosserat solid. The reciprocal of $\underset{\mathfrak{a}}{\mathbf{D}}$ is denoted by $\overset{\mathfrak{a}}{\boldsymbol{\Theta}}$ such that $\underset{\mathfrak{a}}{\mathsf{D}}^A\overset{\mathfrak{a}}{\Theta}_B=\delta^A_B$ and $\underset{\mathfrak{a}}{\mathsf{D}}^A\overset{\mathfrak{b}}{\Theta}_A=\delta^{\mathfrak{b}}_{\mathfrak{a}}$. Similarly, the reciprocal of $\underset{\mathfrak{a}}{\mathbf{d}}$ is denoted by $\overset{\mathfrak{a}}{\boldsymbol{\vartheta}}$ such that $\underset{\mathfrak{a}}{\mathsf{d}}^a\overset{\mathfrak{a}}{\vartheta}_b=\delta^a_b$ and $\underset{\mathfrak{a}}{\mathsf{d}}^a\overset{\mathfrak{b}}{\vartheta}_a=\delta^{\mathfrak{b}}_{\mathfrak{a}}$. The referential directors vary from point to point and $\underset{\mathfrak{a}}{\mathsf{D}}^A{}_{|B}=W^A{}_{CB}\underset{\mathfrak{a}}{\mathsf{D}}^C$, where $W^A{}_{CB}$ is called the wryness of the director field and is defined as

$$W^A{}_{BC}=\underset{\mathfrak{a}}{\mathsf{D}}^A{}_{|C}\overset{\mathfrak{a}}{\Theta}_B=-\underset{\mathfrak{a}}{\mathsf{D}}^A\overset{\mathfrak{a}}{\Theta}_{B|C}\,. \tag{5.69}$$

The director gradient $\underset{\mathfrak{a}}{\mathbf{F}}=\nabla^{\mathbf{G}}\underset{\mathfrak{a}}{\mathbf{d}}$ with components $\underset{\mathfrak{a}}{\mathsf{F}}^a{}_A=\underset{\mathfrak{a}}{\mathsf{d}}^a{}_{|A}$ is related to the relative wryness as

$$\underset{\mathfrak{a}}{\mathsf{F}}^a{}_A=w^a{}_{bA}\ \underset{\mathfrak{a}}{\mathsf{d}}^b\,, \tag{5.70}$$

where

$$w^a{}_{bA}=\underset{\mathfrak{a}}{\mathsf{d}}^a{}_{|A}\overset{\mathfrak{a}}{\vartheta}_b\,. \tag{5.71}$$

One can define the following director metrics: $\underset{\mathfrak{a}\mathfrak{b}}{\mathfrak{G}}=\underset{\mathfrak{a}}{\mathsf{D}}^A\underset{\mathfrak{b}}{\mathsf{D}}^BG_{AB}$, and $\underset{\mathfrak{a}\mathfrak{b}}{\mathfrak{g}}=\underset{\mathfrak{a}}{\mathsf{d}}^a\underset{\mathfrak{b}}{\mathsf{d}}^bg_{ab}$. Using these metrics, $\overset{\mathfrak{a}}{\Theta}_A=\overset{\mathfrak{a}\mathfrak{b}}{\mathfrak{G}}\ G_{AB}\underset{\mathfrak{b}}{\mathsf{D}}^B$, and $\overset{\mathfrak{a}}{\vartheta}_a=\overset{\mathfrak{a}\mathfrak{b}}{\mathfrak{g}}\ g_{ab}\underset{\mathfrak{b}}{\mathsf{d}}^b$.

The kinetic energy density of a Cosserat solid is written as [Toupin, 1964]

$$T=\frac{1}{2}\rho V^aV^bg_{ab}+\frac{1}{2}\overset{\mathfrak{a}\mathfrak{b}}{\nu}\ \underset{\mathfrak{a}}{\dot{\mathsf{d}}}^a\underset{\mathfrak{b}}{\dot{\mathsf{d}}}^bg_{ab}, \tag{5.72}$$

where $\overset{\mathfrak{a}\mathfrak{b}}{\nu}=\overset{\mathfrak{b}\mathfrak{a}}{\nu}$ is the micro-mass moment of inertia, and $\underset{\mathfrak{a}}{\dot{\mathbf{d}}}(X,t)=\frac{\partial}{\partial t}\underset{\mathfrak{a}}{\mathbf{d}}(X,t)$ is the director velocity. The energy density is written as $W=W(X,\mathbf{F},\underset{\mathfrak{a}}{\mathbf{F}},\mathbf{G},\mathbf{g})$.[30]

**Balance of linear momentum.** We next derive the governing equations of generalized Cosserat elasticity using Hamilton's principle of least action. The variation of the kinetic energy is written as

$$\begin{aligned}\delta T&=\rho_0\langle\!\langle\mathbf{V},D^{\mathbf{g}}_t\delta\varphi\rangle\!\rangle_{\mathbf{g}}+\overset{\mathfrak{a}\mathfrak{b}}{\nu}\ \langle\!\langle\underset{\mathfrak{a}}{\dot{\mathbf{d}}},D^{\mathbf{g}}_t\delta\underset{\mathfrak{b}}{\mathbf{d}}\rangle\!\rangle_{\mathbf{g}}\\&=\frac{d}{dt}\left(\rho_0\langle\!\langle\mathbf{V},\delta\varphi\rangle\!\rangle_{\mathbf{g}}+\overset{\mathfrak{a}\mathfrak{b}}{\nu}\ \langle\!\langle\underset{\mathfrak{a}}{\dot{\mathbf{d}}},\delta\underset{\mathfrak{b}}{\mathbf{d}}\rangle\!\rangle_{\mathbf{g}}\right)-\rho_0\langle\!\langle\mathbf{A},\delta\varphi\rangle\!\rangle_{\mathbf{g}}-\overset{\mathfrak{a}\mathfrak{b}}{\nu}\ \langle\!\langle\underset{\mathfrak{a}}{\ddot{\mathbf{d}}},\delta\underset{\mathfrak{b}}{\mathbf{d}}\rangle\!\rangle_{\mathbf{g}},\end{aligned} \tag{5.73}$$

[30] Note that the energy function can have an explicit dependence on the director field, i.e., $W=W(X,\mathbf{F},\underset{\mathfrak{a}}{\mathbf{d}},\underset{\mathfrak{a}}{\mathbf{F}},\mathbf{G},\mathbf{g})$. The partial derivative $\frac{\partial W}{\partial\underset{\mathfrak{a}}{\mathbf{d}}}$ is a micro body force that we do not consider in this paper.

where $\underset{a}{\ddot{\mathbf{d}}} = D_t^{\mathbf{g}} \underset{a}{\dot{\mathbf{d}}}$ is the director acceleration. Assuming that $\delta\varphi(X, t_1) = \delta\varphi(X, t_2) = 0$, and $\delta \underset{b}{\mathbf{d}}(X, t_1) = \delta \underset{b}{\mathbf{d}}(X, t_2) = \mathbf{0}$ one obtains

$$\delta \int_{t_1}^{t_2} \int_{\mathcal{B}} -T dV dt = \int_{t_1}^{t_2} \int_{\mathcal{B}} \left( \rho_0 \langle\!\langle \mathbf{A}, \delta\varphi \rangle\!\rangle_{\mathbf{g}} + \overset{ab}{\nu} \, \langle\!\langle \underset{a}{\ddot{\mathbf{d}}}, \delta \underset{b}{\mathbf{d}} \rangle\!\rangle_{\mathbf{g}} \right) dV dt. \tag{5.74}$$

The variation of the energy density is calculated as

$$\delta W = \nabla^{\varphi_\epsilon}_{\frac{\partial}{\partial \epsilon}} W \Big|_{\epsilon=0} = \frac{\partial W}{\partial \mathbf{F}} : \delta \mathbf{F} + \frac{\partial W}{\partial \underset{a}{\mathsf{F}}} : \delta \underset{a}{\mathsf{F}} = \frac{\partial W}{\partial F^a{}_A} \delta F^a{}_A + \frac{\partial W}{\partial \underset{a}{\mathsf{F}}^a{}_A} \delta \underset{a}{\mathsf{F}}^a{}_A \,. \tag{5.75}$$

Note that $\delta F^a{}_A = U^a{}_{|A}$ and $\delta \underset{a}{\mathsf{F}}^a{}_A = \delta \underset{a}{\mathsf{d}}^a{}_{|A}$. Thus

$$\begin{aligned}
\int_{\mathcal{B}} \delta W dV &= \int_{\mathcal{B}} \left( \frac{\partial W}{\partial F^a{}_A} \delta \varphi^a{}_{|A} + \frac{\partial W}{\partial \underset{a}{\mathsf{F}}^a{}_A} \delta \underset{a}{\mathsf{d}}^a{}_{|A} \right) dV \\
&= - \int_{\mathcal{B}} \left[ \left( \frac{\partial W}{\partial F^a{}_A} \right)_{|A} \delta \varphi^a + \left( \frac{\partial W}{\partial \underset{a}{\mathsf{F}}^a{}_A} \right)_{|A} \delta \underset{a}{\mathsf{d}}^a \right] dV \\
&\quad + \int_{\partial \mathcal{B}} \left( \frac{\partial W}{\partial F^a{}_A} N_A \delta \varphi^a + \frac{\partial W}{\partial \underset{a}{\mathsf{F}}^a{}_A} N_A \delta \underset{a}{\mathsf{d}}^a \right) dA \,.
\end{aligned} \tag{5.76}$$

Therefore, the balance of linear momentum and the balance of micro linear momentum read

$$P^{aA}{}_{|A} = \rho_0 A^a, \quad \overset{a}{\mathsf{H}}{}^{aA}{}_{|A} = \overset{ab}{\nu} \, \underset{b}{\ddot{\mathsf{d}}}{}^a \,, \tag{5.77}$$

where

$$P^{aA} = g^{ab} \frac{\partial W}{\partial F^b{}_A}, \quad \overset{a}{\mathsf{H}}{}^{aA} = g^{ab} \frac{\partial W}{\partial \underset{a}{\mathsf{F}}^b{}_A} \,. \tag{5.78}$$

$\overset{a}{\mathsf{H}}{}^{aA}$ is called the hyperstress tensor. The traction and micro-traction are defined as $T^a = P^{aA} N_A$ and $\overset{a}{\mathsf{T}}{}^a = \overset{a}{\mathsf{H}}{}^{aA} N_A$, respectively.

**Balance of angular momentum.** In order to derive the balance of angular momentum in a Euclidean ambient space consider the flow $\psi_s(x) = x + s\mathbf{\Omega} x$, where $\mathbf{\Omega}$ is an anti-symmetric matrix. As the kinetic energy density (5.72) is invariant under rotations one only needs to require invariance of the energy function under flows of rotations of the ambient space, i.e.,

$$W(X, \mathbf{F}, \underset{a}{\mathsf{F}}, \mathbf{G}, \mathbf{g}) = W(X, \psi_{s*} \mathbf{F}, \psi_{s*} \underset{a}{\mathsf{F}}, \mathbf{G}, \psi_{s*} \mathbf{g}). \tag{5.79}$$

Note that under a rotation $R^a{}_b$ of the Euclidean ambient space the directors are transformed as $\underset{a}{\mathsf{d}}'^a = R^a{}_b \underset{a}{\mathsf{d}}^b$. Taking derivative with respect to $s$ of both sides of (5.79) and evaluating at $s = 0$, one obtains

$$g^{ac} \left[ F^b{}_A \frac{\partial W}{\partial F^c{}_A} + \underset{a}{\mathsf{F}}^b{}_A \frac{\partial W}{\partial \underset{a}{\mathsf{F}}^c{}_A} \right] \Omega_{ab} = 0 \,. \tag{5.80}$$

As $\Omega_{ba} = -\Omega_{ab}$, one concludes that $\Pi^{ab} = \Pi^{ba}$, or $\Pi^{[ab]} = 0$, where

$$\Pi^{ab} = g^{ac} \left[ F^b{}_A \frac{\partial W}{\partial F^c{}_A} + \underset{a}{\mathsf{F}}^b{}_A \frac{\partial W}{\partial \underset{a}{\mathsf{F}}^c{}_A} \right] = F^b{}_A P^{aA} + \underset{a}{\mathsf{F}}^b{}_A \overset{a}{\mathsf{H}}{}^{aA} \,. \tag{5.81}$$

Thus, the balance of angular momentum reads

$$P^{[aA} F^{b]}{}_A + \overset{a}{\mathsf{H}}{}^{[aA} \underset{a}{\mathsf{F}}^{b]}{}_A = 0 \,. \tag{5.82}$$

**Linearized balance of linear momentum.** Linearizing the balance of linear momentum and the balance of micro linear momentum about a pair $(\mathring{\varphi}, \underset{\mathfrak{a}}{\mathring{\mathsf{d}}})$ one obtains

$$(\delta P^{aA})_{|A} = \rho_0 \ddot{U}^a, \quad (\delta \overset{\mathfrak{a}}{\mathsf{H}}{}^{aA})_{|A} = \overset{\mathfrak{a}\mathfrak{b}}{\nu}\, \underset{\mathfrak{b}}{\ddot{\mathfrak{U}}}, \tag{5.83}$$

where $\underset{\mathfrak{a}}{\mathfrak{U}} = \delta \underset{\mathfrak{a}}{\mathsf{d}}$ is the director displacement field. Note that

$$\begin{aligned} \delta P^{aA} &= \mathbb{A}^{aA}{}_{b}{}^{B}\ U^b{}_{|B} + \overset{\mathfrak{a}}{\mathbb{B}}{}^{aA}{}_{b}{}^{B}\ \underset{\mathfrak{a}}{\mathfrak{U}}{}^b{}_{|B}, \\ \delta \overset{\mathfrak{a}}{\mathsf{H}}{}^{aA} &= \overset{\mathfrak{a}}{\mathbb{B}}{}^{aA}{}_{b}{}^{B}\ U^b{}_{|B} + \overset{\mathfrak{a}\mathfrak{b}}{\mathbb{C}}{}^{aA}{}_{b}{}^{B}\ \underset{\mathfrak{b}}{\mathfrak{U}}{}^b{}_{|B}\,, \end{aligned} \tag{5.84}$$

where

$$\mathbb{A}_a{}^A{}_b{}^B = \frac{\partial^2 W}{\partial F^a{}_A \partial F^b{}_B}, \quad \overset{\mathfrak{a}}{\mathbb{B}}_a{}^A{}_b{}^B = \frac{\partial^2 W}{\partial F^a{}_A \partial \underset{\mathfrak{a}}{\mathsf{F}}{}^b{}_B}, \quad \overset{\mathfrak{a}\mathfrak{b}}{\mathbb{C}}_a{}^A{}_b{}^B = \frac{\partial^2 W}{\partial \underset{\mathfrak{a}}{\mathsf{F}}{}^a{}_A \partial \underset{\mathfrak{b}}{\mathsf{F}}{}^b{}_B}\,. \tag{5.85}$$

The elastic constants satisfy the following symmetries: $\mathbb{A}^{aAbB} = \mathbb{A}^{bBaA}$ and $\overset{\mathfrak{a}\mathfrak{b}}{\mathbb{C}}{}^{aAbB} = \overset{\mathfrak{b}\mathfrak{a}}{\mathbb{C}}{}^{bBaA}$.

**Linearized balance of angular momentum.** Suppose the reference motion is an isometric embedding of an initially stress-free body into the Euclidean space, i.e., $\mathring{F}^a{}_A = \delta^a_A$, $\mathring{P}^{aA} = 0$, and $\overset{\mathfrak{a}}{\mathring{\mathsf{H}}}{}^{aA} = 0$. We will linearize the balance of angular momentum about this motion and about a director field $\underset{\mathfrak{a}}{\mathring{\mathsf{d}}}{}^a = \delta^a_A \underset{\mathfrak{a}}{\mathsf{D}}{}^A$. Thus, $\underset{\mathfrak{a}}{\mathring{\mathsf{F}}}{}^a{}_A = \delta^a_B \underset{\mathfrak{a}}{\mathsf{D}}{}^B{}_{|A} = \mathring{w}^a{}_{mA} \underset{\mathfrak{a}}{\mathring{\mathsf{d}}}{}^m = \delta^a_C W^C{}_{MA} \underset{\mathfrak{a}}{\mathsf{D}}{}^M$. Linearization of $\Pi^{ab}$ in (5.81) one obtains

$$\delta \Pi^{ab} = \left( \mathbb{A}^{aA}{}_c{}^B \mathring{F}^b{}_A + \overset{\mathfrak{a}}{\mathbb{B}}{}^{aA}{}_c{}^B \underset{\mathfrak{a}}{\mathring{\mathsf{F}}}{}^b{}_A \right) U^c{}_{|B} + \left( \overset{\mathfrak{b}}{\mathbb{B}}{}^{aA}{}_c{}^B \mathring{F}^b{}_A + \overset{\mathfrak{a}\mathfrak{b}}{\mathbb{C}}{}^{aA}{}_c{}^B \underset{\mathfrak{a}}{\mathring{\mathsf{F}}}{}^b{}_A \right) \underset{\mathfrak{b}}{\mathfrak{U}}{}^c{}_{|B}\,. \tag{5.86}$$

Knowing that $\delta \Pi^{[ab]} = 0$ and that the displacement and director displacement fields are independent, one obtains

$$\begin{aligned} &\mathbb{A}^{[aA}{}_c{}^B \mathring{F}^{b]}{}_A + \overset{\mathfrak{a}}{\mathbb{B}}{}^{[aA}{}_c{}^B \underset{\mathfrak{a}}{\mathring{\mathsf{F}}}{}^{b]}{}_A = 0, \\ &\overset{\mathfrak{b}}{\mathbb{B}}{}^{[aA}{}_c{}^B \mathring{F}^{b]}{}_A + \overset{\mathfrak{a}\mathfrak{b}}{\mathbb{C}}{}^{[aA}{}_c{}^B \underset{\mathfrak{a}}{\mathring{\mathsf{F}}}{}^{b]}{}_A = 0. \end{aligned} \tag{5.87}$$

Or equivalently

$$\begin{aligned} &\mathbb{A}^{[aAbB} \mathring{F}^{c]}{}_A + \overset{\mathfrak{a}}{\mathbb{B}}{}^{[aAbB} \underset{\mathfrak{a}}{\mathring{\mathsf{F}}}{}^{c]}{}_A = 0, \\ &\overset{\mathfrak{a}}{\mathbb{B}}{}^{[aAbB} \mathring{F}^{c]}{}_A + \overset{\mathfrak{b}\mathfrak{a}}{\mathbb{C}}{}^{[aAbB} \underset{\mathfrak{b}}{\mathring{\mathsf{F}}}{}^{c]}{}_A = 0. \end{aligned} \tag{5.88}$$

It is seen that the wryness of the reference configuration enters the linearized balance of angular momentum. In other words, in addition to the elastic constants, one needs some information on the non-uniformity of the director field in the stress-free reference configuration.

**Transformation cloaking in generalized Cosserat elastodynamics.** It is straightforward to show that the issue with the acceleration term observed in classical nonlinear elasticity persists even in nonlinear Cosserat elasticity, and hence, we do not discuss transformation cloaking in nonlinear Cosserat elastodynamics. We start from a virtual body that is made of a homogeneous, isotropic Cosserat elastic solid. We consider a cloaking transformation and try to calculate the elastic constants of the physical body induced from the cloaking transformation such that the balance of linear and angular momenta are respected in both the virtual and physical bodies.

The divergence term $(\delta P^{aA})_{|A}$ in the balance of linear momentum in the physical body is written as

$$\left( \mathbb{A}^{aA}{}_b{}^B\ U^b{}_{|B} + \overset{\mathfrak{a}}{\mathbb{B}}{}^{aA}{}_b{}^B\ \underset{\mathfrak{a}}{\mathfrak{U}}{}^b{}_{|B} \right)_{|A} \frac{\partial}{\partial x^a}. \tag{5.89}$$

Under a cloaking transformation $\Xi : \mathcal{B} \to \tilde{\mathcal{B}}$ and using the shifter map this is transformed to

$$J_\Xi \left( \tilde{\mathbb{A}}{}^{\tilde{a}\tilde{A}}{}_{\tilde{b}}{}^{\tilde{B}}\ \tilde{U}^{\tilde{b}}{}_{|\tilde{B}} + \overset{\mathfrak{a}}{\tilde{\mathbb{B}}}{}^{\tilde{a}\tilde{A}}{}_{\tilde{b}}{}^{\tilde{B}}\ \underset{\mathfrak{a}}{\tilde{\mathfrak{U}}}{}^{\tilde{b}}{}_{|\tilde{B}} \right)_{|\tilde{A}} \frac{\partial}{\partial \tilde{x}^{\tilde{a}}}, \tag{5.90}$$

where

$$
\begin{aligned}
\tilde{U}^{\tilde{a}} &= \mathsf{s}^{\tilde{a}}{}_a U^a\,, \quad \underset{\mathfrak{a}}{\tilde{\mathfrak{U}}}^{\tilde{a}} = \mathsf{s}^{\tilde{a}}{}_a\ \underset{\mathfrak{a}}{\mathfrak{U}}^a, \\
\tilde{\mathbb{A}}^{\tilde{a}\tilde{A}}{}_{\tilde{b}}{}^{\tilde{B}} &= J_\Xi^{-1}\mathsf{s}^{\tilde{a}}{}_a \overset{\equiv}{F}{}^{\tilde{A}}{}_A (\mathsf{s}^{-1})^b{}_{\tilde{b}} \overset{\equiv}{F}{}^{\tilde{B}}{}_B\ \mathbb{A}^{aA}{}_b{}^B\,, \\
\tilde{\overset{\mathfrak{a}}{\mathbb{B}}}{}^{\tilde{a}\tilde{A}}{}_{\tilde{b}}{}^{\tilde{B}} &= J_\Xi^{-1}\mathsf{s}^{\tilde{a}}{}_a \overset{\equiv}{F}{}^{\tilde{A}}{}_A (\mathsf{s}^{-1})^b{}_{\tilde{b}} \overset{\equiv}{F}{}^{\tilde{B}}{}_B\ \overset{\mathfrak{a}}{\mathbb{B}}{}^{aA}{}_b{}^B\,.
\end{aligned}
\tag{5.91}
$$

Or, equivalently

$$
\begin{aligned}
\mathbb{A}^{aA}{}_b{}^B &= J_\Xi(\mathsf{s}^{-1})^a{}_{\tilde{a}}(\overset{\equiv}{F}{}^{-1})^A{}_{\tilde{A}}\mathsf{s}^{\tilde{b}}{}_b(\overset{\equiv}{F}{}^{-1})^B{}_{\tilde{B}}\ \tilde{\mathbb{A}}^{\tilde{a}\tilde{A}}{}_{\tilde{b}}{}^{\tilde{B}}\,, \\
\overset{\mathfrak{a}}{\mathbb{B}}{}^{aA}{}_b{}^B &= J_\Xi(\mathsf{s}^{-1})^a{}_{\tilde{a}}(\overset{\equiv}{F}{}^{-1})^A{}_{\tilde{A}}\mathsf{s}^{\tilde{b}}{}_b(\overset{\equiv}{F}{}^{-1})^B{}_{\tilde{B}}\ \tilde{\overset{\mathfrak{a}}{\mathbb{B}}}{}^{\tilde{a}\tilde{A}}{}_{\tilde{b}}{}^{\tilde{B}}\,.
\end{aligned}
\tag{5.92}
$$

The divergence term $(\delta\overset{\mathfrak{a}}{\mathsf{H}}{}^{aA})_{|A}$ in the balance of micro linear momentum in the physical body is written as

$$
\left(\overset{\mathfrak{a}}{\mathbb{B}}{}^{aA}{}_b{}^B\ U^b{}_{|B} + \overset{\mathfrak{ab}}{\mathbb{C}}{}^{aA}{}_b{}^B\ \underset{\mathfrak{b}}{\mathfrak{U}}^b{}_{|B}\right)_{|A}\frac{\partial}{\partial x^a}. \tag{5.93}
$$

Under the cloaking transformation and using the shifter map this is transformed to read

$$
J_\Xi\left(\tilde{\overset{\mathfrak{a}}{\mathbb{B}}}{}^{\tilde{a}\tilde{A}}{}_{\tilde{b}}{}^{\tilde{B}}\ \tilde{U}^{\tilde{b}}{}_{|\tilde{B}} + \tilde{\overset{\mathfrak{ab}}{\mathbb{C}}}{}^{\tilde{a}\tilde{A}}{}_{\tilde{b}}{}^{\tilde{B}}\ \underset{\mathfrak{a}}{\tilde{\mathfrak{U}}}^{\tilde{b}}{}_{|\tilde{B}}\right)_{|\tilde{A}}\frac{\partial}{\partial \tilde{x}^{\tilde{a}}}, \tag{5.94}
$$

where

$$
\tilde{\overset{\mathfrak{ab}}{\mathbb{C}}}{}^{\tilde{a}\tilde{A}}{}_{\tilde{b}}{}^{\tilde{B}} = J_\Xi^{-1}\mathsf{s}^{\tilde{a}}{}_a \overset{\equiv}{F}{}^{\tilde{A}}{}_A (\mathsf{s}^{-1})^b{}_{\tilde{b}} \overset{\equiv}{F}{}^{\tilde{B}}{}_B\ \overset{\mathfrak{ab}}{\mathbb{C}}{}^{aA}{}_b{}^B\,, \tag{5.95}
$$

and the other transformed quantities are given in (5.97). Equivalently

$$
\overset{\mathfrak{ab}}{\mathbb{C}}{}^{aA}{}_b{}^B = J_\Xi(\mathsf{s}^{-1})^a{}_{\tilde{a}}(\overset{\equiv}{F}{}^{-1})^A{}_{\tilde{A}}\mathsf{s}^{\tilde{b}}{}_b(\overset{\equiv}{F}{}^{-1})^B{}_{\tilde{B}}\ \tilde{\overset{\mathfrak{ab}}{\mathbb{C}}}{}^{\tilde{a}\tilde{A}}{}_{\tilde{b}}{}^{\tilde{B}}\,. \tag{5.96}
$$

The micro-mass moment of inertia is transformed as $\tilde{\overset{\mathfrak{ab}}{\nu}} = J_\Xi^{-1}\overset{\mathfrak{ab}}{\nu}$. Similar to classical linear elasticity the mass density is transformed as $\tilde{\rho}_0 = J_\Xi^{-1}\rho_0$. In summary, the linearized balance of linear momentum and micro linear momentum are form-invariant under the following *cloaking transformations*:

$$
\begin{aligned}
&\tilde{X} = \Xi(X), \quad \tilde{\mathbf{U}} = \mathbf{s}\circ\mathring{\varphi}\ \mathbf{U}\circ\Xi^{-1}, \quad \underset{\mathfrak{a}}{\tilde{\boldsymbol{\mathfrak{U}}}} = \mathbf{s}\circ\mathring{\varphi}\ \underset{\mathfrak{a}}{\boldsymbol{\mathfrak{U}}}\circ\Xi^{-1}, \\
&\tilde{\rho}_0 = J_\Xi^{-1}\rho_0\circ\Xi^{-1}, \quad \tilde{\overset{\mathfrak{ab}}{\nu}} = J_\Xi^{-1}\overset{\mathfrak{ab}}{\nu}\circ\Xi^{-1}, \quad \tilde{\mathbb{A}} = (J_\Xi^{-1}\mathbf{s}^{-1}\circ\mathring{\varphi}\ \overset{\equiv}{\mathbf{F}}\ \mathbf{s}\circ\mathring{\varphi}\ \mathbb{A}\ \overset{\equiv}{\mathbf{F}}{}^\star)\circ\Xi^{-1}, \\
&\tilde{\overset{\mathfrak{a}}{\mathbb{B}}} = (J_\Xi^{-1}\mathbf{s}^{-1}\circ\mathring{\varphi}\ \overset{\equiv}{\mathbf{F}}\ \mathbf{s}\circ\mathring{\varphi}\ \overset{\mathfrak{a}}{\mathbb{B}}\ \overset{\equiv}{\mathbf{F}}{}^\star)\circ\Xi^{-1}, \quad \tilde{\overset{\mathfrak{ab}}{\mathbb{C}}} = (J_\Xi^{-1}\mathbf{s}^{-1}\circ\mathring{\varphi}\ \overset{\equiv}{\mathbf{F}}\ \mathbf{s}\circ\mathring{\varphi}\ \overset{\mathfrak{ab}}{\mathbb{C}}\ \overset{\equiv}{\mathbf{F}}{}^\star)\circ\Xi^{-1}\,.
\end{aligned}
\tag{5.97}
$$

Note that under a cloaking transformation, the stress and hyperstress tensors are transformed as

$$
\tilde{P}^{\tilde{a}\tilde{A}} = J_\Xi^{-1}\mathsf{s}^{\tilde{a}}{}_a\overset{\equiv}{F}{}^{\tilde{A}}{}_A P^{aA}\,, \quad \tilde{\overset{\mathfrak{a}}{\mathsf{H}}}{}^{\tilde{a}\tilde{A}} = J_\Xi^{-1}\mathsf{s}^{\tilde{a}}{}_a\overset{\equiv}{F}{}^{\tilde{A}}{}_A\overset{\mathfrak{a}}{\mathsf{H}}{}^{aA}\,. \tag{5.98}
$$

We assume that in the stress-free reference configuration of the virtual body the director field is uniform, i.e., $\underset{\mathfrak{a}}{\tilde{\mathsf{D}}}{}^{\tilde{A}}{}_{|\tilde{B}} = 0$, and hence, $\underset{\mathfrak{a}}{\tilde{\mathring{\mathsf{F}}}}{}^{\tilde{a}}{}_{\tilde{A}} = 0$. In other words, the wryness vanishes in the virtual structure. Therefore, the balance of angular momentum (5.88) for the virtual body reads

$$
\tilde{\mathbb{A}}^{[\tilde{a}\tilde{A}\tilde{b}\tilde{B}}\tilde{\mathring{F}}{}^{\tilde{c}]}{}_{\tilde{A}} = 0, \qquad \tilde{\overset{\mathfrak{a}}{\mathbb{B}}}{}^{[\tilde{a}\tilde{A}\tilde{b}\tilde{B}}\tilde{\mathring{F}}{}^{\tilde{c}]}{}_{\tilde{A}} = 0. \tag{5.99}
$$

Assuming that in the virtual body the balance of angular momentum (5.99) is satisfied, the balance of angular momentum in the physical body (5.88) requires that

$$
\begin{aligned}
&(\mathsf{s}^{-1})^{[a}{}_{\tilde{a}}(\mathsf{s}^{-1})^b{}_{\tilde{b}}(\overset{\equiv}{F}{}^{-1})^A{}_{\tilde{A}}(\overset{\equiv}{F}{}^{-1})^B{}_{\tilde{B}}\left\{\mathring{F}{}^{c]}{}_A\tilde{\mathbb{A}}^{\tilde{a}\tilde{A}\tilde{b}\tilde{B}} + \underset{\mathfrak{a}}{\mathring{\mathsf{F}}}{}^{c]}{}_A\ \tilde{\overset{\mathfrak{a}}{\mathbb{B}}}{}^{\tilde{a}\tilde{A}\tilde{b}\tilde{B}}\right\} = 0\,, \\
&(\mathsf{s}^{-1})^{[a}{}_{\tilde{a}}(\mathsf{s}^{-1})^b{}_{\tilde{b}}(\overset{\equiv}{F}{}^{-1})^A{}_{\tilde{A}}(\overset{\equiv}{F}{}^{-1})^B{}_{\tilde{B}}\left\{\mathring{F}{}^{c]}{}_A\tilde{\overset{\mathfrak{a}}{\mathbb{B}}}{}^{\tilde{a}\tilde{A}\tilde{b}\tilde{B}} + \underset{\mathfrak{b}}{\mathring{\mathsf{F}}}{}^{c]}{}_A\tilde{\overset{\mathfrak{ba}}{\mathbb{C}}}{}^{\tilde{a}\tilde{A}\tilde{b}\tilde{B}}\right\} = 0\,,
\end{aligned}
\tag{5.100}
$$

where $\underset{\mathfrak{a}}{\mathring{\mathsf{F}}}{}^c{}_A = \delta^c_C W^C{}_{MA}\underset{\mathfrak{a}}{\mathsf{D}}{}^M$. Note that $(\overset{\equiv}{F})^{\tilde{A}}{}_{A|B} = (\overset{\equiv}{F})^{\tilde{A}}{}_{B|A}$.

**Remark 5.4.** Note that $\underset{\mathfrak{a}}{\mathring{\mathsf{F}}}{}^a{}_{A|B} = \underset{\mathfrak{a}}{\mathring{\mathsf{d}}}{}^a{}_{|A|B}$. Knowing that the reference configuration of the physical body is flat, covariant derivatives commute. Therefore, a necessary condition for a field $\underset{\mathfrak{a}}{\mathring{\mathsf{F}}}{}^a{}_A$ to be director gradient of a director field is

$$\underset{\mathfrak{a}}{\mathring{\mathsf{F}}}{}^a{}_{A|B} = \underset{\mathfrak{a}}{\mathring{\mathsf{F}}}{}^a{}_{B|A}\,. \tag{5.101}$$

When designing a cloak the field $\underset{\mathfrak{a}}{\mathring{\mathsf{F}}}{}^a{}_A$ is not known a priori. It must satisfy both (5.100) and the compatibility equations (5.101).

**Remark 5.5.** In the literature it has been assumed that the virtual body is made of a classical linear elastic solid while the cloak is suggested to be made of a Cosserat elastic solid (without any discussion on how to calculate the Cosserat elastic constants). Note, however, that the virtual body cannot be made of a classical solid. If one assumes that the virtual body is classical, i.e., $\tilde{\overset{\mathfrak{a}}{\mathbb{B}}}$ and $\tilde{\overset{\mathfrak{ab}}{\mathbb{C}}}$ vanish, (5.100) reduces to the classical balance of angular momentum for the physical body (4.54), which is an obstruction to transformation cloaking. In other words, both the virtual body and the physical body (even outside the cloak) must be made of Cosserat linear elastic solids in order to achieve elastodynamics transformation cloaking.

**Elastic constants of the virtual body.** We assume that the virtual body is made of an elastic, homogeneous, and isotropic generalized Cosserat solid. The most general form of a fourth-order isotropic tensor is given by $a_1\delta_{ij}\delta_{kl} + a_2\delta_{ik}\delta_{jl} + a_3\delta_{il}\delta_{jk}$ for some scalars $a_1$, $a_2$, and $a_3$. Therefore, the elastic constant of the virtual body are written as

$$\begin{aligned}\tilde{\mathbb{A}}_{\tilde a}{}^{\tilde A}{}_{\tilde b}{}^{\tilde B} = \lambda \tilde G^{\tilde M\tilde A}\tilde G^{\tilde N\tilde B}\mathring F^{\tilde m}{}_{\tilde M}\mathring F^{\tilde n}{}_{\tilde N}g_{\tilde a\tilde m}g_{\tilde b\tilde n} + \mu\Big(&\tilde G^{\tilde M\tilde N}\tilde G^{\tilde A\tilde B}\mathring F^{\tilde m}{}_{\tilde M}\mathring F^{\tilde n}{}_{\tilde N}g_{\tilde a\tilde m}g_{\tilde b\tilde n}\\ &+\tilde G^{\tilde M\tilde B}\tilde G^{\tilde N\tilde A}\mathring F^{\tilde m}{}_{\tilde M}\mathring F^{\tilde n}{}_{\tilde N}g_{\tilde a\tilde m}g_{\tilde b\tilde n}\Big)\,.\end{aligned} \tag{5.102}$$

This is a consequence of the minor symmetries. Also

$$\begin{aligned}\tilde{\overset{\mathfrak{a}}{\mathbb{B}}}_{\tilde a}{}^{\tilde A}{}_{\tilde b}{}^{\tilde B} = \overset{\mathfrak{a}}{b}_1 G^{\tilde M\tilde A}G^{\tilde N\tilde B}\mathring F^{\tilde m}{}_{\tilde M}\mathring F^{\tilde n}{}_{\tilde N}g_{\tilde a\tilde m}g_{\tilde b\tilde n} &+ \overset{\mathfrak{a}}{b}_2 G^{\tilde M\tilde N}G^{\tilde A\tilde B}\mathring F^{\tilde m}{}_{\tilde M}\mathring F^{\tilde n}{}_{\tilde N}g_{\tilde a\tilde m}g_{\tilde b\tilde n}\\ &+ \overset{\mathfrak{a}}{b}_3 G^{\tilde M\tilde B}G^{\tilde N\tilde A}\mathring F^{\tilde m}{}_{\tilde M}\mathring F^{\tilde n}{}_{\tilde N}g_{\tilde a\tilde m}g_{\tilde b\tilde n}\,.\end{aligned} \tag{5.103}$$

But note that $\tilde{\overset{\mathfrak{a}}{\mathbb{B}}}$ has the minor symmetry from $(5.99)_2$, and hence, $\overset{\mathfrak{a}}{b}_2 = \overset{\mathfrak{a}}{b}_3$. Similarly

$$\begin{aligned}\tilde{\overset{\mathfrak{ab}}{\mathbb{C}}}_{\tilde a}{}^{\tilde A}{}_{\tilde b}{}^{\tilde B} = \overset{\mathfrak{ab}}{c}_1 G^{\tilde M\tilde A}G^{\tilde N\tilde B}\mathring F^{\tilde m}{}_{\tilde M}\mathring F^{\tilde n}{}_{\tilde N}g_{\tilde a\tilde m}g_{\tilde b\tilde n} &+ \overset{\mathfrak{ab}}{c}_2 G^{\tilde M\tilde N}G^{\tilde A\tilde B}\mathring F^{\tilde m}{}_{\tilde M}\mathring F^{\tilde n}{}_{\tilde N}g_{\tilde a\tilde m}g_{\tilde b\tilde n}\\ &+ \overset{\mathfrak{ab}}{c}_3 G^{\tilde M\tilde B}G^{\tilde N\tilde A}\mathring F^{\tilde m}{}_{\tilde M}\mathring F^{\tilde n}{}_{\tilde N}g_{\tilde a\tilde m}g_{\tilde b\tilde n}\,,\end{aligned} \tag{5.104}$$

where $\overset{\mathfrak{ab}}{c}_i = \overset{\mathfrak{ba}}{c}_i$, $i = 1,2,3$, because $\tilde{\overset{\mathfrak{ab}}{\mathbb{C}}}{}^{\tilde a\tilde A\tilde b\tilde B} = \tilde{\overset{\mathfrak{ba}}{\mathbb{C}}}{}^{\tilde b\tilde B\tilde a\tilde A}$. Knowing that $\mathring F^a{}_A = \delta^a_A$, one can identify the spatial and material manifolds with the same metric, and thus, with a slight abuse of notation, the elastic constants are simplified to read

$$\begin{aligned}\tilde{\mathbb{A}}^{\tilde a\tilde A\tilde b\tilde B} &= \lambda\tilde G^{\tilde a\tilde A}\tilde G^{\tilde b\tilde B} + \mu(\tilde G^{\tilde a\tilde b}\tilde G^{\tilde A\tilde B} + \tilde G^{\tilde a\tilde B}\tilde G^{\tilde b\tilde A})\,,\\ \tilde{\overset{\mathfrak{a}}{\mathbb{B}}}{}^{\tilde a\tilde A\tilde b\tilde B} &= \overset{\mathfrak{a}}{b}_1 G^{\tilde a\tilde A}G^{\tilde b\tilde B} + \overset{\mathfrak{a}}{b}_2(G^{\tilde a\tilde b}G^{\tilde A\tilde B} + G^{\tilde a\tilde B}G^{\tilde b\tilde A})\,,\\ \tilde{\overset{\mathfrak{ab}}{\mathbb{C}}}{}^{\tilde a\tilde A\tilde b\tilde B} &= \overset{\mathfrak{ab}}{c}_1 G^{\tilde a\tilde A}G^{\tilde b\tilde B} + \overset{\mathfrak{ab}}{c}_2 G^{\tilde a\tilde b}G^{\tilde A\tilde B} + \overset{\mathfrak{ab}}{c}_3 G^{\tilde a\tilde B}G^{\tilde b\tilde A}\,.\end{aligned} \tag{5.105}$$

Therefore, one has 15 and 26 independent elastic constants in dimensions two and three, respectively.

**Positive-definiteness of the energy density.** Knowing that in the initial configuration $\mathring{P}^{aA} = 0$ and $\overset{\mathfrak{a}}{\mathsf{H}}^{aA} = 0$, one can write

$$\begin{aligned}\delta W &= \frac{1}{2}\frac{\partial W}{\partial F^a{}_A \partial F^b{}_B} U^a{}_{|A} U^b{}_{|B} + \frac{\partial W}{\partial F^a{}_A \partial \underset{\mathfrak{a}}{\mathsf{F}}{}^b{}_B} U^a{}_{|A} \underset{\mathfrak{a}}{\mathfrak{U}}{}^b{}_{|B} + \frac{1}{2}\frac{\partial W}{\partial \underset{\mathfrak{a}}{\mathsf{F}}{}^a{}_A \partial \underset{\mathfrak{b}}{\mathsf{F}}{}^b{}_B} \underset{\mathfrak{a}}{\mathfrak{U}}{}^a{}_{|A} \underset{\mathfrak{b}}{\mathfrak{U}}{}^b{}_{|B} \\ &= \frac{1}{2}\mathbb{A}_a{}^A{}_b{}^B U^a{}_{|A} U^b{}_{|B} + \overset{\mathfrak{a}}{\mathbb{B}}_a{}^A{}_b{}^B U^a{}_{|A} \underset{\mathfrak{a}}{\mathfrak{U}}{}^b{}_{|B} + \frac{1}{2}\overset{\mathfrak{a}\mathfrak{b}}{\mathbb{C}}_a{}^A{}_b{}^B \underset{\mathfrak{a}}{\mathfrak{U}}{}^a{}_{|A} \underset{\mathfrak{b}}{\mathfrak{U}}{}^b{}_{|B} \\ &= \frac{1}{2}\mathbb{A}^{aAbB} U_{a|A} U_{b|B} + \overset{\mathfrak{a}}{\mathbb{B}}{}^{aAbB} U_{a|A} \underset{\mathfrak{a}}{\mathfrak{U}}{}_{b|B} + \frac{1}{2}\overset{\mathfrak{a}\mathfrak{b}}{\mathbb{C}}{}^{aAbB} \underset{\mathfrak{a}}{\mathfrak{U}}{}_{a|A} \underset{\mathfrak{b}}{\mathfrak{U}}{}_{b|B}.\end{aligned} \tag{5.106}$$

Let us introduce the new indices $\gamma = \{aA\}$, $\Gamma = \{\mathfrak{a}aA\}$,[31] and define the new variables $X_\gamma = U_{a|A}$ and $Y_\Gamma = \underset{\mathfrak{a}}{\mathfrak{U}}{}_{a|A}$. Thus, energy density is rewritten as

$$\delta W = \frac{1}{2}\mathbb{A}^{\gamma\lambda} X_\gamma X_\lambda + \mathbb{B}^{\gamma\Gamma} X_\gamma Y_\Gamma + \frac{1}{2}\mathbb{C}^{\Gamma\Lambda} Y_\Gamma Y_\Lambda. \tag{5.107}$$

Next let us define a new variable $\mathbf{Z} = \begin{Bmatrix}\mathbf{X} \\ \mathbf{Y}\end{Bmatrix}$. It is straightforward to show that

$$\delta W = \frac{1}{2}\mathbf{Z}^\mathsf{T} \cdot \mathbb{D}\mathbf{Z}, \quad \mathbb{D} = \begin{bmatrix}\mathbb{A} & \mathbb{B} \\ \mathbb{B}^\mathsf{T} & \mathbb{C}\end{bmatrix}. \tag{5.108}$$

Note that $\mathbb{A}^{\gamma\lambda} = \mathbb{A}^{\lambda\gamma}$, and $\mathbb{C}^{\Gamma\Lambda} = \mathbb{C}^{\Lambda\Gamma}$, but $\mathbb{B}^{\gamma\Gamma} \neq \mathbb{B}^{\Gamma\gamma}$, in general. Note that $\mathbb{D}$ is symmetric, and hence, $\delta W > 0$, $\forall \mathbf{Z} \neq \mathbf{0}$ if and only if all the eigenvalues of $\mathbb{D}$ are positive ($\mathbb{D}$ is a square matrix of size 12 and 36 in dimensions two and three, respectively). For the energy density to be positive-definite it is necessary that $\mathbb{A}$ and $\mathbb{C}$ be positive definite because $\mathbf{X}$ and $\mathbf{Y}$ are independent. We note that $\mathbb{A}$ is positive-definite if and only if $\mu > 0$ and $3\lambda + 2\mu > 0$. Note that $\underset{\mathfrak{a}}{\mathfrak{U}}$, $\mathfrak{a} = 1, 2, 3$ are independent, and hence, as a consequence of positive-definiteness of $\mathbb{C}$, $3\overset{\mathfrak{a}\mathfrak{a}}{c}_1 + \overset{\mathfrak{a}\mathfrak{a}}{c}_2 + \overset{\mathfrak{a}\mathfrak{a}}{c}_3 > 0$, $\overset{\mathfrak{a}\mathfrak{a}}{c}_2 + \overset{\mathfrak{a}\mathfrak{a}}{c}_3 > 0$, and $\overset{\mathfrak{a}\mathfrak{a}}{c}_2 - \overset{\mathfrak{a}\mathfrak{a}}{c}_3 > 0$, (no summation on $\mathfrak{a}$). In particular, $\overset{\mathfrak{a}\mathfrak{a}}{c}_2 > 0$. $\mathbb{A}$ and $\mathbb{C}$ are positive-definite, and hence, invertible. From Schur's complement condition [De Klerk, 2006], positive-definiteness of $\mathbb{D}$ is now equivalent to positive-definiteness of either $\mathbb{C} - \mathbb{B}^\mathsf{T}\mathbb{A}^{-1}\mathbb{B}$ or $\mathbb{A} - \mathbb{B}\mathbb{C}^{-1}\mathbb{B}^\mathsf{T}$.

Suppose $\delta W > 0$ for any $(U^a{}_{|A}, \underset{\mathfrak{a}}{\mathfrak{U}}{}^a{}_{|A}) \neq (0, 0)$. In the virtual body

$$\delta\tilde{W} = \frac{1}{2}\tilde{\mathbb{A}}^{\tilde{a}\tilde{A}\tilde{b}\tilde{B}} \tilde{U}_{\tilde{a}|\tilde{A}} \tilde{U}_{\tilde{b}|\tilde{B}} + \overset{\mathfrak{a}}{\tilde{\mathbb{B}}}{}^{\tilde{a}\tilde{A}\tilde{b}\tilde{B}} \tilde{U}_{\tilde{a}|\tilde{A}} \underset{\mathfrak{a}}{\tilde{\mathfrak{U}}}{}_{\tilde{b}|\tilde{B}} + \frac{1}{2}\overset{\mathfrak{a}\mathfrak{b}}{\tilde{\mathbb{C}}}{}^{\tilde{a}\tilde{A}\tilde{b}\tilde{B}} \underset{\mathfrak{a}}{\tilde{\mathfrak{U}}}{}_{\tilde{a}|\tilde{A}} \underset{\mathfrak{b}}{\tilde{\mathfrak{U}}}{}_{\tilde{b}|\tilde{B}}. \tag{5.109}$$

Using (5.97), the relations $\tilde{U}^{\tilde{b}}{}_{|\tilde{B}} = \mathsf{s}^{\tilde{b}}{}_b (\bar{\bar{F}}^{-1})^B{}_{\tilde{B}} U^b{}_{|B}$, $\underset{\mathfrak{a}}{\tilde{\mathfrak{U}}}{}^{\tilde{b}}{}_{|\tilde{B}} = \mathsf{s}^{\tilde{b}}{}_b (\bar{\bar{F}}^{-1})^B{}_{\tilde{B}} \underset{\mathfrak{a}}{\mathfrak{U}}{}^b{}_{|B}$, and the relation $\mathsf{s}^{\tilde{a}}{}_a \mathsf{s}^{\tilde{b}}{}_b \tilde{g}_{\tilde{a}\tilde{b}} = g_{ab}$, one can easily show that $\delta\tilde{W} = J_\Xi^{-1}\delta W > 0$. Therefore, the elastic constants of the physical body are positive-definite if and only if those of the virtual body are positive-definite. In other words, the cloaking transformations (5.97) preserve the positive-definiteness of the elastic constants.

**Boundary conditions in the physical and virtual problems.** Let $\partial\mathcal{B} = \mathcal{H} \cup \partial_o\mathcal{B}$, where $\mathcal{H}$ is the boundary of the hole and $\partial_o\mathcal{B}$ is the outer boundary of $\mathcal{B}$. Suppose $\partial_o\mathcal{B} = \partial_o\mathcal{B}_t \cup \partial_o\mathcal{B}_d$ such that the Neumann and Dirichlet boundary conditions are written as

$$\begin{aligned}&\begin{cases}\mathbf{PN} = \bar{\mathbf{t}}(X,t) \\ \overset{\mathfrak{a}}{\mathbf{H}}\mathbf{N} = \overset{\mathfrak{a}}{\bar{\mathbf{m}}}(X,t)\end{cases} \quad \text{on} \quad \partial_o\mathcal{B}_t, \\ &\begin{cases}\varphi(X,t) = \bar{\varphi}(X,t) \\ \underset{\mathfrak{a}}{\mathbf{d}}(X,t) = \underset{\mathfrak{a}}{\bar{\mathbf{d}}}(X,t)\end{cases} \quad \text{on} \quad \partial_o\mathcal{B}_d\,,\end{aligned} \tag{5.110}$$

where $\mathbf{N}$ is the unit normal vector on $\partial_o\mathcal{B}_t$ and $\bar{\mathbf{t}}$, $\bar{\mathbf{m}}$, $\bar{\varphi}$, and $\underset{\mathfrak{a}}{\bar{\mathbf{d}}}$ are given. Under the mapping $\Xi : \mathcal{B} \to \tilde{\mathcal{B}}$, $\partial_o\tilde{\mathcal{B}} = \Xi(\partial_o\mathcal{B}) = \Xi(\partial_o\mathcal{B}_t) \cup \Xi(\partial_o\mathcal{B}_d)$. Suppose on $\partial\mathcal{H}$, $\mathbf{t} = \mathbf{PN} = \bar{\mathbf{t}}$, and $\overset{\mathfrak{a}}{\mathbf{H}}\mathbf{N} = \overset{\mathfrak{a}}{\bar{\mathbf{m}}}$. Note that

$$\mathbf{st}dA = \mathbf{s}\bar{\mathbf{t}}dA = \tilde{\mathbf{t}}d\tilde{A}, \quad \mathbf{s}\overset{\mathfrak{a}}{\bar{\mathbf{m}}}dA = \mathbf{s}\overset{\mathfrak{a}}{\bar{\mathbf{m}}}dA = \overset{\mathfrak{a}}{\tilde{\mathbf{m}}}d\tilde{A}\,. \tag{5.111}$$

[31] More specifically, $\{11, 12, \cdots, 1n, 21, 22, \cdots, nn\} \to \{1, 2, \cdots, n^2\}$ and $\{\mathfrak{1}11, \mathfrak{1}12, \cdots, \mathfrak{1}nn, \mathfrak{2}11, \cdots, \mathfrak{n}nn\} \to \{1, 2, \cdots, n^3\}$.

Therefore, on $\partial_o\mathcal{B}_t$, $\tilde{\mathbf{t}} = \tilde{\mathbf{P}}\tilde{\mathbf{N}} = (dA/d\tilde{A})\bar{\mathbf{t}}$, and $\tilde{\overset{a}{\mathbf{m}}} = \tilde{\overset{a}{\mathbf{H}}}\tilde{\mathbf{N}} = (dA/d\tilde{A})\bar{\overset{a}{\mathbf{m}}}$. We know that $\tilde{N}_{\tilde{A}}d\tilde{A} = J_\Xi(\bar{\bar{\bar{F}}}^{-1})^A{}_{\tilde{A}}N_A dA$. Therefore (note that $\tilde{\mathbf{N}}$ is a $\tilde{\mathbf{G}}$-unit one-form)

$$\tilde{G}^{\tilde{A}\tilde{B}}\tilde{N}_{\tilde{A}}\tilde{N}_{\tilde{B}}d\tilde{A}^2 = d\tilde{A}^2 = J_\Xi^2\left[(\bar{\bar{\bar{F}}}^{-1})^A{}_{\tilde{A}}(\bar{\bar{\bar{F}}}^{-1})^B{}_{\tilde{B}}\tilde{G}^{\tilde{A}\tilde{B}}N_A N_B\right]dA^2 . \tag{5.112}$$

Thus

$$\frac{dA}{d\tilde{A}} = J_\Xi^{-1}\left[(\bar{\bar{\bar{C}}}^{-1})^{AB}N_A N_B\right]^{-\frac{1}{2}} , \tag{5.113}$$

where $\bar{\bar{\bar{C}}}_{AB} = \bar{\bar{\bar{F}}}^{\tilde{A}}{}_A\bar{\bar{\bar{F}}}^{\tilde{B}}{}_B\tilde{G}_{\tilde{A}\tilde{B}}$, and $(\bar{\bar{\bar{C}}}^{-1})^{AB} = (\bar{\bar{\bar{F}}}^{-1})^A{}_{\tilde{A}}(\bar{\bar{\bar{F}}}^{-1})^B{}_{\tilde{B}}\tilde{G}^{\tilde{A}\tilde{B}}$.

Therefore, the traction boundary condition on both $\partial\tilde{\mathcal{H}}$ and $\partial_o\tilde{\mathcal{B}}_t$ reads

$$\tilde{\mathbf{t}} = J_\Xi^{-1}\left[(\bar{\bar{\bar{C}}}^{-1})^{AB}N_A N_B\right]^{-\frac{1}{2}}\mathbf{s}\bar{\mathbf{t}} , \quad \tilde{\overset{a}{\mathbf{m}}} = J_\Xi^{-1}\left[(\bar{\bar{\bar{C}}}^{-1})^{AB}N_A N_B\right]^{-\frac{1}{2}}\mathbf{s}\bar{\overset{a}{\mathbf{m}}} . \tag{5.114}$$

Knowing that in $\mathcal{B}\setminus\mathcal{C}$, $\Xi = id$, we have

$$\tilde{\mathbf{t}} = \mathbf{s}\bar{\mathbf{t}}, \quad \tilde{\overset{a}{\mathbf{m}}} = \mathbf{s}\bar{\overset{a}{\mathbf{m}}} \quad \text{on} \quad \partial_o\tilde{\mathcal{B}}_t . \tag{5.115}$$

The hole in the physical body is assumed to be traction free, i.e.,

$$\mathbf{PN} = \mathbf{t} = \mathbf{0}, \quad \overset{a}{\mathbf{H}}\mathbf{N} = \overset{a}{\mathbf{m}} = \mathbf{0}, \quad \text{on} \quad \partial\mathcal{H} . \tag{5.116}$$

Using (5.114) in the virtual body one has

$$\tilde{\mathbf{P}}\tilde{\mathbf{N}} = \tilde{\mathbf{t}} = \mathbf{0}, \quad \tilde{\overset{a}{\mathbf{H}}}\tilde{\mathbf{N}} = \tilde{\overset{a}{\mathbf{m}}} = \mathbf{0} \quad \text{on} \quad \partial\tilde{\mathcal{H}} . \tag{5.117}$$

In other words, the traction-free boundary condition on the surface of the hole $\mathcal{H}$ implies that the surface of the transformed (infinitesimal) hole $\tilde{\mathcal{H}}$ in the virtual structure is traction-free as well.

On $\partial_o\tilde{\mathcal{B}}_d = \Xi(\partial_o\mathcal{B}_d)$, one assumes that the virtual and physical problems have the same Dirichlet boundary condition, i.e.,

$$\tilde{\varphi}(\tilde{X},t) = \bar{\varphi}\circ\Xi^{-1}(\tilde{X},t), \quad \tilde{\underset{a}{\mathbf{d}}}(\tilde{X},t) = \bar{\underset{a}{\mathbf{d}}}\circ\Xi^{-1}(\tilde{X},t) \quad \text{on} \quad \partial_o\tilde{\mathcal{B}}_d . \tag{5.118}$$

Moreover, we note that $\partial(\mathcal{B}\setminus\mathcal{C}) = \partial_o\mathcal{B}\cup\partial\mathcal{C}$, and similarly, $\partial(\tilde{\mathcal{B}}\setminus\tilde{\mathcal{C}}) = \partial_o\tilde{\mathcal{B}}\cup\partial\tilde{\mathcal{C}}$. As $\Xi$ is defined to be the identity map in $\mathcal{B}\setminus\mathcal{C}$, from (5.118), $\partial_o\mathcal{B}_d$ and $\partial_o\tilde{\mathcal{B}}_d$ have the same displacement boundary conditions. Notice that $T\Xi|_{\partial_o\mathcal{B}_t} = \bar{\bar{\bar{\mathbf{F}}}}|_{\partial_o\mathcal{B}_t} = id$, and hence $J_\Xi|_{\partial_o\mathcal{B}_t} = 1$, which implies that the traction boundary conditions on $\partial_o\mathcal{B}_t$ and $\partial_o\tilde{\mathcal{B}}_t$ are identical. Thus, $\partial_o\mathcal{B}$ and $\partial_o\tilde{\mathcal{B}}$ have the same traction and displacement boundary conditions. Note that $\partial\mathcal{C} = \partial\mathcal{H}\cup\partial_o\mathcal{C}$, where $\partial\mathcal{H}$ is the boundary of the hole and $\partial_o\mathcal{C}$ is the outer boundary of the cloak. On $\partial_o\mathcal{C}$ one can write

$$\tilde{\mathbf{t}} = J_\Xi^{-1}\left[(\bar{\bar{\bar{C}}}^{-1})^{AB}N_A N_B\right]^{-\frac{1}{2}}\mathbf{st} . \tag{5.119}$$

Let us assume that in addition to $\Xi|_{\partial_o\mathcal{C}} = id$, the tangent map is the identity map as well, i.e., $T\Xi|_{\partial_o\mathcal{C}} = \bar{\bar{\bar{\mathbf{F}}}}|_{\partial_o\mathcal{C}} = id$.[32] For such maps $J_\Xi|_{\partial_o\mathcal{C}} = 1$, and hence

$$\tilde{\mathbf{t}} = \left[\delta^A_{\tilde{A}}\delta^B_{\tilde{B}}\tilde{G}^{\tilde{A}\tilde{B}}N_A N_B\right]^{-\frac{1}{2}}\mathbf{st} \quad \text{on} \quad \partial_o\mathcal{C} . \tag{5.120}$$

Note that $\mathbf{G}$ and $\tilde{\mathbf{G}}$ are induced Euclidean metrics and because $\Xi$ and $T\Xi$ are both identity on $\partial_o\mathcal{C}$, $\delta^A_{\tilde{A}}\delta^B_{\tilde{B}}\tilde{G}^{\tilde{A}\tilde{B}}N_A N_B = G^{AB}N_A N_B = 1$, and hence

$$\tilde{\mathbf{t}} = \mathbf{st} \quad \text{on} \quad \partial_o\tilde{\mathcal{C}} . \tag{5.121}$$

[32] This condition has been ignored in the existing works on elastodynamics transformation cloaking. In particular, borrowing the cloaking transformation of Pendry et al. [2006] from electromagnetism is not acceptable as it does not satisfy this condition.

Similarly

$$\tilde{\underset{a}{\mathbf{m}}} = \mathbf{s}\underset{a}{\mathbf{m}} \quad \text{on} \quad \partial_o \tilde{\mathcal{C}}\,, \tag{5.122}$$

that is, $\partial_o \mathcal{C}$ and $\partial_o \tilde{\mathcal{C}}$ have identical traction boundary conditions. Outside the cloak the elastic constants of the two problems are obviously identical from (5.97). One needs to assume that outside the cloak the two bodies have identical mechanical properties. In particular, outside the cloak $\underset{a}{\mathring{\mathbf{F}}} = \mathbf{0}$, i.e., the director field in the physical body outside the cloak is uniform. This means that on $\mathcal{B} \setminus \mathcal{C}$, $\underset{a}{\mathbf{D}} = \underset{a}{\tilde{\mathbf{D}}}$. This holds on the outer boundary of the cloak as well, i.e., $\underset{a}{\tilde{\mathring{\mathbf{F}}}} = \mathbf{0}$, and $\underset{a}{\mathbf{D}} = \underset{a}{\tilde{\mathbf{D}}}$ on $\partial_o \mathcal{C}$. Therefore, $\mathcal{B} \setminus \mathcal{C}$ and $\tilde{\mathcal{B}} \setminus \tilde{\mathcal{C}}$ are made of the same generalized Cosserat solid (have identical elastic constants and have the same director fields in their undeformed configurations), and are subject to the same body forces and boundary conditions. It then immediately follows that $\varphi(X,t) = \tilde{\varphi}(\tilde{X},t)$ and $\underset{a}{\mathbf{d}}(X,t) = \underset{a}{\tilde{\mathbf{d}}}(\tilde{X},t)$ on $\mathcal{B} \setminus \mathcal{C}$, and hence, the current configurations of the two bodies are identical outside the cloak. This, in turn, renders cloaking possible as the virtual body is isotropic and homogeneous, and contains an infinitesimal hole, which has a low scattering effect on the incident waves.

The following summarizes the construction of an elastic generalized Cosserat cloak in a linear elastic generalized Cosserat solid. Consider a diffeomorphism $\Xi : \mathcal{B} \to \tilde{\mathcal{B}}$ that shrinks a hole in the physical body $\mathcal{B}$ to an infinitesimal hole in the virtual body $\tilde{\mathcal{B}}$. The hole is surrounded by a cloak $\mathcal{C}$ in the physical body $\mathcal{B}$. Assume that $\Xi|_{\mathcal{B}\setminus\mathcal{C}} = id$ and that on the outer boundary of the cloak $T\Xi = id$. Assume that the displacement vectors in the physical and virtual bodies, mass densities, body forces, and the elastic constants are related as given in (5.97). Outside the cloak the physical body is homogeneous and isotropic and has a constitutive equation identical to that of the virtual body. One assumes that the two problems have identical boundary conditions on the outer boundaries of $\mathcal{B}$ and $\tilde{\mathcal{B}}$, and the physical hole and the virtual hole are both traction free. Under the above assumptions the two boundary-value problems are equivalent. In other words, the governing equations of the physical problem are satisfied if and only if those of the virtual problem are satisfied. In addition, the two bodies have identical current configurations outside the cloak $\mathcal{C}$.

**The impossibility of transformation cloaking in generalized Cosserat elasticity.** In this section we prove that in dimension two transformation cloaking is not possible in linear generalized Cosserat elasticity. A corollary of this result is that transformation cloaking cannot be possible in any subclass of generalized Cosserat solids, and in particular, transformation cloaking is not possible in Cosserat elasticity. We start our discussion by first looking at the example of a cylindrical hole covered by a cylindrical cloak. It has been claimed in the literature that a cylindrical cloak would have to be made of a Cosserat solid. We show that this is not possible. In other words, (generalized) Cosserat elasticity allowing for a non-symmetric Cauchy stress does not imply that transformation cloaking can be achieved in (generalized) Cosserat elasticity.

**Example (A generalized Cosserat cylindrical cloak).** Consider an infinitely-long hollow solid cylinder that in its stress-free reference configuration has inner and outer radii $R_i$ and $R_o$, respectively. Let us transform the reference configuration to the reference configuration of another body (virtual body) that is a hollow cylinder with inner and outer radii $\epsilon$ and $R_o$, respectively, using a cloaking map $\Xi(R,\Theta,Z) = (f(R),\Theta,Z)$ such that $f(R_o) = R_o$. For such a map we have

$$\overset{\Xi}{\mathbf{F}} = \begin{bmatrix} f'(R) & 0 \\ 0 & 1 \end{bmatrix}. \tag{5.123}$$

Note that the shifter map is given as

$$\mathbf{s} = \begin{bmatrix} 1 & 0 \\ 0 & \frac{R}{f(R)} \end{bmatrix}. \tag{5.124}$$

Following our discussion in §5.2, we note that $\Xi$ must satisfy $T\Xi|_{\partial_o\mathcal{C}} = \overset{\Xi}{\mathbf{F}}|_{\partial_o\mathcal{C}} = id$, i.e., $f'(R_o) = 1$. Therefore, the simplest form of $f(R)$ is a quadratic polynomial and is given by

$$f(R) = -\frac{R_o^2(R_i-\epsilon)}{(R_o-R_i)^2} + \frac{R_o^2+R_i^2-2R_o\epsilon}{(R_o-R_i)^2}R - \frac{R_i-\epsilon}{(R_o-R_i)^2}R^2\,. \tag{5.125}$$

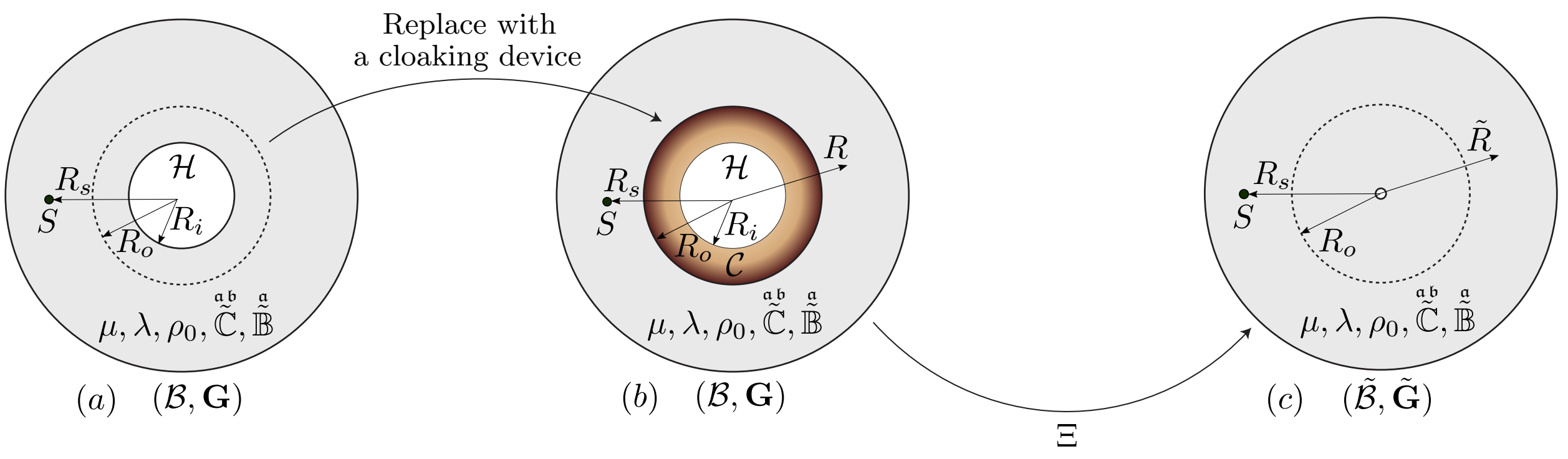

Figure 7: *Cloaking an object inside a hole $\mathcal{H}$ from elastic waves generated by a source $S$ located at a distance $R_s$ from the center of the hole. The system (a) is an isotropic, homogeneous, generalized Cosserat medium with elastic properties $\mu$, $\lambda$, $\rho_0$, $\overset{\mathfrak{a}}{\tilde{\mathbb{B}}}$, $\overset{\mathfrak{ab}}{\tilde{\mathbb{C}}}$, and $\overset{\mathfrak{ab}}{\tilde{\nu}}$ containing a finite hole $\mathcal{H}$. The system (c) is isotropic and homogeneous generalized Cosserat solid with the same elastic properties as the medium in the system (a). The cloak $\mathcal{C}$ in (b) has the elastic properties $\mathbb{A}$, $\overset{\mathfrak{a}}{\mathbb{B}}$, $\overset{\mathfrak{ab}}{\mathbb{C}}$, and $\overset{\mathfrak{ab}}{\nu}$. The configuration (b) is mapped to the new reference configuration (c) such that the hole $\mathcal{H}$ is mapped to an infinitesimal hole with negligible scattering effects on the elastic waves. The cloaking transformation is the identity mapping in $\mathcal{B} \setminus \mathcal{C}$.*

Note that $\rho = J_\Xi\, \tilde{\rho} \circ \Xi$ and $\overset{\mathfrak{ab}}{\nu} = J_\Xi\, \overset{\mathfrak{ab}}{\tilde{\nu}} \circ \Xi$, and therefore, the mass density and the micro-mass moment of inertia in $\mathcal{B} \setminus \mathcal{C}$ are homogeneous and are equal to $\rho_0$ and $\overset{\mathfrak{ab}}{\tilde{\nu}}$, respectively. The mass density in the cloaking device is inhomogeneous and is given by

$$\rho_{\mathcal{C}}(R) = \frac{f(R) f'(R)}{R} \rho_0\,, \qquad R_i \leq R \leq R_o\,. \tag{5.126}$$

Hence, (5.126) implies that

$$\rho_{\mathcal{C}}(R) = \frac{\left[(R_o^2 + R_i^2 - 2R_o\epsilon)R - (R_o^2 + R^2)(R_i - \epsilon)\right]}{\left[R_o^2 + R_i^2 - 2R_o\epsilon - 2R(R_i - \epsilon)\right]^{-1} R(R_o - R_i)^4} \rho_0\,, \quad R_i \leq R \leq R_o\,. \tag{5.127}$$

Similarly ($\mathfrak{a}, \mathfrak{b} \in \{1, 2\}$)

$$\overset{\mathfrak{ab}}{\nu}_{\mathcal{C}}(R) = \frac{\left[(R_o^2 + R_i^2 - 2R_o\epsilon)R - (R_o^2 + R^2)(R_i - \epsilon)\right]}{\left[R_o^2 + R_i^2 - 2R_o\epsilon - 2R(R_i - \epsilon)\right]^{-1} R(R_o - R_i)^4} \overset{\mathfrak{ab}}{\tilde{\nu}}\,, \quad R_i \leq R \leq R_o\,. \tag{5.128}$$

From (5.92), we find the Cosserat elastic constants of the cloak as follows ($\mathfrak{a}, \mathfrak{b} \in \{1, 2\}$)[33]

$$\hat{\mathbb{A}}^{aAbB} = \begin{bmatrix} \begin{bmatrix} \frac{(\lambda+2\mu)f(R)}{Rf'(R)} & 0 \\ 0 & \lambda \end{bmatrix} & \begin{bmatrix} 0 & \frac{\mu R f'(R)}{f(R)} \\ \mu & 0 \end{bmatrix} \\ \begin{bmatrix} 0 & \mu \\ \frac{\mu f(R)}{Rf'(R)} & 0 \end{bmatrix} & \begin{bmatrix} \lambda & 0 \\ 0 & \frac{(\lambda+2\mu)Rf'(R)}{f(R)} \end{bmatrix} \end{bmatrix}, \tag{5.129}$$

$$\overset{\mathfrak{a}}{\hat{\mathbb{B}}}{}^{aAbB} = \begin{bmatrix} \begin{bmatrix} \frac{(\overset{\mathfrak{a}}{b}_1+2\overset{\mathfrak{a}}{b}_2)f(R)}{Rf'(R)} & 0 \\ 0 & \overset{\mathfrak{a}}{b}_1 \end{bmatrix} & \begin{bmatrix} 0 & \frac{\overset{\mathfrak{a}}{b}_2 R f'(R)}{f(R)} \\ \overset{\mathfrak{a}}{b}_2 & 0 \end{bmatrix} \\ \begin{bmatrix} 0 & \overset{\mathfrak{a}}{b}_2 \\ \frac{\overset{\mathfrak{a}}{b}_2 f(R)}{Rf'(R)} & 0 \end{bmatrix} & \begin{bmatrix} \overset{\mathfrak{a}}{b}_1 & 0 \\ 0 & \frac{(\overset{\mathfrak{a}}{b}_1+2\overset{\mathfrak{a}}{b}_2)Rf'(R)}{f(R)} \end{bmatrix} \end{bmatrix}, \tag{5.130}$$

[33] Note that the physical components of the elasticity tensor $\hat{\mathbb{A}}^{aAbB}$ are related to the components of the elasticity tensor as $\hat{\mathbb{A}}^{aAbB} = \sqrt{g_{aa}}\sqrt{G_{AA}}\sqrt{g_{bb}}\sqrt{G_{BB}}\,\mathbb{A}^{aAbB}$ (no summation).

$$
\overset{ab}{\hat{\mathbb{C}}}{}^{aAbB} = \begin{bmatrix} \begin{bmatrix} \frac{(\overset{ab}{c}_1+\overset{ab}{c}_2+\overset{ab}{c}_3)f(R)}{Rf'(R)} & 0 \\ 0 & \overset{ab}{c}_1 \end{bmatrix} & \begin{bmatrix} 0 & \frac{\overset{ab}{c}_2 Rf'(R)}{f(R)} \\ \overset{ab}{c}_3 & 0 \end{bmatrix} \\ \begin{bmatrix} 0 & \overset{ab}{c}_3 \\ \frac{\overset{ab}{c}_2 f(R)}{Rf'(R)} & 0 \end{bmatrix} & \begin{bmatrix} \overset{ab}{c}_1 & 0 \\ 0 & \frac{(\overset{ab}{c}_1+\overset{ab}{c}_2+\overset{ab}{c}_3)Rf'(R)}{f(R)} \end{bmatrix} \end{bmatrix} . \tag{5.131}
$$

We next consider a cylindrical cloak and find the distribution of those (initial) director gradient fields that are compatible with the balance of angular momentum. $(5.100)_1$ gives the following equations for $\overset{\circ}{\underset{a}{\mathsf{F}}}{}^c{}_A = \overset{\circ}{\underset{a}{\mathsf{F}}}{}^c{}_A(R,\Theta)$:

$$
Rf(R)\left[(\overset{1}{b}_1+2\overset{1}{b}_2)\overset{\circ}{\underset{1}{\mathsf{F}}}{}^{\theta}{}_R+(\overset{2}{b}_1+2\overset{2}{b}_2)\overset{\circ}{\underset{2}{\mathsf{F}}}{}^{\theta}{}_R\right]-\left[\overset{1}{b}_1\overset{\circ}{\underset{1}{\mathsf{F}}}{}^{r}{}_\Theta+\overset{2}{b}_1\overset{\circ}{\underset{2}{\mathsf{F}}}{}^{r}{}_\Theta\right]f'(R)=0\,. \tag{5.132}
$$

$$
Rf(R)\left[\overset{1}{b}_1\overset{\circ}{\underset{1}{\mathsf{F}}}{}^{\theta}{}_R+\overset{2}{b}_1\overset{\circ}{\underset{2}{\mathsf{F}}}{}^{\theta}{}_R\right]-\left[(\overset{1}{b}_1+2\overset{1}{b}_2)\overset{\circ}{\underset{1}{\mathsf{F}}}{}^{r}{}_\Theta+(\overset{2}{b}_1+2\overset{2}{b}_2)\overset{\circ}{\underset{2}{\mathsf{F}}}{}^{r}{}_\Theta\right]f'(R)=0\,. \tag{5.133}
$$

$$
Rf'(R)\left[\overset{1}{b}_2\overset{\circ}{\underset{1}{\mathsf{F}}}{}^{\theta}{}_\Theta+\overset{2}{b}_2\overset{\circ}{\underset{2}{\mathsf{F}}}{}^{\theta}{}_\Theta\right]-f(R)\left[\overset{1}{b}_2\overset{\circ}{\underset{1}{\mathsf{F}}}{}^{r}{}_R+\overset{2}{b}_2\overset{\circ}{\underset{2}{\mathsf{F}}}{}^{r}{}_R\right]=\mu[f(R)-Rf'(R)]\,. \tag{5.134}
$$

$(5.100)_2$ gives the following equations

$$
Rf(R)\left[(\overset{11}{c}_1+\overset{11}{c}_2+\overset{11}{c}_3)\overset{\circ}{\underset{1}{\mathsf{F}}}{}^{\theta}{}_R+(\overset{12}{c}_1+\overset{12}{c}_2+\overset{12}{c}_3)\overset{\circ}{\underset{2}{\mathsf{F}}}{}^{\theta}{}_R\right]-\left[\overset{11}{c}_1\overset{\circ}{\underset{1}{\mathsf{F}}}{}^{r}{}_\Theta+\overset{12}{c}_1\overset{\circ}{\underset{2}{\mathsf{F}}}{}^{r}{}_\Theta\right]f'(R)=0\,. \tag{5.135}
$$

$$
Rf(R)\left[(\overset{12}{c}_1+\overset{12}{c}_2+\overset{12}{c}_3)\overset{\circ}{\underset{1}{\mathsf{F}}}{}^{\theta}{}_R+(\overset{22}{c}_1+\overset{22}{c}_2+\overset{22}{c}_3)\overset{\circ}{\underset{2}{\mathsf{F}}}{}^{\theta}{}_R\right]-\left[\overset{12}{c}_1\overset{\circ}{\underset{1}{\mathsf{F}}}{}^{r}{}_\Theta+\overset{22}{c}_1\overset{\circ}{\underset{2}{\mathsf{F}}}{}^{r}{}_\Theta\right]f'(R)=0\,. \tag{5.136}
$$

$$
Rf(R)\left[\overset{11}{c}_1\overset{\circ}{\underset{1}{\mathsf{F}}}{}^{\theta}{}_R+\overset{12}{c}_1\overset{\circ}{\underset{2}{\mathsf{F}}}{}^{\theta}{}_R\right]-f'(R)\left[(\overset{11}{c}_1+\overset{11}{c}_2+\overset{11}{c}_3)\overset{\circ}{\underset{1}{\mathsf{F}}}{}^{r}{}_\Theta+(\overset{12}{c}_1+\overset{12}{c}_2+\overset{12}{c}_3)\overset{\circ}{\underset{2}{\mathsf{F}}}{}^{r}{}_\Theta\right]=0\,. \tag{5.137}
$$

$$
Rf(R)\left[\overset{12}{c}_1\overset{\circ}{\underset{1}{\mathsf{F}}}{}^{\theta}{}_R+\overset{22}{c}_1\overset{\circ}{\underset{2}{\mathsf{F}}}{}^{\theta}{}_R\right]-f'(R)\left[(\overset{12}{c}_1+\overset{12}{c}_2+\overset{12}{c}_3)\overset{\circ}{\underset{1}{\mathsf{F}}}{}^{r}{}_\Theta+(\overset{22}{c}_1+\overset{22}{c}_2+\overset{22}{c}_3)\overset{\circ}{\underset{2}{\mathsf{F}}}{}^{r}{}_\Theta\right]=0\,. \tag{5.138}
$$

$$
Rf'(R)\left[\overset{11}{c}_3\overset{\circ}{\underset{1}{\mathsf{F}}}{}^{\theta}{}_\Theta+\overset{12}{c}_3\overset{\circ}{\underset{2}{\mathsf{F}}}{}^{\theta}{}_\Theta\right]-f(R)\left[\overset{11}{c}_2\overset{\circ}{\underset{1}{\mathsf{F}}}{}^{r}{}_R+\overset{12}{c}_2\overset{\circ}{\underset{2}{\mathsf{F}}}{}^{r}{}_R\right]=\overset{1}{b}_2[f(R)-Rf'(R)]\,. \tag{5.139}
$$

$$
Rf'(R)\left[\overset{11}{c}_2\overset{\circ}{\underset{1}{\mathsf{F}}}{}^{\theta}{}_\Theta+\overset{12}{c}_2\overset{\circ}{\underset{2}{\mathsf{F}}}{}^{\theta}{}_\Theta\right]-f(R)\left[\overset{11}{c}_3\overset{\circ}{\underset{1}{\mathsf{F}}}{}^{r}{}_R+\overset{12}{c}_3\overset{\circ}{\underset{2}{\mathsf{F}}}{}^{r}{}_R\right]=\overset{1}{b}_2[f(R)-Rf'(R)]\,. \tag{5.140}
$$

$$
Rf'(R)\left[\overset{12}{c}_3\overset{\circ}{\underset{1}{\mathsf{F}}}{}^{\theta}{}_\Theta+\overset{22}{c}_3\overset{\circ}{\underset{2}{\mathsf{F}}}{}^{\theta}{}_\Theta\right]-f(R)\left[\overset{12}{c}_2\overset{\circ}{\underset{1}{\mathsf{F}}}{}^{r}{}_R+\overset{22}{c}_2\overset{\circ}{\underset{2}{\mathsf{F}}}{}^{r}{}_R\right]=\overset{2}{b}_2[f(R)-Rf'(R)]\,. \tag{5.141}
$$

$$
Rf'(R)\left[\overset{12}{c}_2\overset{\circ}{\underset{1}{\mathsf{F}}}{}^{\theta}{}_\Theta+\overset{22}{c}_2\overset{\circ}{\underset{2}{\mathsf{F}}}{}^{\theta}{}_\Theta\right]-f(R)\left[\overset{12}{c}_3\overset{\circ}{\underset{1}{\mathsf{F}}}{}^{r}{}_R+\overset{22}{c}_3\overset{\circ}{\underset{2}{\mathsf{F}}}{}^{r}{}_R\right]=\overset{2}{b}_2[f(R)-Rf'(R)]\,. \tag{5.142}
$$

Note that the algebraic equations governing the diagonal and off-diagonal director gradients are uncoupled. We have six equations (5.132)-(5.133), and (5.135)-(5.138) for the four off-diagonal terms and five equations (5.134), and (5.139)-(5.142) for the diagonal terms. We first show that the determinant of the coefficient matrix of the linear system (5.135)-(5.138) is non-zero, and hence, $\overset{\circ}{\underset{1}{\mathsf{F}}}{}^{r}{}_\Theta=\overset{\circ}{\underset{2}{\mathsf{F}}}{}^{r}{}_\Theta=0$, and $\overset{\circ}{\underset{1}{\mathsf{F}}}{}^{\theta}{}_R=\overset{\circ}{\underset{2}{\mathsf{F}}}{}^{\theta}{}_R=0$. Note that the determinant of the coefficient matrix reads

$$
R^2f^2(R){f'}^2(R)\left[-(\overset{12}{c}_2+\overset{12}{c}_3)^2+(\overset{11}{c}_2+\overset{11}{c}_3)(\overset{22}{c}_2+\overset{22}{c}_3)\right]\left[-(2\overset{12}{c}_1+\overset{12}{c}_2+\overset{12}{c}_3)^2+(2\overset{11}{c}_1+\overset{11}{c}_2+\overset{11}{c}_3)(2\overset{22}{c}_1+\overset{22}{c}_2+\overset{22}{c}_3)\right]. \tag{5.143}
$$

It turns out that this determinant cannot vanish. This is a consequence of the positive-definiteness of $\mathbb{C}$. To see this let us assume that $U^a{}_{|A}=0$, and $\underset{a}{\mathfrak{U}}{}^1{}_{|2}=\underset{a}{\mathfrak{U}}{}^2{}_{|1}=0$. For this class of deformations $\delta W>0$ implies positive-definiteness of the following matrix:

$$
\begin{bmatrix}
\overset{11}{c}_1+\overset{11}{c}_2+\overset{11}{c}_3 & \overset{11}{c}_1 & \overset{12}{c}_1+\overset{12}{c}_2+\overset{12}{c}_3 & \overset{12}{c}_1 \\
\overset{11}{c}_1 & \overset{11}{c}_1+\overset{11}{c}_2+\overset{11}{c}_3 & \overset{12}{c}_1 & \overset{12}{c}_1+\overset{12}{c}_2+\overset{12}{c}_3 \\
\overset{12}{c}_1+\overset{12}{c}_2+\overset{12}{c}_3 & \overset{12}{c}_1 & \overset{22}{c}_1+\overset{22}{c}_2+\overset{22}{c}_3 & \overset{22}{c}_1 \\
\overset{12}{c}_1 & \overset{12}{c}_1+\overset{12}{c}_2+\overset{12}{c}_3 & \overset{22}{c}_1 & \overset{22}{c}_1+\overset{22}{c}_2+\overset{22}{c}_3
\end{bmatrix}. \tag{5.144}
$$

In particular, its determinant must be positive, i.e.,

$$
\left[-(\overset{12}{c}_2+\overset{12}{c}_3)^2+(\overset{11}{c}_2+\overset{11}{c}_3)(\overset{22}{c}_2+\overset{22}{c}_3)\right]\left[-(2\overset{12}{c}_1+\overset{12}{c}_2+\overset{12}{c}_3)^2+(2\overset{11}{c}_1+\overset{11}{c}_2+\overset{11}{c}_3)(2\overset{22}{c}_1+\overset{22}{c}_2+\overset{22}{c}_3)\right]>0\,. \tag{5.145}
$$

Therefore

$$\underset{1}{\mathring{\mathsf{F}}}{}^{r}{}_{\Theta} = \underset{2}{\mathring{\mathsf{F}}}{}^{r}{}_{\Theta} = 0\,, \qquad \underset{1}{\mathring{\mathsf{F}}}{}^{\theta}{}_{R} = \underset{2}{\mathring{\mathsf{F}}}{}^{\theta}{}_{R} = 0\,. \tag{5.146}$$

Note that (5.132) and (5.133) are now trivially satisfied.

The determinant of the coefficient matrix of the linear system (5.139)-(5.142) reads

$$-R^2 f^2(R) f'^2(R) \left[(\overset{12}{c}_2 - \overset{12}{c}_3)^2 - (\overset{11}{c}_2 - \overset{11}{c}_3)(\overset{22}{c}_2 - \overset{22}{c}_3)\right] \left[(\overset{12}{c}_2 + \overset{12}{c}_3)^2 - (\overset{11}{c}_2 + \overset{11}{c}_3)(\overset{22}{c}_2 + \overset{22}{c}_3)\right]\,. \tag{5.147}$$

Positive-definiteness of energy requires that this determinant be non-vanishing. To see this let us assume that $U^a{}_{|A} = 0$, and $\underset{a}{\mathfrak{U}}{}^1{}_{|1} = \underset{a}{\mathfrak{U}}{}^2{}_{|2} = 0$. For this class of deformations $\delta W > 0$ implies positive-definiteness of the following matrix.

$$\begin{bmatrix} \overset{11}{c}_2 & \overset{11}{c}_3 & \overset{12}{c}_2 & \overset{12}{c}_3 \\ \overset{11}{c}_3 & \overset{11}{c}_2 & \overset{12}{c}_3 & \overset{12}{c}_2 \\ \overset{12}{c}_2 & \overset{12}{c}_3 & \overset{22}{c}_2 & \overset{22}{c}_3 \\ \overset{12}{c}_3 & \overset{12}{c}_2 & \overset{22}{c}_3 & \overset{22}{c}_2 \end{bmatrix}\,. \tag{5.148}$$

In particular, its determinant must be positive, and hence

$$\left[(\overset{12}{c}_2 - \overset{12}{c}_3)^2 - (\overset{11}{c}_2 - \overset{11}{c}_3)(\overset{22}{c}_2 - \overset{22}{c}_3)\right] \left[(\overset{12}{c}_2 + \overset{12}{c}_3)^2 - (\overset{11}{c}_2 + \overset{11}{c}_3)(\overset{22}{c}_2 + \overset{22}{c}_3)\right] > 0\,. \tag{5.149}$$

Therefore

$$\underset{1}{\mathring{\mathsf{F}}}{}^{r}{}_{R} = -\frac{\left[\overset{2}{b}_2(\overset{12}{c}_2 + \overset{12}{c}_3) - \overset{1}{b}_2(\overset{22}{c}_2 + \overset{22}{c}_3)\right](f(R) - Rf'(R))}{f(R)\left[(\overset{12}{c}_2 + \overset{12}{c}_3)^2 - (\overset{11}{c}_2 + \overset{11}{c}_3)(\overset{22}{c}_2 + \overset{22}{c}_3)\right]}\,, \tag{5.150}$$

$$\underset{1}{\mathring{\mathsf{F}}}{}^{\theta}{}_{\Theta} = \frac{\left[\overset{2}{b}_2(\overset{12}{c}_2 + \overset{12}{c}_3) - \overset{1}{b}_2(\overset{22}{c}_2 + \overset{22}{c}_3)\right](f(R) - Rf'(R))}{Rf'(R)\left[(\overset{12}{c}_2 + \overset{12}{c}_3)^2 - (\overset{11}{c}_2 + \overset{11}{c}_3)(\overset{22}{c}_2 + \overset{22}{c}_3)\right]}\,, \tag{5.151}$$

$$\underset{2}{\mathring{\mathsf{F}}}{}^{r}{}_{R} = \frac{\left[\overset{2}{b}_2(\overset{11}{c}_2 + \overset{11}{c}_3) - \overset{1}{b}_2(\overset{12}{c}_2 + \overset{12}{c}_3)\right](f(R) - Rf'(R))}{f(R)\left[(\overset{12}{c}_2 + \overset{12}{c}_3)^2 - (\overset{11}{c}_2 + \overset{11}{c}_3)(\overset{22}{c}_2 + \overset{22}{c}_3)\right]}\,, \tag{5.152}$$

$$\underset{2}{\mathring{\mathsf{F}}}{}^{\theta}{}_{\Theta} = \frac{\left[\overset{2}{b}_2(\overset{11}{c}_2 + \overset{11}{c}_3) - \overset{1}{b}_2(\overset{12}{c}_2 + \overset{12}{c}_3)\right](Rf'(R) - f(R))}{Rf'(R)\left[(\overset{12}{c}_2 + \overset{12}{c}_3)^2 - (\overset{11}{c}_2 + \overset{11}{c}_3)(\overset{22}{c}_2 + \overset{22}{c}_3)\right]}\,. \tag{5.153}$$

Note that

$$\underset{1}{\mathring{\mathsf{F}}}{}^{r}{}_{R} = -\frac{Rf'(R)}{f(R)}\underset{1}{\mathring{\mathsf{F}}}{}^{\theta}{}_{\Theta}, \quad \underset{2}{\mathring{\mathsf{F}}}{}^{r}{}_{R} = -\frac{Rf'(R)}{f(R)}\underset{2}{\mathring{\mathsf{F}}}{}^{\theta}{}_{\Theta}\,. \tag{5.154}$$

From (5.146), it is immediate to see that (5.132) and (5.133) are automatically satisfied, while (5.134) imposes the following constraint on the elastic constants of the virtual body:

$$2(\overset{1}{b}_2)^2(\overset{22}{c}_2 + \overset{22}{c}_3) - 4\overset{2}{b}_2\overset{1}{b}_2(\overset{12}{c}_2 + \overset{12}{c}_3) + 2(\overset{2}{b}_2)^2(\overset{11}{c}_2 + \overset{11}{c}_3) + \mu\left[(\overset{12}{c}_2 + \overset{12}{c}_3)^2 - (\overset{11}{c}_2 + \overset{11}{c}_3)(\overset{22}{c}_2 + \overset{22}{c}_3)\right] = 0\,. \tag{5.155}$$

Let us assume that $U^a{}_{|A} = \delta^a_1\delta^2_A$, and $\underset{a}{\mathfrak{U}}{}^1{}_{|1} = \underset{a}{\mathfrak{U}}{}^2{}_{|2} = 0$. For this class of deformations $\delta W > 0$ implies positive-definiteness of the following matrix.

$$\begin{bmatrix} \mu & \overset{1}{b}_2 & \overset{1}{b}_2 & \overset{2}{b}_2 & \overset{2}{b}_2 \\ \overset{1}{b}_2 & \overset{11}{c}_2 & \overset{11}{c}_3 & \overset{12}{c}_2 & \overset{12}{c}_3 \\ \overset{1}{b}_2 & \overset{11}{c}_3 & \overset{11}{c}_2 & \overset{12}{c}_3 & \overset{12}{c}_2 \\ \overset{2}{b}_2 & \overset{12}{c}_2 & \overset{12}{c}_3 & \overset{22}{c}_2 & \overset{22}{c}_3 \\ \overset{2}{b}_2 & \overset{12}{c}_3 & \overset{12}{c}_2 & \overset{22}{c}_3 & \overset{22}{c}_2 \end{bmatrix}\,. \tag{5.156}$$

Therefore, its determinant must be positive, and hence

$$\begin{aligned} &\left[(\overset{12}{c}_2 - \overset{12}{c}_3)^2 - (\overset{11}{c}_2 - \overset{11}{c}_3)(\overset{22}{c}_2 - \overset{22}{c}_3)\right] \Big\{\mu\left[(\overset{12}{c}_2 + \overset{12}{c}_3)^2 - (\overset{11}{c}_2 + \overset{11}{c}_3)(\overset{22}{c}_2 + \overset{22}{c}_3)\right] \\ &\quad - 2\left[2\overset{1}{b}_2\overset{2}{b}_2(\overset{12}{c}_2 + \overset{12}{c}_3) - (\overset{2}{b}_2)^2(\overset{11}{c}_2 + \overset{11}{c}_3) - (\overset{1}{b}_2)^2(\overset{22}{c}_2 + \overset{22}{c}_3)\right]\Big\} > 0\,. \end{aligned} \tag{5.157}$$

In particular

$$\mu\left[(\overset{12}{c}_2+\overset{12}{c}_3)^2-(\overset{11}{c}_2+\overset{11}{c}_3)(\overset{22}{c}_2+\overset{22}{c}_3)\right]\;-2\left[2\overset{1}{b}_2\overset{2}{b}_2(\overset{12}{c}_2+\overset{12}{c}_3)-(\overset{2}{b}_2)^2\,(\overset{11}{c}_2+\overset{11}{c}_3)-(\overset{1}{b}_2)^2\,(\overset{22}{c}_2+\overset{22}{c}_3)\right]\neq 0. \tag{5.158}$$

Therefore, (5.155) violates positive-definiteness of the energy, and hence cloaking is not possible.

**Remark 5.6.** Even if one accepts a positive-semidefinite energy, the director gradient field given by (5.150) to (5.153) is not compatible. In other words, the director gradient given by (5.150) to (5.153) does not correspond to a single-valued director field in the physical body, in general. Necessary compatibility equations for the director gradients are written as

$$\begin{aligned}
&\underset{1}{\mathring{\mathsf{F}}}^r{}_{\Theta|R}=\underset{1}{\mathring{\mathsf{F}}}^r{}_{R|\Theta}=0,\quad \underset{1}{\mathring{\mathsf{F}}}^\theta{}_{\Theta|R}=\underset{1}{\mathring{\mathsf{F}}}^\theta{}_{R|\Theta}=0\\
&\underset{2}{\mathring{\mathsf{F}}}^r{}_{\Theta|R}=\underset{2}{\mathring{\mathsf{F}}}^r{}_{R|\Theta}=0,\quad \underset{2}{\mathring{\mathsf{F}}}^\theta{}_{\Theta|R}=\underset{2}{\mathring{\mathsf{F}}}^\theta{}_{R|\Theta}=0\,.
\end{aligned} \tag{5.159}$$

Note that $\underset{\mathfrak{a}}{\mathring{\mathsf{F}}}^a{}_{A|B}=\partial\underset{\mathfrak{a}}{\mathring{\mathsf{F}}}^a{}_A/\partial X^B-\Gamma^C{}_{AB}\underset{\mathfrak{a}}{\mathring{\mathsf{F}}}^a{}_C+\gamma^a{}_{bc}\mathring{F}^b{}_B\underset{\mathfrak{a}}{\mathring{\mathsf{F}}}^c{}_A$. Thus

$$\underset{\mathfrak{a}}{\mathring{\mathsf{F}}}^r{}_{R|\Theta}=\underset{\mathfrak{a}}{\mathring{\mathsf{F}}}^r{}_{R,\Theta}-R\,\underset{\mathfrak{a}}{\mathring{\mathsf{F}}}^\theta{}_R-\frac{1}{R}\underset{\mathfrak{a}}{\mathring{\mathsf{F}}}^r{}_\Theta\,, \tag{5.160}$$

$$\underset{\mathfrak{a}}{\mathring{\mathsf{F}}}^r{}_{\Theta|R}=\underset{\mathfrak{a}}{\mathring{\mathsf{F}}}^r{}_{\Theta,R}-\frac{1}{R}\underset{\mathfrak{a}}{\mathring{\mathsf{F}}}^r{}_\Theta\,, \tag{5.161}$$

$$\underset{\mathfrak{a}}{\mathring{\mathsf{F}}}^\theta{}_{\Theta|R}=\underset{\mathfrak{a}}{\mathring{\mathsf{F}}}^\theta{}_{\Theta,R}+\frac{1}{R}\underset{\mathfrak{a}}{\mathring{\mathsf{F}}}^\theta{}_\Theta-\frac{1}{R}\underset{\mathfrak{a}}{\mathring{\mathsf{F}}}^\theta{}_\Theta=\underset{\mathfrak{a}}{\mathring{\mathsf{F}}}^\theta{}_{\Theta,R}\,, \tag{5.162}$$

$$\underset{\mathfrak{a}}{\mathring{\mathsf{F}}}^\theta{}_{R|\Theta}=\underset{\mathfrak{a}}{\mathring{\mathsf{F}}}^\theta{}_{R,\Theta}+\frac{1}{R}\left[\underset{\mathfrak{a}}{\mathring{\mathsf{F}}}^r{}_R-\underset{\mathfrak{a}}{\mathring{\mathsf{F}}}^\theta{}_\Theta\right]\,. \tag{5.163}$$

Hence, $\underset{\mathfrak{a}}{\mathring{\mathsf{F}}}^r{}_{R|\Theta}=\underset{\mathfrak{a}}{\mathring{\mathsf{F}}}^r{}_{\Theta|R}$, and $\underset{\mathfrak{a}}{\mathring{\mathsf{F}}}^\theta{}_{\Theta|R}=\underset{\mathfrak{a}}{\mathring{\mathsf{F}}}^\theta{}_{R|\Theta}$ imply that

$$\begin{aligned}
&\underset{\mathfrak{a}}{\mathring{\mathsf{F}}}^r{}_{R,\Theta}-\underset{\mathfrak{a}}{\mathring{\mathsf{F}}}^r{}_{\Theta,R}=R\underset{\mathfrak{a}}{\mathring{\mathsf{F}}}^\theta{}_R\,,\\
&\underset{\mathfrak{a}}{\mathring{\mathsf{F}}}^\theta{}_{\Theta,R}-\underset{\mathfrak{a}}{\mathring{\mathsf{F}}}^\theta{}_{R,\Theta}=\frac{1}{R}\left[\underset{\mathfrak{a}}{\mathring{\mathsf{F}}}^r{}_R-\underset{\mathfrak{a}}{\mathring{\mathsf{F}}}^\theta{}_\Theta\right]\,.
\end{aligned} \tag{5.164}$$

Now using (5.146), one obtains

$$\begin{aligned}
&\underset{\mathfrak{a}}{\mathring{\mathsf{F}}}^r{}_{R,\Theta}=0\,,\\
&(R\underset{\mathfrak{a}}{\mathring{\mathsf{F}}}^\theta{}_\Theta)_{,R}=\underset{\mathfrak{a}}{\mathring{\mathsf{F}}}^r{}_R\,.
\end{aligned} \tag{5.165}$$

Note that in this example $(5.165)_1$ is trivially satisfied. From (5.154) and $(5.165)_2$ one obtains

$$\underset{\mathfrak{a}}{\mathring{\mathsf{F}}}^\theta{}_\Theta=\frac{C_{\mathfrak{a}}}{Rf(R)}\,, \tag{5.166}$$

for constants $C_{\mathfrak{a}}$. If $C_{\mathfrak{a}}\neq 0$, (5.166) gives the following ODEs for $f(R)$.

$$Rf(R)f'(R)+k_{\mathfrak{a}}f'(R)-f^2(R)=0\,, \tag{5.167}$$

where $k_{\mathfrak{a}}$ are constant. Knowing that $f(R_o)=R_o$, and $f'(R_o)=1$, one concludes that $k_{\mathfrak{a}}=0$, and thus, $f(R)=R$. If $C_{\mathfrak{a}}=0$, then either $f(R)=R$, or

$$\overset{2}{b}_2(\overset{12}{c}_2+\overset{12}{c}_3)-\overset{1}{b}_2(\overset{22}{c}_2+\overset{22}{c}_3)=0,\quad \overset{2}{b}_2(\overset{11}{c}_2+\overset{11}{c}_3)-\overset{1}{b}_2(\overset{12}{c}_2+\overset{12}{c}_3)=0. \tag{5.168}$$

Knowing that $\overset{11}{c}_2+\overset{11}{c}_3>0$, and $\overset{22}{c}_2+\overset{22}{c}_3>0$, the above relations are written as

$$\overset{1}{b}_2=\frac{\overset{12}{c}_2+\overset{12}{c}_3}{\overset{22}{c}_2+\overset{22}{c}_3}\overset{2}{b}_2,\quad \overset{2}{b}_2=\frac{\overset{12}{c}_2+\overset{12}{c}_3}{\overset{11}{c}_2+\overset{11}{c}_3}\overset{1}{b}_2. \tag{5.169}$$

Thus

$$\overset{1}{b}_2\frac{(\overset{11}{c}_2+\overset{11}{c}_3)(\overset{22}{c}_2+\overset{22}{c}_3)-(\overset{12}{c}_2+\overset{12}{c}_3)^2}{(\overset{11}{c}_2+\overset{11}{c}_3)(\overset{22}{c}_2+\overset{22}{c}_3)}=0,\quad \overset{2}{b}_2\frac{(\overset{11}{c}_2+\overset{11}{c}_3)(\overset{22}{c}_2+\overset{22}{c}_3)-(\overset{12}{c}_2+\overset{12}{c}_3)^2}{(\overset{11}{c}_2+\overset{11}{c}_3)(\overset{22}{c}_2+\overset{22}{c}_3)}=0. \tag{5.170}$$

From (5.145), one concludes that $\overset{1}{b}_2=\overset{2}{b}_2=0$. Substituting this into (5.134), one concludes that $f(R)=R$. Therefore, transformation cloaking is not possible in this example for any linear generalized Cosserat solid.

We next prove that the impossibility of transformation cloaking in dimension two for linear generalized Cosserat elasticity is independent of the shape of the hole (cavity). We write (5.100) in Cartesian coordinates for which the shifter is written as $\mathsf{s}^{\tilde{a}}{}_{a} = \delta^{\tilde{a}}{}_{a}$. Knowing that $\mathring{F}^{c}{}_{A} = \delta^{c}{}_{A}$, (5.100) is simplified to read $\forall\, b, B, c, a \in \{1,2,3\}$

$$
\begin{aligned}
&(\bar{\bar{\bar{F}}}^{-1})^{[c}{}_{\tilde{A}}(\bar{\bar{\bar{F}}}^{-1})^{B}{}_{\tilde{B}}\tilde{\mathbb{A}}^{a]\tilde{A}b\tilde{B}} + (\bar{\bar{\bar{F}}}^{-1})^{A}{}_{\tilde{A}}(\bar{\bar{\bar{F}}}^{-1})^{B}{}_{\tilde{B}}\underset{\mathfrak{a}}{\mathring{\mathsf{F}}}{}^{[c}{}_{A}\,\overset{\mathfrak{a}}{\tilde{\mathbb{B}}}{}^{a]\tilde{A}b\tilde{B}} = 0\,,\\
&(\bar{\bar{\bar{F}}}^{-1})^{[c}{}_{\tilde{A}}(\bar{\bar{\bar{F}}}^{-1})^{B}{}_{\tilde{B}}\overset{\mathfrak{a}}{\tilde{\mathbb{B}}}{}^{a]\tilde{A}b\tilde{B}} + (\bar{\bar{\bar{F}}}^{-1})^{A}{}_{\tilde{A}}(\bar{\bar{\bar{F}}}^{-1})^{B}{}_{\tilde{B}}\underset{\mathfrak{b}}{\mathring{\mathsf{F}}}{}^{[c}{}_{A}\overset{\mathfrak{b}\mathfrak{a}}{\tilde{\mathbb{C}}}{}^{a]\tilde{A}b\tilde{B}} = 0\,.
\end{aligned}
\tag{5.171}
$$

First let us consider an arbitrary cloaking map and director gradient with the following components in 2D

$$
\bar{\bar{\bar{\mathbf{F}}}}^{-1} = \begin{bmatrix} F_{11} & F_{12} \\ F_{21} & F_{22} \end{bmatrix}, \qquad \underset{\mathfrak{a}}{\mathring{\mathsf{F}}} = \begin{bmatrix} \underset{\mathfrak{a}}{\mathsf{F}}{}^{11} & \underset{\mathfrak{a}}{\mathsf{F}}{}^{12} \\ \underset{\mathfrak{a}}{\mathsf{F}}{}^{21} & \underset{\mathfrak{a}}{\mathsf{F}}{}^{22} \end{bmatrix}.
\tag{5.172}
$$

Note that $\bar{\bar{\bar{\mathbf{F}}}}^{-1}$ is compatible, and hence, $F_{11,2} = F_{12,1}$, and $F_{22,1} = F_{21,2}$. Similarly, the compatibility equations for the director gradient are $\underset{\mathfrak{a}}{\mathsf{F}}{}^{11}{}_{,2} = \underset{\mathfrak{a}}{\mathsf{F}}{}^{12}{}_{,1}$, and $\underset{\mathfrak{a}}{\mathsf{F}}{}^{22}{}_{,1} = \underset{\mathfrak{a}}{\mathsf{F}}{}^{21}{}_{,2}$. Expanding $(5.171)_1$, for $i \neq j \in \{1,2\}$ one obtains

$$
\begin{aligned}
&-F_{ii}F_{ij}\underset{1}{\mathsf{F}}{}^{ii}(\overset{1}{b}_1 + \overset{1}{b}_2) + \underset{1}{\mathsf{F}}{}^{jj}\left[F_{ii}F_{ji}(\overset{1}{b}_1 + 2\overset{1}{b}_2) + F_{ij}F_{jj}\overset{1}{b}_2\right] - F_{ii}F_{ij}\underset{2}{\mathsf{F}}{}^{ii}(\overset{2}{b}_1 + \overset{2}{b}_2)\\
&\quad + \underset{2}{\mathsf{F}}{}^{jj}\left[F_{ii}F_{ji}(\overset{2}{b}_1 + 2\overset{2}{b}_2) + F_{ij}F_{jj}\overset{2}{b}_2\right] - \underset{1}{\mathsf{F}}{}^{ij}(F_{ii}F_{jj}\overset{1}{b}_1 + F_{ij}F_{ji}\overset{1}{b}_2) + \underset{1}{\mathsf{F}}{}^{ji}\left[F_{ii}^2(\overset{1}{b}_1 + 2\overset{1}{b}_2) + F_{ij}^2\overset{1}{b}_2\right]\\
&\quad - \underset{2}{\mathsf{F}}{}^{ij}(F_{ii}F_{jj}\overset{2}{b}_1 + F_{ij}F_{ji}\overset{2}{b}_2) + \underset{2}{\mathsf{F}}{}^{ji}\left[F_{ii}^2(\overset{2}{b}_1 + 2\overset{2}{b}_2) + F_{ij}^2\overset{2}{b}_2\right]\\
&\quad = F_{ii}(F_{ij} - F_{ji})\lambda - (F_{ij}F_{jj} - F_{ii}F_{ij} + 2F_{ii}F_{ji})\mu\,,
\end{aligned}
\tag{5.173}
$$

$$
\begin{aligned}
&-\underset{1}{\mathsf{F}}{}^{ii}(F_{ii}F_{jj}\overset{1}{b}_2 + F_{ij}F_{ji}\overset{1}{b}_1) + \underset{1}{\mathsf{F}}{}^{jj}\left[F_{ji}^2(\overset{1}{b}_1 + 2\overset{1}{b}_2) + F_{jj}^2\overset{1}{b}_2\right] - \underset{2}{\mathsf{F}}{}^{ii}(F_{ii}F_{jj}\overset{2}{b}_2 + F_{ij}F_{ji}\overset{2}{b}_1)\\
&\quad + \underset{2}{\mathsf{F}}{}^{jj}\left[F_{ji}^2(\overset{2}{b}_1 + 2\overset{2}{b}_2) + F_{jj}^2\overset{2}{b}_2\right] - F_{ji}F_{jj}\underset{1}{\mathsf{F}}{}^{ij}(\overset{1}{b}_1 + \overset{1}{b}_2) + \underset{1}{\mathsf{F}}{}^{ji}\left[F_{ii}F_{ji}(\overset{1}{b}_1 + 2\overset{1}{b}_2) + F_{ij}F_{jj}\overset{1}{b}_2\right]\\
&\quad - F_{ji}F_{jj}\underset{2}{\mathsf{F}}{}^{ij}(\overset{2}{b}_1 + \overset{2}{b}_2) + \underset{2}{\mathsf{F}}{}^{ji}\left[F_{ii}F_{ji}(\overset{2}{b}_1 + 2\overset{2}{b}_2) + F_{ij}F_{jj}\overset{2}{b}_2\right]\\
&\quad = F_{ji}(F_{ij} - F_{ji})\lambda - (2F_{ji}^2 + F_{jj}^2 - F_{ii}F_{jj})\mu\,,
\end{aligned}
\tag{5.174}
$$

On the other hand, expanding $(5.171)_2$ one obtains

$$
\begin{aligned}
&-F_{11}F_{12}\underset{1}{\mathsf{F}}{}^{11}(\overset{11}{c}_1 + \overset{11}{c}_3) + \underset{1}{\mathsf{F}}{}^{22}\left[F_{11}F_{21}\overset{11}{c}_\Sigma + F_{12}F_{22}\overset{11}{c}_2\right] - F_{11}F_{12}\underset{2}{\mathsf{F}}{}^{11}(\overset{12}{c}_1 + \overset{12}{c}_3) + \underset{2}{\mathsf{F}}{}^{22}\left[F_{11}F_{21}\overset{12}{c}_\Sigma + F_{12}F_{22}\overset{12}{c}_2\right]\\
&\quad - \underset{1}{\mathsf{F}}{}^{12}(F_{11}F_{22}\overset{11}{c}_1 + F_{12}F_{21}\overset{11}{c}_3) + \underset{1}{\mathsf{F}}{}^{21}\left[F_{11}^2\overset{11}{c}_\Sigma + F_{12}^2\overset{11}{c}_2\right] - \underset{2}{\mathsf{F}}{}^{12}(F_{11}F_{22}\overset{12}{c}_1 + F_{12}F_{21}\overset{12}{c}_3)\\
&\quad + \underset{2}{\mathsf{F}}{}^{21}\left[F_{11}^2\overset{12}{c}_\Sigma + F_{12}^2\overset{12}{c}_2\right] = F_{11}(F_{12} - F_{21})\overset{1}{b}_1 + (F_{11}F_{12} - 2F_{11}F_{21} - F_{12}F_{22})\overset{1}{b}_2\,,
\end{aligned}
\tag{5.175}
$$

$$
\begin{aligned}
&-\underset{1}{\mathsf{F}}{}^{11}\left[F_{11}^2\overset{11}{c}_2 + F_{12}^2\overset{11}{c}_\Sigma\right] + \underset{1}{\mathsf{F}}{}^{22}(F_{11}F_{22}\overset{11}{c}_3 + F_{12}F_{21}\overset{11}{c}_1) - \underset{2}{\mathsf{F}}{}^{11}\left[F_{11}^2\overset{12}{c}_2 + F_{12}^2\overset{12}{c}_\Sigma\right] + \underset{2}{\mathsf{F}}{}^{22}(F_{11}F_{22}\overset{12}{c}_3 + F_{12}F_{21}\overset{12}{c}_1)\\
&\quad - \underset{1}{\mathsf{F}}{}^{12}[F_{11}F_{21}\overset{11}{c}_2 + F_{12}F_{22}\overset{11}{c}_\Sigma] + F_{11}F_{12}\underset{1}{\mathsf{F}}{}^{21}(\overset{11}{c}_1 + \overset{11}{c}_3) - \underset{2}{\mathsf{F}}{}^{12}[F_{11}F_{21}\overset{12}{c}_2 + F_{12}F_{22}\overset{12}{c}_\Sigma]\\
&\quad + F_{11}F_{12}\underset{2}{\mathsf{F}}{}^{21}(\overset{12}{c}_1 + \overset{12}{c}_3) = F_{12}(F_{12} - F_{21})\overset{1}{b}_1 + (F_{11}^2 + 2F_{12}^2 - F_{11}F_{22})\overset{1}{b}_2\,,
\end{aligned}
\tag{5.176}
$$

$$
\begin{aligned}
&-\underset{1}{\mathsf{F}}{}^{11}[F_{11}F_{22}\overset{11}{c}_3 + F_{12}F_{21}\overset{11}{c}_1] + \underset{1}{\mathsf{F}}{}^{22}\left[F_{21}^2\overset{11}{c}_\Sigma + F_{22}^2\overset{11}{c}_2\right] - \underset{2}{\mathsf{F}}{}^{11}[F_{11}F_{22}\overset{12}{c}_3 + F_{12}F_{21}\overset{12}{c}_1] + \underset{2}{\mathsf{F}}{}^{22}\left(F_{21}^2\overset{12}{c}_\Sigma + F_{22}^2\overset{12}{c}_2\right)\\
&\quad - F_{21}F_{22}\underset{1}{\mathsf{F}}{}^{12}(\overset{11}{c}_1 + \overset{11}{c}_3) + \underset{1}{\mathsf{F}}{}^{21}[F_{11}F_{21}\overset{11}{c}_\Sigma + F_{12}F_{22}\overset{11}{c}_2] - F_{21}F_{22}\underset{2}{\mathsf{F}}{}^{12}(\overset{12}{c}_1 + \overset{12}{c}_3)\\
&\quad + \underset{2}{\mathsf{F}}{}^{21}[F_{11}F_{21}\overset{12}{c}_\Sigma + F_{12}F_{22}\overset{12}{c}_2] = F_{21}(F_{12} - F_{21})\overset{1}{b}_1 + (F_{11}F_{22} - 2F_{21}^2 - F_{22}^2)\overset{1}{b}_2\,,
\end{aligned}
\tag{5.177}
$$

$$
\begin{aligned}
&-\underset{1}{\mathsf{F}}{}^{11}[F_{11}F_{21}\overset{11}{c}_2 + F_{12}F_{22}\overset{11}{c}_\Sigma] + F_{21}F_{22}\underset{1}{\mathsf{F}}{}^{22}(\overset{11}{c}_1 + \overset{11}{c}_3) - \underset{2}{\mathsf{F}}{}^{11}[F_{11}F_{21}\overset{12}{c}_2 + F_{12}F_{22}\overset{12}{c}_\Sigma] + F_{21}F_{22}\underset{2}{\mathsf{F}}{}^{22}(\overset{12}{c}_1 + \overset{12}{c}_3)\\
&\quad - \underset{1}{\mathsf{F}}{}^{12}\left(F_{21}^2\overset{11}{c}_2 + F_{22}^2\overset{11}{c}_\Sigma\right) + \underset{1}{\mathsf{F}}{}^{21}(F_{11}F_{22}\overset{11}{c}_1 + F_{12}F_{21}\overset{11}{c}_3) - \underset{2}{\mathsf{F}}{}^{12}\left(F_{21}^2\overset{12}{c}_2 + F_{22}^2\overset{12}{c}_\Sigma\right)\\
&\quad + \underset{2}{\mathsf{F}}{}^{21}(F_{11}F_{22}\overset{12}{c}_1 + F_{12}F_{21}\overset{12}{c}_3) = F_{22}(F_{12} - F_{21})\overset{1}{b}_1 + (F_{11}F_{21} - F_{21}F_{22} + 2F_{12}F_{22})\overset{1}{b}_2\,,
\end{aligned}
\tag{5.178}
$$

where $\overset{\mathfrak{ab}}{c}_\Sigma = \overset{\mathfrak{ab}}{c}_1 + \overset{\mathfrak{ab}}{c}_2 + \overset{\mathfrak{ab}}{c}_3$. Another independent set of four equations are generated from (5.175)-(5.178) using the transformation $\{\overset{12}{c} \to \overset{22}{c},\ \overset{11}{c} \to \overset{12}{c},\ \overset{1}{b} \to \overset{2}{b}\}$. The determinant of the coefficient matrix reads

$$\begin{aligned}(F_{12}F_{21} - F_{11}F_{22})^8 \left[(\overset{12}{c}_2 + \overset{12}{c}_3)^2 - (\overset{11}{c}_2 + \overset{11}{c}_3)(\overset{22}{c}_2 + \overset{22}{c}_3)\right]^2 &\left[(\overset{12}{c}_2 - \overset{12}{c}_3)^2 - (\overset{11}{c}_2 - \overset{11}{c}_3)(\overset{22}{c}_2 - \overset{22}{c}_3)\right] \\ &\left[(\overset{12}{c}_1 + \overset{12}{c}_\Sigma)^2 - (\overset{11}{c}_1 + \overset{11}{c}_\Sigma)(\overset{22}{c}_1 + \overset{22}{c}_\Sigma)\right] .\end{aligned} \tag{5.179}$$

Note that from the positive-definiteness of energy, this determinant is non-zero, and hence the system (5.175)-(5.178) (along with four more equations obtained by the transformation) has a unique solution, which reads

$$\underset{1}{\mathsf{F}}^{11} = \frac{F_{22}(F_{11} - F_{22})\overset{1}{\mathsf{t}}_3 - F_{12}F_{21}\overset{1}{\mathsf{t}}_2 + F_{21}^2\overset{1}{\mathsf{t}}_1}{F_{11}F_{22} - F_{12}F_{21}}, \qquad \underset{1}{\mathsf{F}}^{22} = \frac{F_{11}(F_{22} - F_{11})\overset{1}{\mathsf{t}}_3 - F_{12}F_{21}\overset{1}{\mathsf{t}}_2 + F_{12}^2\overset{1}{\mathsf{t}}_1}{F_{11}F_{22} - F_{12}F_{21}}, \tag{5.180}$$

$$\underset{1}{\mathsf{F}}^{12} = \frac{F_{12}F_{22}\overset{1}{\mathsf{t}}_3 - F_{11}F_{21}\overset{1}{\mathsf{t}}_1 + F_{11}F_{12}\overset{1}{\mathsf{t}}_4}{F_{11}F_{22} - F_{12}F_{21}}, \qquad \underset{1}{\mathsf{F}}^{21} = \frac{F_{21}F_{11}\overset{1}{\mathsf{t}}_3 - F_{12}F_{22}\overset{1}{\mathsf{t}}_1 + F_{21}F_{22}\overset{1}{\mathsf{t}}_4}{F_{11}F_{22} - F_{12}F_{21}}, \tag{5.181}$$

$$\underset{2}{\mathsf{F}}^{11} = \frac{F_{22}(F_{11} - F_{22})\overset{2}{\mathsf{t}}_3 - F_{12}F_{21}\overset{2}{\mathsf{t}}_2 + F_{21}^2\overset{2}{\mathsf{t}}_1}{F_{11}F_{22} - F_{12}F_{21}}, \qquad \underset{2}{\mathsf{F}}^{22} = \frac{F_{11}(F_{22} - F_{11})\overset{2}{\mathsf{t}}_3 - F_{12}F_{21}\overset{2}{\mathsf{t}}_2 + F_{12}^2\overset{2}{\mathsf{t}}_1}{F_{11}F_{22} - F_{12}F_{21}}, \tag{5.182}$$

$$\underset{2}{\mathsf{F}}^{12} = \frac{F_{12}F_{22}\overset{2}{\mathsf{t}}_3 - F_{11}F_{21}\overset{2}{\mathsf{t}}_1 + F_{11}F_{12}\overset{2}{\mathsf{t}}_4}{F_{11}F_{22} - F_{12}F_{21}}, \qquad \underset{2}{\mathsf{F}}^{21} = \frac{F_{21}F_{11}\overset{2}{\mathsf{t}}_3 - F_{12}F_{22}\overset{2}{\mathsf{t}}_1 + F_{21}F_{22}\overset{2}{\mathsf{t}}_4}{F_{11}F_{22} - F_{12}F_{21}}, \tag{5.183}$$

where $\overset{a}{\mathsf{t}}_i$, $i = 1, \cdots, 4$, $\mathfrak{a} = 1, 2$ are constants given as

$$\overset{1}{\mathsf{t}}_1 = \frac{(\overset{1}{b}_1 + \overset{1}{b}_2)(\overset{22}{c}_1 + \overset{22}{c}_\Sigma) - (\overset{2}{b}_1 + \overset{2}{b}_2)(\overset{12}{c}_1 + \overset{12}{c}_\Sigma)}{(\overset{12}{c}_1 + \overset{12}{c}_\Sigma)^2 - (\overset{11}{c}_1 + \overset{11}{c}_\Sigma)(\overset{22}{c}_1 + \overset{22}{c}_\Sigma)} + \frac{\overset{2}{b}_2(\overset{12}{c}_2 + \overset{12}{c}_3) - \overset{1}{b}_2(\overset{22}{c}_2 + \overset{22}{c}_3)}{(\overset{12}{c}_2 + \overset{12}{c}_3)^2 - (\overset{11}{c}_2 + \overset{11}{c}_3)(\overset{22}{c}_2 + \overset{22}{c}_3)}, \tag{5.184}$$

$$\overset{2}{\mathsf{t}}_1 = \frac{(\overset{2}{b}_1 + \overset{2}{b}_2)(\overset{11}{c}_1 + \overset{11}{c}_\Sigma) - (\overset{1}{b}_1 + \overset{1}{b}_2)(\overset{12}{c}_1 + \overset{12}{c}_\Sigma)}{(\overset{12}{c}_1 + \overset{12}{c}_\Sigma)^2 - (\overset{11}{c}_1 + \overset{11}{c}_\Sigma)(\overset{22}{c}_1 + \overset{22}{c}_\Sigma)} + \frac{\overset{1}{b}_2(\overset{12}{c}_2 + \overset{12}{c}_3) - \overset{2}{b}_2(\overset{11}{c}_2 + \overset{11}{c}_3)}{(\overset{12}{c}_2 + \overset{12}{c}_3)^2 - (\overset{11}{c}_2 + \overset{11}{c}_3)(\overset{22}{c}_2 + \overset{22}{c}_3)}, \tag{5.185}$$

$$\begin{aligned}\overset{1}{\mathsf{t}}_2 =& \frac{\overset{1}{b}_1(\overset{22}{c}_1 + \overset{22}{c}_\Sigma) - \overset{2}{b}_1(\overset{12}{c}_1 + \overset{12}{c}_\Sigma)}{(\overset{12}{c}_1 + \overset{12}{c}_\Sigma)^2 - (\overset{11}{c}_1 + \overset{11}{c}_\Sigma)(\overset{22}{c}_1 + \overset{22}{c}_\Sigma)} \\ &- 2\overset{2}{b}_2 \frac{\overset{12}{c}_\Sigma(\overset{12}{c}_1 + \overset{12}{c}_\Sigma)(\overset{12}{c}_2 + \overset{12}{c}_3) - (\overset{22}{c}_1\overset{11}{c}_\Sigma + \overset{11}{c}_1\overset{22}{c}_\Sigma)(\overset{12}{c}_2 + \overset{12}{c}_3) - \overset{12}{c}_\Sigma(\overset{11}{c}_2 + \overset{11}{c}_3)(\overset{22}{c}_2 + \overset{22}{c}_3)}{\left[(\overset{12}{c}_2 + \overset{12}{c}_3)^2 - (\overset{11}{c}_2 + \overset{11}{c}_3)(\overset{22}{c}_2 + \overset{22}{c}_3)\right]\left[(\overset{12}{c}_1 + \overset{12}{c}_\Sigma)^2 - (\overset{11}{c}_1 + \overset{11}{c}_\Sigma)(\overset{22}{c}_1 + \overset{22}{c}_\Sigma)\right]} \\ &+ 2\overset{1}{b}_2 \frac{\left[2\overset{12}{c}_1\overset{12}{c}_\Sigma - \overset{11}{c}_\Sigma(\overset{22}{c}_1 + \overset{22}{c}_\Sigma)\right](\overset{22}{c}_2 + \overset{22}{c}_3) + \overset{22}{c}_\Sigma(\overset{12}{c}_2 + \overset{12}{c}_3)^2}{\left[(\overset{12}{c}_2 + \overset{12}{c}_3)^2 - (\overset{11}{c}_2 + \overset{11}{c}_3)(\overset{22}{c}_2 + \overset{22}{c}_3)\right]\left[(\overset{12}{c}_1 + \overset{12}{c}_\Sigma)^2 - (\overset{11}{c}_1 + \overset{11}{c}_\Sigma)(\overset{22}{c}_1 + \overset{22}{c}_\Sigma)\right]},\end{aligned} \tag{5.186}$$

$$\begin{aligned}\overset{2}{\mathsf{t}}_2 =& \frac{\overset{2}{b}_1(\overset{11}{c}_1 + \overset{11}{c}_\Sigma) - \overset{1}{b}_1(\overset{12}{c}_1 + \overset{12}{c}_\Sigma)}{(\overset{12}{c}_1 + \overset{12}{c}_\Sigma)^2 - (\overset{11}{c}_1 + \overset{11}{c}_\Sigma)(\overset{22}{c}_1 + \overset{22}{c}_\Sigma)} \\ &+ 2\overset{2}{b}_2 \frac{\left[2\overset{12}{c}_1\overset{12}{c}_\Sigma - \overset{22}{c}_\Sigma(\overset{11}{c}_1 + \overset{11}{c}_\Sigma)\right](\overset{11}{c}_2 + \overset{11}{c}_3) + \overset{11}{c}_\Sigma(\overset{12}{c}_2 + \overset{12}{c}_3)^2}{\left[(\overset{12}{c}_2 + \overset{12}{c}_3)^2 - (\overset{11}{c}_2 + \overset{11}{c}_3)(\overset{22}{c}_2 + \overset{22}{c}_3)\right]\left[(\overset{12}{c}_1 + \overset{12}{c}_\Sigma)^2 - (\overset{11}{c}_1 + \overset{11}{c}_\Sigma)(\overset{22}{c}_1 + \overset{22}{c}_\Sigma)\right]} \\ &- 2\overset{1}{b}_2 \frac{\overset{12}{c}_\Sigma(\overset{12}{c}_1 + \overset{12}{c}_\Sigma)(\overset{12}{c}_2 + \overset{12}{c}_3) - (\overset{22}{c}_1\overset{11}{c}_\Sigma + \overset{11}{c}_1\overset{22}{c}_\Sigma)(\overset{12}{c}_2 + \overset{12}{c}_3) - \overset{12}{c}_\Sigma(\overset{11}{c}_2 + \overset{11}{c}_3)(\overset{22}{c}_2 + \overset{22}{c}_3)}{\left[(\overset{12}{c}_2 + \overset{12}{c}_3)^2 - (\overset{11}{c}_2 + \overset{11}{c}_3)(\overset{22}{c}_2 + \overset{22}{c}_3)\right]\left[(\overset{12}{c}_1 + \overset{12}{c}_\Sigma)^2 - (\overset{11}{c}_1 + \overset{11}{c}_\Sigma)(\overset{22}{c}_1 + \overset{22}{c}_\Sigma)\right]},\end{aligned} \tag{5.187}$$

$$\overset{1}{\mathsf{t}}_3 = \frac{\overset{1}{b}_2(\overset{22}{c}_2 + \overset{22}{c}_3) - \overset{2}{b}_2(\overset{12}{c}_2 + \overset{12}{c}_3)}{(\overset{12}{c}_2 + \overset{12}{c}_3)^2 - (\overset{11}{c}_2 + \overset{11}{c}_3)(\overset{22}{c}_2 + \overset{22}{c}_3)}, \qquad \overset{2}{\mathsf{t}}_3 = \frac{\overset{2}{b}_2(\overset{11}{c}_2 + \overset{11}{c}_3) - \overset{1}{b}_2(\overset{12}{c}_2 + \overset{12}{c}_3)}{(\overset{12}{c}_2 + \overset{12}{c}_3)^2 - (\overset{11}{c}_2 + \overset{11}{c}_3)(\overset{22}{c}_2 + \overset{22}{c}_3)}, \tag{5.188}$$

$$\overset{1}{\mathsf{t}}_4 = \frac{(\overset{1}{b}_1 + \overset{1}{b}_2)(\overset{22}{c}_1 + \overset{22}{c}_\Sigma) - (\overset{2}{b}_1 + \overset{2}{b}_2)(\overset{12}{c}_1 + \overset{12}{c}_\Sigma)}{(\overset{12}{c}_1 + \overset{12}{c}_\Sigma)^2 - (\overset{11}{c}_1 + \overset{11}{c}_\Sigma)(\overset{22}{c}_1 + \overset{22}{c}_\Sigma)}, \qquad \overset{2}{\mathsf{t}}_4 = \frac{(\overset{2}{b}_1 + \overset{2}{b}_2)(\overset{11}{c}_1 + \overset{11}{c}_\Sigma) - (\overset{1}{b}_1 + \overset{1}{b}_2)(\overset{12}{c}_1 + \overset{12}{c}_\Sigma)}{(\overset{12}{c}_1 + \overset{12}{c}_\Sigma)^2 - (\overset{11}{c}_1 + \overset{11}{c}_\Sigma)(\overset{22}{c}_1 + \overset{22}{c}_\Sigma)}. \tag{5.189}$$

Substituting this solution into (5.173) and (5.174) one obtains the following two conditions to be satisfied by the material elastic constants

$$\mu = 2\left[\frac{2\overset{1}{b}_2\overset{2}{b}_2(\overset{12}{c}_2 + \overset{12}{c}_3) - (\overset{2}{b}_2)^2\,(\overset{11}{c}_2 + \overset{11}{c}_3) - (\overset{1}{b}_2)^2\,(\overset{22}{c}_2 + \overset{22}{c}_3)}{(\overset{12}{c}_2 + \overset{12}{c}_3)^2 - (\overset{11}{c}_2 + \overset{11}{c}_3)(\overset{22}{c}_2 + \overset{22}{c}_3)}\right], \tag{5.190}$$

$$
\begin{aligned}
\lambda = 2 &\left[ \frac{2(\overset{1}{b}_1 + \overset{1}{b}_2)(\overset{2}{b}_1 + \overset{2}{b}_2)(\overset{12}{c}_1 + \overset{12}{c}_\Sigma) - (\overset{1}{b}_1 + \overset{1}{b}_2)^2 \, (\overset{22}{c}_1 + \overset{22}{c}_\Sigma) - (\overset{2}{b}_1 + \overset{2}{b}_2)^2 \, (\overset{11}{c}_1 + \overset{11}{c}_\Sigma)}{(\overset{12}{c}_1 + \overset{12}{c}_\Sigma)^2 - (\overset{11}{c}_1 + \overset{11}{c}_\Sigma)(\overset{22}{c}_1 + \overset{22}{c}_\Sigma)} \right] \\
&+ 2 \left[ \frac{(\overset{2}{b}_2)^2 \, (\overset{11}{c}_2 + \overset{11}{c}_3) + (\overset{1}{b}_2)^2 \, (\overset{22}{c}_2 + \overset{22}{c}_3) - 2\overset{1}{b}_2\overset{2}{b}_2(\overset{12}{c}_2 + \overset{12}{c}_3)}{(\overset{12}{c}_2 + \overset{12}{c}_3)^2 - (\overset{11}{c}_2 + \overset{11}{c}_3)(\overset{22}{c}_2 + \overset{22}{c}_3)} \right] .
\end{aligned}
\tag{5.191}
$$

Note that (5.190) is identical to what we obtained for the cylindrical cloak example in (5.155), i.e., (5.190) violates positive-definiteness of the energy, and hence cloaking is not possible.

The following proposition summarizes the discussions and calculations of this section.

**Proposition 5.7.** *Elastodynamics transformation cloaking is not possible for linear generalized Cosserat elastic solids in dimension two.*

**Corollary 5.8.** *Elastodynamics transformation cloaking is not possible for linear Cosserat elastic solids in dimension two.*

**Example (A generalized Cosserat spherical cloak).** Let us next consider a finite spherical cavity $\mathcal{H}$ with radius $R_i$ embedded in an infinite isotropic homogeneous generalized Cosserat elastic medium with the elastic properties given in (5.105). Let $(R, \Theta, \Phi)$ be the spherical coordinates, for which $R \geq 0$ with $R = 0$ corresponding to the center of the cavity, $0 \leq \Theta \leq \pi$, and $0 \leq \Phi \leq 2\pi$. Suppose that there is a wave source located at $(R_p, \Theta_p, \Phi_p)$. The cloak $\mathcal{C}$ is a spherical shell with inner radius $R_i$ and outer radius $R_o$ surrounding the hole such that the source is located outside the cloaking region, i.e., $R_o < R_p$. Similar to the cylindrical cloak example, the reference configuration of the body $\mathcal{B}$ is mapped to that of $\tilde{\mathcal{B}}$ (the virtual body) using a mapping $\Xi : \mathcal{B} \to \tilde{\mathcal{B}}$. Note that $\Xi$ is the identity outside the cloak, i.e., $R \geq R_o$ and is defined as $(\tilde{R}, \tilde{\Theta}, \tilde{\Phi}) = \Xi(R, \Theta, \Phi) = (f(R), \Theta, \Phi)$ for $R_i \leq R \leq R_o$ such that $f(R_i) = \epsilon$, $f(R_o) = R_o$, and $f'(R_o) = 1$.

The reference configurations of $\mathcal{B}$ and $\tilde{\mathcal{B}}$ are endowed with the induced metrics from $\mathbb{R}^3$, i.e., $\mathbf{G} = \operatorname{diag}(1, R^2, R^2 \sin^2 \Theta)$ and $\tilde{\mathbf{G}} = \operatorname{diag}(1, \tilde{R}^2, \tilde{R}^2 \sin^2 \tilde{\Theta})$, respectively. In the coordinates $(r, \theta, \phi)$, the ambient space metric reads $\mathbf{g} = \operatorname{diag}(1, r^2, r^2 \sin^2 \theta)$. We assume that the virtual body $\tilde{\mathcal{B}}$ is isotropic and has the same elastic properties as the primary medium in the region $\mathcal{B} \setminus \mathcal{C}$. For the cloaking map $\overset{\Xi}{\mathbf{F}} = \operatorname{diag}(f'(R), 1, 1)$ in $R_i \leq R \leq R_o$. Note also that

$$
\mathsf{s} = \begin{bmatrix} 1 & 0 & 0 \\ 0 & \frac{R}{f(R)} & 0 \\ 0 & 0 & \frac{R}{f(R)} \end{bmatrix} .
\tag{5.192}
$$

Noting that $\rho = J_\Xi \tilde{\rho} \circ \Xi = J_\Xi \rho_0$, the mass density of the cloak is obtained as

$$
\rho_{\mathcal{C}}(R) = \frac{f(R)^2 f'(R)}{R^2} \rho_0 \,, \qquad R_i \leq R \leq R_o \,.
\tag{5.193}
$$

Also

$$
\overset{\mathfrak{ab}}{\nu}_{\mathcal{C}}(R) = \frac{f(R)^2 f'(R)}{R^2} \overset{\mathfrak{ab}}{\tilde{\nu}} \,, \qquad R_i \leq R \leq R_o \,.
\tag{5.194}
$$

Using (5.92) and (5.96), the elastic constants of the cloak are found as ($\mathfrak{a}, \mathfrak{b} \in \{1, 2, 3\}$)

$$
\hat{\mathbb{A}} = \begin{bmatrix}
\begin{bmatrix} \frac{(\lambda+2\mu) f(R)^2}{R^2 f'(R)} & 0 & 0 \\ 0 & \frac{\lambda f(R)}{R} & 0 \\ 0 & 0 & \frac{\lambda f(R)}{R} \end{bmatrix} &
\begin{bmatrix} 0 & \mu f'(R) & 0 \\ \frac{\mu f(R)}{R} & 0 & 0 \\ 0 & 0 & 0 \end{bmatrix} &
\begin{bmatrix} 0 & 0 & \mu f'(R) \\ 0 & 0 & 0 \\ \frac{\mu f(R)}{R} & 0 & 0 \end{bmatrix} \\
\begin{bmatrix} 0 & \frac{\mu f(R)}{R} & 0 \\ \frac{\mu f(R)^2}{R^2 f'(R)} & 0 & 0 \\ 0 & 0 & 0 \end{bmatrix} &
\begin{bmatrix} \frac{\lambda f(R)}{R} & 0 & 0 \\ 0 & (\lambda+2\mu) f'(R) & 0 \\ 0 & 0 & \lambda f'(R) \end{bmatrix} &
\begin{bmatrix} 0 & 0 & 0 \\ 0 & 0 & \mu f'(R) \\ 0 & \mu f'(R) & 0 \end{bmatrix} \\
\begin{bmatrix} 0 & 0 & \frac{\mu f(R)}{R} \\ 0 & 0 & 0 \\ \frac{\mu f(R)^2}{R^2 f'(R)} & 0 & 0 \end{bmatrix} &
\begin{bmatrix} 0 & 0 & 0 \\ 0 & 0 & \mu f'(R) \\ 0 & \mu f'(R) & 0 \end{bmatrix} &
\begin{bmatrix} \frac{\lambda f(R)}{R} & 0 & 0 \\ 0 & \lambda f'(R) & 0 \\ 0 & 0 & (\lambda+2\mu) f'(R) \end{bmatrix}
\end{bmatrix} ,
\tag{5.195}
$$

$$
\hat{\overset{a}{\mathbb{B}}} = \begin{bmatrix}
\begin{bmatrix} \frac{\overset{a}{b}_{\Sigma} f(R)^2}{R^2 f'(R)} & 0 & 0 \\ 0 & \frac{\overset{a}{b}_1 f(R)}{R} & 0 \\ 0 & 0 & \frac{\overset{a}{b}_1 f(R)}{R} \end{bmatrix} &
\begin{bmatrix} 0 & \overset{a}{b}_2 f'(R) & 0 \\ \frac{\overset{a}{b}_2 f(R)}{R} & 0 & 0 \\ 0 & 0 & 0 \end{bmatrix} &
\begin{bmatrix} 0 & 0 & \overset{a}{b}_2 f'(R) \\ 0 & 0 & 0 \\ \frac{\overset{a}{b}_2 f(R)}{R} & 0 & 0 \end{bmatrix} \\
\begin{bmatrix} 0 & \frac{\overset{a}{b}_2 f(R)}{R} & 0 \\ \frac{\overset{a}{b}_2 f(R)^2}{R^2 f'(R)} & 0 & 0 \\ 0 & 0 & 0 \end{bmatrix} &
\begin{bmatrix} \frac{\overset{a}{b}_1 f(R)}{R} & 0 & 0 \\ 0 & \overset{a}{b}_{\Sigma} f'(R) & 0 \\ 0 & 0 & \overset{a}{b}_1 f'(R) \end{bmatrix} &
\begin{bmatrix} 0 & 0 & 0 \\ 0 & 0 & \overset{a}{b}_2 f'(R) \\ 0 & \overset{a}{b}_2 f'(R) & 0 \end{bmatrix} \\
\begin{bmatrix} 0 & 0 & \frac{\overset{a}{b}_2 f(R)}{R} \\ 0 & 0 & 0 \\ \frac{\overset{a}{b}_2 f(R)^2}{R^2 f'(R)} & 0 & 0 \end{bmatrix} &
\begin{bmatrix} 0 & 0 & 0 \\ 0 & 0 & \overset{a}{b}_2 f'(R) \\ 0 & \overset{a}{b}_2 f'(R) & 0 \end{bmatrix} &
\begin{bmatrix} \frac{\overset{a}{b}_1 f(R)}{R} & 0 & 0 \\ 0 & \overset{a}{b}_1 f'(R) & 0 \\ 0 & 0 & \overset{a}{b}_{\Sigma} f'(R) \end{bmatrix}
\end{bmatrix}, \tag{5.196}
$$

$$
\hat{\overset{ab}{\mathbb{C}}} = \begin{bmatrix}
\begin{bmatrix} \frac{\overset{ab}{c}_{\Sigma} f(R)^2}{R^2 f'(R)} & 0 & 0 \\ 0 & \frac{\overset{ab}{c}_1 f(R)}{R} & 0 \\ 0 & 0 & \frac{\overset{ab}{c}_1 f(R)}{R} \end{bmatrix} &
\begin{bmatrix} 0 & \overset{ab}{c}_2 f'(R) & 0 \\ \frac{\overset{ab}{c}_3 f(R)}{R} & 0 & 0 \\ 0 & 0 & 0 \end{bmatrix} &
\begin{bmatrix} 0 & 0 & \overset{ab}{c}_2 f'(R) \\ 0 & 0 & 0 \\ \frac{\overset{ab}{c}_3 f(R)}{R} & 0 & 0 \end{bmatrix} \\
\begin{bmatrix} 0 & \frac{\overset{ab}{c}_3 f(R)}{R} & 0 \\ \frac{\overset{ab}{c}_2 f(R)^2}{R^2 f'(R)} & 0 & 0 \\ 0 & 0 & 0 \end{bmatrix} &
\begin{bmatrix} \frac{\overset{ab}{c}_1 f(R)}{R} & 0 & 0 \\ 0 & \overset{ab}{c}_{\Sigma} f'(R) & 0 \\ 0 & 0 & \overset{ab}{c}_1 f'(R) \end{bmatrix} &
\begin{bmatrix} 0 & 0 & 0 \\ 0 & 0 & \overset{ab}{c}_2 f'(R) \\ 0 & \overset{ab}{c}_3 f'(R) & 0 \end{bmatrix} \\
\begin{bmatrix} 0 & 0 & \frac{\overset{ab}{c}_3 f(R)}{R} \\ 0 & 0 & 0 \\ \frac{\overset{ab}{c}_2 f(R)^2}{R^2 f'(R)} & 0 & 0 \end{bmatrix} &
\begin{bmatrix} 0 & 0 & 0 \\ 0 & 0 & \overset{ab}{c}_3 f'(R) \\ 0 & \overset{ab}{c}_2 f'(R) & 0 \end{bmatrix} &
\begin{bmatrix} \frac{\overset{ab}{c}_1 f(R)}{R} & 0 & 0 \\ 0 & \overset{ab}{c}_1 f'(R) & 0 \\ 0 & 0 & \overset{ab}{c}_{\Sigma} f'(R) \end{bmatrix}
\end{bmatrix}, \tag{5.197}
$$

where $\overset{ab}{c}_{\Sigma} = \overset{ab}{c}_1 + \overset{ab}{c}_2 + \overset{ab}{c}_3$, and $\overset{a}{b}_{\Sigma} = \overset{a}{b}_1 + 2\overset{a}{b}_2$.

Similar to the cylindrical cloak example, the balance of angular momentum Eq.(5.100) for the diagonal and off-diagonal components of the director gradients are uncoupled. We consider the equations for the diagonal components of the director gradients in $(5.100)_2$ and show that they force the cloaking map to be the identity. For the choice $(a, c) = (2, 3)$, $(5.100)_2$ is expanded and gives the following equations for the (circumferential and azimuthal) diagonal director gradient components

$$\overset{11}{c}_2 \underset{1}{\mathsf{F}}^{\phi}{}_{\Phi} + \overset{12}{c}_2 \underset{2}{\mathsf{F}}^{\phi}{}_{\Phi} + \overset{13}{c}_2 \underset{3}{\mathsf{F}}^{\phi}{}_{\Phi} - \overset{11}{c}_3 \underset{1}{\mathsf{F}}^{\theta}{}_{\Theta} - \overset{12}{c}_3 \underset{2}{\mathsf{F}}^{\theta}{}_{\Theta} - \overset{13}{c}_3 \underset{3}{\mathsf{F}}^{\theta}{}_{\Theta} = 0\,, \tag{5.198}$$

$$\overset{11}{c}_3 \underset{1}{\mathsf{F}}^{\phi}{}_{\Phi} + \overset{12}{c}_3 \underset{2}{\mathsf{F}}^{\phi}{}_{\Phi} + \overset{13}{c}_3 \underset{3}{\mathsf{F}}^{\phi}{}_{\Phi} - \overset{11}{c}_2 \underset{1}{\mathsf{F}}^{\theta}{}_{\Theta} - \overset{12}{c}_2 \underset{2}{\mathsf{F}}^{\theta}{}_{\Theta} - \overset{13}{c}_2 \underset{3}{\mathsf{F}}^{\theta}{}_{\Theta} = 0\,, \tag{5.199}$$

$$\overset{12}{c}_2 \underset{1}{\mathsf{F}}^{\phi}{}_{\Phi} + \overset{22}{c}_2 \underset{2}{\mathsf{F}}^{\phi}{}_{\Phi} + \overset{23}{c}_2 \underset{3}{\mathsf{F}}^{\phi}{}_{\Phi} - \overset{12}{c}_3 \underset{1}{\mathsf{F}}^{\theta}{}_{\Theta} - \overset{22}{c}_3 \underset{2}{\mathsf{F}}^{\theta}{}_{\Theta} - \overset{23}{c}_3 \underset{3}{\mathsf{F}}^{\theta}{}_{\Theta} = 0\,, \tag{5.200}$$

$$\overset{12}{c}_3 \underset{1}{\mathsf{F}}^{\phi}{}_{\Phi} + \overset{22}{c}_3 \underset{2}{\mathsf{F}}^{\phi}{}_{\Phi} + \overset{23}{c}_3 \underset{3}{\mathsf{F}}^{\phi}{}_{\Phi} - \overset{12}{c}_2 \underset{1}{\mathsf{F}}^{\theta}{}_{\Theta} - \overset{22}{c}_2 \underset{2}{\mathsf{F}}^{\theta}{}_{\Theta} - \overset{23}{c}_2 \underset{3}{\mathsf{F}}^{\theta}{}_{\Theta} = 0\,, \tag{5.201}$$

$$\overset{13}{c}_2 \underset{1}{\mathsf{F}}^{\phi}{}_{\Phi} + \overset{23}{c}_2 \underset{2}{\mathsf{F}}^{\phi}{}_{\Phi} + \overset{33}{c}_2 \underset{3}{\mathsf{F}}^{\phi}{}_{\Phi} - \overset{13}{c}_3 \underset{1}{\mathsf{F}}^{\theta}{}_{\Theta} - \overset{23}{c}_3 \underset{2}{\mathsf{F}}^{\theta}{}_{\Theta} - \overset{33}{c}_3 \underset{3}{\mathsf{F}}^{\theta}{}_{\Theta} = 0\,, \tag{5.202}$$

$$\overset{13}{c}_3 \underset{1}{\mathsf{F}}^{\phi}{}_{\Phi} + \overset{23}{c}_3 \underset{2}{\mathsf{F}}^{\phi}{}_{\Phi} + \overset{33}{c}_3 \underset{3}{\mathsf{F}}^{\phi}{}_{\Phi} - \overset{13}{c}_2 \underset{1}{\mathsf{F}}^{\theta}{}_{\Theta} - \overset{23}{c}_2 \underset{2}{\mathsf{F}}^{\theta}{}_{\Theta} - \overset{33}{c}_2 \underset{3}{\mathsf{F}}^{\theta}{}_{\Theta} = 0\,. \tag{5.203}$$

Choosing a class of deformations for which $U^a{}_{|A} = 0$, and $\underset{a}{\mathfrak{U}}^1{}_{|1} = \underset{a}{\mathfrak{U}}^2{}_{|2} = \underset{a}{\mathfrak{U}}^3{}_{|3} = 0$, the positive-definiteness of the energy function implies the positive definiteness of the following matrix

$$
\begin{bmatrix}
\overset{11}{c}_2 & \overset{11}{c}_3 & \overset{12}{c}_2 & \overset{12}{c}_3 & \overset{13}{c}_2 & \overset{13}{c}_3 \\
\overset{11}{c}_3 & \overset{11}{c}_2 & \overset{12}{c}_3 & \overset{12}{c}_2 & \overset{13}{c}_3 & \overset{13}{c}_2 \\
\overset{12}{c}_2 & \overset{12}{c}_3 & \overset{22}{c}_2 & \overset{22}{c}_3 & \overset{23}{c}_2 & \overset{23}{c}_3 \\
\overset{12}{c}_3 & \overset{12}{c}_2 & \overset{22}{c}_3 & \overset{22}{c}_2 & \overset{23}{c}_3 & \overset{23}{c}_2 \\
\overset{13}{c}_2 & \overset{13}{c}_3 & \overset{23}{c}_2 & \overset{23}{c}_3 & \overset{33}{c}_2 & \overset{33}{c}_3 \\
\overset{13}{c}_3 & \overset{13}{c}_2 & \overset{23}{c}_3 & \overset{23}{c}_2 & \overset{33}{c}_3 & \overset{33}{c}_2
\end{bmatrix}. \tag{5.204}
$$

In particular, the determinant of (5.204) is non-vanishing, and thus, so is determinant of the coefficient matrix of the system (5.198)-(5.203). Therefore, $\underset{1}{\mathsf{F}}^{\phi}{}_{\Phi} = \underset{2}{\mathsf{F}}^{\phi}{}_{\Phi} = \underset{3}{\mathsf{F}}^{\phi}{}_{\Phi} = 0$ and $\underset{1}{\mathsf{F}}^{\theta}{}_{\Theta} = \underset{2}{\mathsf{F}}^{\theta}{}_{\Theta} = \underset{3}{\mathsf{F}}^{\theta}{}_{\Theta} = 0$. Using this, $(5.100)_2$ for choices $(a, c) = (1, 2)$ and $(a, c) = (1, 3)$ gives one the following equations $\underset{1}{\mathsf{F}}^{r}{}_{R}$, $\underset{2}{\mathsf{F}}^{r}{}_{R}$, and $\underset{3}{\mathsf{F}}^{r}{}_{R}$:

$$f(R)\left(\overset{11}{c}_3\underset{1}{\mathsf{F}}^{r}{}_{R} + \overset{12}{c}_3\underset{2}{\mathsf{F}}^{r}{}_{R} + \overset{13}{c}_3\underset{3}{\mathsf{F}}^{r}{}_{R}\right) = \overset{1}{b}_2\left(Rf'(R) - f(R)\right) , \tag{5.205}$$

$$f(R)\left(\overset{11}{c}_2\underset{1}{\mathsf{F}}^{r}{}_{R} + \overset{12}{c}_2\underset{2}{\mathsf{F}}^{r}{}_{R} + \overset{13}{c}_2\underset{3}{\mathsf{F}}^{r}{}_{R}\right) = \overset{1}{b}_2\left(Rf'(R) - f(R)\right) , \tag{5.206}$$

$$f(R)\left(\overset{12}{c}_3\underset{1}{\mathsf{F}}^{r}{}_{R} + \overset{22}{c}_3\underset{2}{\mathsf{F}}^{r}{}_{R} + \overset{23}{c}_3\underset{3}{\mathsf{F}}^{r}{}_{R}\right) = \overset{2}{b}_2\left(Rf'(R) - f(R)\right) , \tag{5.207}$$

$$f(R)\left(\overset{12}{c}_2\underset{1}{\mathsf{F}}^{r}{}_{R} + \overset{22}{c}_2\underset{2}{\mathsf{F}}^{r}{}_{R} + \overset{23}{c}_2\underset{3}{\mathsf{F}}^{r}{}_{R}\right) = \overset{2}{b}_2\left(Rf'(R) - f(R)\right) , \tag{5.208}$$

$$f(R)\left(\overset{13}{c}_3\underset{1}{\mathsf{F}}^{r}{}_{R} + \overset{23}{c}_3\underset{2}{\mathsf{F}}^{r}{}_{R} + \overset{33}{c}_3\underset{3}{\mathsf{F}}^{r}{}_{R}\right) = \overset{3}{b}_2\left(Rf'(R) - f(R)\right) , \tag{5.209}$$

$$f(R)\left(\overset{13}{c}_2\underset{1}{\mathsf{F}}^{r}{}_{R} + \overset{23}{c}_2\underset{2}{\mathsf{F}}^{r}{}_{R} + \overset{33}{c}_2\underset{3}{\mathsf{F}}^{r}{}_{R}\right) = \overset{3}{b}_2\left(Rf'(R) - f(R)\right) . \tag{5.210}$$

The coefficient matrix of the system of homogeneous algebraic equations that can be obtained from (5.205)-(5.206), (5.207)-(5.208), and (5.209)-(5.210) reads

$$\begin{bmatrix} \overset{11}{c}_3 - \overset{11}{c}_2 & \overset{12}{c}_3 - \overset{12}{c}_2 & \overset{13}{c}_3 - \overset{13}{c}_2 \\ \overset{12}{c}_3 - \overset{12}{c}_2 & \overset{22}{c}_3 - \overset{22}{c}_2 & \overset{23}{c}_3 - \overset{23}{c}_2 \\ \overset{13}{c}_3 - \overset{13}{c}_2 & \overset{23}{c}_3 - \overset{23}{c}_2 & \overset{33}{c}_3 - \overset{33}{c}_2 \end{bmatrix} . \tag{5.211}$$

It is straightforward to see that the determinant of (5.204) being non-zero implies that the determinant of (5.211) is non-zero as well. Thus, $\underset{1}{\mathsf{F}}^{r}{}_{R} = \underset{2}{\mathsf{F}}^{r}{}_{R} = \underset{3}{\mathsf{F}}^{r}{}_{R} = 0$, and from (5.205)-(5.210), one concludes that $f(R) = R$, which, in turn, means that cloaking is not possible. Therefore, we have proved the following result.

**Proposition 5.9.** *Elastodynamics transformation cloaking is not possible for a spherical cavity using a spherical cloak in linear generalized Cosserat elastic solids.*

**Corollary 5.10.** *Elastodynamics transformation cloaking is not possible for a spherical cavity using a spherical cloak in linear Cosserat elastic solids.*

We suspect that transformation cloaking in dimension three is not possible for a cavity of any shape. The idea of proof is similar to that of 2D. However, in this case there are 108 equations for 27 unknown director gradients $\underset{a}{\mathsf{F}}^{a}{}_{A}$. We have not been able to solve this large system of equations but expect that they would violate positive-definiteness of the energy and also force the cloaking map to be the identity.

**Conjecture 5.11.** Elastodynamics transformation cloaking is not possible for linear generalized Cosserat elastic solids in dimension three.

# Acknowledgement

This research was supported by ARO W911NF-16-1-0064 and ARO W911NF-18-1-0003 (Dr. David Stepp).

# Appendix A Riemannian Geometry

To make the paper self-contained, in this appendix some basic concepts of Riemannian geometry are tersely reviewed. It should be emphasized that only what has been used in the paper is discussed here.

For a smooth $n$-dimensional manifold $\mathcal{B}$, the tangent space of $\mathcal{B}$ at a point $X \in \mathcal{B}$ is denoted by $T_X\mathcal{B}$. Assume that $\mathcal{S}$ is another $n$-dimensional manifold and $\varphi : \mathcal{B} \to \mathcal{S}$ is a diffeomorphism (smooth and invertible map with a smooth inverse) between the two manifolds. A smooth vector field $\mathbf{W}$ on $\mathcal{B}$ assigns a vector $\mathbf{W}_X \in T_X\mathcal{B}$ for every $X \in \mathcal{B}$ such that the mapping $X \mapsto \mathbf{W}_X$ is smooth. If $\mathbf{W}$ is a vector field on $\mathcal{B}$, then the push-forward of $\mathbf{W}$ by $\varphi$ is a vector field on $\varphi(\mathcal{B})$ defined as $\varphi_*\mathbf{W} = T\varphi \cdot \mathbf{W} \circ \varphi^{-1}$. Similarly, if $\mathbf{w}$ is a vector field on $\varphi(\mathcal{B}) \subset \mathcal{S}$, the pull-back of $\mathbf{w}$ by $\varphi$ is a vector field on $\mathcal{B}$ defined as $\varphi^*\mathbf{w} = T(\varphi^{-1}) \cdot \mathbf{w} \circ \varphi$. Let us denote the tangent map of $\varphi$ by $\mathbf{F}$, i.e., $\mathbf{F} = T\varphi$. Let $\{X^A\}$ and $\{x^a\}$ be the local charts for $\mathcal{B}$ and $\mathcal{S}$, respectively. More specifically, a local chart for $\mathcal{B}$ at $X \in \mathcal{B}$ is a homeomorphism from an open subset $\mathcal{U} \subset \mathcal{B}$ ($X \in \mathcal{U}$) to an open subset $\mathcal{V} \subset \mathbb{R}^n$. $\{X^A\}$ are components of this map. The derivative map $\mathbf{F}$ is a two-point tensor with the following representation in the local charts: $\mathbf{F} = F^a{}_A \frac{\partial}{\partial X^A} \otimes dx^a$, $F^a{}_A = \frac{\partial \varphi^a}{\partial X^A}$, where $\{\frac{\partial}{\partial X^A}\}$ and $\{dx^a\}$ are bases for $T_X\mathcal{B}$ and $T^*_{\varphi(X)}\varphi(\mathcal{B})$, respectively. Recall that $T^*_{\varphi(X)}\varphi(\mathcal{B})$ denotes the cotangent space (or the dual space) of $T_{\varphi(X)}\varphi(\mathcal{B})$. The push-forward and pull-back of vectors have the following coordinate representations: $(\varphi_*\mathbf{W})^a = F^a{}_A W^A$, and $(\varphi^*\mathbf{w})^A = (F^{-1})^A{}_a w^a$.

A type $\binom{0}{2}$-tensor at $X \in \mathcal{B}$ is a bilinear map $\mathbf{T} : T_X\mathcal{B} \times T_X\mathcal{B} \to \mathbb{R}$, where in a local coordinate chart $\{X^A\}$ for $\mathcal{B}$ reads $\mathbf{T}(\mathbf{U},\mathbf{V}) = T_{AB}U^AV^B$, $\forall\, \mathbf{U},\mathbf{V} \in T_X\mathcal{B}$. A Riemannian manifold $(\mathcal{B},\mathbf{G})$ is a smooth manifold $\mathcal{B}$ endowed with an inner product $\mathbf{G}_X$ (a symmetric $\binom{0}{2}$-tensor field) on the tangent space $T_X\mathcal{B}$ that smoothly varies in the sense that if $\mathbf{U}$ and $\mathbf{V}$ are smooth vector fields on $\mathcal{B}$, then $X \mapsto \mathbf{G}_X(\mathbf{U}_X,\mathbf{V}_X) =: \langle\!\langle \mathbf{U}_X,\mathbf{V}_X\rangle\!\rangle_{\mathbf{G}_X}$, is a smooth function.

Let $(\mathcal{B},\mathbf{G})$ and $(\mathcal{S},\mathbf{g})$ be Riemannian manifolds and let $\varphi : \mathcal{B} \to \mathcal{S}$ be a diffeomorphism (smooth map with smooth inverse). The push-forward of the metric $\mathbf{G}$ is a metric on $\varphi(\mathcal{B}) \subset \mathcal{S}$, which is denoted by $\varphi_*\mathbf{G}$ defined as

$$(\varphi_*\mathbf{G})_{\varphi(X)}\left(\mathbf{u}_{\varphi(X)},\mathbf{v}_{\varphi(X)}\right) := \mathbf{G}_X\left((\varphi^*\mathbf{u})_X,(\varphi^*\mathbf{v})_X\right). \tag{A.1}$$

In components, $(\varphi_*\mathbf{G})_{ab} = (F^{-1})^A{}_a(F^{-1})^B{}_b G_{AB}$. Similarly, the pull-back of the metric $\mathbf{g}$ is a metric in $\mathcal{B}$, which is denoted by $\varphi^*\mathbf{g}$ defined as

$$(\varphi^*\mathbf{g})_X(\mathbf{U}_X,\mathbf{V}_X) := \mathbf{g}_{\varphi(X)}((\varphi_*\mathbf{U})_{\varphi(X)},(\varphi_*\mathbf{V})_{\varphi(X)}). \tag{A.2}$$

In components, $(\varphi^*\mathbf{g})_{AB} = F^a{}_AF^b{}_B g_{ab}$. The diffeomorphism $\varphi$ is an isometry between two Riemannian manifolds $(\mathcal{B},\mathbf{G})$ and $(\mathcal{S},\mathbf{g})$ if $\mathbf{g} = \varphi_*\mathbf{G}$, or equivalently, $\mathbf{G} = \varphi^*\mathbf{g}$. An isometry, by definition, preserves distances.

**Affine connections, and their torsion and curvature tensors.** A linear (affine) connection on a manifold $\mathcal{B}$ is an operation $\nabla : \mathcal{X}(\mathcal{B}) \times \mathcal{X}(\mathcal{B}) \to \mathcal{X}(\mathcal{B})$, where $\mathcal{X}(\mathcal{B})$ is the set of vector fields on $\mathcal{B}$, such that $\forall\, \mathbf{X},\mathbf{Y},\mathbf{X}_1,\mathbf{X}_2,\mathbf{Y}_1,\mathbf{Y}_2 \in \mathcal{X}(\mathcal{B}), \forall\, f, f_1, f_2 \in C^\infty(\mathcal{B}), \forall\, a_1,a_2 \in \mathbb{R}$: i) $\nabla_{f_1\mathbf{X}_1+f_2\mathbf{X}_2}\mathbf{Y} = f_1\nabla_{\mathbf{X}_1}\mathbf{Y} + f_2\nabla_{\mathbf{X}_2}\mathbf{Y}$, ii) $\nabla_{\mathbf{X}}(a_1\mathbf{Y}_1 + a_2\mathbf{Y}_2) = a_1\nabla_{\mathbf{X}}(\mathbf{Y}_1) + a_2\nabla_{\mathbf{X}}(\mathbf{Y}_2)$, iii) $\nabla_{\mathbf{X}}(f\mathbf{Y}) = f\nabla_{\mathbf{X}}\mathbf{Y} + (\mathbf{X}f)\mathbf{Y}$. $\nabla_{\mathbf{X}}\mathbf{Y}$ is called the covariant derivative of $\mathbf{Y}$ along $\mathbf{X}$. In a local coordinate chart $\{X^A\}$, $\nabla_{\partial_A}\partial_B = \Gamma^C{}_{AB}\partial_C$, where $\Gamma^C{}_{AB}$ are Christoffel symbols of the connection, and $\partial_A = \frac{\partial}{\partial x^A}$ are natural bases for the tangent space corresponding to a coordinate chart $\{x^A\}$. A linear connection is said to be compatible with a metric $\mathbf{G}$ on the manifold if

$$\nabla_{\mathbf{X}}\langle\!\langle \mathbf{Y},\mathbf{Z}\rangle\!\rangle_{\mathbf{G}} = \langle\!\langle \nabla_{\mathbf{X}}\mathbf{Y},\mathbf{Z}\rangle\!\rangle_{\mathbf{G}} + \langle\!\langle \mathbf{Y},\nabla_{\mathbf{X}}\mathbf{Z}\rangle\!\rangle_{\mathbf{G}}, \tag{A.3}$$

where $\langle\!\langle .,.\rangle\!\rangle_{\mathbf{G}}$ is the inner product induced by the metric $\mathbf{G}$. It can be shown that $\nabla$ is compatible with $\mathbf{G}$ if and only if $\nabla\mathbf{G} = \mathbf{0}$, or in components

$$G_{AB|C} = \frac{\partial G_{AB}}{\partial X^C} - \Gamma^S{}_{CA}G_{SB} - \Gamma^S{}_{CB}G_{AS} = 0. \tag{A.4}$$

Suppose $\mathbf{V},\mathbf{W} \in \mathcal{X}(\mathcal{B})$ are vector fields and $\alpha : I \to \mathcal{B}$ is a smooth curve. The restriction of the vector fields to $\alpha$, i.e., $\mathbf{V}\circ\alpha$ and $\mathbf{W}\circ\alpha$ are called vector fields along the curve $\alpha$. The set of all vector fields along $\alpha$ is denoted by $\mathcal{X}(\alpha)$. Covariant derivative along the curve $\alpha$ is a map $D_t : \mathcal{X}(\alpha) \to \mathcal{X}(\alpha)$ with the following properties: $D_t(\mathbf{V}+\mathbf{W}) = D_t\mathbf{V} + D_t\mathbf{W}$, and $D_t(f\mathbf{W}) = \frac{df}{dt}\mathbf{W} + fD_t\mathbf{W}$. If $\mathbf{W} \in \mathcal{X}(\alpha)$ is the restriction of $\widetilde{\mathbf{W}} \in \mathcal{X}(\mathcal{B})$ to $\alpha$, then, $D_t\mathbf{W} = \nabla_{\alpha'(t)}\widetilde{\mathbf{W}}$. If the connection $\nabla$ is $\mathbf{G}$-compatible, then

$$\frac{d}{dt}\langle\!\langle \mathbf{X},\mathbf{Y}(X,t)\rangle\!\rangle_{\mathbf{G}} = \langle\!\langle D_t\mathbf{X},\mathbf{Y}\rangle\!\rangle_{\mathbf{G}} + \langle\!\langle \mathbf{X},D_t\mathbf{Y}\rangle\!\rangle_{\mathbf{G}}. \tag{A.5}$$

The covariant derivative of a two-point tensor $\mathbf{T}$ is given by

$$\begin{aligned} T^{AB\cdots F}{}_{G\cdots Q}{}^{ab\cdots f}{}_{g\cdots q|K} =& \frac{\partial}{\partial X^k}T^{AB\cdots F}{}_{G\cdots Q}{}^{ab\cdots f}{}_{g\cdots q} \\ &+ T^{RB\cdots F}{}_{G\cdots Q}{}^{ab\cdots f}{}_{g\cdots q}\Gamma^A{}_{RK} + (\text{all upper referential indices}) \\ &- T^{AB\cdots F}{}_{R\cdots Q}{}^{ab\cdots f}{}_{g\cdots q}\Gamma^R{}_{GK} - (\text{all lower referential indices}) \\ &+ T^{RB\cdots F}{}_{G\cdots Q}{}^{lb\cdots f}{}_{g\cdots q}\gamma^a{}_{lr}F^r{}_K + (\text{all upper spatial indices}) \\ &- T^{AB\cdots F}{}_{G\cdots Q}{}^{ab\cdots f}{}_{l\cdots q}\gamma^l{}_{gr}F^r{}_K - (\text{all lower spatial indices})\,. \end{aligned} \tag{A.6}$$

The torsion of a connection is defined as $\boldsymbol{T}(\mathbf{X},\mathbf{Y})=\nabla_{\mathbf{X}}\mathbf{Y}-\nabla_{\mathbf{Y}}\mathbf{X}-[\mathbf{X},\mathbf{Y}]$, where $[\mathbf{X},\mathbf{Y}](F)=\mathbf{X}(\mathbf{Y}(F))-\mathbf{Y}(\mathbf{X}(F))$, $\forall\ F\in C^\infty(\mathcal{S})$, is the commutator of $\mathbf{X}$ and $\mathbf{Y}$. In components, in a local chart $\{X^A\}$, $T^A{}_{BC}=\Gamma^A{}_{BC}-\Gamma^A{}_{CB}$, and $[\mathbf{X},\mathbf{Y}]^a=\frac{\partial Y^a}{\partial x^b}X^b-\frac{\partial X^a}{\partial x^b}Y^b$. $\nabla$ is symmetric if it is torsion-free, i.e., $\nabla_{\mathbf{X}}\mathbf{Y}-\nabla_{\mathbf{Y}}\mathbf{X}=[\mathbf{X},\mathbf{Y}]$. On any Riemannian manifold $(\mathcal{B},\mathbf{G})$ there is a unique linear connection $\nabla^{\mathbf{G}}$ that is compatible with $\mathbf{G}$ and is torsion-free. This is the Levi-Civita connection. If the Levi-Civita connection $\nabla^{\mathbf{G}}$ is used, the covariant time derivative is denoted by $D_t^{\mathbf{G}}$. In a manifold with a connection the curvature is a map $\boldsymbol{\mathcal{R}}:\mathcal{X}(\mathcal{B})\times\mathcal{X}(\mathcal{B})\times\mathcal{X}(\mathcal{B})\to\mathcal{X}(\mathcal{B})$ defined by $\boldsymbol{\mathcal{R}}(\mathbf{X},\mathbf{Y},\mathbf{Z})=\nabla_{\mathbf{X}}\nabla_{\mathbf{Y}}\mathbf{Z}-\nabla_{\mathbf{Y}}\nabla_{\mathbf{X}}\mathbf{Z}-\nabla_{[\mathbf{X},\mathbf{Y}]}\mathbf{Z}$, or in components $\mathcal{R}^A{}_{BCD}=\frac{\partial\Gamma^A{}_{CD}}{\partial X^B}-\frac{\partial\Gamma^A{}_{BD}}{\partial X^C}+\Gamma^A{}_{BM}\Gamma^M{}_{CD}-\Gamma^A{}_{CM}\Gamma^M{}_{BD}$. The Riemannian curvature is the curvature tensor of the Levi-Civita connection $\nabla^{\mathbf{G}}$ and is denoted by $\boldsymbol{\mathcal{R}}_{\mathbf{G}}$. The Ricci identity for a vector field $\mathbf{U}$ with components $W^A$ reads $U^A{}_{|BC}-U^A{}_{|CB}=\mathcal{R}^A{}_{BCD}U^D$. Ricci identity for a 1-form $\boldsymbol{\alpha}$ with components $\alpha_A$ reads $\alpha_{A|BC}-\alpha_{A|CB}=\mathcal{R}^D{}_{BCA}\alpha_D$. The Ricci curvature $\textsf{Ric}$ is defined as $\textsf{Ric}_{CD}=\mathcal{R}^A{}_{ACD}$, and is a symmetric tensor. The Ricci curvature of the Levi-Civita connection $\nabla^{\mathbf{G}}$ is denoted by $\textsf{Ric}_{\mathbf{G}}$.

**Vector bundles.** Suppose $\mathcal{E}$ and $\mathcal{B}$ are sets and consider a map $\pi:\mathcal{E}\to\mathcal{B}$. The fiber over $X\in\mathcal{B}$ is the set $\mathcal{E}_X:=\pi^{-1}(X)\subset\mathcal{E}$. For an onto map $\pi$ fibers are non-empty and $\mathcal{E}=\sqcup_{X\in\mathcal{B}}\mathcal{E}_X$, where $\sqcup$ denoted disjoint union of sets. Now suppose $\mathcal{E}$ and $\mathcal{B}$ are manifolds and assume that for any $X\in\mathcal{B}$, there exists a neighborhood $\mathcal{U}\subset\mathcal{B}$ of $X$, a manifolds $\mathcal{F}$, and a diffeomorphism $\psi:\pi^{-1}(\mathcal{U})\to\mathcal{U}\times\mathcal{F}$ such that $\pi=\mathrm{pr}_1\circ\psi$, where $\mathrm{pr}_1:\mathcal{U}\times\mathcal{F}\to\mathcal{U}$ is projection onto the first factor. $(\mathcal{E},\pi,\mathcal{B})$ is called a fiber bundle and $\mathcal{E}$, $\pi$, and $\mathcal{B}$ are called the total space, the projection, and the base space, respectively. If for any $X\in\mathcal{B}$, $\pi^{-1}(X)$ is a vector space, $(\mathcal{E},\pi,\mathcal{B})$ is called a vector bundle. The set of sections of this bundle $\Gamma(\mathcal{E})$ is the set of all smooth maps $\sigma:\mathcal{B}\to\mathcal{E}$ such that $\sigma(X)\in\mathcal{E}_X$, $\forall\ X\in\mathcal{B}$. An important example of a vector bundle is the tangent bundle of a manifold for which $\mathcal{E}=T\mathcal{B}$.

**Induced bundle and connection.** Consider a map between Riemannian manifolds $\varphi:\mathcal{B}\to\mathcal{S}$. The tangent bundles of $\mathcal{B}$ and $\mathcal{S}$ are denoted by $T\mathcal{B}=\sqcup_{X\in\mathcal{B}}T_X\mathcal{B}$ and $T\mathcal{S}=\sqcup_{x\in\mathcal{S}}T_x\mathcal{S}$, respectively. We define an induced vector bundle $\varphi^{-1}T\mathcal{S}$, which is a vector bundle over $\mathcal{B}$ whose fiber over $X\in\mathcal{B}$ is $T_{\varphi(X)}\mathcal{S}$ [Nishikawa, 2002]. The connection $\nabla^{\mathbf{g}}$ induces a unique connection $\nabla^\varphi$ on $\varphi^{-1}T\mathcal{S}$ defined as

$$\nabla^\varphi_{\mathbf{W}}\mathbf{w}\circ\varphi=\nabla^{\mathbf{g}}_{\varphi_*\mathbf{W}}\mathbf{w},\quad \mathbf{W}\in T_X\mathcal{B},\ \mathbf{w}\in\Gamma(T\mathcal{S})\,. \tag{A.7}$$

$\nabla^\varphi$ is called the induced connection. It can be shown that its connection coefficients with respect to the coordinate charts $\{X^A\}$ and $\{x^a\}$ of $\mathcal{B}$ and $\mathcal{S}$, respectively, are $\frac{\partial\varphi^b}{\partial X^A}\gamma^a{}_{bc}$. In particular, the variation field $\delta\varphi$ defined in §3 is a section of $\Gamma(\varphi^{-1}T\mathcal{S})$, i.e., $\delta\varphi$ defines a vector fields in $\mathcal{S}$ along the map $\varphi$. For a two-point tensor, e.g., deformation gradient, covariant derivative involves both $\nabla^{\mathbf{g}}$ and $\nabla^{\mathbf{G}}$: $F^a{}_{A|B}=\frac{\partial F^a{}_A}{\partial X^B}+(F^b{}_B\gamma^a{}_{bc})F^c{}_A-\Gamma^C{}_{AB}F^a{}_C=\frac{\partial F^a{}_A}{\partial X^B}+\gamma^a{}_{bc}F^b{}_BF^c{}_A-\Gamma^C{}_{AB}F^a{}_C$. We denote the covariant derivative of the deformation gradient by $\nabla\mathbf{F}=F^a{}_{A|B}dX^B\otimes dX^A\otimes\frac{\partial}{\partial x^a}$. It is straightforward to show that [Nishikawa, 2002]

$$\nabla^\varphi\mathbf{F}(\mathbf{X},\mathbf{Y})=\nabla^\varphi_{\mathbf{X}}\varphi_*\mathbf{Y}-\varphi_*\nabla^{\mathbf{G}}_{\mathbf{W}}\mathbf{Y},\quad \nabla^\varphi_{\mathbf{X}}\varphi_*\mathbf{Y}-\nabla^\varphi_{\mathbf{Y}}\varphi_*\mathbf{X}=\varphi_*[\mathbf{X},\mathbf{Y}]\,. \tag{A.8}$$

The metrics $\mathbf{G}$ and $\mathbf{g}$ induce an inner product $\langle,\rangle_X$ in $T_{\varphi(X)}\mathcal{S}\otimes T^*_X\mathcal{B}$. This is defined first for the basis $\left\{\frac{\partial}{\partial x^a}\otimes dX^A,\ 1\le a\le n,\ 1\le A\le n\right\}$ as $\left\langle\frac{\partial}{\partial x^a}\otimes dX^A,\frac{\partial}{\partial x^b}\otimes dX^B\right\rangle_X=g_{ab}G^{AB}$, and then one extends it linearly to arbitrary elements in $T_{\varphi(X)}\mathcal{S}\otimes T^*_X\mathcal{B}$. $\varphi^{-1}T\mathcal{S}\otimes T^*\mathcal{B}$ is the vector bundle whose fiber at $X\in\mathcal{B}$ is $T_{\varphi(X)}\mathcal{S}\otimes T^*_X\mathcal{B}$. The two-point tensor $\mathbf{F}=T\varphi:\mathcal{B}\to\varphi^{-1}T\mathcal{S}\otimes T^*\mathcal{B}$, i.e., $\mathbf{F}\in\Gamma(\varphi^{-1}T\mathcal{S}\otimes T^*\mathcal{B})$. One can define a fiber metric $\langle\!\langle,\rangle\!\rangle$ on $\varphi^{-1}T\mathcal{S}\otimes T^*\mathcal{B}$ using the inner product $\langle,\rangle_X$ in $T_{\varphi(X)}\mathcal{S}\otimes T^*_X\mathcal{B}$ as follows. For $\sigma,\tau\in\Gamma(\varphi^{-1}T\mathcal{S}\otimes T^*\mathcal{B})$, define $\langle\!\langle\sigma,\tau\rangle\!\rangle(X)=\langle\sigma(X),\tau(X)\rangle_X$, $X\in\mathcal{B}$. One can define a connection $\nabla$ in $\varphi^{-1}T\mathcal{S}\otimes T^*\mathcal{B}$ using the Levi-Civita connections $\nabla^{\mathbf{G}}$ and $\nabla^{\mathbf{g}}$: consider a section $\mathbf{W}\otimes\boldsymbol{\alpha}\in\Gamma(\varphi^{-1}T\mathcal{S}\otimes T^*\mathcal{B})$ and let $\nabla(\mathbf{W}\otimes\boldsymbol{\alpha})=\nabla^\varphi\mathbf{W}\otimes\boldsymbol{\alpha}+\mathbf{W}\otimes\nabla^{\mathbf{G}}\boldsymbol{\alpha}$. This connection is compatible with the fiber metric $\langle\!\langle,\rangle\!\rangle$ in $\varphi^{-1}T\mathcal{S}\otimes T^*\mathcal{B}$.

**The Piola transform.** The Piola transform of a vector $\mathbf{w}\in T_{\varphi(X)}\mathcal{S}$ is a vector $\mathbf{W}\in T_X\mathcal{B}$ given by $\mathbf{W}=J\varphi^*\mathbf{w}=J\mathbf{F}^{-1}\mathbf{w}$. In coordinates, $W^A=J(F^{-1})^A{}_bw^b$, where $J=\sqrt{\frac{\det\mathbf{g}}{\det\mathbf{G}}}\det\mathbf{F}$ is the Jacobian of $\varphi$ with $\mathbf{G}$ and $\mathbf{g}$ the Riemannian metrics of $\mathcal{B}$ and $\mathcal{S}$, respectively. It can be shown that $\operatorname{Div}\mathbf{W}=J(\operatorname{div}\mathbf{w})\circ\varphi$.

In coordinates, ${W^A}_{|A} = J{w^a}_{|a}$. This is also known as the Piola identity. Another way of writing the Piola identity is in terms of the unit normal vectors of a surface in $\mathcal{B}$ and its corresponding surface in $\mathcal{S}$ and the area elements. It is written as $\hat{\mathbf{n}}da = J\mathbf{F}^{-\star}\hat{\mathbf{N}}dA$, or in components, $n_a da = J(F^{-1})^A{}_a N_A dA$. In the literature of continuum mechanics, this is called Nanson's formula.

**Lie derivative.** Let $\mathbf{w} : \mathcal{U} \to T\mathcal{S}$ be a vector field, where $\mathcal{U} \subset \mathcal{S}$ is open. A curve $\alpha : I \to \mathcal{S}$, where $I$ is an open interval, is an integral curve of $\mathbf{w}$ if $\frac{d\alpha(t)}{dt} = \mathbf{w}(\alpha(t))$, $\forall\ t \in I$. For a time-dependent vector field $\mathbf{w} : \mathcal{S} \times I \to T\mathcal{S}$, where $I$ is some open interval, the collection of maps $\psi_{\tau,t}$ is the flow of $\mathbf{w}$ if for each $t$ and $x$, $\tau \mapsto \psi_{\tau,t}(x)$ is an integral curve of $\mathbf{w}_t$, i.e., $\frac{d}{d\tau}\psi_{\tau,t}(x) = \mathbf{w}(\psi_{\tau,t}(x), \tau)$, and $\psi_{t,t}(x) = x$. Let $\mathbf{t}$ be a time-dependent tensor field on $\mathcal{S}$, i.e., $\mathbf{t}_t(x) = \mathbf{t}(x,t)$ is a tensor. The Lie derivative of $\mathbf{t}$ with respect to $\mathbf{w}$ is defined as $\mathbf{L}_{\mathbf{w}}\mathbf{t} = \frac{d}{d\tau}\psi^*_{\tau,t}\mathbf{t}_\tau\Big|_{\tau=t}$. Note that $\psi_{\tau,t}$ maps $\mathbf{t}_t$ to $\mathbf{t}_\tau$. Hence, to calculate the Lie derivative one drags $\mathbf{t}$ along the flow of $\mathbf{w}$ from $\tau$ to $t$ and then differentiates the Lie dragged tensor with respect to $\tau$. The autonomous Lie derivative of $\mathbf{t}$ with respect to $\mathbf{w}$ is defined as $\mathfrak{L}_{\mathbf{w}}\mathbf{t} = \frac{d}{d\tau}\psi^*_{\tau,t}\mathbf{t}_t\Big|_{\tau=t}$. Thus, $\mathbf{L}_{\mathbf{w}}\mathbf{t} = \partial\mathbf{t}/\partial t + \mathfrak{L}_{\mathbf{w}}\mathbf{t}$.

For a scalar $f$, $\mathbf{L}_{\mathbf{w}}f = \partial f/\partial t + \mathbf{w}[f]$. In a coordinate chart $\{x^a\}$ this reads, $\mathbf{L}_{\mathbf{w}}f = \frac{\partial f}{\partial t} + \frac{\partial f}{\partial x^a}w^a$. For a vector $\mathbf{u}$, one can show that $\mathbf{L}_{\mathbf{w}}\mathbf{u} = \frac{\partial \mathbf{w}}{\partial t} + [\mathbf{w},\mathbf{u}]$. If $\nabla$ is a torsion-free connection, then $[\mathbf{w},\mathbf{u}] = \nabla_{\mathbf{w}}\mathbf{u} - \nabla_{\mathbf{u}}\mathbf{w}$. Thus, $\mathbf{L}_{\mathbf{w}}\mathbf{u} = \frac{\partial \mathbf{w}}{\partial t} + \nabla_{\mathbf{w}}\mathbf{u} - \nabla_{\mathbf{u}}\mathbf{w}$.

When linearizing nonlinear elasticity one starts with a one-parameter family of motions $\varphi_{t,\epsilon} : \mathcal{B} \to \mathcal{S}$. By definition of the variation field $\mathbf{U}_t = \delta\varphi_t$, $\varphi_{t,\epsilon}$ is the flow of the variation field. Given the tensor field $\mathbf{t}$ in $\mathcal{S}$, $\bar{\mathbf{T}}_\epsilon = \varphi^*_{t,\epsilon}\mathbf{t} \circ \varphi_{t,\epsilon}$ is a vector field on $\mathcal{B}$. Its linearization is defined as

$$\delta\bar{\mathbf{T}} = \frac{d}{d\epsilon}\bar{\mathbf{T}}_\epsilon\Big|_{\epsilon=0} = \left(\frac{d}{d\epsilon}\varphi^*_{t,\epsilon}\mathbf{t} \circ \varphi_{t,\epsilon}\right)\Big|_{\epsilon=0} = \left(\varphi^*_{t,\epsilon}\mathbf{L}_{\mathbf{U}_{t,\epsilon}}\mathbf{t} \circ \varphi_{t,\epsilon}\right)\Big|_{\epsilon=0} = \mathring{\varphi}^*_t\left(\mathbf{L}_{\mathbf{U}_t}\mathbf{t} \circ \mathring{\varphi}_t\right). \tag{A.9}$$

Thus, $\delta\mathbf{t} = \mathbf{L}_{\mathbf{u}_t}\mathbf{t}$, where $\mathbf{u}_t = \mathbf{U}_t \circ \mathring{\varphi}_t^{-1}$.